\documentclass[12pt]{article}
\usepackage[margin=.88in]{geometry}
\usepackage[utf8]{inputenc}
\usepackage{amsfonts}
\usepackage{amsgen,amsmath,amstext,amsbsy,amsopn,amssymb,subcaption,mathrsfs}
\usepackage[dvips]{graphicx}
\usepackage{comment}
\usepackage{placeins}

\usepackage{enumitem} 
\usepackage{natbib}
\usepackage{booktabs}
\usepackage{amsthm}
\usepackage{hyperref}
\usepackage{blkarray}
\usepackage{algorithm}
\usepackage{algorithmic}
\usepackage{url}
\usepackage{longtable}
\usepackage{stmaryrd}
\usepackage{mwe}
\usepackage{mathtools}
\usepackage{bbm}
\usepackage{threeparttable}
\DeclareMathAlphabet\mathbfcal{OMS}{cmsy}{b}{n}

\makeatletter
\newcommand*{\rom}[1]{\expandafter\@slowromancap\romannumeral #1@}
\makeatother

\usepackage[dvipsnames]{xcolor}

\newcommand\fro[1]{\| #1 \|_{\rm{F}}}

\newcommand{\inp}[2]{\langle #1,#2\rangle}

\newcommand{\argmin}{\mathop{\rm arg\min}}
\newcommand{\argmax}{\mathop{\rm arg\max}}

\newcommand{\PP}{\mathbb{P}}

\def\ubs{\textsf{unbs}}
\def\init{\textsf{init}}

\def\PMQ{\mathcal{P}_M(Q)}

\def\calD{{\mathcal D}}

\def\calF{{\mathcal F}}

\def\calM{{\mathcal M}}

\def\calP{{\mathcal P}}

\def\EE{{\mathbb E}}

\def\NN{{\mathbb N}}
\def\OO{{\mathbb O}}
\def\PP{{\mathbb P}}

\def\RR{{\mathbb R}}

\def\indicator{\mathbf{1}}

\def\init{\textsf{init}}

\def\oto{\textsf{oto}}
\def\otm{\textsf{otm}}

\def\hat{\widehat}
\def\tside{\textsf{tside}}

\newtheorem{Theorem}{Theorem}

\newtheorem{Lemma}{Lemma}
\newtheorem{Remark}{Remark}
\theoremstyle{plain}

\newtheorem{Proposition}{Proposition}

\title{Statistical Inference for Matching Decisions via Matrix Completion under Dependent Missingness}

 \author{Congyuan Duan$^1$, Wanteng Ma$^2$, Dong Xia$^1$ and Kan Xu$^3$\\
 ~ \\
 {\small $^{1}$Department of Statistics and Data Science, The Chinese University of Hong Kong, Hong Kong SAR}\\
 {\small $^2$Department of Statistics and Data Science, University of Pennsylvania, USA}\\
 {\small $^3$W. P. Carey School of Business, Arizona State University, USA}\\
 }
\date{}

\begin{document}

\maketitle
 \footnotetext[1]{Dong Xia’s research was partially supported by Hong Kong RGC Grant GRF 16303224.} 
\footnotetext[3]{Dong Xia and Kan Xu are co-corresponding authors.}

\begin{abstract}
This paper studies decision-making and statistical inference for two-sided matching markets under \emph{unknown preferences} via matrix completion. In contrast to the independent sampling assumed in classical matrix completion literature, the observed entries, which arise from past matching data, are constrained by matching capacity. This matching-induced dependence poses new challenges for both estimation and inference in the matrix completion framework. We propose a non-convex algorithm based on Grassmannian gradient descent and establish entrywise convergence rates for three canonical mechanisms, i.e., one-to-one matching, one-to-many matching with one-sided random arrival, and two-sided random arrival. To facilitate valid uncertainty quantification and hypothesis testing on matching decisions, we further develop a general debiasing and projection framework for arbitrary linear forms of the reward matrix, deriving asymptotic normality with finite-sample guarantees under matching-induced dependent sampling. Our empirical experiments demonstrate that the proposed approach provides accurate estimation, valid confidence intervals, and efficient evaluation of matching policies.
\end{abstract}

\begin{sloppypar}

\section{Introduction}  \label{sec:intro}
The centralized two-sided matching market refers to the platform or system that facilitates connections between two distinct groups of agents, where agents from one side can only be matched with agents from the other side \citep{roth1992two}. In such platforms, matching any two agents from different sides produces a matching score or reward, which typically reflects the preferences or satisfaction of the agents and indicates how well the two agents are suited to each other. The primary objective of these platforms is to identify an optimal matching policy that creates mutually beneficial matches between agents on both sides, thereby maximizing the total reward.

Centralized two-sided matching platforms exhibit several characteristic features. First, the term ``centralized" implies that the allocation process is automated and governed by a centralized matching scheme, meaning agents on either side do not directly approach one another to form allocations. Additionally, these platforms typically operate on a recurring basis, with all agents from both sides participating at regular intervals. Second, the preferences of agents are often unknown due to the large volume of participants or privacy concerns, as preferences are considered as sensitive information \citep{wu2019bptm, li2023two}. Consequently, the platform may occasionally resort to random allocation for matching learning. Third, there are capacity constraints on both sides of the market. For instance, each agent can only be matched with one or a limited number of agents from the other side at any given time. Below, we provide examples to further illustrate the centralized two-sided matching market.
\begin{itemize}
    \item[(i)] \emph{Crowdsourcing platform.} Crowdsourcing refers to the practice of obtaining services, ideas, or content from a large group of people, typically via a centralized platform that assigns tasks to participants. For example, volunteer platforms, such as VolunteerMatch and Food Rescue Hero, connect volunteers with events and activities \citep{alam2017temporal, lo2024commitment}; human intelligence platforms, like Amazon Mechanical Turk and CrowdFlower, facilitate tasks such as large-scale translation, information collection, and image acquisition \citep{van2012designing, strickland2019use}. Each task is usually assigned to a limited number of participants, while each participant handles a fixed number of tasks. The key to improve the overall efficiency and profitability is to optimize the matching between participants and tasks, thereby increasing participants' engagement and improving task performance.  
    \item[(ii)] \emph{Major League Baseball (MLB) umpire matching.} During each MLB regular season, there is a fixed set of teams that play a series of games against one another, as well as a fixed set of umpires assigned to officiate these games. After each game, the umpire’s accuracy is recorded as a measure of umpiring quality. Because different teams have different technique characteristics, while umpires may vary in their familiarity with these characteristics, appropriately matching
    teams with umpires can improve the overall umpiring accuracy and quality of the games \citep{deshpande2017hierarchical}. Capacity constraints arise naturally since on any given day, each team can appear in at most one game and each umpire can officiate at most one game.
    \item[(iii)] \emph{Emergency departments.} In hospital emergency departments, patients are typically assigned to physicians on a random basis to minimize waiting times and ensure prompt access to medical care \citep{gowrisankaran2023physician, jameson2025impact}.  While this randomized assignment mechanism supports operational efficiency, it may lead to overuse or underuse of resources and substantial variations in treatment outcomes since physicians differ in their expertise and practice styles. As a result, emergency departments provide a natural setting to study the matching system performance of physicians on patient health outcomes. It is crucial to develop an accurate and efficient data-driven patient-to-physician assignment system to improve the overall quality of care and optimize operational performance in healthcare delivery.
\end{itemize}

Since the random allocation in the aforementioned platforms may result in a suboptimal total reward below expectations, it is crucial to identify the optimal matching to achieve the highest possible total reward. This is possible only if all the matching scores are available. 
However, in complex matching platforms, the data often exhibits high dimensionality due to the large number of participants and tasks, each with varying matching preferences and constraints. Incorporating every individual’s preferences and capacity constraints would result in an extremely intricate model and matching system, requiring the solution of a highly complex optimization problem. Thus, it is imperative to address the question: \emph{How can we identify the optimal matching under high dimensional data without explicitly modeling the complex preferences of either side?} Furthermore, in practice, uncertainty quantification is even more critical for matching decision making. That is, we need to perform policy evaluation on a carefully designed matching or a new matching policy that deviates significantly from the one currently in use. This is because, even if we learn a better decision in expectation from the data, directly switching to a different policy often involves friction costs. As a result, it is essential to quantify the worst-case performance of the new policy to ensure a high level of confidence in its improvement before implementation \citep{bertsimas2006robust, hayashi2025off}. Moreover, in scenarios where multiple matching policy options are available, we need to determine which policy is significantly better than the others. Ultimately, we aim to answer the key questions: \emph{How confident are we that a certain matching can achieve a total reward that meets our expectations? How confident are we that the newly designed matching is better than the existing one?}


Motivated by these questions, in this work, we take a novel insight into the centralized two-sided matching learning and evaluation with capacity constraints through low rank matrix completion. 
For convenience, we will use the aforementioned crowdsourcing platform as an illustrative example. Suppose there are $d_1$ participants and $d_2$ tasks, then we represent the rewards for all possible allocations via a low rank matrix $M\in \RR^{d_1\times d_2}$. In particular, the $(i,j)$-th entry of $M$ denotes the reward (e.g., satisfaction or task completion quality) when task $j$ is assigned to participant $i$. While accurately identifying the relationship between rewards and preferences of both sides is highly challenging, the low-rank assumption of $M$ eliminates the need to know the preferences explicitly or to model the relationship between rewards and preference covariates. Instead, the rewards for different allocations are captured by the inherent similarities between participants and tasks. Intuitively, some participants may share similar backgrounds and strengths, while certain tasks may belong to the same category or have similar requirements. As a result, even with only a few observed random allocations and no prior knowledge of preferences, we can estimate the reward for any allocation by leveraging these similarities, ultimately identifying the optimal matching. 

In classical  matrix completion literature, the observed entries are typically assumed to be sampled uniformly or independently \citep{candes2009exact, candes2010power, candes2010matrix}. Formally, within the trace regression framework, the collected data is represented as $\{(y_t, X_t)\}_{t=1}^T$, where each $X_t$ is sampled uniformly from the orthonormal basis 
$\mathcal{M}^{\textsf{iid}}:=\left\{e_{j_1}e_{j_2}^{\top}: j_1\in [d_1], j_2\in [d_2] \right\},$
where $e_{j_1}$'s and $e_{j_2}$'s are the canonical basis vectors in $\mathbb{R}^{d_1}$ and $\mathbb{R}^{d_2}$, respectively. The response $y_t$ is related to $X_t$ via 
$ y_t=\inp{M}{X_t} + \xi_t,$ 
where $\xi_t$'s are i.i.d. sub-Gaussian noise. In other words, we observe $T$ random noisy entries of the underlying matrix $M$.   

In matching problems, the observed data consist of $T$ random \emph{matchings} and the corresponding noisy rewards. Here a \emph{matching} refers to a complete allocation of all participants to tasks, meaning that we observe multiple entries across different rows and columns simultaneously. Consider the one-to-one matching mechanism as an example, which has been studied in the literature under different contexts. We assume throughout the paper that there are $d_1$ participants and $d_2$ tasks with $d_2 \geq d_1$, i.e., the number of tasks is at least the number of participants. For one-to-one matching, each participant is assigned with exactly one task and each task is assigned to at most one participant. In particular, each $X_t$ is uniformly sampled from the set of all possible one-to-one matchings,
\begin{align}\label{eq:one-to-one}
    \mathcal{M}^{\textsf{oto}}:=\Big\{X\in \{0,1\}^{d_1\times d_2}\Big|e_i^{\top} X 1_{d_2}=1, 1_{d_1}^{\top}X e_j\leq 1, \forall (i,j)\in [d_1]\times [d_2] \Big\},
\end{align}
where $1_{d}$ is the $d$-dimensional all-ones vector, i.e., $1_{d}=(1,\cdots,1)^{\top}$. 
Each entry of the observed reward $y_t\in\RR^{d_1}$ represents the reward for the corresponding participant under this matching scheme, see Section~\ref{sec:problem} for more details. The one-to-one matching problem can be conveniently viewed as matrix completion with {\it dependent sampling}, where the dependence arises from the matching pattern and its capacity constraints on both sides, i.e., the entries observed in a one-to-one matching cannot overlap in rows or columns.

Furthermore, we can formulate matching evaluation as hypothesis testing for  linear forms of $M$. Suppose we are interested in whether a specific matching scheme $Q\in \mathcal{M}^{\textsf{oto}}$ achieves a total reward at least $V_0$. This can be addressed through testing
\begin{align}\label{eq:test-one}
    H_0: \langle M, Q\rangle\leq V_0\quad {\rm v.s.}\quad H_1: \langle M, Q\rangle>V_0.
\end{align}
Additionally, comparing two matching schemes $Q_1, Q_2\in \calM^{\textsf{oto}}$ can be formulated as testing against the null hypothesis $H_0: \langle M, Q_1-Q_2\rangle= 0$.

However, the matching learning and evaluation via matrix completion is not that straightforward. The majority of existing literature on matrix completion and inference focus on i.i.d. missingness, where $X_t$ is uniformly sampled from $\mathcal{M}^{\textsf{iid}}$. The challenges in matrix completion under matching observations mainly arise from three aspects. First, a rate-optimal initial estimator $\hat M^{\init}$ plays a critical role in inference for $M$. In particular, $\hat M^{\init}$ needs to achieve an entrywise error rate proportional to the noise level $\sigma$ to provide valid inference. Unfortunately, most existing algorithms are either inapplicable due to the dependent sampling in matching problems or only achieve an entrywise error rate proportional to $(\sigma+\|M\|_{\max})$ (see Section~\ref{sec:estimation} for further discussion). Second, statistical inference for matrix completion requires a specifically designed debiasing procedure and an exact, non-asymptotic characterization of the bias and variance. The dependent sampling and capacity constraints inherent in matrix problems further complicate the analysis, particularly in one-to-many and two-sided random arrival matching problems (mentioned in later sections), where special technique is needed to handle the unique dependence structure of the observed entries. Lastly, on the technical front, classical tools such as the Central Limit Theorem and Berry-Esseen Theorem rely on the assumption of independent samples. For instance, although there are $T$ independently sampled matchings in a one-to-one matching problem, there are $Td_1$ observed entries in total. While the Berry-Esseen bounds might suggest that the effective sample size is $T$, we show that it is actually $Td_1$ when determining the variance, as if the entries were observed independently. Moreover, the dependence among entries in one matching poses significant challenges to controlling the third-order moment when applying the Berry-Esseen Theorem, where we must carefully examine the dependence structure and exploit novel techniques to achieve sharp upper bounds.

\subsection{Main contribution}
We investigate the centralized two-sided matching problem via matrix completion under dependent missingness. We focus on three common matching mechanisms: one-to-one matching, one-to-many matching, and two-sided random arrival. We propose a matching learning algorithm and a general matching evaluation framework that are applicable to all of the aforementioned matching mechanisms. To the best of our knowledge, this is the first work to explore statistical inference for matrix completion under matching-induced dependent sampling. Our contributions can be summarized as follows:

\textit{Matching learning algorithm}. We propose a non-convex matching learning algorithm that employs gradient descent on the Grassmannian with sample splitting. This algorithm generalizes the approach introduced in \citep{xia2021statistical} to matrix completion under dependent sampling. Our key contribution lies in establishing convergence under dependent missingness and deriving a sharp entrywise error rate of $O_p\big(\sigma/\sqrt{Td_1\nu}\big)$ for all three matching schemes through a careful analysis of the dependence among the observed entries. Here, $\nu$ denotes the probability that any given entry is observed under the corresponding matching scheme. The resulting estimator provides a valid initialization for the subsequent matching evaluation procedure. More importantly, our matching learning algorithm achieves an error rate with the same dependence on the noise level, sample size, and matrix dimensions as the optimal rate under i.i.d. entry sampling. The required sample size and SNR conditions are also comparable to those imposed under i.i.d. sampling.

\textit{General matching evaluation framework and martingale technique for dependent sampling.} We introduce a general framework for matching evaluation based on debiasing and low-rank projection. Although our framework builds on existing statistical inference methods for matrix completion under independent sampling, extending these methods to matching problems is highly nontrivial because of the dependence among entries observed within each matching. Specifically, establishing asymptotic normality requires sophisticated spectral analysis and careful control of higher-order terms. Such analysis further requires precise characterization and control of the covariance among entries observed within each matching sample. To address this challenge, we develop a novel martingale technique that represents the dependent matching process as sequential sampling without replacement. To clarify this technical contribution, we provide a proof sketch in Section \ref{sec: proof sketch}. Our framework is both flexible and straightforward to use. It permits the evaluation of any target matching pattern, including arbitrary subsets of a matching, complete matchings, weighted matchings, and comparisons between different matchings, regardless of the matching scheme under which the data are observed.

\textit{Real data analysis}. We evaluate the practical performance of our matching learning and evaluation framework using the Major League Baseball team-umpire matching data introduced in the beginning. We apply our matching learning algorithm to identify optimal matchings for game days with different sets of arriving teams and then conduct matching evaluations comparing their rewards with those of the originally observed matchings. Our findings indicate that the identified optimal matchings significantly outperform the benchmark matchings and can improve overall umpiring accuracy and quality. These findings are also supported by historical evidence.

\subsection{Related works}

\textit{Matrix completion and inference under independent missingness}. The pioneering work of \citet{candes2009exact} established exact recovery of low rank matrices from a uniformly sampled subset of entries through convex nuclear-norm minimization in the noiseless setting. The subsequent foundational work of \citet{candes2010power} improved the sample complexity toward the information-theoretic limit. The performance of this convex approach under arbitrary noise was further investigated by \citet{candes2010matrix}, although only a suboptimal estimation error rate was established. By contrast, \citet{keshavan2010matrix} proposed an early nonconvex matrix completion method combining spectral initialization with manifold optimization and established order-optimal estimation guarantees under subGaussian noise and certain regimes. This has inspired numerous follow-up works on nonconvex matrix completion \citep{chen2015fast, ma2018implicit} focusing on designing more efficient optimization algorithms and establishing more profound statistical theories. 

Statistical inference for noisy matrix completion is considerably more challenging, since deriving the asymptotic behavior of low-rank estimators involves sophisticated spectral perturbation analyses.  
See, for instance, \citet{chen2019inference, xia2021statistical, yan2021inference, chernozhukov2023inference}. In particular, \citet{chen2019inference} and \citet{yan2021inference} employ a leave-one-out analysis for entrywise inference, while \citet{xia2021statistical} utilizes double-sample debiasing and low-rank projection for inference of general linear forms. All these works assume independent missingness, and it is far from straightforward to extend their methods to dependent missingness.

\textit{Matrix completion and inference under structured missingness}. Motivated by various applications, matrix completion and inference have been studied under non-uniform sampling and missing-not-at-random scenarios \citep{ma2019missing, foucart2020weighted, wang2021matrix,duan2024online}, where entries are observed with unequal probabilities. Notably, matrix completion with structured missingness has been extensively applied to estimating treatment effects over time \citep{athey2021matrix, agarwal2023causal, xiong2023large, choi2024matrix, yan2024entrywise}, where propensity scores vary across individuals and time. In particular, \cite{choi2024matrix} investigated causal matrix completion using convex nuclear norm optimization and derived entrywise inference results via a leave-one-out technique; \cite{yan2024entrywise} focused on the staggered adoption pattern and developed optimal entrywise confidence intervals that meet a non-asymptotic instance-wise lower bound. Nevertheless, these settings differ from matching problems, as their methods do not address the correlation structure inherent in the data and cannot provide inference for linear forms.

\textit{Matching with capacity constraints}. 
There have been extensive interests in operations research studying matching strategies for centralized platforms through learning \citep[see, e.g.,][]{massoulie2016capacity,bimpikis2019learning,shah2020adaptive,johari2021matching,hsu2022integrated}.
The majority of this literature focus on learning unknown types of market participants given knowledge of matching rewards, with a few exceptions. For example, \cite{hsu2022integrated} consider an optimal matching problem in a queuing system with both unknown types and rewards; \cite{liu2020competing} investigate dynamic stable matching with unknown rewards learned through matching outcomes. Recently, \cite{tang2026match} provide the first study of reward learning using matrix completion in the matching markets, considering the reward estimation problem under dependent missingness of one-to-one matching. However, their approach relied on a nuclear norm regularization convex estimation method, which failed to achieve the convergence rate only proportional to the noise level $\sigma$. In this work, we propose a general non-convex algorithm and statistical inference framework that can be applied across various matching patterns, accommodating different types of dependencies among the observed entries.

\section{The Matching Problem}\label{sec:problem}


Throughout this paper, we assume that the reward matrix $M$ has rank $r\ll\min(d_1,d_2)$ and admits the singular value decomposition $M=U\Lambda V^{\top}$, where $U\in\OO^{d_1\times r}$ and $V\in\OO^{d_2\times r}$. Here, $\mathbb O^{d\times r}$ denotes the set of $d\times r$ matrices with orthonormal columns. Let $\lambda_{\max}$ and $\lambda_{\min}$ denote the largest and smallest nonzero singular values of $M$, respectively, and define the condition number as $\kappa:=\lambda_{\max}/\lambda_{\min}$. We define the aspect ratio of $M$ as $\alpha_d:=d_2/d_1$. We use $M(i,j)$ to denote the $(i,j)$-th entry of $M$, and $M(i,\cdot)$ and $M(\cdot,j)$ to denote its $i$-th row and $j$-th column, respectively. We also define the matrix max norm as $\|M\|_{\max}=\max_{i,j}|M(i,j)|$. Matrix completion becomes ill-posed when the underlying matrix is highly spiked, meaning that one or a few entries are substantially larger than the others. To address this issue, we assume that $M$ is \emph{incoherent} with parameter $\mu$, meaning that $\max\big\{(d_1/r)^{1/2}\|U\|_{2,\max}, (d_2/r)^{1/2}\|V\|_{2,\max} \big\}\leq \mu$. When $\mu$ is bounded, this condition prevents a small number of entries from dominating the matrix. Here, the maximum row-wise $\ell_2$-norm is defined as $\|A\|_{2,\max} := \max_{j} \|e_j^{\top}A\|$. For simplicity, we treat $\mu, \kappa, r$ as constants throughout the paper.                  

For all mechanisms considered in this paper, we assume that the observed matchings $X_t$ are sampled i.i.d. from the corresponding matching set. In what follows, we introduce three matching mechanisms commonly used in practice.

\begin{figure} [h]
	\centering
	\begin{subfigure}[t] {0.3\linewidth}
		\includegraphics[scale=0.5]{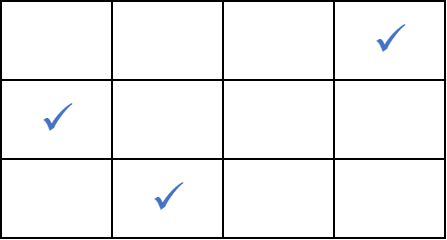}
        \caption{One-to-one}
    \end{subfigure}
    \begin{subfigure}[t] {0.3\linewidth} 
		\includegraphics[scale=0.5]{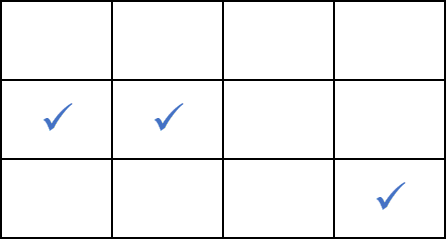}
        \caption{One-to-many with \\ one-sided random arrival}
    \end{subfigure}
    \begin{subfigure}[t] {0.3\linewidth} 
		\includegraphics[scale=0.5]{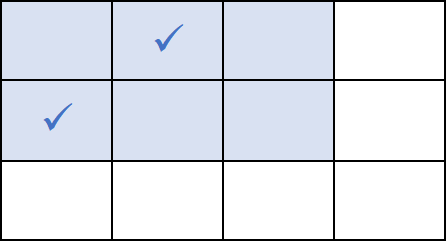}
        \caption{Two-sided random arrival}
    \end{subfigure}
    \caption{The three matching mechanism: rows and columns represent participants and tasks, respectively. The shaded area in (c) denotes the arrived participants and available tasks. }
    \label{matching pattern} 
\end{figure}

\textbf{One-to-one matching}. The leading example is the one-to-one matching discussed in Section \ref{sec:intro} and illustrated in Figure \ref{matching pattern} (a). The matching sample $X_t$ is drawn i.i.d. from the set $\calM^{\oto}$. Note that each one-to-one matching reveals $d_1$ entries of $M$ (with noise). We can decompose $X_t=\sum_{i=1}^{d_1}X_t^i$ row-wise, where $X_t^i=e_ie_{j_t(i)}^{\top}$ and the index $j_t(i)\in[d_2]$ such that $X_t(i,j_t(i))=1$. This representation encodes the task assignment for each participant $i\in[d_1]$, and the index set $\{j_t(i): i\in[d_1]\}$ contains no repeated elements, since every participant is assigned exactly one task and every task is assigned to at most one participant. The observed noisy rewards is denoted as $y_t=(y_t^1,\cdots, y_t^{d_1})^{\top}$, with each entry satisfying $y_t^{i}=\inp{M}{X_t^i} + \xi_t^i$. For simplicity, we assume that the noise $\xi_t^i$ are i.i.d. centered sub-Gaussian across both $t$ and $i$ and are independent of $X_t$, with $\EE[\xi_t^{i2}]=\sigma^2$ and $\|\xi_t^i\|_{\psi_2}\leq C\sigma$ for some $C>0$.

\textbf{One-to-many matching with one-sided random arrival}. In one-to-many matching, the system allows no task assignments to some participants and more than one task to other participants, as illustrated in Figure \ref{matching pattern} (b). It is one-sided random meaning that only participants arrive randomly and each task is assigned to at most one participant. For simplicity, we assume that the number of tasks chosen by each participant follows a Binomial distribution with parameters $K$ and $p_0$. Specifically, we define the following matching set:
\begin{align}\label{eq:one-to-many}
    \mathcal{M}_{K,p_0}^{\otm}:=\Big\{X\in \{0,1\}^{d_1\times d_2}\Big|e_i^{\top}X 1_{d_2}\sim {\rm Bin}(K,p_0),  1_{d_1}^{\top}X e_j\leq 1, \forall (i,j)\in [d_1]\times [d_2] \Big\}.
\end{align}
The special case $K=p_0=1$ reduces to the one-to-one matching mechanism. If $K=1$ and $0<p_0<1$, participants are allowed to be randomly absent. Participants may be assigned multiple tasks when $K>1$. Denote $h_{it}:= \{j\in [d_2] \big| X_t(i, j)=1 \}$ the set of tasks assigned to participant $i$ by the matching $X_t\in \calM_{K, p_0}^{\otm}$. Similarly, the matching $X_t$ can be decomposed row-wise and column-wise as
$X_t=\sum_{i=1}^{d_1}\sum_{q\in h_{it}} X_t^{iq}$ where the matrix $X_t^{iq}=e_ie_q^{\top}$ encodes task $q$ being assigned to participant $i$.
The corresponding observed reward satisfies $y_t^{iq}=\inp{M}{X_t^{iq}} + \xi_t^{iq}$, where the noise $\xi_t^{iq}$ are i.i.d. centered sub-Gaussian with $\EE[\xi_t^{iq2}]=\sigma^2$ and $\|\xi_t^{iq}\|_{\psi_2}\leq C\sigma$ for some $C>0$.

\textbf{Two-sided random arrival}. In some scenarios, both sides of the market may experience random absences. For example, in the Major League Baseball team–umpire matching problem introduced earlier, each team can appear in at most one game on any given day, while each umpire can officiate at most one game. This kind of two-sided random arrival matching scheme is illustrated in Figure \ref{matching pattern} (c). For simplicity, we focus on random allocation and formulate the two-sided random arrival as the following matching set:
\begin{equation}\label{eq: two-random}
\begin{aligned}
    &\mathcal{M}_{p_1,p_2}^{\tside}=\Big\{X\in \{0,1\}^{d_1\times d_2}\big| e_i^{\top}X 1_{d_2}\leq 1,  1_{d_1}^{\top}X e_j\leq 1, 1_{d_1}^{\top}X 1_{d_2}=\min(B_r,B_s), \forall (i,j)\Big\},
\end{aligned}
\end{equation}
where $B_r$ and $B_s$ are ${\rm Bin}(d_1, p_1)$ and ${\rm Bin}(d_2, p_2)$ random variables, respectively.  This matching scheme allows both participants and tasks to be randomly absent. In particular, each participant tends to be absent with probability $1-p_1$, while each task tends to be absent with probability $1-p_2$. The arrived participants and present tasks still follow specific matching rules. Specifically, if the number of arrived participants exceeds the number of available tasks, each task is assigned to exactly one participant, and each participant can be assigned at most one task. Conversely, if the number of available tasks exceeds the number of arrived participants, the roles are reversed, ensuring all matches meet the one-to-one constraints. Note that our subsequent theoretical results for this model require $(B_r,B_s)$ to follow the truncated joint binomial distribution specified in Section \ref{sec:estimation}. Define the set of matched pairs given by $X_t$ as $ g_t:= \left\{(i,j)\in [d_1]\times [d_2] \big| X_t(i, j)=1 \right\}$. Similarly, we can decompose $X_t$ as
$X_t=\sum_{(i,j)\in g_t} X_t^{ij}$, where $X_t^{ij}=e_ie_j^{\top}$. 
For each pair $(i,j)\in g_t$, the observed reward satisfies $y_t^{ij}=\inp{M}{X_t^{ij}} + \xi_t^{ij}$, assuming the noise are i.i.d. as in the aforementioned cases. 

Finally, in order to study the aforementioned three matching mechanisms in a unified framework, we slightly abuse the notation and define the response matrix $Y_t$ with elements
\begin{equation*}
    Y_t(i,j)= \begin{cases}
        y_t^i, \quad &\textrm{ if } X_t(i,j)=1 \ (\text{one-to-one matching}); \\
        y_t^{ij}, \quad &\textrm{ if } X_t(i,j)=1 \ (\text{one-to-many matching and two-sided random arrival}); \\
        0, \quad & \textrm{ if } X_t(i,j)=0 \ (\textrm{all three matching schemes});
    \end{cases}
\end{equation*}

\section{Learning Matching Rewards}\label{sec:estimation}
One of the classical matrix completion methods relies on convex optimization using nuclear norm regularization \citep{candes2010matrix}. Under a uniform sampling scheme, \cite{chen2020noisy} proved that convex relaxation achieves near-optimal estimation error rates in both the Frobenius and max norms by leveraging a sophisticated leave-one-out technique. Unfortunately, the leave-one-out technique is not suitable for matching observations because it relies heavily on the independence among the entries. 
Recently, \cite{tang2026match} investigated the nuclear-norm penalized least squares estimator for one-to-one matching. Together with a novel double enhancement approach, they derived convergence rates in both the Frobenius and max norms. However, these rates are not proportional to the noise level, making them inappropriate for statistical inference or matching evaluation.

We propose to learn the matching score matrix via non-convex methods, inspired by existing literature  \citep{keshavan2010matrix, ma2018implicit, xia2021statistical,li2023online}, by solving the following optimization problem:
\begin{equation}\label{eq:optimization}
    \min_{U,G,V} \mathscr{L}_{\calD}(U,G,V)=\sum_{t\in \calD} \big\|Y_t - X_t\circ (UGV^{\top})\big\|_{\rm F}^2 \quad \text{s.t.} \quad U\in \OO^{d_1\times r}, V\in \OO^{d_2\times r}, G\in \RR^{r\times r},
\end{equation}
where $\circ$ denotes the matrix Hadamard product. This optimization problem is highly non-convex, so we adapt a gradient descent algorithm on Grassmannians, previously proposed in \cite{xia2021statistical} for matrix completion under uniform sampling.  Note that each matching observation $X_t$ contains multiple entries that are mutually dependent and may be subject to constrained sampling. Consequently, the performance of most existing non-convex algorithms must be carefully reexamined under matching observations. 

Furthermore, statistical inference for matching evaluation requires an initial estimator that achieves a sharp entrywise error rate. Establishing entrywise error rates is technically much more challenging than deriving Frobenius-norm error rates for matrix completion. Most existing results \citep{ xia2019polynomial, ma2018implicit,li2023online} rely on either the leave-one-out analytic framework or data splitting under uniform sampling. For convenience, we employ a gradient descent algorithm equipped with rotation calibration and sample splitting. We demonstrate that, under reasonable conditions on the sample size and signal-to-noise ratio, our proposed algorithm delivers estimators with sharp entrywise error rates across the aforementioned three matching schemes. It is worth pointing out that while sample splitting may cause a loss of efficiency, the resulting estimators serve only as initial estimators; the subsequent debiasing procedure (Section~\ref{sec:inference}) recovers full data efficiency.

Without loss of generality, we assume $T=2mN_0$, and partition the data into $2m$ disjoint subsets $\mathcal{D}_p=\{t\in \NN| (p-1)N_0 +1\leq t\leq pN_0 \}$ for $p\in[2m]$. The challenge of optimization problem (\ref{eq:optimization}) lies in enforcing the orthogonality constraint on $U$ and $V$, which we view as points on Grassmannians, i.e., the sets of $r$-dimensional linear subspaces of $\RR^{d_1}$ and $\RR^{d_2}$, respectively. Note that the matrix $G$ is constraint-free and admits a closed-form solution via linear regression for fixed $U$ and $V$. We employ rotation calibration to maintain row-wise accuracy of $\hat U$ and $\hat V$, as detailed in Algorithm~\ref{alg:gd}. We denote by $\textsf{SVD}_r(M)$ the operator that returns the top-$r$ left and singular vectors of $M$ and define $(\partial/\partial M)\mathscr{L}_{\calD}(U,G,V):=2\sum_{t\in\calD}\big(X_t\circ (UGV^{\top})-Y_t\big)$. We introduce the entrywise sampling probability $\nu$ to study all three matching schemes in a unified algorithm. For simplicity, we assume that the true rank $r$ is known; otherwise, it may be determined in a data-driven manner by examining the scree plots of $\sum_{t\in \mathcal{D}_1}  Y_t \circ X_t$.  If the scree plots do not clearly reveal an elbow point, we recommend selecting a slightly larger rank. Underestimating the rank can cause significant information loss, leading to an estimator that deviates substantially from the true matrix.

\begin{algorithm}
    \caption{Matching-Based Rotation-Calibrated Gradient Descent on Grassmannians}\label{alg:gd}
\begin{algorithmic}
\STATE{\textbf{Input}: stepsize $\eta$; matching observations $\bigcup_{p=1}^{2m} \calD_p$; entrywise sampling probability $\nu$. }
\STATE{\textbf{Output}: $\widehat{U}^{(m)}, \widehat{G}^{(m)}, \widehat{V}^{(m)}$ and $M^{(m)}=\widehat{U}^{(m)}\widehat{G}^{(m)}\widehat{V}^{(m)\top}$.}
\STATE{\textbf{Initialization}:} Compute $(\widehat{U}^{(1)},\widehat{V}^{(1)})\leftarrow \textsf{SVD}_r\big((\nu N_0)^{-1}\sum_{t\in \mathcal{D}_1} (Y_t\circ X_t)\big)$ and $\widehat{G}^{(1)}=\argmin_{G} \mathscr{L}_{\mathcal{D}_2}(\widehat{U}^{(1)},G,\widehat{V}^{(1)})$ with $(\hat L_G^{(1)}, \hat R_G^{(1)})\leftarrow \textsf{SVD}_r(\hat G^{(1)})$.  
\FOR{$p\in[m-1]$}  
    \STATE Update by rotation calibrated gradient descent
    \begin{equation*}
    \begin{aligned}
        \widehat{U}^{(p+0.5)}&=\bigg(\widehat{U}^{(p)}-\frac{\eta}{2N_0\nu}\cdot\frac{\partial \mathscr{L}_{\calD_{2p+1}}}{\partial M}\big(\hat U^{(p)}, \hat G^{(p)}, \hat V^{(p)}\big)\hat V^{(p)}(\hat G^{(p)})^{-1}\bigg)\hat L_G^{(p)}; \\
        \widehat{V}^{(p+0.5)}&=\bigg(\widehat{V}^{(p)}-\frac{\eta}{2N_0\nu}\cdot \frac{\partial \mathscr{L}_{\calD_{2p+1}}^{\top}}{\partial M}\big(\hat U^{(p)}, \hat G^{(p)}, \hat V^{(p)}\big)\hat U^{(p)}(\hat G^{(p)\top})^{-1}\bigg)\widehat{R}_G^{(p)}.
    \end{aligned}
    \vspace{-0.2cm}
    \end{equation*}
    \STATE Compute $(\hat U^{(p+1)}, \hat K_U^{(p+1)})\leftarrow \textsf{SVD}_r(\hat U^{(p+0.5)})$ and $(\hat V^{(p+1)}, \hat K_V^{(p+1)})\leftarrow \textsf{SVD}_r(\hat V^{(p+0.5)})$.
    \STATE Compute $\widehat{G}^{(p+1)}=\argmin_{G} \mathscr{L}_{\mathcal{D}_{2p+2}}(\widehat{U}^{(p+1)},G,\widehat{V}^{(p+1)})$ and $(\hat L_G^{(p+1)}, \hat R_G^{(p+1)})\leftarrow \textsf{SVD}_r(\hat G^{(p+1)})$.
    \ENDFOR  
\end{algorithmic}
\end{algorithm}

\textbf{One-to-one matching}. The total number of observed entries under one-to-one matching is $Td_1$. Note that, at each time $t$, the probability of observing any entry of $M$ is $\nu=\EE[X_t(i,j)]=1/d_2$.  The entrywise error rate of $\widehat{M}^{(m)}$ is provided in Proposition \ref{thm:estimation}.  
\begin{Proposition}[one-to-one matching]\label{thm:estimation}
    Denote $\alpha_d=d_2/d_1$. There exist absolute constants $C_1,\cdots, C_4>0$ such that $T\geq C_1\alpha_d^2r^2\log^2 d_2$, $\lambda_{\min}/\sigma\geq C_2\sqrt{\alpha_d rd_2^2/T}\log d_2$, $\|M\|_{\max}/\sigma\leq d_1^{C_3}$, and set $m=\lceil C_3\log d_2 \rceil$ and $\eta=0.75$ in Algorithm~\ref{alg:gd}, then with probability at least $1-5d_2^{-10}$,
    \begin{align*}
        \|\widehat{M}^{(m)} - M\|_{\max}\leq C_4\sigma\sqrt{\frac{r^2d_2}{Td_1}}\log d_2.
    \end{align*}
\end{Proposition}
The sample size and SNR conditions are comparable to those in existing literature \citep{koltchinskii2011nuclear, ma2018implicit} up to the logarithmic factors, if assuming $\alpha_d=O(1)$. Indeed, since each matching reveals $d_1$ entries, the sample size condition requires that the number of observed entries, namely $Td_1$, is at least $O\big(r^2d_2\cdot{\rm polylog}(d_2)\big)$. The max-norm error rate is proportional to the noise level $\sigma$, which has the same dependence on the number of observed entries and the matrix dimensions as the optimal rate under i.i.d. observed entries, up to rank-dependent and logarithmic factors. 

After obtaining such a consistent estimator of the reward matrix with the entrywise error guarantee, we can identify the optimal matching under an additional optimality-gap condition, which ensures that the gap between the optimal and suboptimal matchings dominates the entrywise estimation error. We also provide a practical algorithm for computing an optimal matching from the estimated reward matrix. Further details are provided in Appendix \ref{sec: optimal matching}.

\begin{Remark}
     The condition $\|M\|_{\max}/\sigma\leq d_1^{C_3}$ with $\sigma>0$ is for technical convenience. It ensures that the contraction error arising from the computation convergence arising from gradient descent iterations is dominated by the statistical error after $m=O(\log d_2)$ iterations, yielding the desired error rate proportional to $\sigma$. Since $C_3$ can be chosen to be a sufficiently large constant, this condition is not particularly restrictive. However, it does not imply that our method cannot handle the noiseless case. Following the same arguments as in our proof, we can show that, when $\sigma=0$, the error bound decays polynomially to zero as $d_2$ increases after $O(\log d_2)$ iterations, implying almost exact recovery.
\end{Remark}

\textbf{One-to-many matching with one-sided random arrival}. If the matching $X_t$ is sampled uniformly from $\calM_{K,p_0}^{\otm}$ defined in (\ref{eq:one-to-many}), then the probability of observing the $(i,j)$-th entry of $M$ is $\nu=\EE[X_t(i,j)]=Kp_0/d_2$. As a result, given $T$ independently and uniformly sampled matchings from $\calM_{K,p_0}^{\otm}$, the expected total number of observed entries is $Td_1Kp_0$. Under this matching mechanism, the performance of Algorithm~\ref{alg:gd} is guaranteed by the following proposition. Note the natural requirement $d_2\geq Kd_1$ under the one-to-many matching mechanism. 
\begin{Proposition}[one-to-many matching]\label{thm:estimation2}
There exist absolute constants $C_1,\cdots, C_4>0$ such that if $T\geq C_1(Kp_0)^{-1}\alpha_d^2r^2\log^2 d_2$, $\lambda_{\min}/\sigma\geq C_2\sqrt{\alpha_d rd_2^2/(TKp_0)}\log d_2$, $\|M\|_{\max}/\sigma\leq d_1^{C_3}$, and set $m=\lceil C_3\log d_2 \rceil$ and  $\eta=0.75$ in Algorithm~\ref{alg:gd}, then with probability at least $1-5d_2^{-10}$,
    \begin{align*}
        \|\widehat{M}^{(m)} - M\|_{\max}\leq C_4\sigma\sqrt{\frac{r^2d_2}{Td_1Kp_0}}\log d_2.
    \end{align*}
\end{Proposition}
The dependence among the entries of $X_t\in\calM_{K,p_0}^{\otm}$ is more complicate than that in one-to-one matching. Each matching may reveal multiple  entries on every row, but reveals at most one entry on every column.  Since $Td_1Kp_0\gg 1$, the Chernoff bound implies that the total number of observed entries is of order $O(Td_1Kp_0)$ with high probability. This result shows that, under the one-to-many matching mechanism, our error rate also has the same dependence on the number of observed entries, the matrix dimensions, and the noise level, up to logarithmic factors as that under i.i.d. observed entries. The corresponding sample size and SNR conditions are likewise comparable.

\textbf{Two-sided random arrival}. We focus on {\it truncated Binomial distributions} under the two-sided random arrival matching scheme. For fixed constants $c_r, c_s\in(0,1)$ and $\gamma>0$, define the bivariate truncated Binomial distribution $\textsf{tBin}_{c_r,c_s,\gamma}(d_1,p_1,d_2,p_2)$ with p.m.f. given by $\PP(B_r=k_1, B_s=k_2)\propto {d_1\choose k_1}{d_2\choose k_2}p_1^{k_1}p_2^{k_2}(1-p_1)^{d_1-k_1}(1-p_2)^{d_2-k_2}$ if $k_1\geq c_rd_1, k_2\geq c_sd_2$ and $k_1\geq (1+\gamma)k_2$ or $k_2\geq (1+\gamma)k_1$; and zero otherwise. The truncated binomial distribution is adopted for technical convenience, as it guarantees that each observation $X_t$ contains $O(d_1)$ matching pairs. Otherwise, a more complicated analysis than our current theoretical framework would be required. The conditions $k_1\geq(1+\gamma)k_2$ or $k_2\geq(1+\gamma)k_1$ are also imposed for technical convenience when applying martingale arguments to control the higher-order remainder terms in establishing asymptotic normality (see Section \ref{sec:inference}).

We redefine the sampling set under two-sided random arrival matching mechanism as
\begin{equation}\label{eq:two-random-complete}
    {\mathcal{M}}_{p_1, p_2, c_r, c_s}^{\tside}=\Big\{X\in \{0,1\}^{d_1\times d_2}\big| e_i^{\top}X 1_{d_2}\leq 1,  1_{d_1}^{\top}X e_j\leq 1, 1_{d_1}^{\top}X 1_{d_2}=\min(B_r, B_s), \forall (i,j)\Big\},
\end{equation}
where $(B_r, B_s)^{\top}\sim \textsf{tBin}_{c_r,c_s,\gamma}(d_1,p_1,d_2,p_2)$. For notational brevity, we omit its dependence on $c_r,c_s,\gamma$.  If a matching is sampled uniformly from ${\mathcal{M}}_{p_1, p_2, c_r, c_s}^{\tside}$, the probability of revealing the $(i,j)$-th entry is given by $\nu:=\EE[X_t(i,j)]=\EE\left[\min(B_r,B_s)\right]/(d_1d_2)$. 
In practice, $\nu$ can be estimated through Monte Carlo sampling if $p_1, p_2, c_r, c_s$ are known or estimated directly from the observed data as $\widehat{\nu}=1/(Td_1d_2)\sum_t\sum_{(i,j)\in g_t}X_t(i,j)$. Given $T$ independently sampled matchings from ${\mathcal{M}}_{p_1, p_2,c_r,c_s}^{\tside}$, the expected total number of observed entries is  $T\nu d_1d_2$.
\begin{Proposition}[two-sided random arrival]\label{thm:estimation3}
There exist absolute constants $C_1,\cdots, C_4>0$ such that if $T\geq C_1(d_2\nu)^{-1}\alpha_d^2r^2\log^2 d_2$, $\lambda_{\min}/\sigma\geq C_2\sqrt{\alpha_d rd_2/(T\nu)}\log d_2$, $\|M\|_{\max}/\sigma\leq d_1^{C_3}$, and set $m=\lceil C_3\log d_2 \rceil$ and $\eta=0.75$ in Algorithm~\ref{alg:gd}, then with probability at least $1-5d_2^{-10}$,
    \begin{align*}
        \|\widehat{M}^{(m)} - M\|_{\max}\leq C_4\sigma\sqrt{\frac{r^2}{Td_1\nu}}\log d_2.
    \end{align*}
\end{Proposition}
Note that under the two-sided random arrival matching mechanism (\ref{eq:two-random-complete}), each participant is assigned only one task, and each task is assigned to at most one participant. This assignment is identical to that under the one-to-one matching scheme, except that  available participants and tasks arrive at random.  If $c_r, c_s, \gamma, p_1, p_2$ are constants, then the entrywise sampling probability $\nu$ is of order $d_2^{-1}$ under the assumption $d_2\geq d_1$.  As a result, the max-norm error rate attained under two-sided random arrival is comparable to that under the one-to-one matching mechanism, up to a constant factor.

\section{Evaluating Matching Decisions}\label{sec:inference}
For matching evaluation, we aim to quantify the uncertainty in the estimated reward matrix, which facilitates the construction of confidence intervals, hypothesis testing, and decision making subject to a specified tolerance of mistakes. The high-probability bounds derived for the estimator in Section~\ref{sec:estimation} cannot provide a precise characterization of its inherent uncertainty. For instance, one may be interested in constructing a $95\%$ confidence interval for the expected reward when assigning task $j$ to participant $i$, or for the expected total rewards of a complete matching policy when assigning multiple tasks to a group of participants. Achieving this essentially requires precisely characterizing the distributions of entries of the estimator $\hat M$ or its linear form $\langle \hat M, Q\rangle$, where $Q$ is typically a predetermined matrix. It is particularly challenging under matching observations due to the dependence among revealed entries, as most existing literature rely on the classical central limit or Berry-Esseen theorem under independent sampling. Therefore, we need more sophisticated tools to precisely characterize the bias, variance, and asymptotic distribution under the three matching mechanisms.



While the estimators obtained from Algorithm~\ref{alg:gd} achieve a sharp max-norm error rate, they are not directly applicable for statistical inference due to the bias introduced by implicit regularization, and their asymptotic distributions are not analytically tractable. To address these challenges, we employ a debiasing procedure with sample splitting that uses the outputs of Algorithm~\ref{alg:gd} as initial estimators. Without loss of generality, we assume $T=2T_0$. We split the collected data $\{(Y_t, X_t)\}_{t\in[T]}$ into two subsets with indices $\widetilde{\mathcal{D}}_1=\{(Y_t, X_t)\}_{t=T_0+1}^T$ and $\widetilde{\mathcal{D}}_2=\{(Y_t, X_t)\}_{t=1}^{T_0}$. The initial estimators $\widehat{M}_1^{\init}$ and $\widehat{M}_2^{\init}$ are obtained by applying Algorithm \ref{alg:gd} to $\widetilde{\mathcal{D}}_1$ and $\widetilde{\mathcal{D}}_2$, respectively. By cross-sample debiasing, we obtain the unbiased estimator $\widehat{M}_1^{\ubs}=\widehat{M}_1^{\init} + (T_0\nu)^{-1}\sum_{t\in\widetilde{\calD}_2}\big(Y_t - X_t \circ \widehat{M}_1^{\init}\big)$, where $\nu$ represents the (marginal) entrywise sampling probability under the specified matching mechanisms. The unbiased estimator $\hat M_2^{\ubs}$ is defined similarly. It is straightforward to verify that $\EE \hat M_1^{\ubs}=\EE \hat M_2^{\ubs}=M$. 

While $\hat M_1^{\ubs}$ and $\hat M_2^{\ubs}$ are unbiased, they are generally full-rank and have large variances. 
To exploit the low-rank structure and variance reduction, we compute the best rank-$r$ approximations of $\widehat{M}_1^{\ubs}$ and $\widehat{M}_2^{\ubs}$, denoted by $\widehat{M}_1$ and $\widehat{M}_2$, respectively. In particular, we define
$\widehat{M}_1=\widehat{U}_1\widehat{U}_1^{\top}\widehat{M}_1^{\ubs} \widehat{V}_1\widehat{V}_1^{\top}$, 
where $\widehat{U}_1$ and $\widehat{V}_1$ consist of the top-$r$  left and right singular vectors of $\widehat{M}_1^{\ubs}$, respectively.  The final estimator is then given by
\begin{align*}
    \widehat{M}=\frac{1}{2}(\widehat{M}_1+\widehat{M}_2).
\end{align*}
The low-rank approximation introduces negligible bias but substantially reducing the variance. To test hypothesis related to the linear form $\langle M, Q\rangle$, we use the plug-in estimator $\langle \hat M, Q\rangle$.

\begin{Remark}
Sample splitting ensures the independence between $\hat M_1^{\init}$ and $\widetilde{\calD}_2$, which in turn guarantees the unbiasedness of the debiased estimator $\hat M_1^{\ubs}$. This property is particularly useful for establishing asymptotic normality because the final estimator is obtained through a low-rank projection, requiring a nontrivial spectral perturbation analysis. The unbiasedness provide technical convenience to control higher-order projection remainders. Moreover, the final  estimator combines  $\widehat M_1$ and $\widehat M_2$, which prevents the loss of asymptotic efficiency from sample splitting. The potential cost of sample splitting is greater finite-sample variability due to the reduced accuracy of the initial estimators. In the classic i.i.d. setting, \citet{ma2023multiple} show that removing sample splitting does not affect asymptotic normality. We provide additional simulation results in Appendix \ref{sec: sample splitting},  suggesting that sample splitting may be a limitation of technical tools rather than a practical necessity.
\end{Remark}

We establish the asymptotic normality of $\langle \hat M, Q\rangle$ under three matching mechanisms and derive finite-sample convergence rates. It turns out that the asymptotic variance of $\langle \hat M, Q\rangle$ is determined by the magnitude of the projection of $Q$ onto the tangent space at $M$. Specifically, we define $\mathcal{P}_M(Q)=UU^{\top}QV_{\perp}V_{\perp}^{\top} + U_{\perp}U_{\perp}^{\top}QVV^{\top} + UU^{\top}QVV^{\top}=Q-U_{\perp}U_{\perp}^{\top}QV_{\perp}V_{\perp}^{\top}$, where $U_{\perp}\in\OO^{d_1\times (d_1-r)}$ and $V_{\perp}\in\OO^{d_2\times (d_2-r)}$ are chosen such that $(U, U_{\perp})$ and $(V, V_{\perp})$ form orthogonal matrices. Denote by $\Phi(t)$ the c.d.f. of the standard normal distribution and $\|\cdot\|_{\ell_1}$ the vectorized $\ell_1$-norm of a matrix.

\textbf{One-to-one matching}. Using the Berry-Esseen theorem, we derive the finite-sample error rate for the normal approximation of the plug-in estimator $\langle \hat M, Q\rangle$. For technical reasons, we assume that the aspect ratio $\alpha_d$ is bounded away from one, which is not overly restrictive. If the data does not satisfy this condition, we can split the matrix into two parts, each containing half of the rows, and perform estimation and inference separately for each part.  

\begin{Theorem}[one-to-one matching]\label{thm:inference-main}
Suppose the conditions in Proposition \ref{thm:estimation} hold and $\alpha_d\geq 1+\gamma$ for some constant $\gamma>0$. Then there exist a constant $C_0>0$ depending only on $\gamma$ such that
\begin{align*}
\sup_{t\in\RR}\left|\PP\left(\frac{\inp{\widehat{M}}{Q}-\inp{M}{Q}}{\sigma\fro{\mathcal{P}_M(Q)}\sqrt{d_2/T}}\leq t\right) - \Phi(t)\right|\leq C_0\left(\sqrt{\frac{r^2\alpha_d\log^3 d_2}{T}} + \frac{\sigma}{\lambda_{\min}}\sqrt{\frac{r^2d_2^2\log d_2}{Td_1}}\frac{\|Q\|_{\ell 1}}{\fro{\mathcal{P}_M(Q)}}\right).
\end{align*}
\end{Theorem}

Therefore, if the sample size and SNR are such that the right-hand side converges to zero, Theorem~\ref{thm:inference-main} shows that 
\begin{align*}
\frac{\inp{\widehat{M}}{Q}-\inp{M}{Q}}{\sigma\fro{\mathcal{P}_M(Q)}\sqrt{d_2/T}} \overset{{\rm d.}}{\rightarrow} N(0,1),\quad \textrm{ as } T\to\infty.
\end{align*}
The asymptotic variance depends on $Q$ through the squared Frobenius norm of its projection onto the tangent space at $M$, given by $\fro{\mathcal{P}_M(Q)}^2 = \fro{U^{\top}Q}^2 + \fro{QV}^2 - \fro{U^{\top}QV}^2$. This quantity measures how strongly $Q$ aligns with the left singular space spanned by $U$ or the right singular space spanned by $V$, with the overlap counted only once. Thus, the variance reflects the alignment between the testing direction $Q$ and the latent low-rank structure of $M$. 

For entrywise inference with $Q=e_ie_j^{\top}$, Theorem~\ref{thm:inference-main} suggests that $\EE \big(\hat M(i,j)-M(i,j)\big)^2=\big(1+o(1)\big)\|\calP_M(e_ie_j^{\top})\|_{\rm F}^2\cdot \sigma^2d_2/T$, which implies that 
$$
\EE \|\hat M-M\|_{\rm F}^2=\big(1+o(1)\big)\cdot \frac{\sigma^2rd_2(d_1+d_2)}{T},
$$
with a sharp constant factor.


\textbf{One-to-many matching with one-sided random arrival}.  In the more general case of one-to-many matching, our estimator attains the following asymptotic normality. 
\begin{Theorem}[one-to-many matching]\label{thm:inference-main-2}
Suppose that the conditions in Proposition \ref{thm:estimation2} hold and $\alpha_d\geq (1+\gamma)K$ for some constant $\gamma>0$. Then there exists a constant $C_0>0$ depending only on $\gamma$ such that
\begin{align*}
        &\sup_{t\in\RR}\left|\PP\left(\frac{\inp{\widehat{M}}{Q} - \inp{M}{Q}}{\sigma\fro{\mathcal{P}_M(Q)}\sqrt{d_2/(TKp_0)}}\leq t\right) - \Phi(t)\right|\leq C_0\left(\sqrt{\frac{r^2\alpha_d^2\log^3 d_2}{TKp_0}} + \frac{\sigma}{\lambda_{\min}}\sqrt{\frac{r^2d_2^2\log d_2}{Td_1Kp_0}}\frac{\|Q\|_{\ell 1}}{\fro{\mathcal{P}_M(Q)}}\right).
\end{align*}
\end{Theorem}

Therefore, the plug-in estimator $\langle \hat M, Q\rangle$ is asymptotic normal provided that the sample size satisfies $T\gg r^2\alpha_d^2(Kp_0)^{-1}\log^3 d_2$ and SNR meets $\lambda_{\min}/\sigma \gg \sqrt{(r^2d_2^2\log d_2)/(Td_1Kp_0)}\cdot \|Q\|_{\ell_1}/\|\calP_M(Q)\|_{\rm F}$. Moreover, Theorem~\ref{thm:inference-main-2} shows that 
$$
\EE \|\hat M-M\|_{\rm F}^2=\big(1+o(1)\big)\cdot \frac{\sigma^2rd_2(d_1+d_2)}{TKp_0}.
$$


\textbf{Two-sided random arrival}. Lastly, we consider the two-sided random arrival matching mechanism. 
Recall that the entrywise sampling probability $\nu$ is defined in Section~\ref{sec:estimation}.
\begin{Theorem}[two-sided random arrival]\label{thm:inference-main-3}
Suppose that the conditions in Proposition \ref{thm:estimation3} hold and $\alpha_d\geq 1+\gamma$ for some constant $\gamma$. Then there exists a constant $C_0>0$ depending only on $\gamma$ such that
\begin{align*}
\sup_{t\in\RR}\left|\PP\left(\frac{\inp{\widehat{M}}{Q} - \inp{M}{Q}}{\sigma\fro{\mathcal{P}_M(Q)}\sqrt{1/(T\nu)}}\leq t\right) - \Phi(t)\right|\leq C_0\left(\sqrt{\frac{r^2\log^3 d_2}{Td_1\nu}} +  \frac{\sigma}{\lambda_{\min}}\sqrt{\frac{r^2d_2\log d_2}{Td_1\nu}}\frac{\|Q\|_{\ell 1}}{\fro{\mathcal{P}_M(Q)}}\right).
\end{align*}
\end{Theorem}


\begin{Remark}
It is worth noting that the target $Q$ is not constrained by the observed matching pattern. In other words, even if the observed data consists of one-to-one matching pairs, we are not limited to conducting hypothesis tests solely on single-entry rewards or specific one-to-one matchings. Instead, we can perform hypothesis tests on any general linear form or matching pattern. For example, we can test whether a certain one-to-many matching yields higher rewards, compare the rewards of two different one-to-many matchings, or examine the impact of missing users.  Therefore, our framework is highly flexible and adaptable for matching learning and inference.  
\end{Remark}

\textbf{Variance estimation and confidence intervals}. The asymptotic results in Theorems~\ref{thm:inference-main}-\ref{thm:inference-main-3} are not immediately applicable for hypothesis testing or constructing confidence intervals as the variance terms depend on unknown quantities $\sigma^2$ and $\fro{\mathcal{P}_M(Q)}^2$. We propose the following data-drive estimates 
: $\widehat{\fro{\mathcal{P}_M(Q)}^2}=\fro{\mathcal{P}_{\widehat{M}}(Q)}^2$ and 
\begin{align*}
    &\widehat{\sigma}^2= \frac{1}{T}\sum_{t\in\widetilde{\calD}_2}  \frac{1}{\textsf{sum}(X_t)}\fro{Y_t - X_t\circ \widehat{M}_1^{\init}}^2 + \frac{1}{T}\sum_{t\in \widetilde{D}_1} \frac{1}{\textsf{sum}(X_t)}\fro{Y_t - X_t\circ \widehat{M}_2^{\init}}^2,
\end{align*}
where $\textsf{sum}(\cdot)$ denotes the sum of all entries of a matrix. 
\begin{Theorem}\label{thm:variance} 
    Under the conditions stated in one of Theorems \ref{thm:inference-main}-\ref{thm:inference-main-3},
   \begin{align*}
        \widehat{\sigma}^2\overset{{\rm p.}}{\rightarrow} \sigma^2 \quad \textrm{and} \quad \fro{\mathcal{P}_{\widehat{M}}(Q)}^2\overset{{\rm p.}}{\rightarrow} \fro{\mathcal{P}_M(Q)}^2,
   \end{align*}
   as $T\to\infty$.  
\end{Theorem}

Based on Theorems \ref{thm:inference-main} and \ref{thm:variance}, we can construct the $100(1-\alpha)\%$ confidence interval of the linear form $\langle M, Q\rangle$ by 
$$
{\rm CI}_{\alpha, Q}^{\oto}=\bigg(\langle \hat M, Q\rangle-z_{\alpha/2}\hat\sigma \|\calP_{\hat M}(Q)\|_{\rm F}\cdot \sqrt{\frac{d_2}{T}},\ 
\langle \hat M, Q\rangle+z_{\alpha/2}\hat\sigma \|\calP_{\hat M}(Q)\|_{\rm F}\cdot \sqrt{\frac{d_2}{T}}\bigg)
$$
under the one-to-one matching mechanism, where $z_{\alpha}=\Phi^{-1}(1-\alpha)$. The confidence intervals under other matching mechanisms can be constructed similarly.

\section{Numerical Experiments}

\subsection{Simulation studies}\label{sec:simulation}
We now present several numerical studies to evaluate the practical performance of our matching learning and evaluation methods across all three matching mechanisms.

Throughout all experiments, we set the dimensions of the true matrix to $d_1=100$ and $d_2=750$, with rank $r=2$. The matrix $M_{\mathrm{base}}$ is generated by applying an SVD to a random matrix whose entries are drawn uniformly from $\operatorname{Unif}[-20,20]$. The sample size is fixed at $T=2000\approx 0.2\alpha_d^2\log^2(d_2)r^2$, and the noise level is set to $\sigma=0.2$. For the one-to-many matching scheme, we set $K=5$ and the arrival probability $p_0=0.8$. For the two-sided random arrival scheme, we set $p_1=p_2=0.9$.

\begin{figure}[htbp]
    \centering

    \begin{subfigure}[t]{0.32\textwidth}
        \centering
        \includegraphics[width=\linewidth]{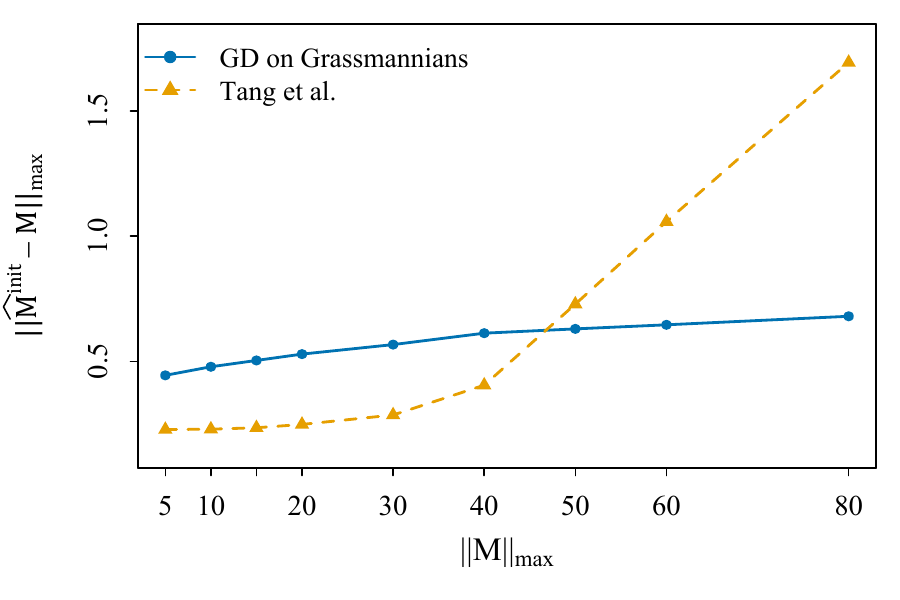}
        \caption{One-to-one matching}
        \label{fig: simu-oto}
    \end{subfigure}
    \hfill
    \begin{subfigure}[t]{0.32\textwidth}
        \centering
        \includegraphics[width=\linewidth]{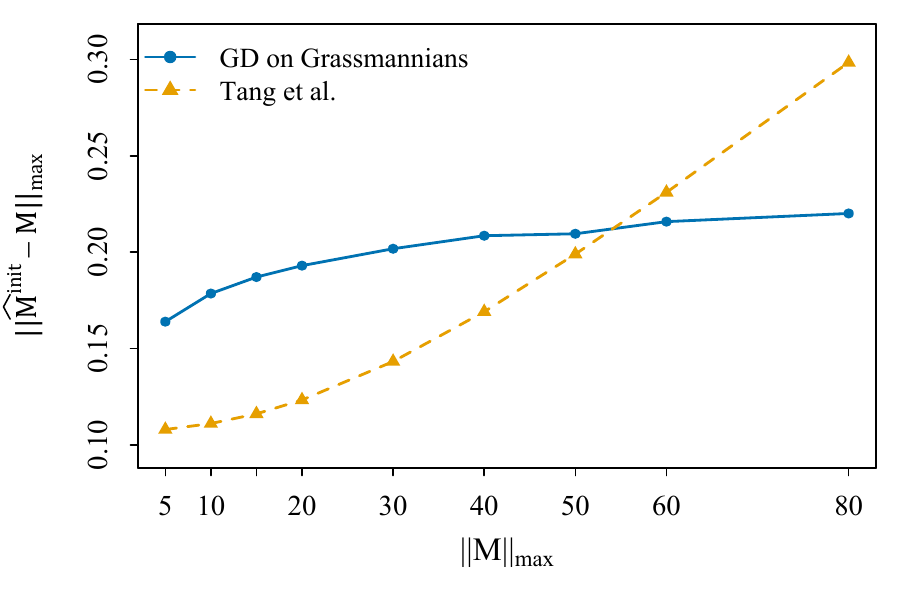}
        \caption{One-to-many matching}
        \label{fig: simu-otm}
    \end{subfigure}
    \hfill
    \begin{subfigure}[t]{0.32\textwidth}
        \centering
        \includegraphics[width=\linewidth]{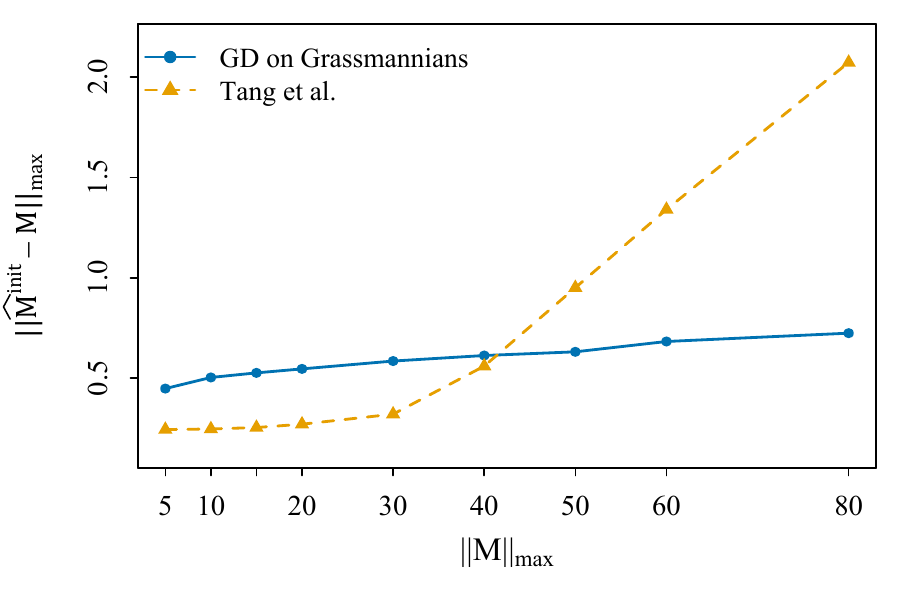}
        \caption{Two-sided random arrival}
        \label{fig: simu-tside}
    \end{subfigure}

    \caption{Comparison of the entrywise estimation errors of Algorithm \ref{alg:gd} and the method of \citet{tang2026match} across different signal magnitudes under the three matching schemes.}
    \label{fig: simu-comparison}
\end{figure}

We first examine the entrywise performance of the initial estimator obtained using Algorithm \ref{alg:gd} and compare it with the double-enhancement algorithm of \citet{tang2026match}. Specifically, we use all $T=2000$ observations to construct the initial estimators and examine how the entrywise estimation error $\|\widehat{M}-M\|_{\max}$ behave w.r.t. the signal magnitude $\|M\|_{\max}$. We rescale the original matrix $M_{\mathrm{base}}$ to obtain different signal levels satisfying $\|M\|_{\max}\in \{5,10,15,20,30, 40, 50, 60, 80\}$. For each matching scheme and signal level, we conduct 500 independent trials. Note that the original algorithm of \citet{tang2026match} applies only to one-to-one matching and not to the other two matching schemes. We therefore extend their algorithm to the other two schemes for comparison. The results are presented in Figure \ref{fig: simu-comparison}. The method of \citet{tang2026match} performs better when $\|M\|_{\max}$ is small, but its estimation error grows more rapidly as $\| M\|_{\max}$ increases. By contrast, our method exhibits relatively stable estimation error, consistent with our theoretical finding that the error is proportional to $\sigma$ and does not increase with $\| M\|_{\max}$.

We then evaluate the performance of our matching evaluation framework. To demonstrate the flexibility of our method, we consider four choices for $Q$ for each observed matching scheme: 1) the single entry, $Q=e_1e_1^{\top}$; 2) the single one-to-one matching, where $Q\in \mathcal{M}^{\oto}$; 3) the difference between two one-to-one matching, $Q=Q_1-Q_2$, where $Q_1$ and $Q_2$ are randomly sampled from $\mathcal{M}^{\oto}$; 4) the single one-to-many matching, where $Q\in \mathcal{M}_{K,p_0}^{\otm}$. We use the original matrix $M_{\mathrm{base}}$ in all settings.

\begin{figure}[t!]
	\centering
	\begin{subfigure} {0.24\linewidth} 
		\includegraphics[width=1.1\linewidth]{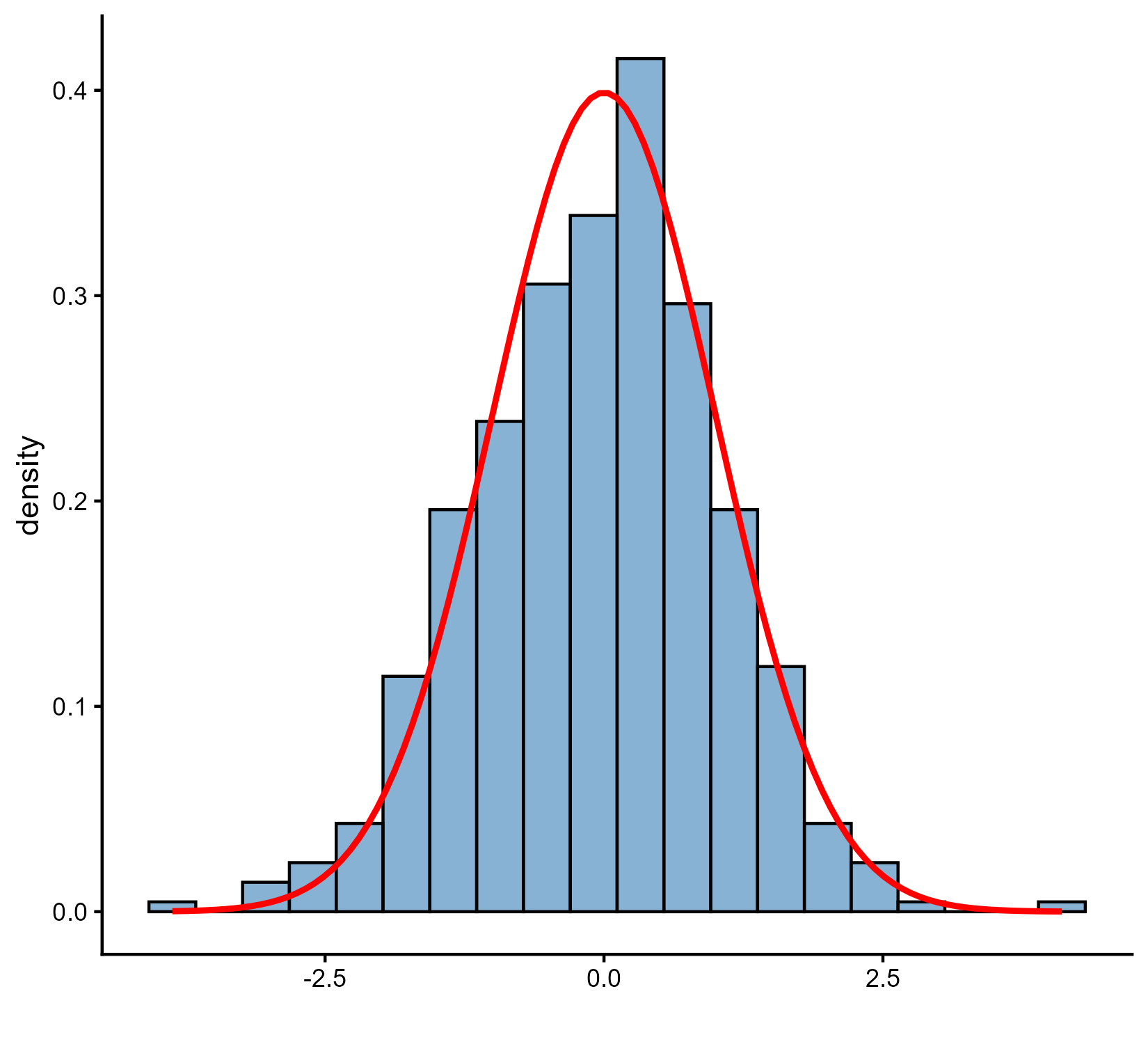}
        \caption{$Q=e_1e_1^{\top}$}
    \end{subfigure}
    \begin{subfigure} {0.24\linewidth} 
		\includegraphics[width=1.1\linewidth]{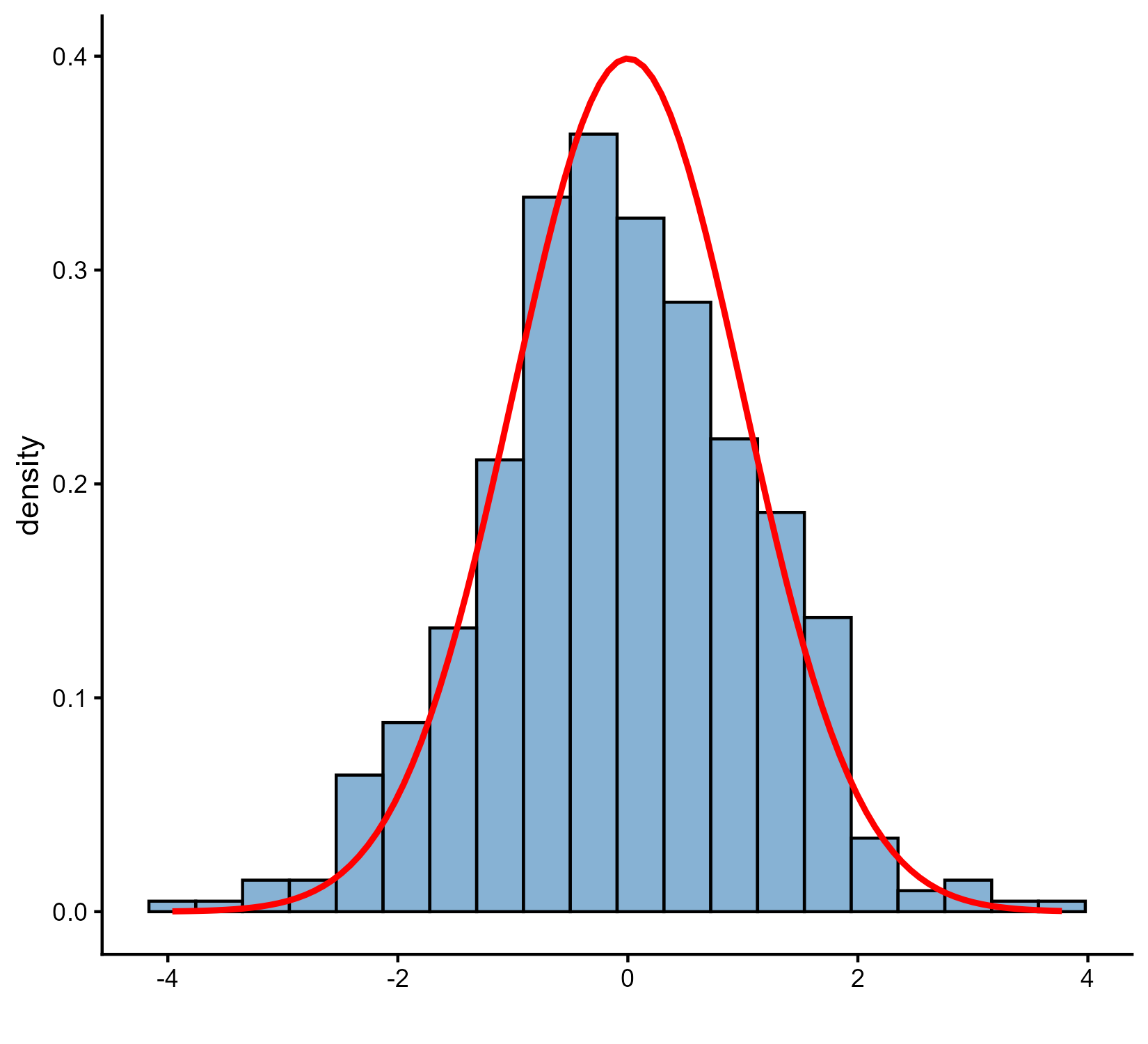}
        \caption{$Q\in \mathcal{M}^{oto}$}
    \end{subfigure}
    \begin{subfigure} {0.24\linewidth} 
		\includegraphics[width=1.1\linewidth]{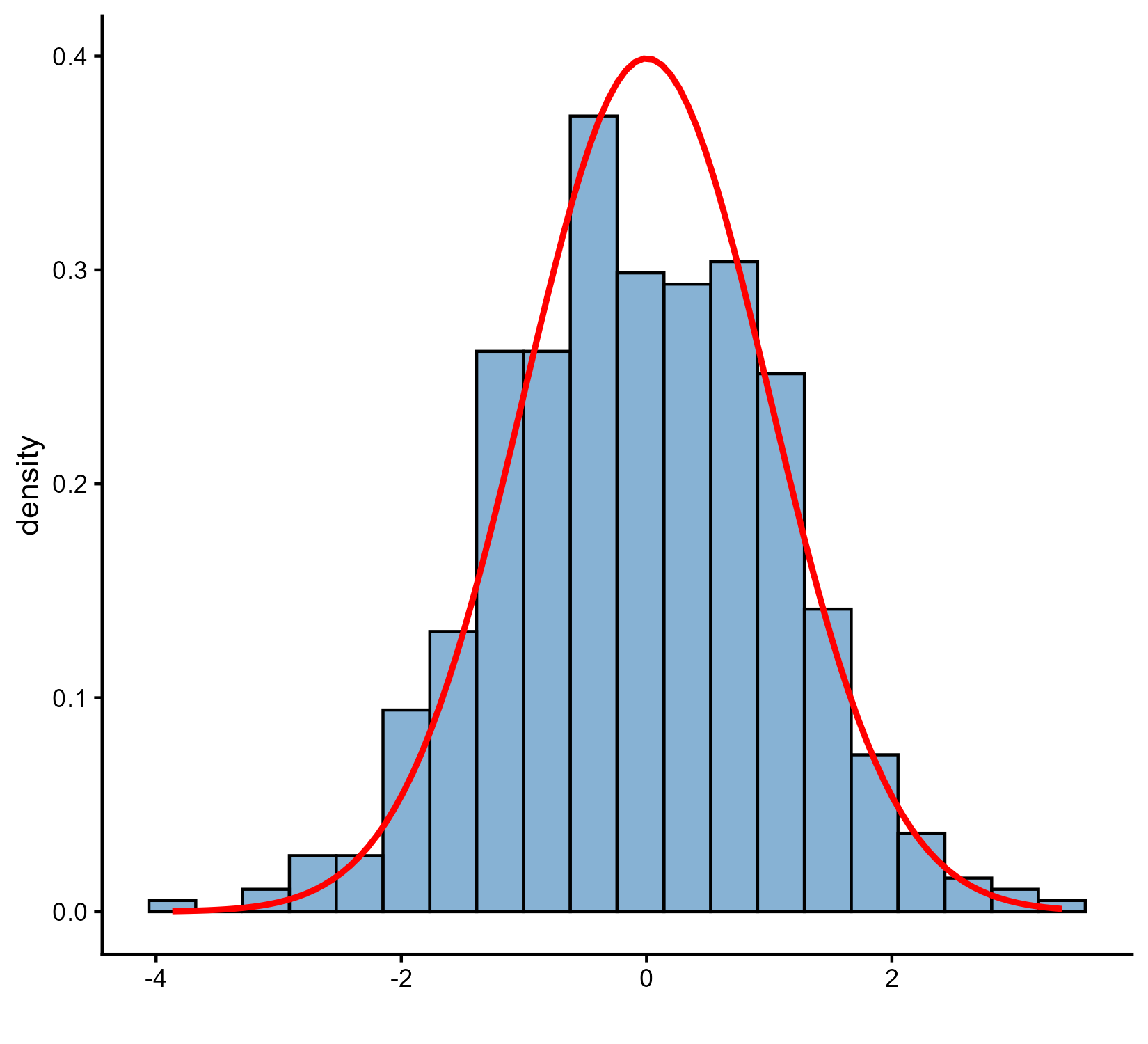}
        \caption{$Q=Q_1-Q_2$}
    \end{subfigure}
    \begin{subfigure} {0.24\linewidth} 
		\includegraphics[width=1.1\linewidth]{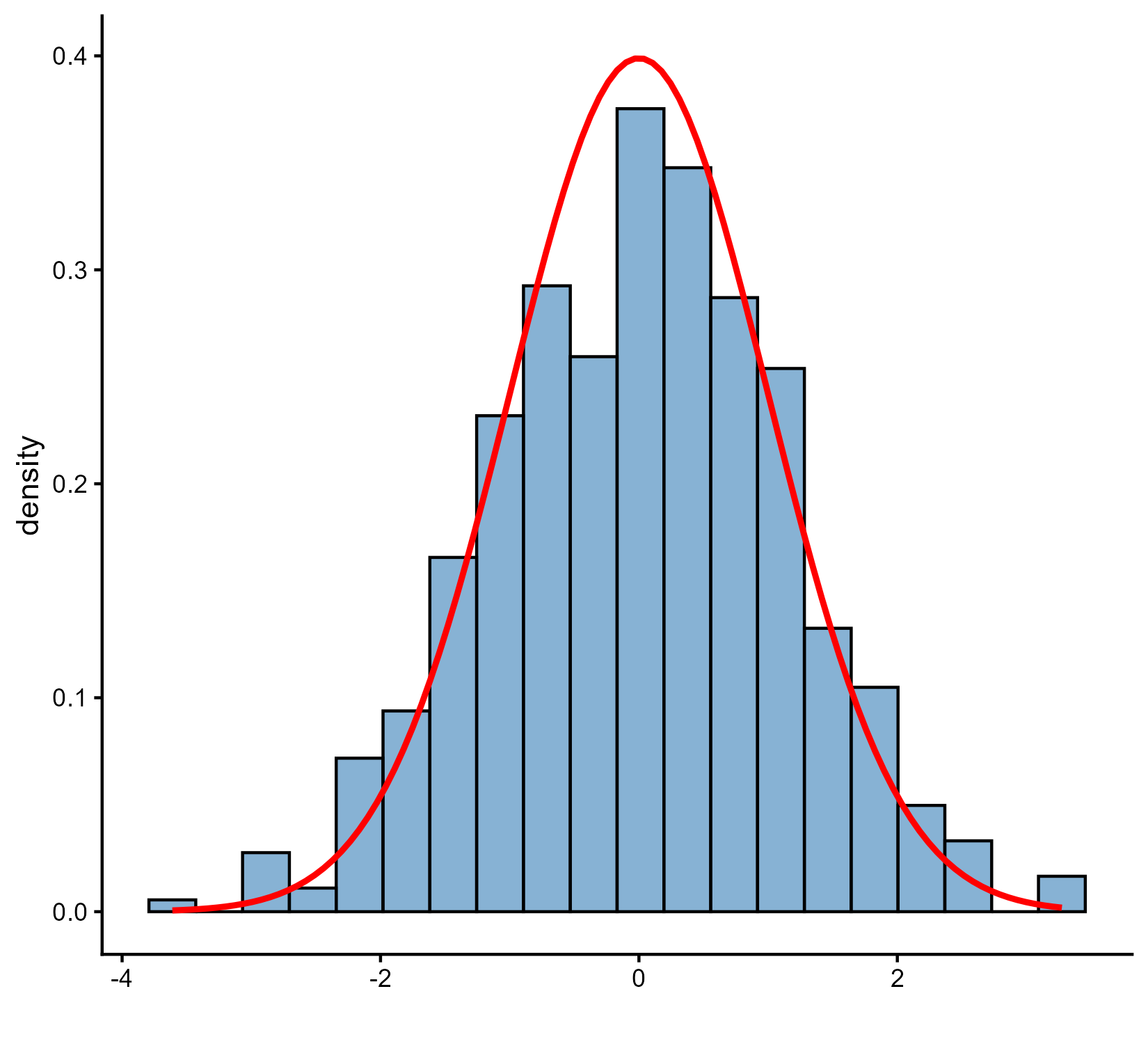}
        \caption{$Q\in \mathcal{M}_{K,p_0}^{\otm}$}
    \end{subfigure}

	\begin{subfigure} {0.24\linewidth} 
		\includegraphics[width=1.1\linewidth]{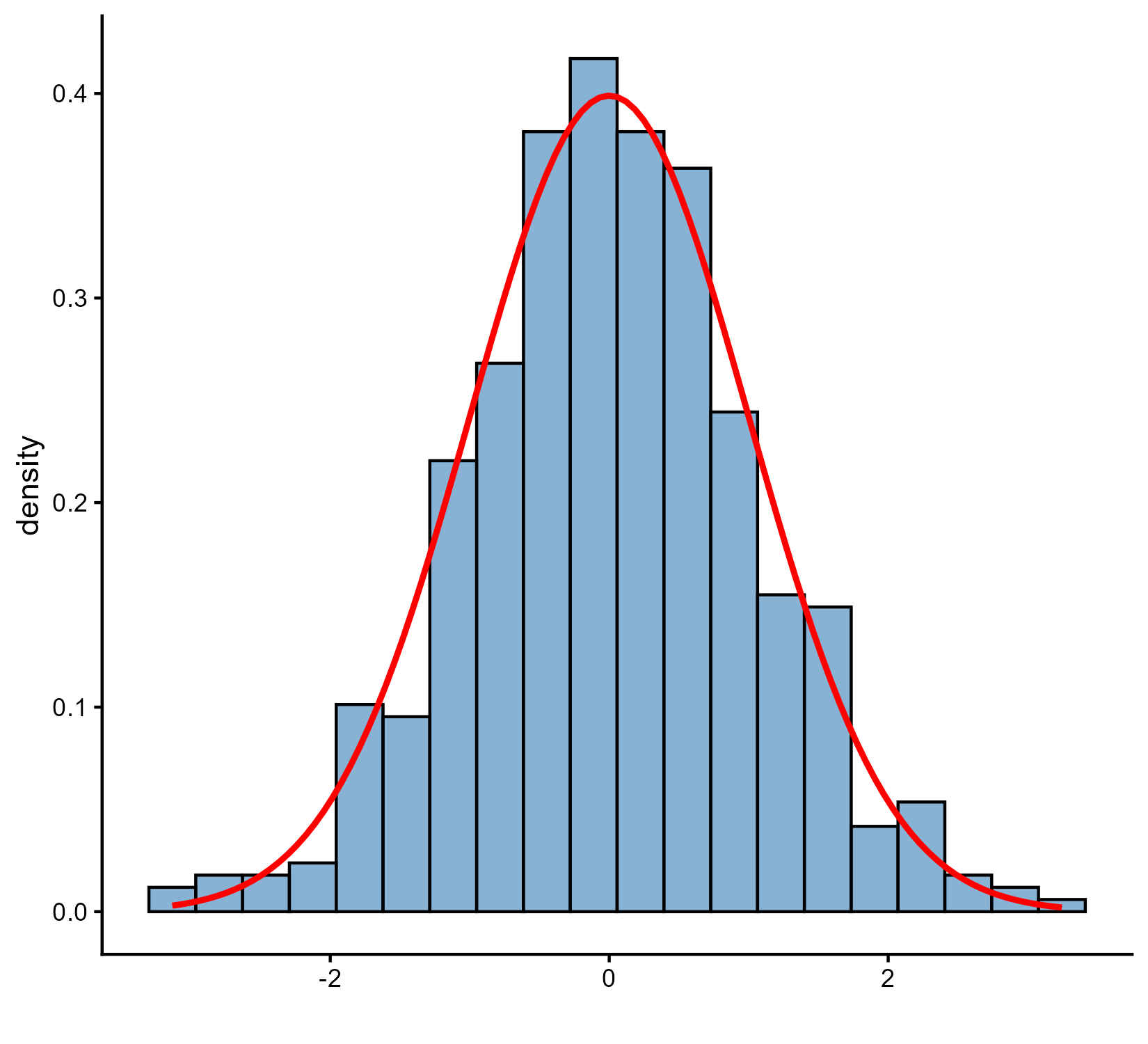}
        \caption{$Q=e_1e_1^{\top}$}
    \end{subfigure}
    \begin{subfigure} {0.24\linewidth} 
		\includegraphics[width=1.1\linewidth]{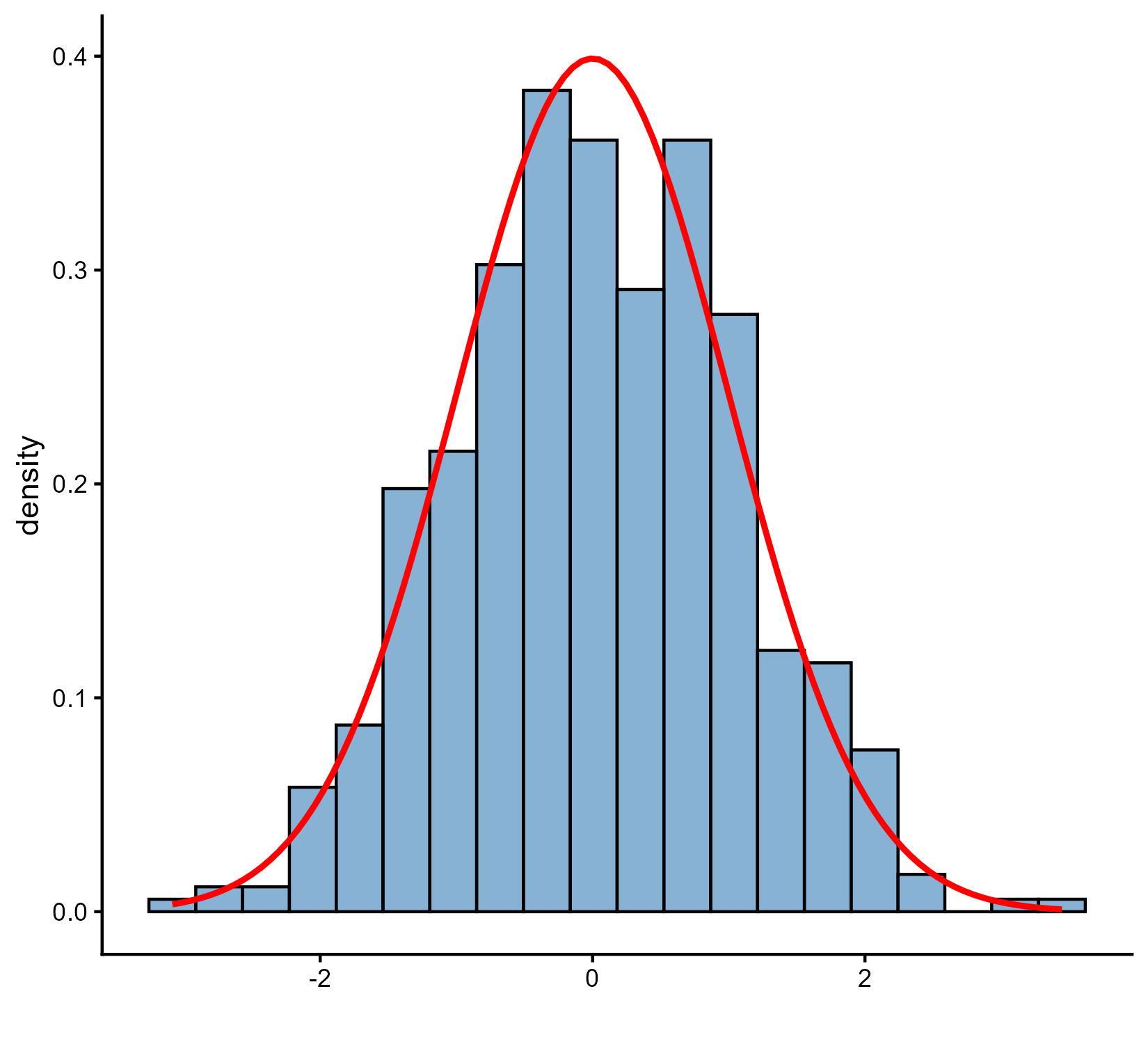}
        \caption{$Q\in \mathcal{M}^{oto}$}
    \end{subfigure}
    \begin{subfigure} {0.24\linewidth} 
		\includegraphics[width=1.1\linewidth]{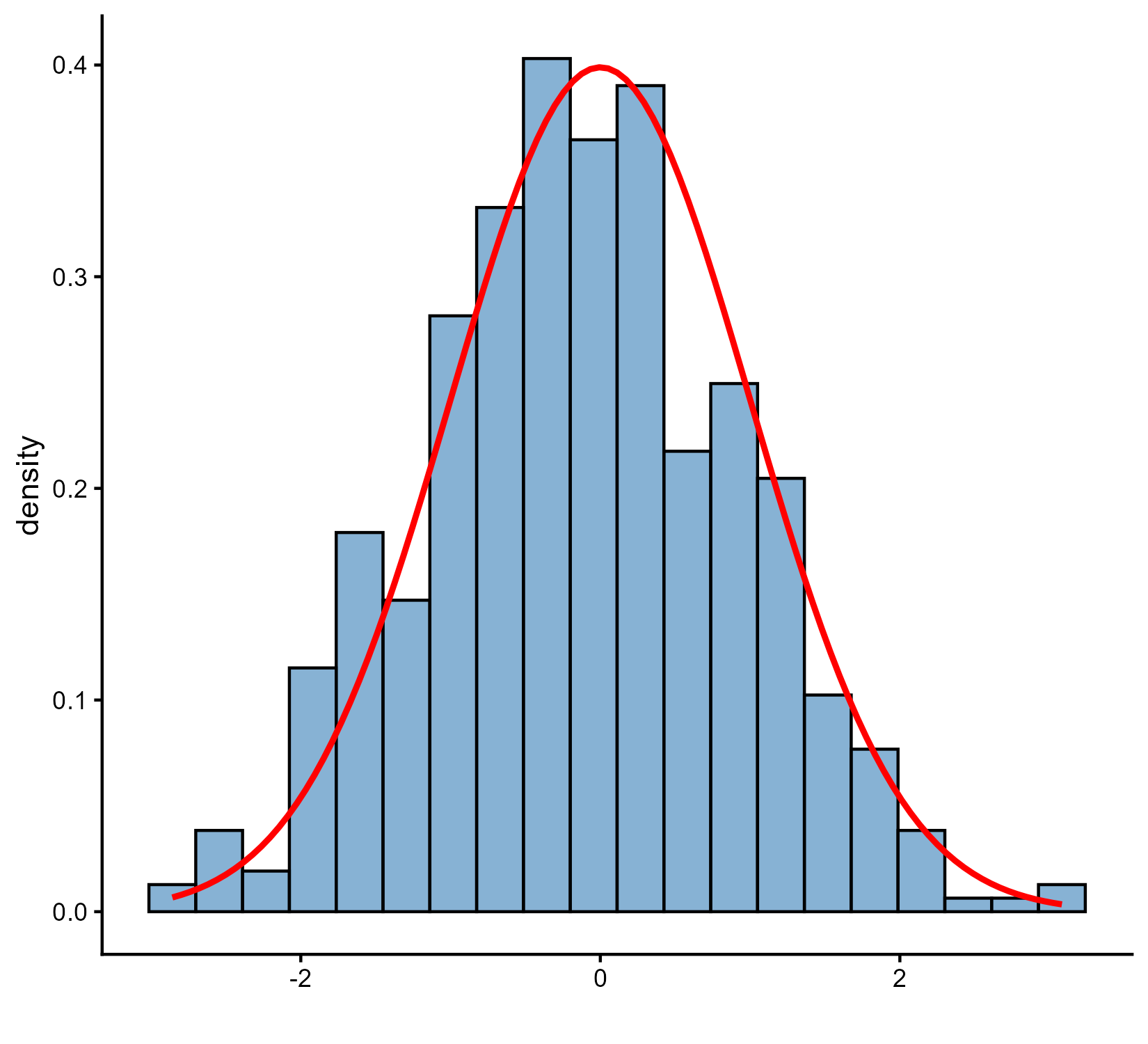}
        \caption{$Q=Q_1-Q_2$}
    \end{subfigure}
    \begin{subfigure} {0.24\linewidth} 
		\includegraphics[width=1.1\linewidth]{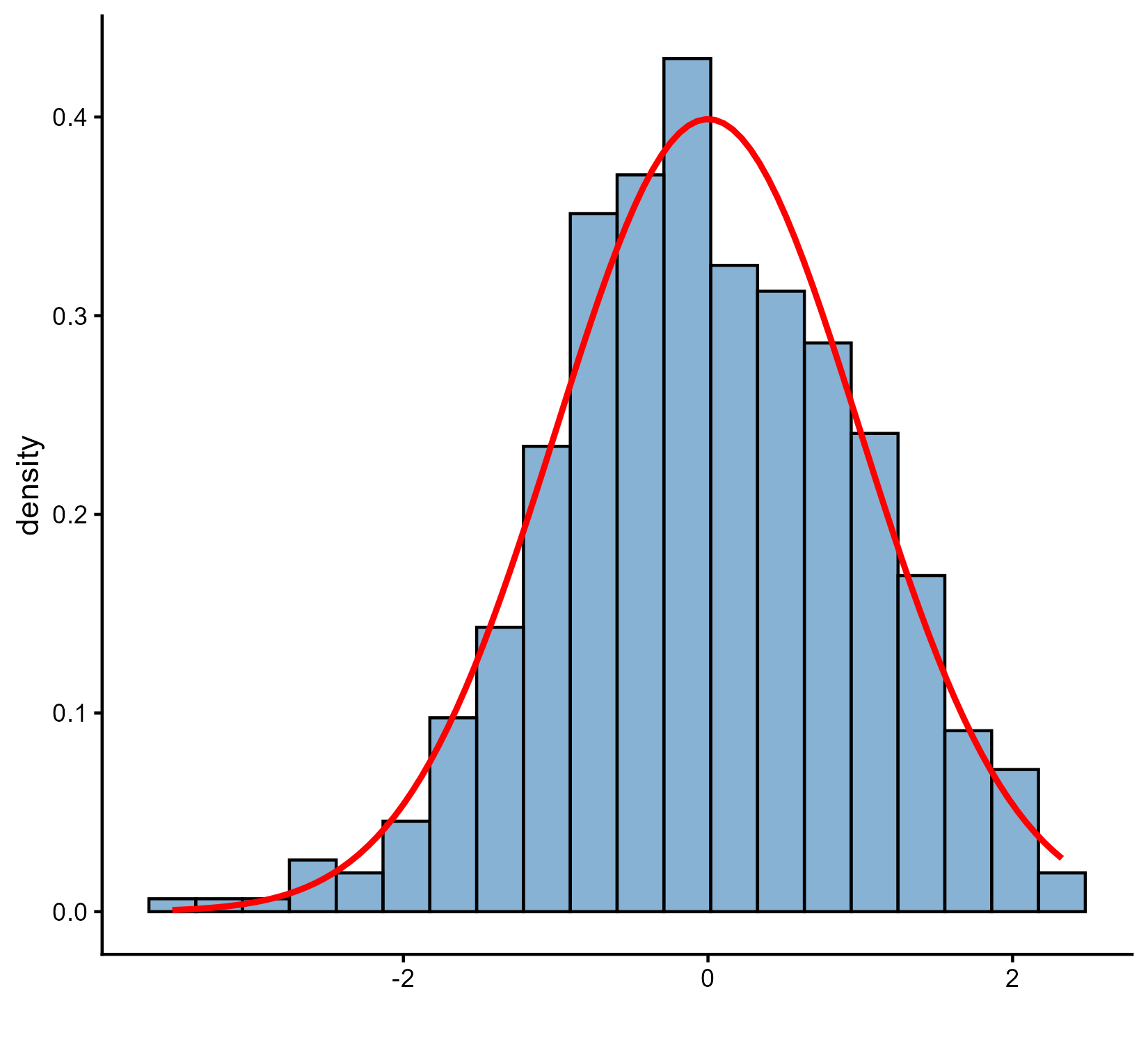}
        \caption{$Q\in \mathcal{M}_{K,p_0}^{\otm}$}
    \end{subfigure}
	
	\begin{subfigure} {0.24\linewidth} 
		\includegraphics[width=1.1\linewidth]{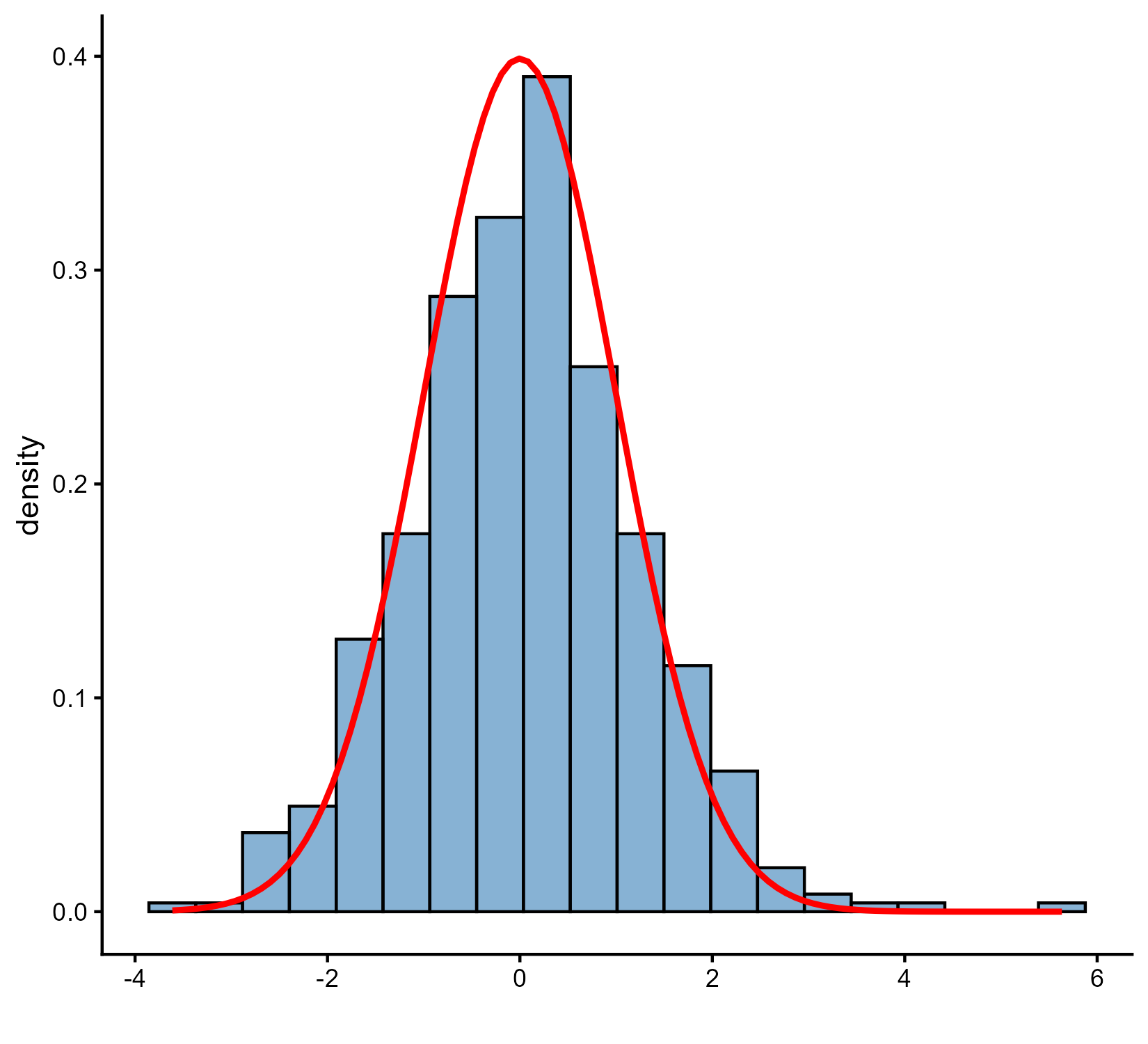}
        \caption{$Q=e_1e_1^{\top}$}
    \end{subfigure}
    \begin{subfigure} {0.24\linewidth} 
		\includegraphics[width=1.1\linewidth]{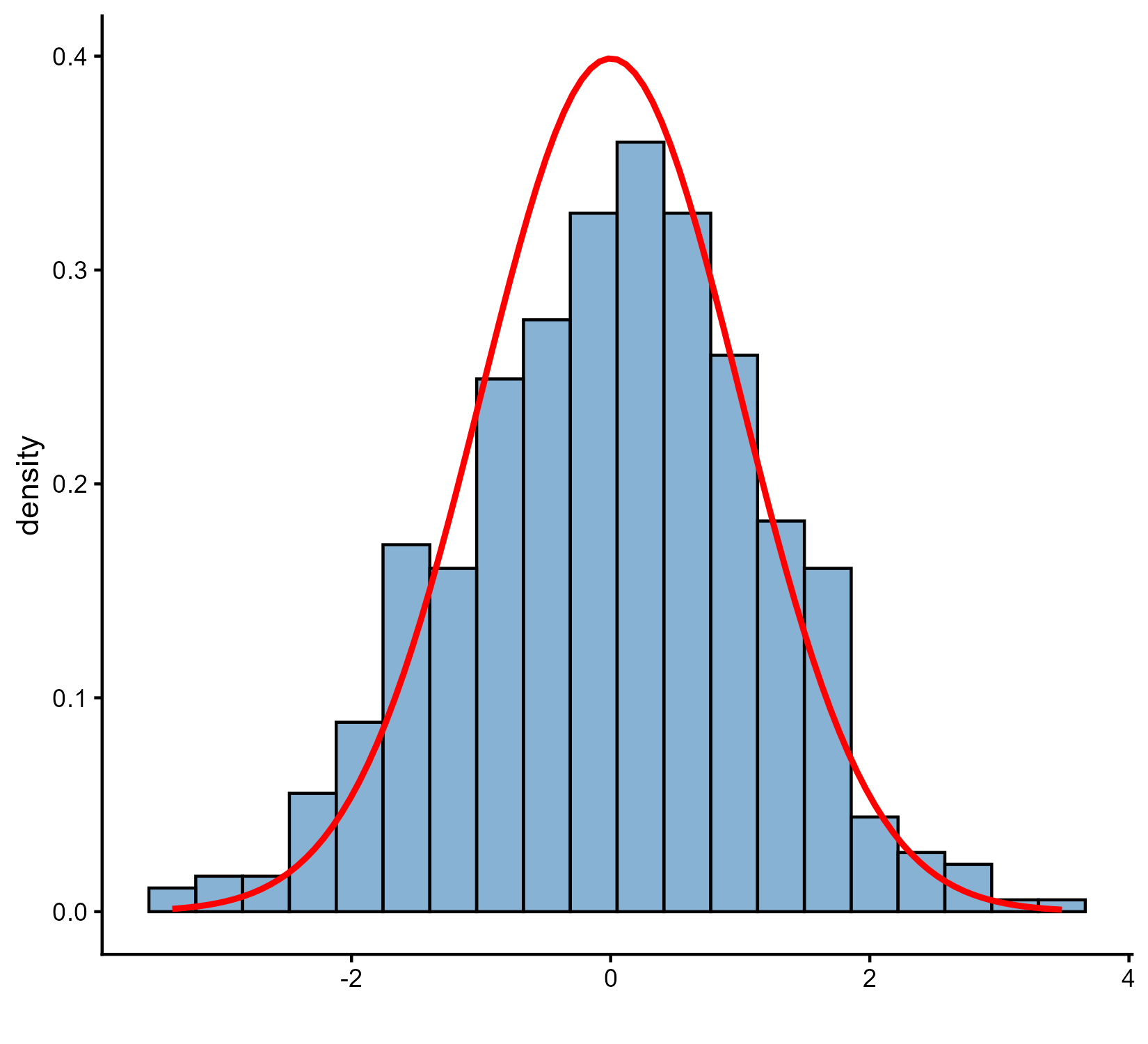}
        \caption{$Q\in \mathcal{M}^{oto}$}
    \end{subfigure}
    \begin{subfigure} {0.24\linewidth} 
		\includegraphics[width=1.1\linewidth]{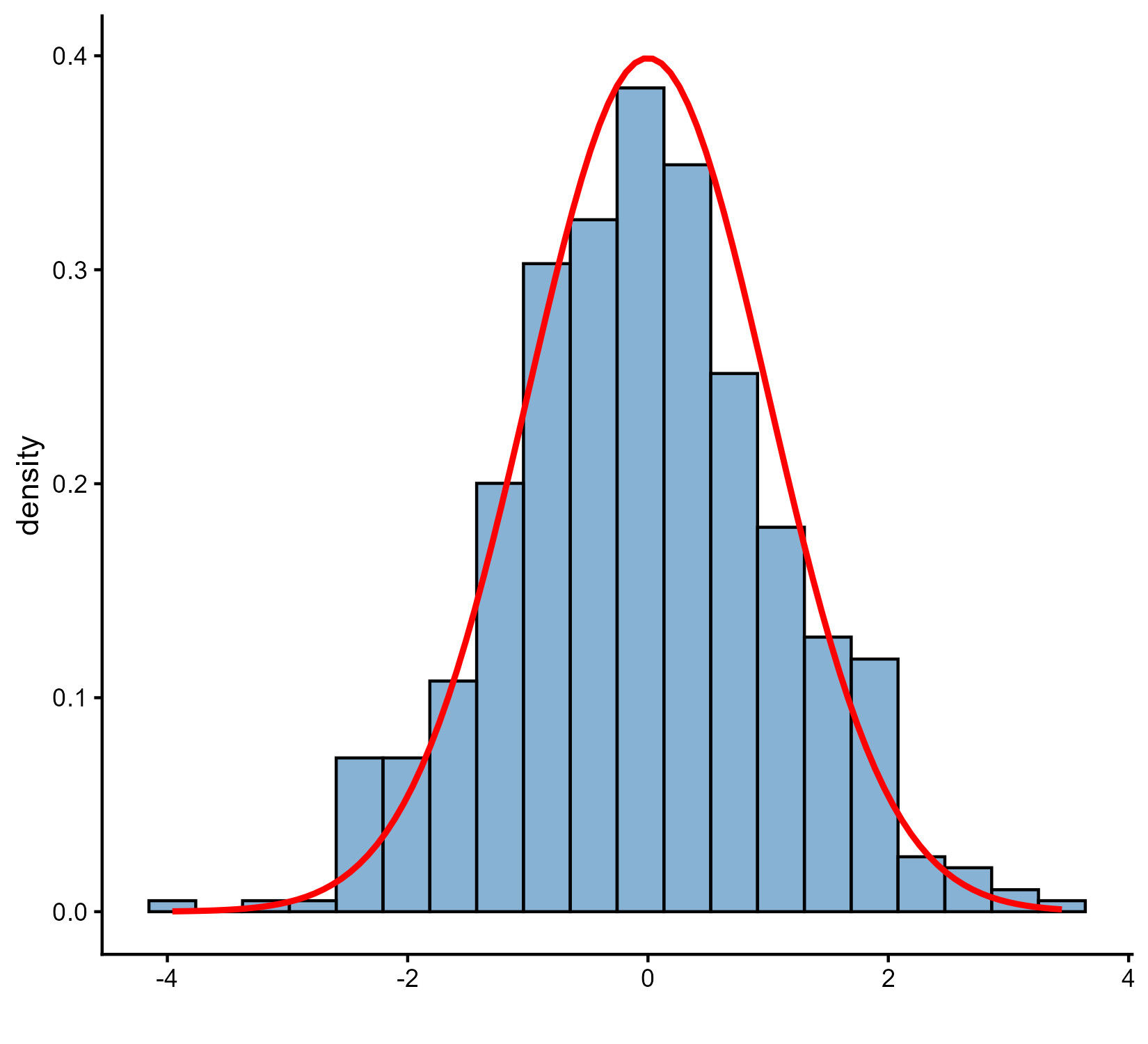}
        \caption{$Q=Q_1-Q_2$}
    \end{subfigure}
    \begin{subfigure} {0.24\linewidth} 
		\includegraphics[width=1.1\linewidth]{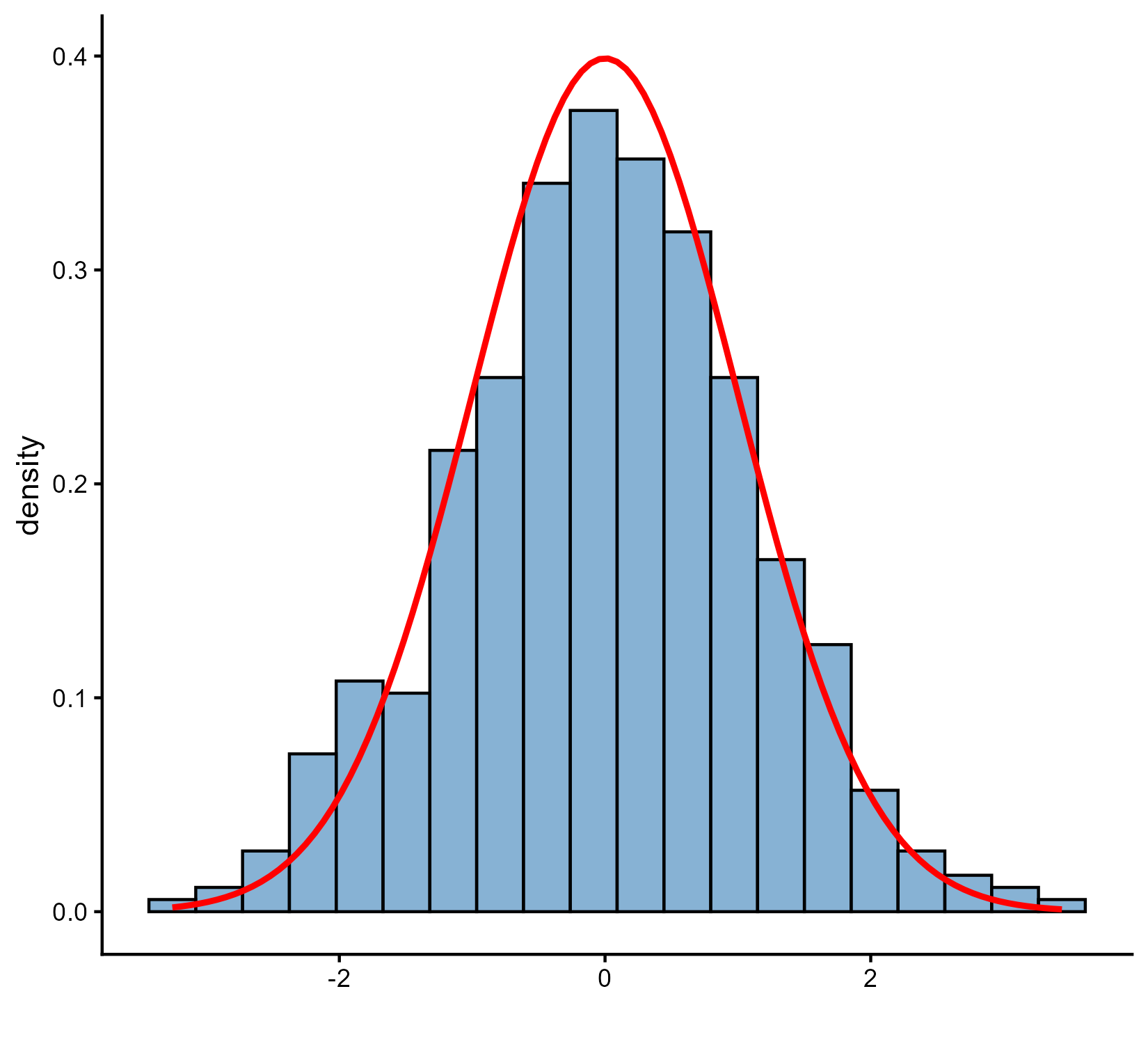}
        \caption{$Q\in \mathcal{M}_{K,p_0}^{\otm}$}
    \end{subfigure}

    \caption{Empirical distributions of $(\inp{\widehat{M}}{Q}-\inp{M}{Q})/(\widehat{\sigma}\fro{\mathcal{P}_{\widehat{M}}(Q)}\sqrt{1/T\nu})$ for four different choices of $Q$ under one-to-one matching  (top four plots),one-to-many matching(middle four plots) and two-sided random arrival observed data (bottom four plots). The red curve represents the p.d.f. of standard normal distributions.}
    \label{fig: pdf}
\end{figure}

Figure \ref{fig: pdf} presents the histograms of $(\inp{\widehat{M}}{Q}-\inp{M}{Q})/(\widehat{\sigma}\fro{\mathcal{P}_{\widehat{M}}(Q)}\sqrt{1/T\nu})$ for all four scenarios under the three observed matching patterns. As shown, the empirical distributions closely align with the standard normal distribution, represented by the red curve, demonstrating the accuracy of our method.

\subsection{Real data analysis}\label{sec:real data}
We now apply our matching learning and evaluation framework to a publicly available dataset on Major League Baseball (MLB). The dataset is available from \url{https://umpscorecards.com/data/games} and contains information on every game during an MLB regular season, including the participating teams, assigned umpires, and umpires’ performance. More specifically, each regular season involves a fixed set of teams and a series of games among them, with multiple games taking place at different ballparks on most game days. The set of umpires is also fixed within a season, and umpires are assigned to officiate these games. These assignments are centrally scheduled by MLB’s Umpiring Department rather than chosen by the teams or umpires.

Modern pitch-tracking equipment and systems can automatically measure both the difficulty of the pitches an umpire is required to judge and the proportion of ball and strike calls made correctly, making it possible to evaluate each umpire’s performance quantitatively. Accordingly, each game record includes an \emph{Accuracy Above Expected} metric that measures umpiring quality by adjusting the raw accuracy with the difficulty of the pitches being judged. Because different home teams have different technique characteristics and play in different ballparks, while umpires may vary in their familiarity with these characteristics and venues, appropriately matching home teams with umpires can improve the overall umpiring accuracy and quality \citep{deshpande2017hierarchical}.

\begin{figure}[htbp]
    \centering

    \begin{subfigure}[t]{0.49\textwidth}
        \centering
        \includegraphics[width=\linewidth]{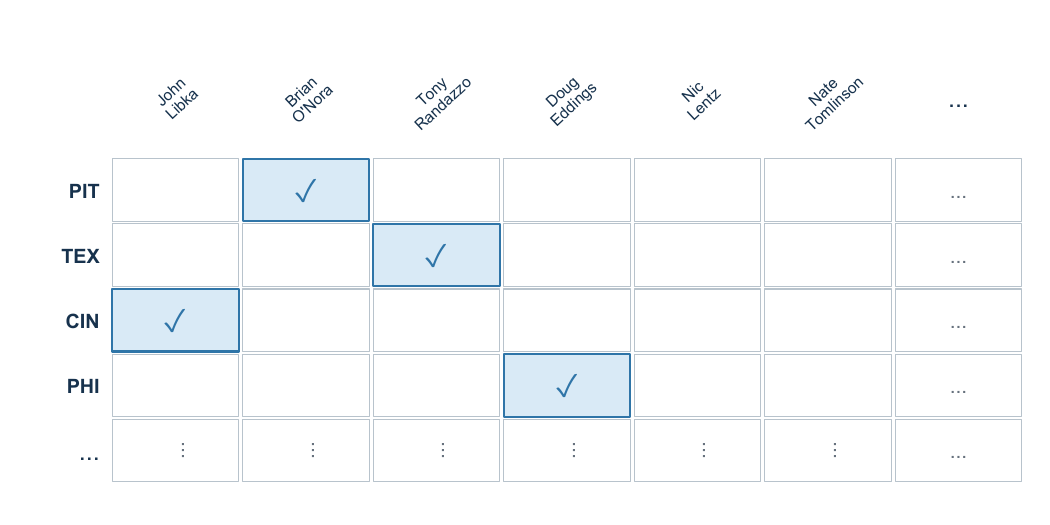}
        \caption{Observed matching.}
        \label{fig: MLB-observed-matching}
    \end{subfigure}
    \hfill
    \begin{subfigure}[t]{0.49\textwidth}
        \centering
        \includegraphics[width=\linewidth]{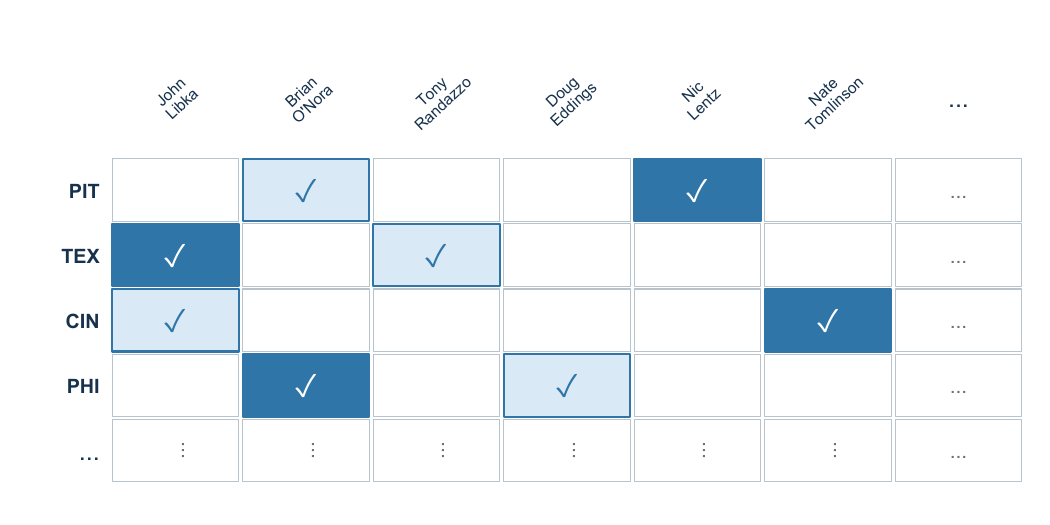}
        \caption{Observed and identified optimal matchings.}
        \label{fig: MLB-optimal-matching}
    \end{subfigure}

    \par\vspace{0.4em}
    \includegraphics[scale=0.8]{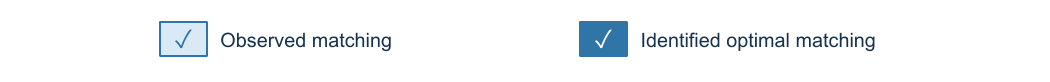}

    \caption{
        Partial illustration of the observed and identified optimal matchings on 2025-05-02. Only 4 of the 15 home teams are displayed.
    }
    \label{fig:MLB-matching-comparison}
\end{figure}

Because the set of active MLB umpires varies across seasons, we restrict our analysis to the 2025 regular season so that both sides of the matching market remain fixed throughout the study period. The dataset contains $d_1=30$ teams and $d_2=75$ umpires observed over $T=180$ game days. We treat each game day as one matching round $t$.  Each row of the reward matrix corresponds to a team in its role as the home team. Thus, although all 30 MLB teams may play either at home or away, we model only assignments involving team $i$ when it is the home team. On each game day, only a subset of the teams play at home, and only a subset of the umpires are assigned to these games. Each home team can appear in at most one game that day, while each umpire can officiate at most one game. Therefore, the team–umpire assignments on each game day naturally form the two-sided random-arrival matching considered in our framework. Figure \ref{fig: MLB-observed-matching} illustrates this pattern using the observed matching on 2025-05-02, where different rows represent different teams, the different columns represent different umpires, and the observed assignments form a matching between the two sides.

We use the recorded umpire’s \emph{Accuracy Above Expected} as the observed reward. The latent entry $M(i,j)$ represents the expected Accuracy Above Expected when umpire $j$ officiates a game hosted by team $i$. The parameters $p_1$, $p_2$, $c_r$, and $c_s$ in the two-sided random-arrival model in \eqref{eq:two-random-complete} are unknown in practice. We therefore estimate the average entrywise sampling probability as $\widehat{\nu}=(Td_1d_2)^{-1}\sum_t\sum_i X_t(i,j_t(i))$, where the numerator is the total number of observed team–umpire assignments across all $T$ matching rounds.

We examine the performance of our matching learning and evaluation procedures. By League regulation, at most 15 teams can play as the home team on any given day $t$. During the 2025 regular season, there are 10 game days with $B_r=15$ home teams. We select all 10 of these game days, each of which has a different set of arriving teams. For each of them, we first identify the optimal matching $\widehat Q_{\mathrm{optimal}}$ and then conduct matching evaluation by comparing its reward with that of the corresponding matching observed in the data, denoted by $Q_{\mathrm{observed}}$.

It is important to note that, if the same final projected estimator $\widehat M$ is used both to identify the optimal matching and to conduct inference on its value, the selected matching will be inherently dependent with $\widehat M$, leading to selection bias. To address this issue, we slightly modify the sample splitting procedure. We first use the $T/2$ observations in $\widetilde{\mathcal D}_1$ to construct the initial estimator $\widehat M^{\init}$ and identify the optimal matching $\widehat Q_{\mathrm{optimal}}$. We then use only the observations in $\widetilde{\mathcal D}_2$ to construct the debiased estimator $\widehat{M}^{\ubs}=\widehat{M}^{\init} + \frac{2}{T\widehat{\nu}}\sum_{t\in \widetilde{D}_2} (Y_t-X_t\circ \widehat{M}^{\init})$ and obtain the corresponding projected estimator $\widehat{M}$. Finally, we conduct the hypothesis tests 
\begin{align*}
    H_0:
\left\langle M,
\widehat Q_{\mathrm{optimal}}-Q_{\mathrm{observed}}
\right\rangle\leq 0
\quad\text{v.s.}\quad
H_1:
\left\langle M,
\widehat Q_{\mathrm{optimal}}-Q_{\mathrm{observed}}
\right\rangle>0.
\end{align*}
for matching evaluation. 

The corresponding $p$-values are reported in Table~\ref{tab: MLB test}. For 8 of the 10 game days with 15 arrived home teams, the estimated total Accuracy Above Expected of the identified optimal matching is significantly higher than that of the matching actually observed in the data, based on the one-sided tests at the 5\% significance level. This finding suggests that our identified optimal matching can significantly improve the overall umpiring accuracy and quality. Figure \ref{fig: MLB-optimal-matching} illustrates how the assignments under the identified optimal matching differ from those under the observed matching. Each team displayed in the figure is assigned to a different umpire under the optimal matching. For example, whereas team TEW was matched with umpire Tony Randazzo in the observed data, our algorithm matches it with John Libka. A previous report also identified John Libka as one of the most outstanding and accurate umpires in a past season \citep{williams2019mlb}.

Our findings are also consistent with evidence from historical data covering the 2015–2024 seasons [\citeauthor{umpscorecards_historical_data}]. For each of the 15 home teams that played on 2025-05-02, we calculate the average Accuracy Above Expected  for the team–umpire pairs under both the identified optimal matching and the observed matching, using all historical observations of the same pairs from 2015 to 2024. Among the 15 teams, 9 have a higher average historical AAE under our identified optimal matching. In particular, Table \ref{tab: MLB historical} reports the differences in historical mean AAE between the alternative assignments of the teams displayed in Figure \ref{fig: MLB-optimal-matching}. For example, our algorithm reassigns team PIT from Brian O’Nora to Nic Lentz, whose historical AAE for this team is 3.692 percentage points higher. Similarly, team TEX is reassigned from Tony Randazzo to John Libka, whose historical AAE for TEX is 3.384 percentage points higher than that of Tony Randazzo. These results further indicate that certain umpires consistently achieve greater accuracy for specific teams across seasons, potentially because of their greater knowledge of and familiarity with those teams' playing characteristics. Properly identifying and implementing such matchings can significantly improve the overall accuracy and quality of the games.

\section{Extension: non-uniform allocation}\label{sec:extension}
In the preceding sections, we focus on uniform sampling from the matching sets  $\mathcal{M}^{\oto}$, $\mathcal{M}_{K,p_0}^{\otm}$ and $\mathcal{M}_{p_1,p_2, c_r, c_s}^{\tside}$.  In practice, however, the data may be collected under a designed matching policy in which certain participants are more likely to be assigned specific tasks, or certain types of patients are more likely to be matched with specific types of physicians. In what follows, we show that our proposed matching learning and evaluation framework can be extended to accommodate non-uniform allocation.

For simplicity, we focus on non-uniform one-to-one matching, where each matching observation $X_t$ is still sampled from ${\calM}^{\oto}$, but different matchings may be sampled with different known probabilities $P(X_t=q), \forall q\in {\calM}^{\oto}$. We define the probability that the $(i,j)$-th entry of $M$ is observed as
\begin{align*}
    p_{ij}:=\mathbb P\bigl(X_t(i,j)=1\bigr)=
\sum_{q\in\mathcal M^{\oto}}
\indicator\bigl(q(i,j)=1\bigr)\mathbb P(X_t=q).
\end{align*}
Under one-to-one matching mechanism, we still have $\sum_j p_{ij}=1, \forall i\in [d_1]$ and $\sum_i p_{ij}\leq 1, \forall j\in [d_2]$. Assuming $p_{ij}>0$ for all $(i,j)$, we define the \emph{inverse probability matrix} $P_{-1}\in \RR_+^{d_1\times d_2}$ such that its $(i,j)$-th entry is $P_{-1}(i,j)=p_{ij}^{-1}$, and let $P_{-1}^{1/2}$ denote its entrywise square root, i.e., $P_{-1}^{1/2}(i,j)=p_{ij}^{-1/2}$. For ease of notation, we further define $p_{\min}:=\min_{i,j}p_{ij}, p_{\max}:=\max_{i,j}p_{ij},
$, and the joint sampling probability for pairs of distinct positions $\pi_{ij,pq}=\PP(X_t(i,j)=X_t(p,q)=1)$. 

For matching learning, based on the inverse probability matrix, we aim to minimize the weighted square loss
$
\mathscr{L}_{\calD}(U,G,V)=\sum_{t\in\calD} \big\|\big(Y_t - X_t\circ (UGV^{\top})\big)\circ P_{-1}^{1/2}\big\|_{\rm F}^2.
$
All steps in Algorithm \ref{alg:gd} remain the same except that the rotation calibrated gradient descent step should be modified to:
\begin{align*}
       \widehat{U}^{(p+0.5)}&=\widehat{U}^{(p)}\widehat{L}_G^{(p)} - \eta\frac{1}{N_0}\sum_{t\in \mathcal{D}_{2p+1}} P_{-1}\circ\left(X_t \circ (\widehat{U}^{(p)}\widehat{G}^{(p)}\widehat{V}^{(p)\top}) - Y_t\right)\widehat{V}^{(p)}\widehat{R}_G^{(p)}(\widehat{\Lambda}^{(p)})^{-1} \\
       \widehat{V}^{(p+0.5)}&=\widehat{V}^{(p)}\widehat{R}_G^{(p)} - \eta\frac{1}{N_0}\sum_{t\in \mathcal{D}_{2p+1}} P_{-1}\circ\left(X_t \circ (\widehat{U}^{(p)}\widehat{G}^{(p)}\widehat{V}^{(p)\top}) - Y_t\right)^{\top}\widehat{U}^{(p)}\widehat{L}_G^{(p)}(\widehat{\Lambda}^{(p)})^{-1}.
\end{align*}
Then we can derive the entrywise error rate for $\widehat{M}^{(m)}$ under non-uniform one-to-one matching.
\begin{Proposition}\label{thm:estimation-non}
    There exist absolute constants $C_1,\cdots, C_5>0$ such that if $T\geq C_1\alpha_d^2(d_1d_2^2)^{-1}(\sum_{i,j} p_{ij}^{-1})\log d_2$, $\lambda_{\min}/\sigma\geq C_2\sqrt{\alpha_d (\sum_{i,j} p_{ij}^{-1})/(d_1T)}\log d_2$, $p_{\min}^{-1}\leq C_3(d_1d_2)^{-1}(\sum_{i,j}p_{ij}^{-1})$ and $\|M\|_{\max}/\sigma\leq d_1^{C_4}$, and set $m=\lceil C_4\log d_2 \rceil$ and stepsize $\eta=0.75$, then with probability at least $1-5d_2^{-10}$,
    \begin{align*}
        \|\widehat{M}^{(m)} - M\|_{\max}\leq C_5\sigma\bigg(\frac{1}{d_1d_2}\sum_{i,j}\frac{1}{p_{ij}}\bigg)^{1/2}\sqrt{\frac{r^2}{Td_1}}\log d_2.
    \end{align*}
\end{Proposition}
The error rate depends on the average inverse sampling probability $1/(d_1d_2)\sum_{i,j}p_{ij}^{-1}$.  When $p_{ij}=1/d_2$ for every $(i,j)$, the setting reduces to uniform one-to-one matching, and the resulting error rate coincides with that in Proposition \ref{thm:estimation}. The condition $p_{\min}^{-1}\leq C_3(d_1d_2)^{-1}(\sum_{i,j}p_{ij}^{-1})$ imposes a lower bound on $p_{\min}$, preventing the problem from becoming ill-posed when some rows or columns are observed too infrequently. This condition is weaker than those commonly imposed in the non-uniform matrix completion literature, which typically require $p_{\max}\asymp p_{\min}$. By contrast, our framework allows $p_{\max}\gtrsim 1$.

For matching inference, the debiasing procedure requires a slight modification using inverse propensity weighting. Specifically, we propose the following debiased estimator
\begin{align*}
    \widehat{M}_1^{\ubs}=\widehat{M}_1^{\init} + T_0^{-1}\sum_{t\in\widetilde{D}_2}\left(Y_t - X_t \circ \widehat{M}_1^{\init}\right)\circ P_{-1},
\end{align*}
The estimator $\widehat M_2^{\mathrm{ubs}}$ is modified analogously. After applying the same projection step as in Section \ref{sec:inference}, we establish the asymptotic normality of $\langle\widehat M,Q\rangle$ in the following theorem:
\begin{Theorem}
    \label{thm:inference-main-non}
    Suppose that the conditions in Proposition \ref{thm:estimation-non} hold and there exists $\kappa_2=O(d_1^{-1})$ such that $|\pi_{ip,jq}-p_{ip}p_{jq}|\leq \kappa_2p_{ip}p_{jq}$ for any $i\neq j$ and $p\neq q$, then as long as
    \begin{align*}
        \max\left\{\frac{p_{\max}}{Tp_{\min}^2d_1}, \bigg(\frac{1}{d_1d_2}\sum_{i,j}\frac{1}{p_{ij}}\bigg)^{1/2}\sqrt{\frac{r^2}{Td_1}}\log d_2, \frac{\sigma}{\lambda_{\min}}\sqrt{\frac{r^2\log d_2}{Td_1^2}\left(\sum_{i=1}^{d_1}\sum_{j=1}^{d_2}\frac{1}{p_{ij}}\right)}\frac{\|Q\|_{\ell_1}}{\fro{\PMQ}}\right\}\rightarrow 0, 
    \end{align*}
    we have 
    \begin{align*}
        \frac{\inp{\widehat{M}}{Q}-\inp{M}{Q}}{\sigma\fro{P_{-1}^{1/2}\circ\mathcal{P}_M(Q)}\sqrt{1/T}} \overset{{\rm d.}}{\rightarrow} N(0,1).
    \end{align*}
\end{Theorem}
The pairwise regularity condition $|\pi_{ip,jq}-p_{ip}p_{jq}|\leq \kappa_2p_{ip}p_{jq}$ is equivalent to $|\text{Cov}(\indicator(X_t(i,p)=1), \indicator(X_t(j,q)=1)|\leq \kappa_2p_{ip}p_{jq}$, which uniformly controls the dependence within a matching.

Besides non-uniform matching, another interesting direction is to extend our framework to online decision making models, such as contextual bandits. In real world applications, rather than performing many i.i.d. matchings, matching platforms often make online and adaptive decisions. It is desirable to use previously observed matchings and rewards to update the reward matrix estimator online and improve future matching decisions. There is already a growing literature on online matrix completion and statistical inference under i.i.d. or adaptive sampling, and extending these methods to matching models is worth exploring. Another challenge arises from the number of candidate arms. If every valid matching in $\mathcal M^{\mathrm{oto}}$ is treated as a candidate arm, the number of arms becomes $A_{d_2}^{d_1}$, which may be far larger than the decision horizon $T$. Therefore, controlling regret under such a large action space and complex model remains an important direction for future research.

\section{Proof Sketch of Theorem \ref{thm:inference-main}} \label{sec: proof sketch}

In this section, we provide a proof sketch of Theorem \ref{thm:inference-main} to illustrate how a specific martingale argument is used to handle the within-matching dependence and establish asymptotic normality.

\emph{Step 1: Basic decomposition.}
Denote $\widehat{\Delta}_{l}=\widehat{M}_l^{\init}-M$. For $l=1,2$, the debiased estimator can be decomposed as $\widehat{M}_l^{\ubs}= M + \underbrace{\frac{d_2}{T_0}\sum_{t\in \widetilde{\mathcal{D}}_{l}}\sum_{i=1}^{d_1} \xi_t^i X_t^i}_{=:\widehat{Z}_1^{(l)}} + \underbrace{\frac{1}{T_0}\sum_{t\in \widetilde{\mathcal{D}}_{l}} \left(\sum_{i=1}^{d_1}d_2\inp{\widehat{\Delta}_{l}}{X_t^i}X_t^i-\widehat{\Delta}_{l}\right)}_{=:\widehat{Z}_2^{(l)}}$. 
The projected estimator is defined as $\widehat{M}=1/2(\widehat{M}_1+\widehat{M}_2)$, and $\inp{\widehat{M}}{Q}-\inp{M}{Q}$ can be further decomposed into $\inp{\widehat{M}}{Q}-\inp{M}{Q}= \underbrace{\frac{1}{2}\sum_{l=1}^{2}\inp{\widehat{Z}_1^{(l)}}{\PMQ}}_{L_T} + \underbrace{\frac{1}{2}\sum_{l=1}^{2}\inp{\widehat{Z}_2^{(l)}}{\PMQ}}_{B_T} + R_T$, where $L_T$ is the leading asymptotically normal term, and $B_T$ and $R_T$ are the higher order remainders.

\emph{Step 2: Asymptotic normality of $L_T$.} By definition, $L_T=\frac{d_2}{T}\sum_{t=1}^T\sum_{i=1}^{d_1} \xi_t^i\inp{X_t^i}{\PMQ}.$ We prove its asymptotic normality by Berry-Esseen Theorem \citep{berry1941accuracy}. Denote $W_t=\sum_i\xi_t^i\inp{X_t^i}{\PMQ}$ and $S_t=\sum_i\inp{X_t^i}{\PMQ}^2$. Then we need to derive the variance and third order moment bound for $W_t$. Conditional on $X_t$, $W_t$ is a weighted sum of independent centered sub-Gaussian variables.  Hence $\EE [W_t]=0$, $\EE[W_t^2\mid X_t]=\sigma^2S_t$, and $\EE[|W_t|^3\mid X_t]\lesssim\sigma^3S_t^{3/2}$.  Moreover, $\EE[S_t]=(d_2)^{-1}\fro{\PMQ}^2$, and expanding $S_t^2$ gives
\begin{align*}
    \EE[S_t^2]
    &=\frac{1}{d_2}\sum_{i,j}\PMQ(i,j)^4
      +\sum_{(i,j)\neq(k,l)}
      \PP\big(X_t(i,j)=X_t(k,l)=1)\PMQ(i,j)^2\PMQ(k,l)^2 \\
    &\leq \frac{1}{d_2}\|\PMQ\|_{\max}^2\fro{\PMQ}^2
      +\frac{1}{d_2(d_2-1)}\fro{\PMQ}^4. 
\end{align*}
Finally by $\EE [S_t^{3/2}]\leq (\EE[S_t])^{1/2}(\EE[S_t^2])^{1/2}$, this directly implies the upper bound of the third order moment for $W_t$. 


\emph{Step 3: Control of $B_T$.}
It suffices to bound $B_{T,1}=\inp{\widehat Z_2^{(1)}}{\PMQ}$, and the other part can be proved following the same arguments.  By definition, $B_{T,1}= 1/T_0\sum_{t\in\widetilde{\mathcal D}_2}
 \{d_2\sum_{i=1}^{d_1}\inp{\widehat\Delta_1}{X_t^i} \inp{X_t^i}{\PMQ}-\inp{\widehat\Delta_1}{\PMQ}\}.$ Conditional on $\widehat\Delta_1$, the summands are i.i.d. and centered. To apply Bernstein's inequality, we need to control both the variance and a uniform bound for each summand. The former follows from a relatively direct calculation, so in the following, we focus on establishing the uniform bound via a martingale trick.

For a fixed $t$, denote
$R:=\sum_{i=1}^{d_1}\inp{\widehat\Delta_1}{X_t^i}
\inp{X_t^i}{\PMQ}$, and we construct a martingale by revealing the matching
row by row.  The key intuition is that, although all $X_t^i$ are dependent,
their joint distribution is symmetric across rows and admits a sequential
sampling-without-replacement representation.  Specifically, $j_t(1)$ is
first sampled uniformly from $[d_2]$, thereby determining $X_t^1$.
Conditional on this assignment, $j_t(2)$ is sampled uniformly from the
remaining $d_2-1$ columns, thereby determining $X_t^2$.  More generally,
after $j_t(1),\ldots,j_t(i-1)$ have been determined, $j_t(i)$ is sampled
uniformly from $[d_2]/\{j_t(1),\ldots,j_t(i-1)\}$, and each
candidate column is selected with probability $1/(d_2-i+1)$.  Continuing
this procedure until the last row generates exactly a uniform one-to-one
matching and naturally induces the row-by-row filtration used below.

Therefore, we can define $\calF_i=\sigma(X_t^1,\ldots,X_t^i)$, and the Doob martingale
\begin{align*}
R_i
&=\EE[R|\calF_{i}]=\sum_{j=1}^{i}\inp{\widehat\Delta_1}{X_t^j}
\inp{X_t^j}{\PMQ}
+\sum_{j=i+1}^{d_1}\frac1{d_2-i}
\sum_{k\in[d_2]/\{j_t(1),\ldots,j_t(i)\}}
\widehat\Delta_1(j,k)\PMQ(j,k).
\end{align*}
Averaging this expression over the uniform draw
$j_t(i)\mid\calF_{i-1}$ gives
$\EE[R_i\mid\calF_{i-1}]=R_{i-1}$. Therefore,
\begin{align*}
R_i-R_{i-1}
&=\underbrace{\inp{\widehat\Delta_1}{X_t^i}\inp{X_t^i}{\PMQ}
-\EE\!\left[\inp{\widehat\Delta_1}{X_t^i}\inp{X_t^i}{\PMQ}
\mid\calF_{i-1}\right]}_{=:A_i}\\
&\quad+\underbrace{\sum_{j=i+1}^{d_1}\left\{
\EE\!\left[\inp{\widehat\Delta_1}{X_t^j}\inp{X_t^j}{\PMQ}
\mid\calF_i\right]
-\EE\!\left[\inp{\widehat\Delta_1}{X_t^j}\inp{X_t^j}{\PMQ}
\mid\calF_{i-1}\right]\right\}}_{=:B_i}.
\end{align*}
The uniformity of $j_t(i)$ gives
\begin{align*}
\EE[A_i^2\mid\calF_{i-1}]
&\leq\frac1{d_2-i+1}
\sum_{k\in[d_2]/ \{j_t(1),\ldots,j_t(i-1)\}}
\widehat\Delta_1(i,k)^2\PMQ(i,k)^2 \lesssim \frac{1}{d_1}\|\widehat{\Delta}_1\|_{\max}^2\|\PMQ(i,)\|^2, 
\end{align*}
where the second inequality follows from the assumption $d_2\geq(1+\gamma)d_1$, which ensures that $1/(d_2-i+1)\lesssim 1/d_1$ uniformly over $i\in[d_1]$.

Similarly, Cauchy Schwarz inequality gives
  \begin{align*}
B_i^2
&\leq 2(d_1-i)\sum_{j=i+1}^{d_1}
\bigg(
\frac{1}
{(d_2-i+1)^2}\widehat{\Delta}_1(j,j_t(i))^2 \PMQ(j,j_t(i))^2 \\
&\quad +\frac{1}{(d_2-i)(d_2-i+1)^2}
\sum_{\substack{k\in[d_2] / \\
\{j_t(1),\ldots,j_t(i)\}}}
\widehat{\Delta}_1(j,k)^2
\PMQ(j,k)^2
\bigg). 
\end{align*}
\begin{align*}
\EE[B_i^2|\calF_{i-1}]&\lesssim \frac{1}{d_1}\|\widehat{\Delta}_1\|_{\max}^2\sum_{j=i+1}^{d_1}\frac{1}{d_2-i+1}\sum_{\substack{k\in [d_2] / \\ \{j_t(1),\cdots,j_t(i-1)\}}}{\PMQ(j,k)}^2 + \frac{1}{d_1^2}\|\widehat{\Delta}_1\|_{\max}^2\fro{\PMQ}^2 \\ &\lesssim \frac{1}{d_1^2}\|\widehat{\Delta}_1\|_{\max}^2\fro{\PMQ}^2.
\end{align*}
    Consequently, we can derive $\sum_{i=1}^{d_1}\operatorname{Var}(R_i\mid\calF_{i-1})$ and $\max_i|R_i-R_{i-1}|$. Apply martingale Bernstein inequality to 
$R-\EE[R]=\sum_i(R_i-R_{i-1})$ gives the high probability bound for $|R|$.  

Finally, $R_T$ can be controlled through the classic pectral decomposition and perturbation expansion. Combine all the terms proves the Theorem. 

\begin{table}[htbp]
\centering
\begin{tabular}{c|cc}
\toprule
Date & $\left\langle \widehat{M},
\widehat Q_{\mathrm{optimal}}-Q_{\mathrm{observed}}
\right\rangle$ & $p$-value \\
\midrule
2025-03-29 & 21.085 & {0.0001} \\
2025-05-02 & 22.422 & {0.0002} \\
2025-05-14 & 21.652 & {0.0002} \\
2025-05-27 & 20.820 & {0.0008} \\
2025-06-21 & 17.702 & {0.0001} \\
2025-07-18 & 10.517 & {0.0490} \\
2025-08-16 &  3.740 & 0.2247 \\
2025-08-19 &  0.203 & 0.4849 \\
2025-09-20 & 12.417 & {0.0449} \\
2025-09-23 & 11.380 & {0.0175} \\
\bottomrule
\end{tabular}
\caption{One-sided tests of the reward differences between the identified optimal and observed matchings for the ten dates with 15 arriving teams.}
\label{tab: MLB test}
\end{table}

\begin{table}[htbp]
\centering
\begin{tabular}{lrrr}
\toprule
& \multicolumn{2}{c}{Historical mean AAE} & \\
\cmidrule(lr){2-3}
Team & Identified optimal pair & Observed pair & Difference \\
\midrule
CIN &  2.238 &  1.569 &  0.668 \\
PHI & -0.408 & -0.771 &  0.363 \\
PIT &  2.149 & -1.543 &  3.692 \\
TEX &  1.205 & -2.178 &  3.384 \\
\bottomrule
\end{tabular}
\caption{Historical mean AAE (2015–2024) for team–umpire pairs under the identified optimal and observed matchings for four home teams on 2025-05-02.}
\label{tab: MLB historical}
\end{table}

\section*{Data Availability}
The code and data for this article is available at \href{https://github.com/yuan0629/Matching-Inference}{https://github.com/yuan0629/Matching-Inference}.

\section*{Acknowledgements}
ChatGPT was used to assist with writing code and processing data for the numerical analyses. All AI-assisted outputs were reviewed and verified by the authors.

\bibliography{main}
\bibliographystyle{plainnat}

\newpage
\appendix

\section{Finite-Sample Performance With and Without Sample Splitting} \label{sec: sample splitting}

We conduct additional numerical experiments to compare the finite-sample variability and asymptotic normality of the estimators with and without sample splitting. The results show that, although sample splitting does not affect first-order asymptotic efficiency, it can lead to greater finite-sample variance when the sample size is small. By contrast, the estimators without sample splitting also exhibit good Gaussian approximation, suggesting that sample splitting may be unnecessary.
    
    Specifically, we consider one-to-one matching observations generated under the same mechanism as in Section \ref{sec:simulation}, with dimensions $d_1=100$ and $d_2=750$ and noise level $\sigma=1$. We consider three sample sizes, $T\in\{1000,2000,5000\}$, and conduct 500 replications for each setting. We consider four choices of $Q$: 1) the single entry, $Q=e_1e_1^{\top}$; 2) the single one-to-one matching, where $Q\in \mathcal{M}^{\oto}$; 3) the difference between two one-to-one matching, $Q=Q_1-Q_2$, where $Q_1$ and $Q_2$ are randomly sampled from $\mathcal{M}^{\oto}$; 4) the single one-to-many matching, where $Q\in \mathcal{M}_{K,p_0}^{\otm}$. For each choice of $Q$, we construct estimators of $\langle M,Q\rangle$ with and without sample splitting.

    To compare their finite-sample variability, we report the empirical variances of the two estimators across the 500 replications in Table \ref{tab: empirical variance}. The variance ratio is defined as the empirical variance of the estimator with sample splitting divided by that of the estimator without sample splitting. For every choice of $Q$ and every sample size $T$, the variance ratio is greater than one, indicating that the estimator without sample splitting has a smaller empirical variance. The variance ratios are largest for the smallest sample size, $T=1000$, and decrease toward one as $T$ increases. This finding is consistent with the theoretical result that, although sample splitting does not change the leading asymptotic variance, it can enlarge the higher-order remainder terms through the greater errors of the initial estimators. These terms are asymptotically negligible but may be non-negligible when $T$ is small.

\begin{figure}[htbp]
  \centering
  \begin{subfigure}{0.48\textwidth}
    \centering
    \includegraphics[width=\linewidth]{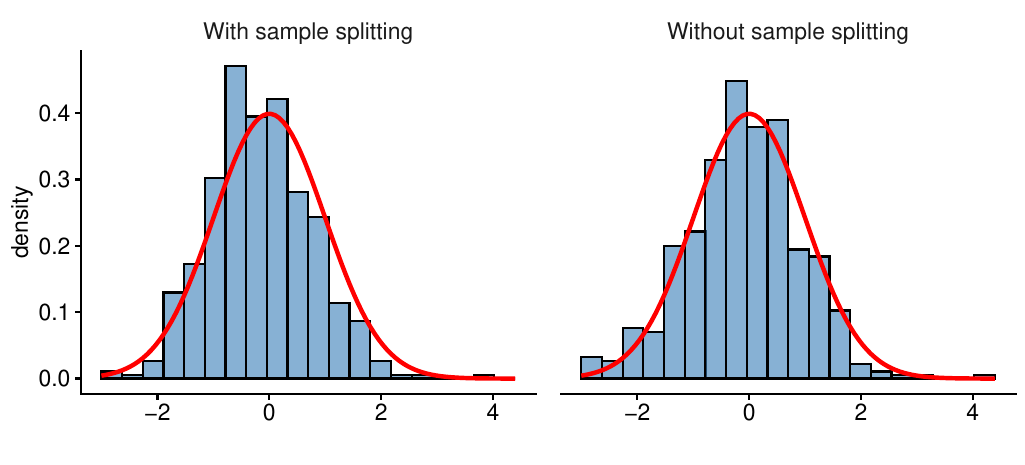}
    \caption{$Q=e_1e_1^{\top}$}
    \label{fig:T1000-Q1}
  \end{subfigure}\hfill
  \begin{subfigure}{0.48\textwidth}
    \centering
    \includegraphics[width=\linewidth]{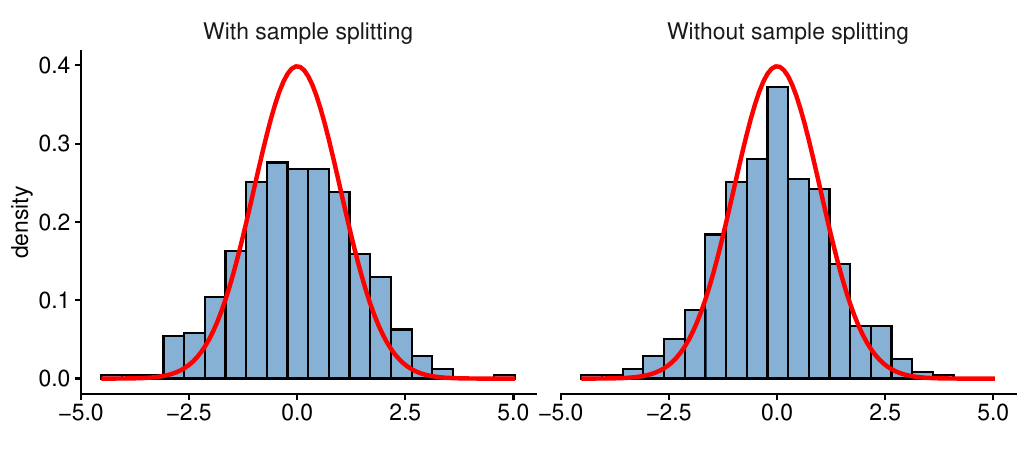}
    \caption{$Q\in\mathcal{M}^{\mathrm{oto}}$}
    \label{fig:T1000-Q2}
  \end{subfigure}

  \medskip
  \begin{subfigure}{0.48\textwidth}
    \centering
    \includegraphics[width=\linewidth]{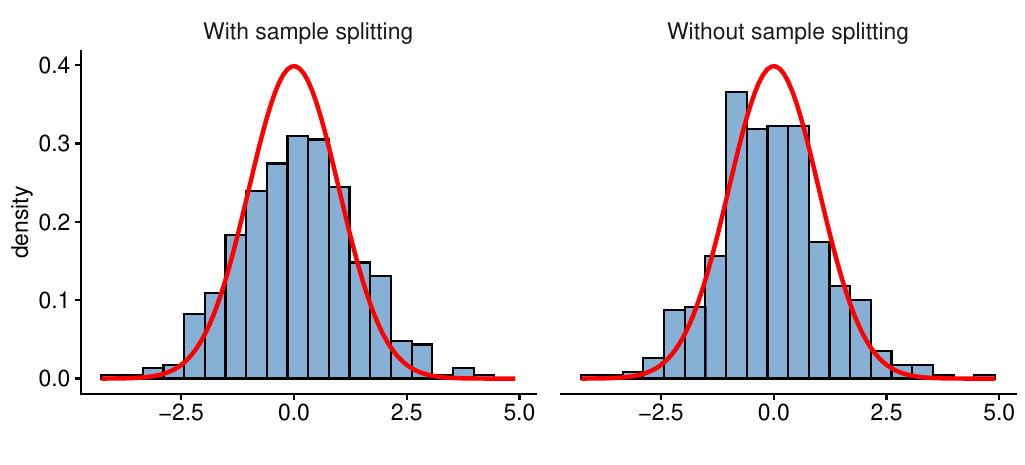}
    \caption{$Q=Q_1-Q_2$}
    \label{fig:T1000-Q3}
  \end{subfigure}\hfill
  \begin{subfigure}{0.48\textwidth}
    \centering
    \includegraphics[width=\linewidth]{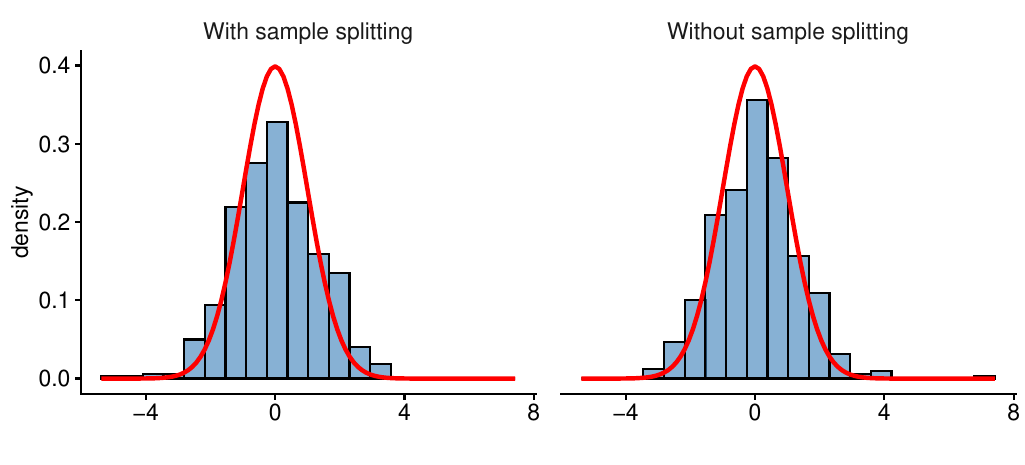}
    \caption{$Q\in\mathcal{M}_{K,p_0}^{\mathrm{otm}}$}
    \label{fig:T1000-Q4}
  \end{subfigure}
  \caption{Empirical distributions of the standardized estimators with and without sample splitting under the one-to-one matching for $T=1000$, based on 500 Monte Carlo replications. The red curves represent the standard normal density. }
  \label{fig:T1000}
\end{figure}

\begin{figure}[htbp]
  \centering
  \begin{subfigure}{0.48\textwidth}
    \centering
    \includegraphics[width=\linewidth]{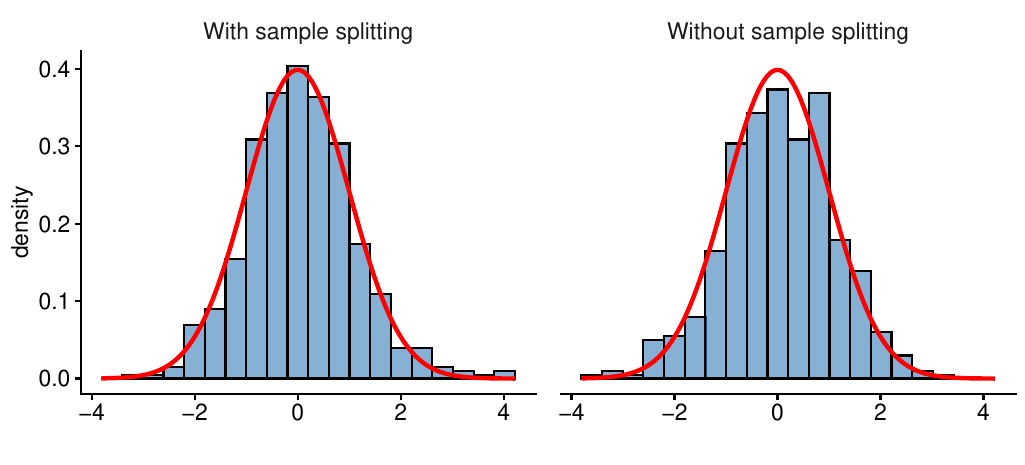}
    \caption{$Q=e_1e_1^{\top}$}
    \label{fig:T2000-Q1}
  \end{subfigure}\hfill
  \begin{subfigure}{0.48\textwidth}
    \centering
    \includegraphics[width=\linewidth]{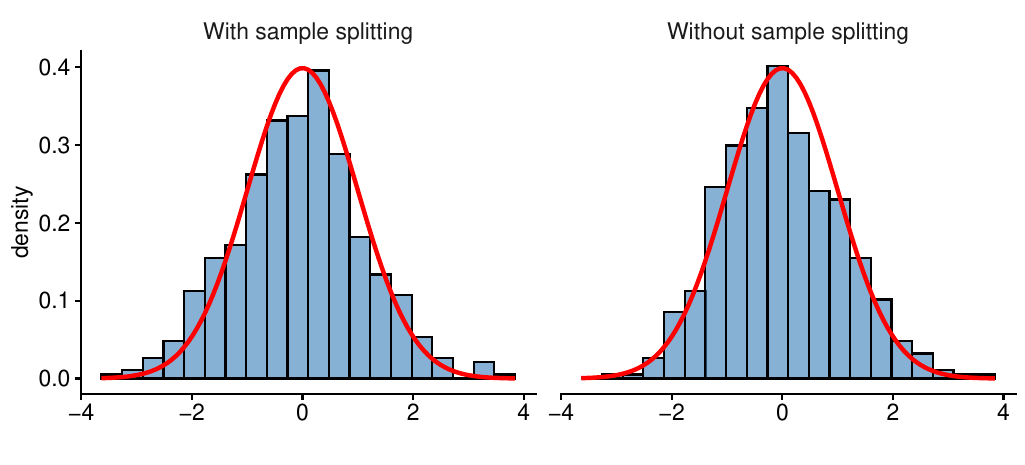}
    \caption{$Q\in\mathcal{M}^{\mathrm{oto}}$}
    \label{fig:T2000-Q2}
  \end{subfigure}

  \medskip
  \begin{subfigure}{0.48\textwidth}
    \centering
    \includegraphics[width=\linewidth]{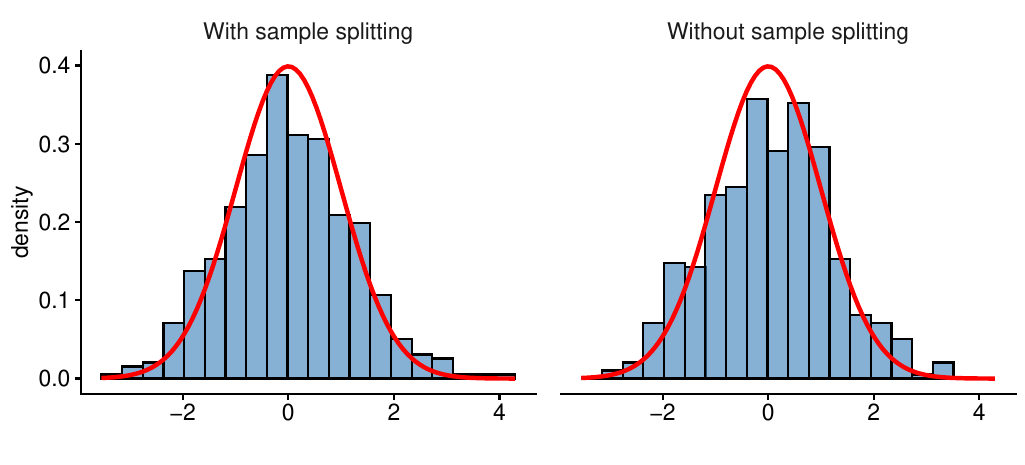}
    \caption{$Q=Q_1-Q_2$}
    \label{fig:T2000-Q3}
  \end{subfigure}\hfill
  \begin{subfigure}{0.48\textwidth}
    \centering
    \includegraphics[width=\linewidth]{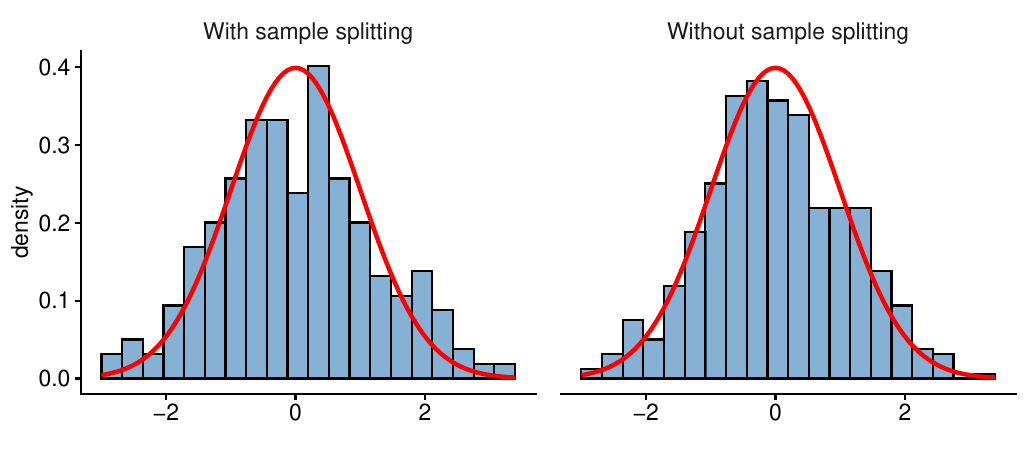}
    \caption{$Q\in\mathcal{M}_{K,p_0}^{\mathrm{otm}}$}
    \label{fig:T2000-Q4}
  \end{subfigure}
  \caption{Empirical distributions of the standardized estimators with and without sample splitting under the one-to-one matching for $T=2000$, based on 500 Monte Carlo replications. The red curves represent the standard normal density.}
  \label{fig:T2000}
\end{figure}

\begin{figure}[htbp]
  \centering
  \begin{subfigure}{0.48\textwidth}
    \centering
    \includegraphics[width=\linewidth]{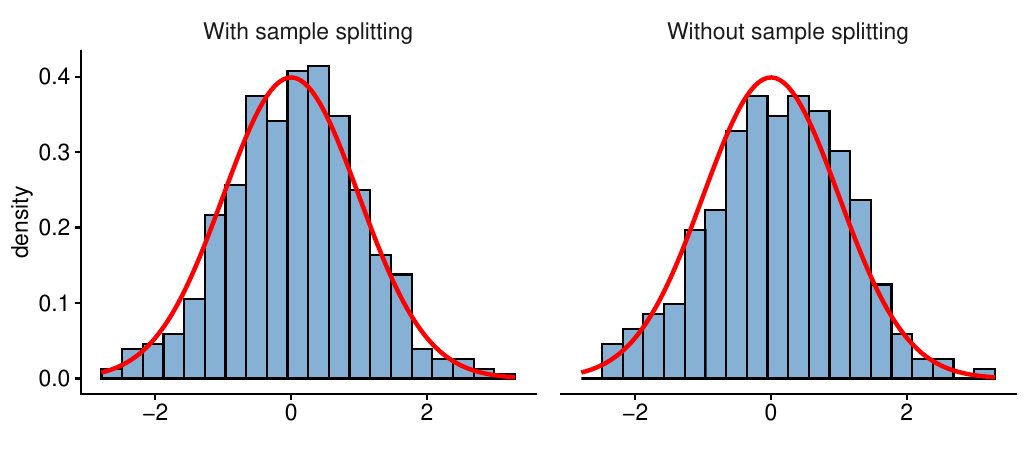}
    \caption{$Q=e_1e_1^{\top}$}
    \label{fig:T5000-Q1}
  \end{subfigure}\hfill
  \begin{subfigure}{0.48\textwidth}
    \centering
    \includegraphics[width=\linewidth]{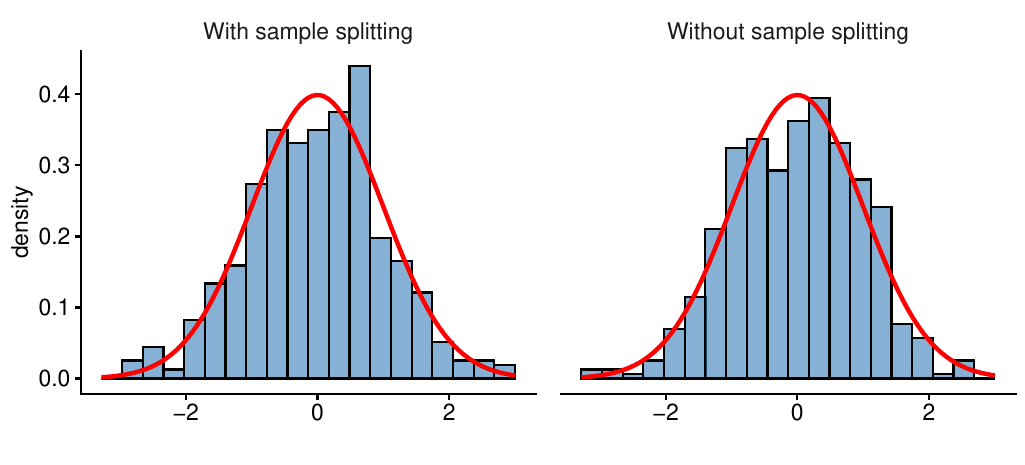}
    \caption{$Q\in\mathcal{M}^{\mathrm{oto}}$}
    \label{fig:T5000-Q2}
  \end{subfigure}

  \medskip
  \begin{subfigure}{0.48\textwidth}
    \centering
    \includegraphics[width=\linewidth]{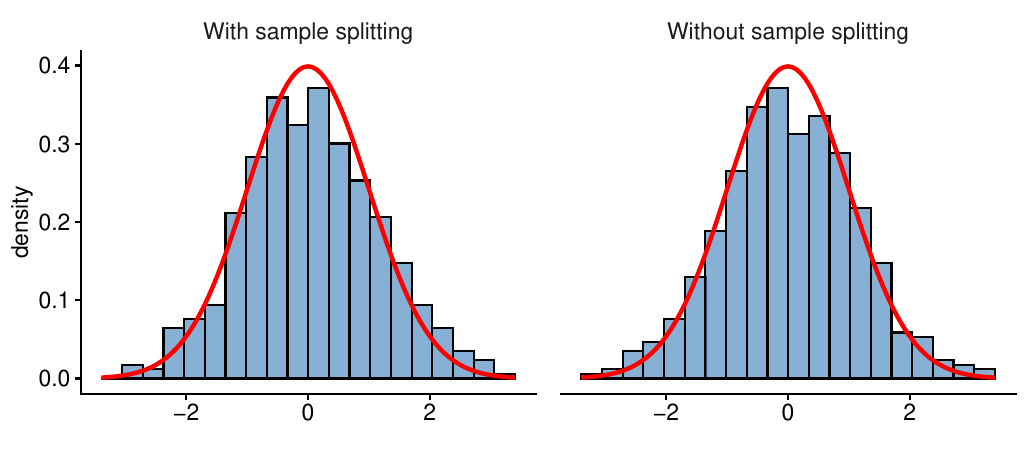}
    \caption{$Q=Q_1-Q_2$}
    \label{fig:T5000-Q3}
  \end{subfigure}\hfill
  \begin{subfigure}{0.48\textwidth}
    \centering
    \includegraphics[width=\linewidth]{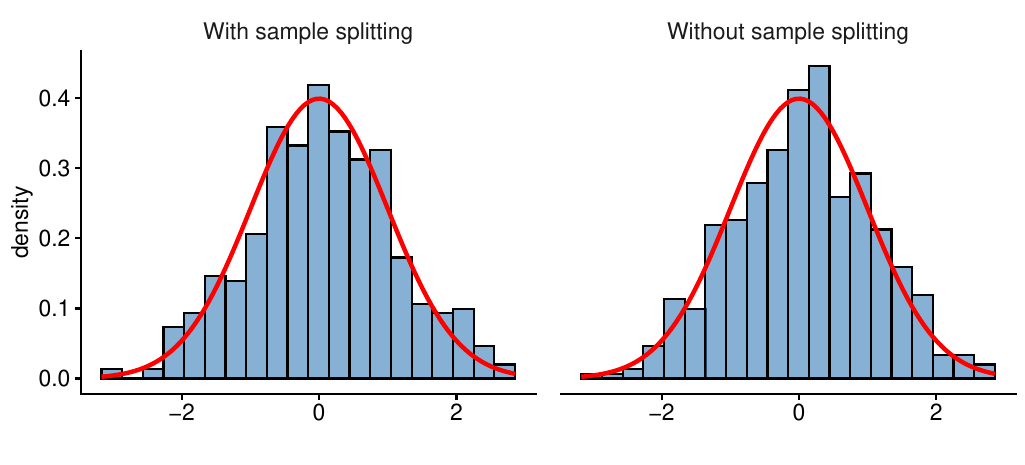}
    \caption{$Q\in\mathcal{M}_{K,p_0}^{\mathrm{otm}}$}
    \label{fig:T5000-Q4}
  \end{subfigure}
  \caption{Empirical distributions of the standardized estimators with and without sample splitting under the one-to-one matching for $T=5000$, based on 500 Monte Carlo replications. The red curves represent the standard normal density.}
  \label{fig:T5000}
\end{figure}

    We also compare the empirical densities of the standardized estimators, as shown in Figures \ref{fig:T1000}-\ref{fig:T5000}. Across all choices of $Q$, the empirical distributions of both estimators are generally close to the standard normal distribution, with improved performance as $T$ increases. For the smallest sample size, $T=1000$, some deviations from the standard normal distribution remain, particularly for the estimator with sample splitting, due to the imprecise initial estimators. 

    Overall, the simulations suggest that removing sample splitting can improve finite-sample efficiency while preserving a comparable Gaussian approximation in the settings considered here. Establishing the asymptotic normality of the estimator without sample splitting under the dependent matching setting remains an interesting question for future research.

        \begin{table}[htbp]
  \centering
  \begin{threeparttable}
    \begin{tabular}{ccccc}
      \toprule
      $T$ & $Q$ & Var (splitting) & Var (no splitting) & Ratio \\
      \midrule
      1000 & $e_1e_1^{\top}$                         & 0.096137  & 0.043783  & 2.1958 \\
      1000 & $\mathcal{M}^{\mathrm{oto}}$          & 17.456    & 5.4780    & 3.1865 \\
      1000 & $Q_1-Q_2$                               & 30.762    & 9.8991    & 3.1075 \\
      1000 & $\mathcal{M}_{K,p_0}^{\mathrm{otm}}$ & 65.678    & 22.212    & 2.9569 \\
      \midrule
      2000 & $e_1e_1^{\top}$                         & 0.024189  & 0.015197  & 1.5917 \\
      2000 & $\mathcal{M}^{\mathrm{oto}}$          & 2.3337    & 1.2543    & 1.8605 \\
      2000 & $Q_1-Q_2$                               & 4.7250    & 2.8315    & 1.6687 \\
      2000 & $\mathcal{M}_{K,p_0}^{\mathrm{otm}}$ & 10.042    & 5.0133    & 2.0030 \\
      \midrule
      5000 & $e_1e_1^{\top}$                         & 0.0050535 & 0.0047681 & 1.0599 \\
      5000 & $\mathcal{M}^{\mathrm{oto}}$          & 0.43763   & 0.35714   & 1.2254 \\
      5000 & $Q_1-Q_2$                               & 0.98684   & 0.86407   & 1.1421 \\
      5000 & $\mathcal{M}_{K,p_0}^{\mathrm{otm}}$ & 1.8272    & 1.5387    & 1.1875 \\
      \bottomrule
    \end{tabular}
  \end{threeparttable}
   \caption{Empirical variances of the estimators with and without sample splitting.}
   \label{tab: empirical variance}
\end{table}

\section{Calculating the Optimal Matching} \label{sec: optimal matching}
      In this section, we first establish a theoretical guarantee for identifying the optimal matching based on our max-norm convergence result under an additional optimality-gap condition. We then present a practical algorithm based on the classical Hungarian algorithm \citep{kuhn1955hungarian} for computing the optimal matching.

    \begin{itemize}
\item [(1)] \emph{Theoretical guarantees.}
We use one-to-one matching as an illustrative example. Let $V^*:= \max_{Q\in\mathcal M^{\oto}} \langle M,Q\rangle$ denote the reward of the optimal matching, and define
\begin{align*}
\mathcal Q^* = \{Q\in\mathcal M^{\oto}: \langle M,Q\rangle = V^*\},
\quad
\widehat{\mathcal Q}^*
=
\argmax_{Q\in\mathcal M^{\oto}}
\langle \widehat M^{\init},Q\rangle,
\end{align*}
where $\mathcal Q^*$ denotes the set of oracle optimal matchings. This means that we allow multiple optimal matchings.

For two one-to-one matchings $Q_1$ and $Q_2$, define
\begin{align*}
k(Q_1,Q_2)
:=
\sum_{i=1}^{d_1}\sum_{j=1}^{d_2}
\left|Q_1(i,j)-Q_2(i,j)\right|,
\end{align*}
to be the number of different assignments in $Q_1$ and $Q_2$. For any matching $Q$, further define its distance from the oracle-optimal set as
\begin{align*}
k(Q,\mathcal Q^*)
:=
\min_{Q^*\in\mathcal Q^*}
k(Q,Q^*).
\end{align*}
Then we define the \emph{normalized optimality gap} as
\begin{align*}
\delta_{\rm match}
:=
\min_{Q\in\mathcal M^{\oto},\,Q\notin\mathcal Q^*}
\frac{
V^*-\langle M,Q\rangle
}{
k(Q,\mathcal Q^*)
}.
\end{align*}

For any $Q\notin\mathcal Q^*$, let
\begin{align*}
Q^\circ
\in
\argmin_{Q^*\in\mathcal Q^*}
k(Q,Q^*).
\end{align*}
Then
\begin{align*}
\left\langle
\widehat M^{\init},
Q^\circ-Q
\right\rangle
&=
\left\langle M,Q^\circ-Q\right\rangle
+
\left\langle
\widehat M^{\init}-M,
Q^\circ-Q
\right\rangle \\
&\geq
V^*-\langle M,Q\rangle
-
k(Q,Q^\circ)
\left\|\widehat M^{\init}-M\right\|_{\max} \\
&\geq
k(Q,\mathcal Q^*)
\left\{
\delta_{\rm match}
-
\left\|\widehat M^{\init}-M\right\|_{\max}
\right\}.
\end{align*}
Consequently, under the \emph{optimality gap condition} that 
\begin{align*}
\delta_{\rm match}
>
\left\|\widehat M^{\init}-M\right\|_{\max},
\end{align*}
where $\left\|\widehat M^{\init}-M\right\|_{\max}$ is provided in Proposition \ref{thm:estimation}, we have
\begin{align*}
    \left\langle
\widehat M^{\init},
Q^\circ-Q
\right\rangle
>0
\qquad
\text{for every }Q\notin\mathcal Q^*.
\end{align*}
Thus, no matching outside $\mathcal Q^*$ can maximize the estimated reward, which directly implies
\begin{align*}
    \widehat Q^*=Q^\circ \in\mathcal Q^*.
\end{align*}
Equivalently, every estimated optimal matching is oracle optimal. Therefore, our max-norm convergence result guarantees the identification of an oracle optimal matching under the additional optimality gap condition.

\item [(2)] \emph{Practical algorithm.}
Next, we provide a practical algorithm for finding an optimal matching based on the estimated reward matrix $\widehat M^{\init}$. Before solving 
\begin{align*}
    \widehat Q^* =
\argmax_{Q\in\mathcal M^{\oto}}
\left\langle \widehat M^{\init},Q\right\rangle,
\end{align*}
we first reduce the number of candidate matching pairs, which is particularly useful when $d_2$ is large. For each participant $i$, let $T_i$ be a set of the $d_1$ tasks with the largest estimated rewards $\widehat M^{\init}(i,j)$, i.e., $T_i := \arg\max_{\substack{S\subseteq[d_2], |S|=d_1}} \sum_{j\in S}\widehat M^{\init}(i,j).$ We claim that there exists an optimal matching under $\widehat M^{\init}$ that uses only pairs $(i,j)$ satisfying $j\in T_i$.

To see this, suppose that an optimal matching assigns participant $i$ to task $j\notin T_i$. Because $T_i$ contains $d_1$ tasks, whereas the other $d_1-1$ participants occupy at most $d_1-1$ tasks, at least one task $j'\in T_i$ must be unmatched. By the definition of $T_i$,
\begin{align*}
    \widehat M^{\init}(i,j') \geq \widehat M^{\init}(i,j).
\end{align*}
Therefore, replacing the assignment $(i,j)$ by $(i,j')$ preserves the one-to-one matching constraint without decreasing the total estimated reward. The resulting matching remains optimal and no longer uses the pair $(i,j)$. Repeating this argument for every pair outside the corresponding candidate set shows that there exists an optimal matching supported entirely on
\begin{align*}
    \left\{(i,j):j\in T_i,\ i\in[d_1]\right\}.
\end{align*}
Hence, it is sufficient to search over at most $d_1^2$ candidate edges rather than the original $d_1d_2$ edges. Each set $T_i$ can be identified in $O(d_2)$ time using a linear-time selection algorithm, so all candidate sets can be constructed in $O(d_1d_2)$ time.

Finally, we solve the maximum-weight bipartite matching problem on the resulting graph, which contains $d_1$ user vertices and at most $d_1^2$ candidate edges. An exact solution can be obtained using a standard maximum-weight bipartite matching algorithm, such as the Hungarian algorithm \citep{kuhn1955hungarian}.
\end{itemize}

\section{Proofs of Main Results}\label{sec: proof-inference}

We first prove a key Lemma for deriving the Berry-Esseen bound that will be used throughout the proof of Theorems \ref{thm:inference-main}-\ref{thm:inference-main-3}.  

\begin{Lemma}
    \label{lemma: pairwise}
    Suppose there exist some $\nu>0$ and $\kappa_1<\infty$ that, for any distinct pairs $(i,j), (k,l) \in [d_1]\times [d_2]$,
    \begin{align*}
        \PP\big(X_t(i,j)=1\big)=\nu,
        \qquad
        \PP\big(X_t(i,j)=X_t(k,l)=1\big)\leq \kappa_1\nu^2
    \end{align*}
    for every two distinct positions $(i,j)\neq(k,l)$. Denote $\Omega_t=\{(i,j): X_t(i,j)=1\}$ and 
    \begin{align*}
        \rho_Q:=\frac{d_1\|\PMQ\|_{\max}^2}{\fro{\PMQ}^2},
            \qquad
        W_t:=\sum_{(i,j)\in\Omega_t}\xi_t(i,j)\PMQ(i,j).
    \end{align*}
    Then
    \begin{align*}
        \EE W_t=0,\qquad
        \operatorname{Var}(W_t)=\sigma^2\nu\fro{\PMQ}^2,
        \qquad
    \EE|W_t|^3
    \lesssim \sigma^3\nu^{3/2}\fro{\PMQ}^3
    \left(\frac{\rho_Q}{d_1\nu}+\kappa_1\right)^{1/2}.
    \end{align*}
Consequently, for i.i.d. matching matrices $X_1,\ldots,X_T$ sampled from (\refeq{eq:one-to-one}), (\refeq{eq:one-to-many}), (\refeq{eq:two-random-complete}),
\begin{align*}
    \sup_{x\in\RR}\left|
    \PP\left(
    \frac{(T\nu)^{-1}\sum_{t=1}^T W_t}
    {\sigma\fro{\PMQ}/\sqrt{T\nu}}\leq x\right)-\Phi(x)
    \right|
    \lesssim
    \sqrt{\frac{\rho_Q/(d_1\nu)+\kappa_1}{T}}.
\end{align*}     
\end{Lemma}

\begin{proof}
Since all $\xi_t(i,j)$ are i.i.d. zero mean subGaussian random variables with variance $\sigma^2$, it is obvious that 
\begin{align*}
    \EE\left[W_t\right]=0, \EE\left[W_t^2\right]=\nu\sigma^2\fro{\PMQ}^2
\end{align*}
for some constant $C_1>0$. Notice that condition on $X_t$, $W_t$ is a weighted sum of independent centered subGaussian random variables. Denote
\begin{align*}
    S_t:=\sum_{(i,j)\in\Omega_t}\PMQ(i,j)^2,
\end{align*}
then $\EE[|W_t|^3\mid X_t]=\sigma^3S_t^{3/2}$. Moreover, $\EE[S_t]=\nu\fro{\PMQ}^2$, and expanding $S_t^2$ gives
\begin{align*}
    \EE[S_t^2]
    &=\nu\sum_{i,j}\PMQ(i,j)^4
      +\sum_{(i,j)\neq(k,l)}
      \PP\big(X_t(i,j)=X_t(k,l)=1\big)\PMQ(i,j)^2\PMQ(k,l)^2 \\
    &\leq \nu\|\PMQ\|_{\max}^2\fro{\PMQ}^2
      +\kappa_1\nu^2\fro{\PMQ}^4 \\
    &=\nu^2\fro{\PMQ}^4
      \left(\frac{\rho_Q}{d_1\nu}+\kappa_1\right).
\end{align*}
Then by Cauchy Schwarz inequality,
\begin{align*}
    \EE[S_t^{3/2}]
    \leq (\EE[S_t])^{1/2}(\EE[S_t^2])^{1/2}
    \leq \nu^{3/2}\fro{\PMQ}^3
    \left(\frac{\rho_Q}{d_1\nu}+\kappa_1\right)^{1/2},
\end{align*}
which directly leads to 
\begin{align*}
    \EE\left[|W_t|^3\right]=\EE\left[\left[W_t| X_t\right]\right]=\sigma^3\EE[S_t^{3/2}]\leq \nu^{3/2}\fro{\PMQ}^3 \left(\frac{\rho_Q}{d_1\nu}+\kappa_1\right)^{1/2}.
\end{align*}
Finally, applying the classical Berry-Esseen theorem \citep{berry1941accuracy,esseen1956moment} to $W_1,\ldots,W_T$ proves the claimed bound.
\end{proof}

Moreover, for any $Q\in \RR^{d_1\times d_2}$, we have $\rho_Q\lesssim\mu^2r$. By the definition 
$\PMQ=UU^{\top}QV_{\perp}V_{\perp}^{\top} + U_{\perp}U_{\perp}^{\top}QVV^{\top} + UU^{\top}QVV^{\top}=UU^{\top}Q + U_{\perp}U_{\perp}^{\top}QVV^{\top}$, for any $i\in [d_1]$ and $j\in [d_2]$, 
\begin{align*}
    |\PMQ(i,j)|&\leq \left|e_i^\top UU^\top Qe_j\right| + \left|e_i^\top U_{\perp}U_{\perp}^{\top}QVV^\top e_j\right| \\
    &\leq  \|U^\top e_i\|_2\,\|U^\top Qe_j\|_2 + \|e_i^\top U_{\perp}U_{\perp}^{\top}QV\|_2\,\|V^\top e_j\|_2 \\
    &\leq \|U\|_{2,\max}\,\|U^\top Q\|_{\mathrm F} + \|V\|_{2,\max}\|U_{\perp}U_{\perp}^{\top} QV\|_{\mathrm F} \\
    &= \|U\|_{2,\max}\,\|UU^{\top}Q\|_{\mathrm F} + \|V\|_{2,\max}\|U_{\perp}U_{\perp}^{\top} QVV^{\top}\|_{\mathrm F}. 
\end{align*}
Also notice that 
\begin{align*}
\left\langle UU^{\top}Q,U_{\perp}U_{\perp}^{\top} QVV^{\top}\right\rangle
&=
\operatorname{tr}\!\left(
(UU^{\top}Q)^\top U_{\perp}U_{\perp}^{\top} QVV^{\top}
\right)=
\operatorname{tr}\!\left(
Q^\top UU^{\top}U_{\perp}U_{\perp}^{\top} QVV^{\top}
\right)
=0.
\end{align*}
Consequently,
\[
\|\PMQ\|_{\mathrm F}^2
=
\|UU^{\top}Q\|_{\mathrm F}^2
+
\|U_{\perp}U_{\perp}^{\top} QVV^{\top}\|_{\mathrm F}^2.
\]
This directly implies
\begin{align*}
    \|\PMQ\|_{\max}
    \leq \big(\|U\|_{2,\max}+\|V\|_{2,\max}\big)\fro{\PMQ}
    \lesssim \mu\sqrt{\frac{r}{d_1}}\fro{\PMQ}
\end{align*}
and further proves the bound for $\rho_Q$.

\subsection{Proof of Theorem \ref{thm:inference-main}}
\begin{proof}
    We start by recalling that $\widehat{M}_l^{\ubs}$ for $l=1,2$ can be decomposed into 
    \begin{align*}
        \widehat{M}_l^{\ubs}= M + \underbrace{\frac{d_2}{T_0}\sum_{t\in \widetilde{\mathcal{D}}_{3-l}}\sum_{i=1}^{d_1} \xi_t^i X_t^i}_{=:\widehat{Z}_1^{(l)}} + \underbrace{\frac{1}{T_0}\sum_{t\in \widetilde{\mathcal{D}}_{3-l}} \left(\sum_{i=1}^{d_1}d_2\inp{\widehat{\Delta}_{l}}{X_t^i}X_t^i-\widehat{\Delta}_{l}\right)}_{=:\widehat{Z}_2^{(l)}},
    \end{align*}
    where $\widehat{\Delta}_{l}=\widehat{M}_l^{\init}-M$. Moreover, notice that
    \begin{align*}
        \inp{\widehat{M}_l}{Q}-\inp{M}{Q}=\inp{\widehat{U}_l\widehat{U}_l^{\top}\widehat{Z}\widehat{V}_l\widehat{V}_l^{\top}}{Q} + \inp{\widehat{U}_l\widehat{U}_l^{\top}M\widehat{V}_l\widehat{V}_l^{\top}-M}{Q}.
    \end{align*}
    Define $\widehat{U}_l$ and $\widehat{V}_l$ to be the top-$r$ left and right singular vectors of $\widehat{M}_l^{\ubs}$. Define $(d_1+d_2)\times (2r)$ matrices 
    $$
        {\Theta}= \begin{pmatrix}
            U & 0 \\
            0 & V
        \end{pmatrix} \quad 
        \widehat{\Theta}_l= \begin{pmatrix}
            \widehat{U}_l & 0 \\
            0 & \widehat{V}_l
        \end{pmatrix} \quad 
        A = \begin{pmatrix}
            0 & M \\
            M^{\top} & 0
        \end{pmatrix}.
    $$
    Then, we can write
    $$
        \widehat{\Theta}_l\widehat{\Theta}_l^{\top}A\widehat{\Theta}_l\widehat{\Theta}_l^{\top}-\Theta\Theta^{\top}A\Theta\Theta^{\top}= \begin{pmatrix}
            0 & \widehat{U}_l\widehat{U}_l^{\top}M\widehat{V}_l\widehat{V}_l^{\top}-M \\
            (\widehat{U}_l\widehat{U}_l^{\top}M\widehat{V}_l\widehat{V}_l^{\top}-M)^{\top} & 0
        \end{pmatrix}.
    $$
    We further define
    $$
        \widetilde{Q}=\begin{pmatrix}
            0 & Q \\
            0 & 0
        \end{pmatrix} \quad \text{and} \quad 
        \widehat{E}^{(l)}= \begin{pmatrix}
            0 & \widehat{Z}^{(l)} \\
            \widehat{Z}^{(l)\top} & 0
        \end{pmatrix},
    $$
    where $\widehat{Z}^{(l)}=\widehat{Z}_1^{(l)}+\widehat{Z}_2^{(l)}$. Therefore, we have
    \begin{align*}
        \inp{\widehat{U}_l\widehat{U}_l^{\top}M\widehat{V}_l\widehat{V}_l^{\top}-M}{Q}=\inp{\widehat{\Theta}_l\widehat{\Theta}_l^{\top}A\widehat{\Theta}_l\widehat{\Theta}_l^{\top}-\Theta\Theta^{\top}A\Theta\Theta^{\top}}{\widetilde{Q}}.
    \end{align*}
    Define
        \begin{align*}
            \mathfrak{P}^{-s}=\left\{
            \begin{aligned}
                \begin{pmatrix}
                    U\Lambda^{-s}U^{\top} & 0 \\
                    0 & V\Lambda^{-s}V^{\top}
                \end{pmatrix} \quad \text{if $s$ is even}  \\
                \begin{pmatrix}
                    0 &  U\Lambda^{-s}U^{\top}\\
                    V\Lambda^{-s}V^{\top} & 0
                \end{pmatrix} \quad \text{if $s$ is odd,} 
            \end{aligned} \right.
        \end{align*}
        and
        \begin{align*}
            \mathfrak{P}^0=\mathfrak{P}^{\perp}=\begin{pmatrix}
                U_{\perp}U_{\perp}^{\top} & 0 \\
                0 & V_{\perp}V_{\perp}^{\top}
            \end{pmatrix}.
        \end{align*}
    By Theorem 1 in \cite{xia2021normal}, $\widehat{\Theta}_l\widehat{\Theta}_l^{\top}-\Theta\Theta^{\top}$ has explicit representation formula of empirical spectral projector, in the form of 
    \begin{align*}
        &\widehat{\Theta}\widehat{\Theta}^{\top}-\Theta\Theta^{\top}=\sum_{k=1}^{\infty}\mathcal{S}_{A,k}(\widehat{E}) \\
        &\mathcal{S}_{A,k}(\widehat{E}^{(l)})=\sum_{s_1+\cdots+s_{k+1}=k}^{k} (-1)^{1+\tau(s)} \mathfrak{P}^{-s_1}\widehat{E}\mathfrak{P}^{-s_2}\cdots\mathfrak{P}^{-s_k}\widehat{E}\mathfrak{P}_{-s_{k+1}}
    \end{align*}
    where $s_1,\cdots,s_{k+1}\geq 0$ are integers, $\tau(\mathbf{s})=\sum_{i=1}^{k+1}\mathbbm{1}(s_i>0)$. \\
    As a result, 
    \begin{align*}
        \widehat{\Theta}_l\widehat{\Theta}_l^{\top}A\widehat{\Theta}\widehat{\Theta}^{\top}&-\Theta\Theta^{\top}A\Theta\Theta^{\top}= (\mathcal{S}_{A,1}(\widehat{E}^{(l)})A\Theta\Theta^{\top}+\Theta\Theta^{\top}A\mathcal{S}_{A,1}(\widehat{E}^{(l)}))\\
        +& \sum_{k=2}^{\infty} (\mathcal{S}_{A,k}(\widehat{E})A\Theta\Theta^{\top}+\Theta\Theta^{\top}A\mathcal{S}_{A,k}(\widehat{E}^{(l)})) \quad + (\widehat{\Theta}\widehat{\Theta}^{\top}-\Theta\Theta^{\top})A(\widehat{\Theta}_l\widehat{\Theta}^{\top}-\Theta\Theta^{\top}).
    \end{align*}
    By definition of $\mathcal{S}_{A,1}(\widehat{E})$, 
    \begin{align*}
        &\inp{\mathcal{S}_{A,1}(\widehat{E}^{(l)})A\Theta\Theta^{\top}+\Theta\Theta^{\top}A\mathcal{S}_{A,1}(\widehat{E}^{(l)})}{\widetilde{Q}}= \inp{UU^{\top}\widehat{Z}^{(l)}V_{\perp}V_{\perp}^{\top}}{Q} + \inp{U_{\perp}U_{\perp}^{\top}\widehat{Z}^{(l)}VV^{\top}}{Q} \\
        &\quad =\inp{UU^{\top}\widehat{Z}_1^{(l)}V_{\perp}V_{\perp}^{\top}}{Q} + \inp{U_{\perp}U_{\perp}^{\top}\widehat{Z}_1^{(l)}VV^{\top}}{Q}  + \inp{UU^{\top}\widehat{Z}_2^{(l)}V_{\perp}V_{\perp}^{\top}}{Q} + \inp{U_{\perp}U_{\perp}^{\top}\widehat{Z}_2^{(l)}VV^{\top}}{Q}.
    \end{align*}
    Combine all the terms above, $\inp{\widehat{M}}{Q}-\inp{M}{Q}$ can be decomposed into
    \begin{align}
        \inp{\widehat{M}}{Q}-&\inp{M}{Q}= \frac{1}{2}\sum_{l=1}^{2} \inp{UU^{\top}\widehat{Z}_1^{(l)}V_{\perp}V_{\perp}^{\top}}{Q} + \inp{U_{\perp}U_{\perp}^{\top}\widehat{Z}_1^{(l)}VV^{\top}}{Q} + \inp{UU^{\top}\widehat{Z}_1^{(l)}VV^{\top}}{Q} \label{main1} \\ &\quad + \frac{1}{2}\sum_{l=1}^{2}\inp{UU^{\top}\widehat{Z}_2^{(l)}V_{\perp}V_{\perp}^{\top}}{Q} + \inp{U_{\perp}U_{\perp}^{\top}\widehat{Z}_2^{(l)}VV^{\top}}{Q} + \inp{UU^{\top}\widehat{Z}_2^{(l)}VV^{\top}}{Q}\label{neg1} \\ &\quad + \frac{1}{2}\sum_{l=1}^{2}\inp{\widehat{U}_l\widehat{U}_l^{\top}\widehat{Z}^{(l)}\widehat{V}_l\widehat{V}_l^{\top}}{Q} - \inp{UU^{\top}\widehat{Z}^{(l)}VV^{\top}}{Q} \label{neg2} \\ &\quad + \frac{1}{2}\sum_{l=1}^{2}\inp{\sum_{k=2}^{\infty} (\mathcal{S}_{A,k}(\widehat{E}^{(l)})A\Theta\Theta^{\top}+\Theta\Theta^{\top}A\mathcal{S}_{A,k}(\widehat{E}^{(l)}))}{\widetilde{Q}} + \inp{(\widehat{\Theta}\widehat{\Theta}^{\top}-\Theta\Theta^{\top})A(\widehat{\Theta}_l\widehat{\Theta}_l^{\top}-\Theta\Theta^{\top})}{\widetilde{Q}}. \label{neg3}
    \end{align}
    In the following, we first apply Berry Esseen Theorem to show equation (\ref{main1}) is the main term which is asymptotic normal. Next, we calculate the upper bounds of (\ref{neg1}), (\ref{neg2}) and (\ref{neg3}) to show they are negligible higher order terms, respectively. 

    \noindent\textbf{Asymptotic normality of (\ref{main1})}
    By definition, \eqref{main1} can be written as
    \begin{align*}
        L_T^{\oto}:=\frac{1}{2}\sum_{l=1}^{2}\Big(
        &\inp{UU^{\top}\widehat{Z}_1^{(l)}V_{\perp}V_{\perp}^{\top}}{Q}
        +\inp{U_{\perp}U_{\perp}^{\top}\widehat{Z}_1^{(l)}VV^{\top}}{Q}
        +\inp{UU^{\top}\widehat{Z}_1^{(l)}VV^{\top}}{Q}\Big)\\
        &=\frac{d_2}{T}\sum_{t=1}^T\sum_{i=1}^{d_1}
        \xi_t^i\inp{X_t^i}{\PMQ}
        :=\frac{1}{T\nu}\sum_{t=1}^T W_t,
    \end{align*}
    where $\nu=\frac{1}{d_2}$.

    It remains to calculate the $\kappa_1$ in Lemma \ref{lemma: pairwise}. For any $(i,j)\neq (k,l)$,
    \begin{align*}
        \PP\big((i,j),(k,l)\in\Omega_t\big)
        =
        \begin{cases}
            \displaystyle\frac{1}{d_2(d_2-1)}, & i\neq k,\ j\neq l,\\[6pt]
            0, & i=k\ \text{or}\ j=l.
        \end{cases}
    \end{align*}
    Consequently,
    $\PP((i,j),(k,l)\in\Omega_t)\leq \frac{d_2}{d_2-1}\nu^2\leq2\nu^2$.  The moment bound above therefore gives
    \begin{align*}
        \sup_{x\in\RR}\left|
        \PP\left(
        \frac{L_T^{\oto}}
        {\sigma\fro{\mathcal{P}_M(Q)}\sqrt{d_2/T}}\leq x
        \right)-\Phi(x)\right|
        \lesssim
        \sqrt{\frac{\rho_Q\alpha_d+2}{T}}.
    \end{align*}

    \noindent\textbf{Bounding (\ref{neg1})}

     Without loss of generality, we only prove the upper bound for $|\inp{UU^{\top}\widehat{Z}_2^{(1)}V_{\perp}V_{\perp}^{\top}}{Q} + \inp{U_{\perp}U_{\perp}^{\top}\widehat{Z}_2^{(1)}VV^{\top}}{Q} + \inp{UU^{\top}\widehat{Z}_2^{(1)}VV^{\top}}{Q}|$. The the upper bound for $|\inp{UU^{\top}\widehat{Z}_2^{(2)}V_{\perp}V_{\perp}^{\top}}{Q} + \inp{U_{\perp}U_{\perp}^{\top}\widehat{Z}_2^{(2)}VV^{\top}}{Q} + \inp{UU^{\top}\widehat{Z}_2^{(2)}VV^{\top}}{Q}|$ can be derived following the same arguments. By definition, 
    \begin{align*}
        &|\inp{UU^{\top}\widehat{Z}_2^{(1)}V_{\perp}V_{\perp}^{\top}}{Q} + \inp{U_{\perp}U_{\perp}^{\top}\widehat{Z}_2^{(1)}VV^{\top}}{Q} + \inp{UU^{\top}\widehat{Z}_2^{(1)}VV^{\top}}{Q}|\\ 
        &\quad =\frac{1}{T_0}\sum_{t=1}^{T_0} \sum_{i=1}^{d_1} d_2\inp{\widehat{\Delta}_1}{X_t^i}\inp{X_t^i}{\PMQ} -\inp{\widehat{\Delta}_1}{\PMQ}.
    \end{align*}
    We first derive the uniform bound for $(1/T_0)|d_2\sum_{i=1}^{d_1}\inp{\widehat{\Delta}_1}{X_t^i}\inp{X_t^i}{\PMQ} -\inp{\widehat{\Delta}_1}{\PMQ}|$. Define $R:= \sum_{i=1}^{d_1}\inp{\widehat{\Delta}_1}{X_t^i}\inp{X_t^i}{\PMQ}$, then $E(R)=(1/d_2)\inp{\widehat{\Delta}_1}{\PMQ}$. Let \(\mathcal F_i\) denote the \(\sigma\)-algebra generated by
\(\{X_t^1,\ldots,X_t^i\}\), i.e., $\mathcal F_i=\sigma(X_t^1,\ldots,X_t^i), \mathcal F_0=\emptyset.$ Then \(R\) induces the Doob martingale
\[
R_i:=\mathbb E[R\mid\mathcal F_i].
\]More explicitly,
\[
\begin{aligned}
R_i
&=
\sum_{j=1}^{i}
\left\langle\widehat\Delta_1,X_t^j\right\rangle
\left\langle X_t^j,\mathcal P_M(Q)\right\rangle +
\sum_{j=i+1}^{d_1}
\mathbb E\!\left[
\left\langle\widehat\Delta_1,X_t^j\right\rangle
\left\langle X_t^j,\mathcal P_M(Q)\right\rangle
\mid\mathcal F_i
\right].
\end{aligned}
\]In particular, $R_0=\mathbb E[R],
R_{d_1}=R.$ Notice that
\[
\begin{aligned}
\mathbb E[R_i\mid\mathcal F_{i-1}]
&=
\sum_{j=1}^{i-1}
\left\langle\widehat\Delta_1,X_t^j\right\rangle
\left\langle X_t^j,\mathcal P_M(Q)\right\rangle
+
\sum_{j=i}^{d_1}
\mathbb E\!\left[
\left\langle\widehat\Delta_1,X_t^j\right\rangle
\left\langle X_t^j,\mathcal P_M(Q)\right\rangle
\mid\mathcal F_{i-1}
\right]=
R_{i-1}.
\end{aligned}
\]Therefore,
\[
\begin{aligned}
R_i-R_{i-1}
&=
\left\langle\widehat\Delta_1,X_t^i\right\rangle
\left\langle X_t^i,\mathcal P_M(Q)\right\rangle
-
\mathbb E\!\left[
\left\langle\widehat\Delta_1,X_t^i\right\rangle
\left\langle X_t^i,\mathcal P_M(Q)\right\rangle
\mid\mathcal F_{i-1}
\right]
\\ &\quad +
\sum_{j=i+1}^{d_1}
\Bigg(
\mathbb E\!\left[
\left\langle\widehat\Delta_1,X_t^j\right\rangle
\left\langle X_t^j,\mathcal P_M(Q)\right\rangle
\mid\mathcal F_i
\right]-
\mathbb E\!\left[
\left\langle\widehat\Delta_1,X_t^j\right\rangle
\left\langle X_t^j,\mathcal P_M(Q)\right\rangle
\mid\mathcal F_{i-1}
\right]
\Bigg).
\end{aligned}
\]Obviously,$\mathbb E[R_i-R_{i-1}\mid\mathcal F_{i-1}]=0$. By \((a+b)^2\leq 2(a^2+b^2)\),
\[
\begin{aligned}
\operatorname{Var}[R_i\mid\mathcal F_{i-1}]
&=
\mathbb E
\left[
\left(
R_i-\mathbb E[R_i\mid\mathcal F_{i-1}]
\right)^2
\mid\mathcal F_{i-1}
\right]=
\mathbb E[(R_i-R_{i-1})^2\mid\mathcal F_{i-1}]
\\
&\leq
2\mathbb E\!\left[
\left\langle\widehat\Delta_1,X_t^i\right\rangle^2
\left\langle X_t^i,\mathcal P_M(Q)\right\rangle^2
\mid\mathcal F_{i-1}
\right]
+
2\mathbb E\!\left[
\left(
\sum_{j=i+1}^{d_1}
\mathbb E\!\left[
\left\langle\widehat\Delta_1,X_t^j\right\rangle
\left\langle X_t^j,\mathcal P_M(Q)\right\rangle
\mid\mathcal F_i
\right]
\right.\right.
\\
&\hspace{32mm}\left.\left.
-
\mathbb E\!\left[
\left\langle\widehat\Delta_1,X_t^j\right\rangle
\left\langle X_t^j,\mathcal P_M(Q)\right\rangle
\mid\mathcal F_{i-1}
\right]
\right)^2
\Biggm|\mathcal F_{i-1}
\right].
\end{aligned}
\]
By the property of the one-to-one matching observation pattern and the definition of \(\mathcal F_i\), conditional on \(\mathcal F_i\), \(X_t^j\) is uniformly sampled from
\[
\left\{
e_je_k^{\top}:
k\in[d_2]\setminus
\{j_t(1),\ldots,j_t(i)\}
\right\},
\qquad i<j\leq d_1.
\]Therefore,
\[
\begin{aligned}
&
\left|
\mathbb E\!\left[
\left\langle\widehat\Delta_1,X_t^j\right\rangle
\left\langle X_t^j,\mathcal P_M(Q)\right\rangle
\mid\mathcal F_i
\right]
-
\mathbb E\!\left[
\left\langle\widehat\Delta_1,X_t^j\right\rangle
\left\langle X_t^j,\mathcal P_M(Q)\right\rangle
\mid\mathcal F_{i-1}
\right]
\right|
\\
&=
\left|
\frac{1}{d_2-i}
\sum_{\substack{
k\in[d_2]\setminus
\{j_t(1),\ldots,j_t(i)\}
}}
\left\langle e_je_k^{\top},\widehat\Delta_1\right\rangle
\left\langle e_je_k^{\top},\mathcal P_M(Q)\right\rangle
\right.
\\
&\hspace{13mm}\left.
-
\frac{1}{d_2-i+1}
\sum_{\substack{
k\in[d_2]\setminus
\{j_t(1),\ldots,j_t(i-1)\}
}}
\left\langle e_je_k^{\top},\widehat\Delta_1\right\rangle
\left\langle e_je_k^{\top},\mathcal P_M(Q)\right\rangle
\right|
\\
&\leq
\frac{1}{d_2-i+1}
\|\widehat\Delta_1\|_{\max}
\left|
\left\langle
e_je_{j_t(i)}^{\top},
\mathcal P_M(Q)
\right\rangle
\right|
\\
&\quad+
\frac{1}{(d_2-i)(d_2-i+1)}
\sum_{\substack{
k\in[d_2]\setminus
\{j_t(1),\ldots,j_t(i)\}
}}
\|\widehat\Delta_1\|_{\max}
\left|
\left\langle e_je_k^{\top},\mathcal P_M(Q)\right\rangle
\right|.
\end{aligned}
\]By Cauchy–Schwarz inequality,
\[
\begin{aligned}
&
\left(
\sum_{j=i+1}^{d_1}
\mathbb E\!\left[
\left\langle\widehat\Delta_1,X_t^j\right\rangle
\left\langle X_t^j,\mathcal P_M(Q)\right\rangle
\mid\mathcal F_i
\right]
-
\mathbb E\!\left[
\left\langle\widehat\Delta_1,X_t^j\right\rangle
\left\langle X_t^j,\mathcal P_M(Q)\right\rangle
\mid\mathcal F_{i-1}
\right]
\right)^2
\\
&\leq
2d_1
\sum_{j=i+1}^{d_1}
\frac{1}{(d_2-i+1)^2}
\|\widehat\Delta_1\|_{\max}^2
\left\langle
e_je_{j_t(i)}^{\top},
\mathcal P_M(Q)
\right\rangle^2
\\
&\quad+
\frac{1}{(d_2-i)(d_2-i+1)^2}
\sum_{j=i+1}^{d_1}
\sum_{\substack{
k\in[d_2]\setminus
\{j_t(1),\ldots,j_t(i)\}
}}
\|\widehat\Delta_1\|_{\max}^2
\left\langle e_je_k^{\top},\mathcal P_M(Q)\right\rangle^2.
\end{aligned}
\]Again, conditional on \(\mathcal F_{i-1}\),
\[
j_t(i)
\sim
\operatorname{Unif}
\left(
[d_2]\setminus
\{j_t(1),\ldots,j_t(i-1)\}
\right).
\]Hence, along with $d_2\geq (1+\gamma)d_1$, 
\[
\begin{aligned}
&
\mathbb E\!\left[
\left(
\sum_{j=i+1}^{d_1}
\mathbb E\!\left[
\left\langle\widehat\Delta_1,X_t^j\right\rangle
\left\langle X_t^j,\mathcal P_M(Q)\right\rangle
\mid\mathcal F_i
\right]
-
\mathbb E\!\left[
\left\langle\widehat\Delta_1,X_t^j\right\rangle
\left\langle X_t^j,\mathcal P_M(Q)\right\rangle
\mid\mathcal F_{i-1}
\right]
\right)^2
\Biggm|\mathcal F_{i-1}
\right]
\\
&\leq
2d_1
\sum_{j=i+1}^{d_1}
\frac{1}{(d_2-i+1)^3}
\sum_{\substack{
k\in[d_2]\setminus
\{j_t(1),\ldots,j_t(i-1)\}
}}
\|\widehat\Delta_1\|_{\max}^2
\left\langle
e_je_k^{\top},
\mathcal P_M(Q)
\right\rangle^2
\\
&\quad+
\frac{1}{(d_2-i)(d_2-i+1)^2}
\sum_{j=i+1}^{d_1}
\sum_{\substack{
k\in[d_2]\setminus
\{j_t(1),\ldots,j_t(i)\}
}}
\|\widehat\Delta_1\|_{\max}^2
\left\langle
e_je_k^{\top},
\mathcal P_M(Q)
\right\rangle^2
\\
&\lesssim
\frac{1}{d_1^2}
\|\widehat\Delta_1\|_{\max}^2
\sum_{j=i+1}^{d_1}
\|\mathcal P_M(Q)(j,\cdot)\|^2\leq
\frac{1}{d_1^2}
\|\widehat\Delta_1\|_{\max}^2
\|\mathcal P_M(Q)\|_{\mathrm F}^2 .
\end{aligned}
\]Similarly,
\[
\begin{aligned}
&
\mathbb E\!\left[
\left\langle\widehat\Delta_1,X_t^i\right\rangle^2
\left\langle X_t^i,\mathcal P_M(Q)\right\rangle^2
\mid\mathcal F_{i-1}
\right]\lesssim
\frac{1}{d_1}
\|\widehat\Delta_1\|_{\max}^2
\|\mathcal P_M(Q)(i,\cdot)\|^2 .
\end{aligned}
\]
Combining the above terms,
\[
\sum_{i=1}^{d_1}
\operatorname{Var}[R_i\mid\mathcal F_{i-1}]
\lesssim
\frac{1}{d_1}
\|\widehat\Delta_1\|_{\max}^2
\|\mathcal P_M(Q)\|_{\mathrm F}^2.
\]
Moreover,
\[
\mathbb E[R]
=
\frac{1}{d_2}
\left\langle
\widehat\Delta_1,\mathcal P_M(Q)
\right\rangle ,
\]and for any \(i\),
\[
\begin{aligned}
|R_i-R_{i-1}|
&\leq
2
\left|
\left\langle\widehat\Delta_1,X_t^i\right\rangle
\left\langle X_t^i,\mathcal P_M(Q)\right\rangle
\right|
 \\ &\quad +
\left|
\sum_{j=i+1}^{d_1}
\mathbb E\!\left[
\left\langle\widehat\Delta_1,X_t^j\right\rangle
\left\langle X_t^j,\mathcal P_M(Q)\right\rangle
\mid\mathcal F_i
\right]
-
\mathbb E\!\left[
\left\langle\widehat\Delta_1,X_t^j\right\rangle
\left\langle X_t^j,\mathcal P_M(Q)\right\rangle
\mid\mathcal F_{i-1}
\right]
\right|
\\
&\leq
2\|\widehat\Delta_1\|_{\max}
\|\mathcal P_M(Q)\|_{\max}
+
\frac{2d_1}{d_2-i}
\|\widehat\Delta_1\|_{\max}
\|\mathcal P_M(Q)\|_{\max}
\lesssim
\frac{1}{\sqrt{d_1}}
\|\widehat\Delta_1\|_{\max}
\|\mathcal P_M(Q)\|_{\mathrm F}.
\end{aligned}
\]Therefore, by martingale Bernstein inequality (Theorem 18 in \cite{chung2006concentration}), with probability at least \(1-d_1^{-10}\),
\[
\begin{aligned}
\frac{1}{T_0}
\left|
d_2\sum_{i=1}^{d_1}
\left\langle\widehat\Delta_1,X_t^i\right\rangle
\left\langle X_t^i,\mathcal P_M(Q)\right\rangle
-
\left\langle\widehat\Delta_1,\mathcal P_M(Q)\right\rangle
\right|&\lesssim
\frac{d_2}{T\sqrt{d_1}}
\log d_1\,
\|\widehat\Delta_1\|_{\max}
\|\mathcal P_M(Q)\|_{\mathrm F}
\\
&\lesssim
\sqrt{\frac{r^2d_2^3}{T^3d_1^2}}
\log^2d_1\,
\sigma
\|\mathcal P_M(Q)\|_{\mathrm F},
\end{aligned}
\]where the last inequality comes from Proposition \ref{thm:estimation}.

    Next we calculate the variance of $\frac{1}{T_0}\bigg|d_2\sum_{i=1}^{d_1}\inp{\widehat{\Delta}_1}{X_t^i}\inp{X_t^i}{\PMQ} -\inp{\widehat{\Delta}_1}{\PMQ}\bigg|$. Notice that $\widehat{\Delta}_1$ is independent of $X_t^i$, then
    \begin{align*}
        &\frac{1}{T_0^2}\EE\left[\bigg(d_2\sum_{i=1}^{d_1}\inp{\widehat{\Delta}_1}{X_t^i}\inp{X_t^i}{\PMQ} -\inp{\widehat{\Delta}_1}{\PMQ} \bigg)^2\right] \\
        &\quad = \frac{1}{T_0^2}\EE\left[\bigg(d_2\sum_{i=1}^{d_1}\inp{\widehat{\Delta}_1}{X_t^i}\inp{X_t^i}{\PMQ} \bigg)^2\right]-\frac{1}{T_0^2}\inp{\widehat{\Delta}_1}{\PMQ}^2 \\
        &\quad = \frac{d_2^2}{T_0^2}\underbrace{\EE\left[\sum_{i=1}^{d_1}\inp{\widehat{\Delta}_1}{X_t^i}^2\inp{X_t^i}{\PMQ}^2 \right]}_{\mathcal{I}_1} \\ &\quad\quad + \frac{d_2^2}{T_0^2}\underbrace{\EE\left[\sum_{i=1}^{d_1}\sum_{j\neq i} \inp{\widehat{\Delta}_1}{X_t^i}\inp{X_t^i}{\PMQ}\inp{\widehat{\Delta}_1}{X_t^j}\inp{X_t^j}{\PMQ} \right]}_{\mathcal{I}_2} - \frac{1}{T_0^2}\inp{\widehat{\Delta}_1}{\PMQ}^2.
    \end{align*}
    Denote $A=\widehat{\Delta}_1\odot \PMQ$ the Hadamard product of $\widehat{\Delta}_1$ and $\PMQ$, then 
    \begin{align*}
        \mathcal{I}_1=\EE[\sum_{i=1}^{d_1}\inp{X_t^i}{A}^2]=\frac{1}{d_2}\fro{A}^2\leq \frac{1}{d_2}\|\widehat{\Delta}_1\|_{\max}^2\fro{\PMQ}^2.
    \end{align*}
    It is obvious that 
    \begin{align*}
        \mathcal{I}_2=\frac{1}{d_2(d_2-1)}\sum_{i\neq j}\sum_{p\neq q} A(i,p)A(j,q).
    \end{align*}
    Since 
    \begin{align*}
        \inp{\widehat{\Delta}_1}{\PMQ}^2&=\big(\sum_{i,p} A(i,p) \big)^2= \sum_{i,p} A(i,p)^2 + \sum_i \sum_{p\neq q} A(i,p)A(i,q) \\ &\quad + \sum_{i\neq j}\sum_p A(i,p)A(j,p) + \sum_{i\neq j}\sum_{p\neq q} A(i,p)A(j,q),
    \end{align*}
    where 
    \begin{align*}
        &|\sum_i \sum_{p\neq q} A(i,p)A(i,q)|\leq \frac{1}{2} \sum_i \sum_{p\neq q} A(i,p)^2 + A(i,q)^2 \\ &\quad \leq \frac{1}{2} \sum_i \sum_{p, q} A(i,p)^2 + A(i,q)^2  \leq \frac{d_2}{2}\sum_{i,p} A(i,p)^2= \frac{d_2}{2}\fro{A}^2
    \end{align*}
    and similarly, 
    \begin{align*}
        |\sum_{i\neq j}\sum_p A(i,p)A(j,p)|\leq \frac{d_1}{2}\fro{A}^2.
    \end{align*}
   As a result,
   \begin{align*}
        d_2(d_2-1)\mathcal{I}_2&\leq \inp{\widehat{\Delta}_1}{\PMQ}^2 - \fro{A}^2 + \frac{1}{2}(d_1+d_2)\fro{A}^2 \\
        &\leq \inp{\widehat{\Delta}_1}{\PMQ}^2 + d_2\|\widehat{\Delta}_1\|_{\max}^2\fro{\PMQ}^2.
   \end{align*}
   Overall, the variance can be upper bounded by
   \begin{align*}
        &\frac{1}{T_0^2}\EE\left[\bigg(d_2\sum_{i=1}^{d_1}\inp{\widehat{\Delta}_1}{X_t^i}\inp{X_t^i}{\PMQ} -\inp{\widehat{\Delta}_1}{\PMQ} \bigg)^2\right]\leq \frac{d_2}{T_0^2}\|\widehat{\Delta}_1\|_{\max}^2\fro{\PMQ}^2 \\ &\quad\quad + \bigg(\frac{d_2}{T_0^2(d_2-1)}-\frac{1}{T_0^2}\bigg)\inp{\widehat{\Delta}_1}{\PMQ}^2 + \frac{d_2^2}{T_0^2(d_2-1)}\|\widehat{\Delta}_1\|_{\max}^2\fro{\PMQ}^2 \\ 
        &\quad\leq \frac{3d_2}{T_0^2}\|\widehat{\Delta}_1\|_{\max}^2\fro{\PMQ}^2 + \frac{2}{T_0^2d_2}d_1d_2\|\widehat{\Delta}_1\|_{\max}^2\fro{\PMQ}^2\leq \frac{5d_2}{T_0^2}\|\widehat{\Delta}_1\|_{\max}^2\fro{\PMQ}^2.
   \end{align*}
   Combine with Proposition \ref{thm:estimation} and the previous uniform bound, by Bernstein inequality, with probability at least $1-2d_1^{-10}$, 
   \begin{align*}
        &\left|\frac{1}{T_0}\sum_{t=1}^{T_0} \sum_{i=1}^{d_1} d_2\inp{\widehat{\Delta}_1}{X_t^i}\inp{X_t^i}{\PMQ} -\inp{\widehat{\Delta}_1}{\PMQ}\right|\\ &\quad \lesssim \sqrt{\frac{r^2d_2^3}{T^3d_1^2}}\log^3d_1\sigma\fro{\PMQ} + \sqrt{\frac{r^2d_2^2\log^3 d_1}{T^2d_1}}\sigma\fro{\PMQ}.
   \end{align*}

    Finally, the following Lemmas show the upper bound of (\ref{neg2}) and (\ref{neg3}).

    \begin{Lemma}
        \label{lemmaneg2}
        Under the conditions in Theorem \ref{thm:inference-main}, 
        \begin{align*}
            \frac{\big|\frac{1}{2}\sum_{l=1}^{2}\inp{\widehat{U}_l\widehat{U}_l^{\top}\widehat{Z}^{(l)}\widehat{V}_l\widehat{V}_l^{\top}}{Q} - \inp{UU^{\top}\widehat{Z}^{(l)}VV^{\top}}{Q}\big|}{\sigma\fro{\mathcal{P}_M(Q)}\sqrt{d_2/T}}\lesssim \frac{\sigma\mu^2}{\lambda_{\min}}\sqrt{\frac{r^2d_2^2\log d_1}{Td_1}}\frac{\|Q\|_{\ell_1}}{\fro{\PMQ}}. 
        \end{align*}
    \end{Lemma}

    \begin{Lemma}
        \label{lemmaneg3}
        Under the conditions in Theorem \ref{thm:inference-main}, 
        \begin{align*}
            &\frac{\big|\sum_{k=2}^{\infty} \inp{(\mathcal{S}_{A,k}(\widehat{E})A\Theta\Theta^{\top}+\Theta\Theta^{\top}A\mathcal{S}_{A,k}(\widehat{E}))}{\widetilde{Q}}\big|}{\sigma\fro{\mathcal{P}_M(Q)}\sqrt{d_2/T}}\lesssim \frac{\sigma\mu^2}{\lambda_{\min}}\sqrt{\frac{r^2d_2^2\log d_1}{Td_1}}\frac{\|Q\|_{\ell_1}}{\fro{\PMQ}}, \\
            &\frac{\big|\inp{(\widehat{\Theta}\widehat{\Theta}^{\top}-\Theta\Theta^{\top})A(\widehat{\Theta}\widehat{\Theta}^{\top}-\Theta\Theta^{\top})}{\widetilde{Q}}\big|}{\sigma\fro{\mathcal{P}_M(Q)}\sqrt{d_2/T}} \lesssim \frac{\sigma\mu^2\kappa}{\lambda_{\min}}\sqrt{\frac{r^2d_2^2\log d_1}{Td_1}}\frac{\|Q\|_{\ell_1}}{\fro{\PMQ}}.
        \end{align*}
    \end{Lemma}
\end{proof}

\subsection{Proof of Theorem \ref{thm:inference-main-2}}
\begin{proof}
    We start by recalling that $\widehat{M}_l^{\ubs}$ for $l=1,2$ can be decomposed into 
    \begin{align*}
        \widehat{M}_l^{\ubs}= M + \underbrace{\frac{d_2}{T_0Kp_0}\sum_{t\in \widetilde{\mathcal{D}}_{l}}\sum_{i=1}^{d_1}\sum_{q\in h_{it}} \xi_t^i X_t^i}_{=:\widehat{Z}_1^{(l)}} + \underbrace{\frac{1}{T_0}\sum_{t\in \widetilde{\mathcal{D}}_{l}} \left(\sum_{i=1}^{d_1}\sum_{q\in h_{it}} \frac{d_2}{Kp_0}\inp{\widehat{\Delta}_{l}}{X_t^{iq}}X_t^{iq}-\widehat{\Delta}_{l}\right)}_{=:\widehat{Z}_2^{(l)}},
    \end{align*}
    where $\widehat{\Delta}_{l}=\widehat{M}_l^{\init}-M$. Following the proof of Theorem \ref{thm:inference-main} and define $\Theta$, $\widehat{\Theta}_l$, $\widehat{E}^{(l)}$, $\mathfrak{P}^{-s}$ and all other variables in the same way, we again can obtain the decomposition
    \begin{align}
        \inp{\widehat{M}}{Q}-&\inp{M}{Q}= \frac{1}{2}\sum_{l=1}^{2} \inp{UU^{\top}\widehat{Z}_1^{(l)}V_{\perp}V_{\perp}^{\top}}{Q} + \inp{U_{\perp}U_{\perp}^{\top}\widehat{Z}_1^{(l)}VV^{\top}}{Q} + \inp{UU^{\top}\widehat{Z}_1^{(l)}VV^{\top}}{Q} \label{main1-otm} \\ &\quad + \frac{1}{2}\sum_{l=1}^{2}\inp{UU^{\top}\widehat{Z}_2^{(l)}V_{\perp}V_{\perp}^{\top}}{Q} + \inp{U_{\perp}U_{\perp}^{\top}\widehat{Z}_2^{(l)}VV^{\top}}{Q} + \inp{UU^{\top}\widehat{Z}_2^{(l)}VV^{\top}}{Q} \label{neg1-otm}\\ &\quad + \frac{1}{2}\sum_{l=1}^{2}\inp{\widehat{U}_l\widehat{U}_l^{\top}\widehat{Z}^{(l)}\widehat{V}_l\widehat{V}_l^{\top}}{Q} - \inp{UU^{\top}\widehat{Z}^{(l)}VV^{\top}}{Q} \label{neg2-otm}\\ &\quad + \frac{1}{2}\sum_{l=1}^{2}\inp{\sum_{k=2}^{\infty} (\mathcal{S}_{A,k}(\widehat{E}^{(l)})A\Theta\Theta^{\top}+\Theta\Theta^{\top}A\mathcal{S}_{A,k}(\widehat{E}^{(l)}))}{\widetilde{Q}} + \inp{(\widehat{\Theta}\widehat{\Theta}^{\top}-\Theta\Theta^{\top})A(\widehat{\Theta}_l\widehat{\Theta}_l^{\top}-\Theta\Theta^{\top})}{\widetilde{Q}} \label{neg3-otm}.
    \end{align}
    In the following Lemmas, we still first apply Berry Esseen Theorem to prove the asymptotic normality of the main term, while showing all the other terms are negligible. Then we can conclude the proof following Theorem \ref{thm:inference-main}.

    \noindent\textbf{Asymptotic normality of (\ref{main1-otm})}

    By definition, \eqref{main1-otm} can be written as
    \begin{align*}
        L_T^{\otm}:=\frac{1}{2}\sum_{l=1}^{2}\Big(
        &\inp{UU^{\top}\widehat{Z}_1^{(l)}V_{\perp}V_{\perp}^{\top}}{Q}
        +\inp{U_{\perp}U_{\perp}^{\top}\widehat{Z}_1^{(l)}VV^{\top}}{Q}
        +\inp{UU^{\top}\widehat{Z}_1^{(l)}VV^{\top}}{Q}\Big)\\
        &=\frac{d_2}{TKp_0}\sum_{t=1}^{T}\sum_{i=1}^{d_1}
        \sum_{q\in h_{it}}\xi_t^{iq}\inp{X_t^{iq}}{B}
        =\frac{1}{T\nu}\sum_{t=1}^T W_t,
        \qquad \nu=\frac{Kp_0}{d_2}.
    \end{align*}

    It remains to calculate the $\kappa_1$ in Lemma \ref{lemma: pairwise}. Since $|h_{it}|$ are i.i.d. sampled from $\operatorname{Bin}(K,p_0)$, and condition on $|h_{it}|$, the disjoint column sets are uniformly allocated. Therefore, for
    $j\neq l$,
    \begin{align*}
        \PP\big((i,j),(k,l)\in\Omega_t\big)
        &=
        \begin{cases}
        \displaystyle\frac{(Kp_0)^2}{d_2(d_2-1)}, & i\neq k,\\[7pt]
        \displaystyle\frac{K(K-1)p_0^2}{d_2(d_2-1)}, & i=k. 
        \end{cases}
    \end{align*}
    Since $\nu=Kp_0/d_2$, each of these probabilities is at most
    $\frac{d_2}{d_2-1}\nu^2\leq2\nu^2$, hence $\kappa_1=2$. Together with the the fact that $\rho_Q\lesssim \mu^2r$, we have
    \begin{align*}
        \sup_{x\in\RR}\left|
        \PP\left(
        \frac{L_T^{\otm}}
        {\sigma\fro{\mathcal{P}_M(Q)}\sqrt{d_2/(TKp_0)}}\leq x
        \right)-\Phi(x)\right|
        \lesssim
        \sqrt{\frac{\mu^2r\alpha_d/(Kp_0)+2}{T}}.
    \end{align*}

    \noindent\textbf{Bounding (\ref{neg1-otm})}
    
     Without loss of generality, we only prove the upper bound for $|\inp{UU^{\top}\widehat{Z}_2^{(1)}V_{\perp}V_{\perp}^{\top}}{Q} + \inp{U_{\perp}U_{\perp}^{\top}\widehat{Z}_2^{(1)}VV^{\top}}{Q} + \inp{UU^{\top}\widehat{Z}_2^{(1)}VV^{\top}}{Q}|$. The the upper bound for $|\inp{UU^{\top}\widehat{Z}_2^{(2)}V_{\perp}V_{\perp}^{\top}}{Q} + \inp{U_{\perp}U_{\perp}^{\top}\widehat{Z}_2^{(2)}VV^{\top}}{Q} + \inp{UU^{\top}\widehat{Z}_2^{(2)}VV^{\top}}{Q}|$ can be derived following the same arguments. By definition, 
    \begin{align*}
        &|\inp{UU^{\top}\widehat{Z}_2^{(1)}V_{\perp}V_{\perp}^{\top}}{Q} + \inp{U_{\perp}U_{\perp}^{\top}\widehat{Z}_2^{(1)}VV^{\top}}{Q} + \inp{UU^{\top}\widehat{Z}_2^{(1)}VV^{\top}}{Q}|\\ 
        &\quad =\frac{1}{T_0}\sum_{t=1}^{T_0} \sum_{i=1}^{d_1}\sum_{q\in h_{it}} \frac{d_2}{Kp_0}\inp{\widehat{\Delta}_1}{X_t^{iq}}\inp{X_t^{iq}}{\PMQ} -\inp{\widehat{\Delta}_1}{\PMQ}.
    \end{align*}
    We first derive the uniform bound for $(1/T_0)|(d_2/Kp_0) \sum_{i=1}^{d_1}\sum_{q\in h_{it}}\inp{\widehat{\Delta}_1}{X_t^{iq}}\inp{X_t^{iq}}{\PMQ} -\inp{\widehat{\Delta}_1}{\PMQ}|$, using a similar martingale type method as Theorem \ref{thm:inference-main}. Let $\mathcal F_i$ denote the $\sigma$-algebra generated by $\{\{X_t^{1q}\}_{q\in h{1t}},\ldots,\{X_t^{iq}\}_{q\in h{it}}\}$, i.e.,
$\mathcal F_i
=
\sigma\!\left(
\{X_t^{1q}\}_{q\in h_{1t}},\ldots,
\{X_t^{iq}\}_{q\in h_{it}}
\right)$ and $\mathcal F_0=\emptyset$. Then $R$ is a martingale associated with filter
$
\mathcal F_0\subset\mathcal F_1\subset\cdots\subset\mathcal F_{d_1}.$ Let
\[
\begin{aligned}
R_i:=\mathbb E[R\mid\mathcal F_i]
={}&
\sum_{j=1}^{i}\sum_{q\in h_{jt}}
\left\langle\widehat\Delta_1,X_t^{jq}\right\rangle
\left\langle X_t^{jq},\mathcal P_M(Q)\right\rangle +
\sum_{j=i+1}^{d_1}\sum_{q\in h_{jt}}
\mathbb E\!\left[
\left\langle\widehat\Delta_1,X_t^{jq}\right\rangle
\left\langle X_t^{jq},\mathcal P_M(Q)\right\rangle
\,\middle|\,\mathcal F_i
\right].
\end{aligned}
\]In particular,
\[
R_0=\mathbb E[R]
=\frac{Kp_0}{d_2}
\left\langle\widehat\Delta_1,\mathcal P_M(Q)\right\rangle,
\qquad
R_{d_1}=R.
\]Notice that,
\[
\begin{aligned}
\mathbb E[R_i\mid\mathcal F_{i-1}]
={}&
\sum_{j=1}^{i-1}\sum_{q\in h_{jt}}
\left\langle\widehat\Delta_1,X_t^{jq}\right\rangle
\left\langle X_t^{jq},\mathcal P_M(Q)\right\rangle+
\sum_{j=i}^{d_1}\sum_{q\in h_{jt}}
\mathbb E\!\left[
\left\langle\widehat\Delta_1,X_t^{jq}\right\rangle
\left\langle X_t^{jq},\mathcal P_M(Q)\right\rangle
\,\middle|\,\mathcal F_{i-1}
\right]
=R_{i-1},
\end{aligned}
\]then
\[
\begin{aligned}
R_i-R_{i-1}&=
\sum_{q\in h_{it}}
\left\langle\widehat\Delta_1,X_t^{iq}\right\rangle
\left\langle X_t^{iq},\mathcal P_M(Q)\right\rangle-
\mathbb E\!\left[
\left\langle\widehat\Delta_1,X_t^{iq}\right\rangle
\left\langle X_t^{iq},\mathcal P_M(Q)\right\rangle
\,\middle|\,\mathcal F_{i-1}
\right] \\ &\quad +
\left(
\sum_{j=i+1}^{d_1}\sum_{q\in h_{jt}}
\mathbb E\!\left[
\left\langle\widehat\Delta_1,X_t^{jq}\right\rangle
\left\langle X_t^{jq},\mathcal P_M(Q)\right\rangle
\,\middle|\,\mathcal F_i
\right]-
\mathbb E\!\left[
\left\langle\widehat\Delta_1,X_t^{jq}\right\rangle
\left\langle X_t^{jq},\mathcal P_M(Q)\right\rangle
\,\middle|\,\mathcal F_{i-1}
\right]
\right).
\end{aligned}
\]Obviously, $\mathbb E[R_i-R_{i-1}\mid\mathcal F_{i-1}]=0$ and by $(a+b)^2\leq2(a^2+b^2)$,
\[
\begin{aligned}
&\operatorname{Var}[R_i\mid\mathcal F_{i-1}]
=
\mathbb E\!\left[
\left(R_i-\mathbb E[R_i\mid\mathcal F_{i-1}]\right)^2
\,\middle|\,\mathcal F_{i-1}
\right]=
\mathbb E[(R_i-R_{i-1})^2\mid\mathcal F_{i-1}]\\
&\leq
2\mathbb E\!\left[
\left(
\sum_{q\in h_{it}}
\left\langle X_t^{iq},\widehat\Delta_1\right\rangle
\left\langle X_t^{iq},\mathcal P_M(Q)\right\rangle
\right)^2
\,\middle|\,\mathcal F_{i-1}
\right]\\
&\quad+
2\mathbb E\!\left[
\left(
\sum_{j=i+1}^{d_1}\sum_{q\in h_{jt}}
\mathbb E\!\left[
\left\langle\widehat\Delta_1,X_t^{jq}\right\rangle
\left\langle X_t^{jq},\mathcal P_M(Q)\right\rangle
\,\middle|\,\mathcal F_i
\right]-
\mathbb E\!\left[
\left\langle\widehat\Delta_1,X_t^{jq}\right\rangle
\left\langle X_t^{jq},\mathcal P_M(Q)\right\rangle
\,\middle|\,\mathcal F_{i-1}
\right]
\right)^2
\,\middle|\,\mathcal F_{i-1}
\right].
\end{aligned}
\]By definition of $\mathcal F_i$ and $\mathcal M_{K,p}^{\otm}$, condition on $\mathcal F_i$, $|h_{jq}|\sim B(K,p)$ and
$
h_{jt}\subset[d_2]/\{h_{1t},\ldots,h_{it}\}
$ for any $i<j\leq d_1$. Therefore, denote
$
m_{it}=\sum_{j=1}^{i}|h_{jt}|,
$ we have
\[
\begin{aligned}
&\left|
\sum_{q\in h_{jt}}
\mathbb E\!\left[
\left\langle\widehat\Delta_1,X_t^j\right\rangle
\left\langle X_t^j,\mathcal P_M(Q)\right\rangle
\,\middle|\,\mathcal F_i
\right]
-
\mathbb E\!\left[
\left\langle\widehat\Delta_1,X_t^j\right\rangle
\left\langle X_t^j,\mathcal P_M(Q)\right\rangle
\,\middle|\,\mathcal F_{i-1}
\right]
\right|\\
&=
\left|
\sum_{s=0}^{K}
\sum_{\substack{
\Omega\subset[d_2]/\{h_{1t},\ldots,h_{it}\}\\
|\Omega|=s}}
\sum_{q\in\Omega}
\left\langle e_je_q^\top,\widehat\Delta_1\right\rangle
\left\langle e_je_q^\top,\mathcal P_M(Q)\right\rangle
C_K^sp_0^s(1-p_0)^{K-s}
\frac{1}{C_{d_2-m_{it}}^s}
\right.\\
&\hspace{10mm}\left.-
\sum_{s=0}^{K}
\sum_{\substack{
\Omega\subset[d_2]/\{h_{1t},\ldots,h_{i-1,t}\}\\
|\Omega|=s}}
\sum_{q\in\Omega}
\left\langle e_je_q^\top,\widehat\Delta_1\right\rangle
\left\langle e_je_q^\top,\mathcal P_M(Q)\right\rangle
C_K^sp_0^s(1-p_0)^{K-s}
\frac{1}{C_{d_2-m_{i-1,t}}^s}
\right|\\
&\leq
\left|
\sum_{s=0}^{K}
\sum_{\substack{
\Omega\subset[d_2]/\{h_{1t},\ldots,h_{it}\}\\
|\Omega|=s}}
\sum_{q\in\Omega}
\left\langle e_je_q^\top,\widehat\Delta_1\right\rangle
\left\langle e_je_q^\top,\mathcal P_M(Q)\right\rangle
C_K^sp_0^s(1-p_0)^{K-s}
\left(
\frac{1}{C_{d_2-m_{it}}^s}
-\frac{1}{C_{d_2-m_{i-1,t}}^s}
\right)
\right|\\
&\quad+
\left|
\sum_{s=0}^{K}
\sum_{\substack{
\Omega\subset[d_2]/\{h_{1t},\ldots,h_{i-1,t}\}\\
|\Omega|=s,\ \Omega\cap h_{it}\neq\emptyset}}
\sum_{q\in\Omega}
\left\langle e_je_q^\top,\widehat\Delta_1\right\rangle
\left\langle e_je_q^\top,\mathcal P_M(Q)\right\rangle
C_K^sp_0^s(1-p_0)^{K-s}
\frac{1}{C_{d_2-m_{i-1,t}}^s}
\right|.
\end{aligned}
\]We first analyze the first term. For a fixed $s\geq1$, by Cauchy Schwarz inequality,
\[
\begin{aligned}
&\left|
\sum_{\substack{
\Omega\subset[d_2]/\{h_{1t},\ldots,h_{it}\}\\
|\Omega|=s}}
\sum_{q\in\Omega}
\left\langle e_je_q^\top,\widehat\Delta_1\right\rangle
\left\langle e_je_q^\top,\mathcal P_M(Q)\right\rangle
\right| \\ &\quad \leq
C_{d_2-m_{it}-1}^{s-1}
\sqrt{d_2-m_{it}}\,
\|\widehat\Delta_1\|_{\max}
\sqrt{
\sum_{q\in[d_2]/\{h_{1t},\ldots,h_{it}\}}
\left\langle e_je_q^\top,\mathcal P_M(Q)\right\rangle^2
}\\
&\leq
C_{d_2-m_{it}-1}^{s-1}
\sqrt{d_2-m_{it}}\,
\|\widehat\Delta_1\|_{\max}
\|\mathcal P_M(Q)(j,)\|.
\end{aligned}
\]By simple calculation,
\[
\begin{aligned}
C_{d_2-m_{it}-1}^{s-1}
\sqrt{d_2-m_{it}}
\left(
\frac{1}{C_{d_2-m_{it}}^s}
-\frac{1}{C_{d_2-m_{i-1,t}}^s}
\right)&=
\frac{s}{\sqrt{d_2-m_{it}}}
\left(
1-\prod_{l=0}^{s}
\frac{d_2-m_{it}-l}{d_2-m_{i-1,t}-l}
\right)\\
&\leq
\frac{s}{\sqrt{d_2-m_{it}}}
\left(
1-
\left(
\frac{d_2-m_{it}-s+1}
     {d_2-m_{i-1,t}-s+1}
\right)^s
\right).
\end{aligned}
\]When $d_2\geq(1+\gamma)Kd_1$ and $d_1\gg K$, by $1-r^s\leq s(1-r)$ for any $s\geq1$ and $r\in[0,1]$,
\[
\left(
\frac{d_2-m_{it}-s+1}
     {d_2-m_{i-1,t}-s+1}
\right)^s
\leq
s\left(
1-
\frac{d_2-m_{it}-s+1}
     {d_2-m_{i-1,t}-s+1}
\right)
\leq
\frac{sK}{\gamma d_1}.
\]As a result,
\[
\begin{aligned}
&\left|
\sum_{s=0}^{K}
\sum_{\substack{
\Omega\subset[d_2]/\{h_{1t},\ldots,h_{it}\}\\
|\Omega|=s}}
\sum_{q\in\Omega}
\left\langle e_je_q^\top,\widehat\Delta_1\right\rangle
\left\langle e_je_q^\top,\mathcal P_M(Q)\right\rangle
C_K^sp_0^s(1-p_0)^{K-s}
\left(
\frac{1}{C_{d_2-m_{it}}^s}
-\frac{1}{C_{d_2-m_{i-1,t}}^s}
\right)
\right|\\
&\leq
\frac{K}{\gamma d_1}
\frac{1}{\sqrt{d_2-m_{it}}}
\|\widehat\Delta_1\|_{\max}
\|\mathcal P_M(Q)(j,)\|.
\end{aligned}
\]We then analyze the second term. Again, for any fixed $s\geq1$,
\[
\begin{aligned}
&\left|
\sum_{\substack{
\Omega\subset[d_2]/\{h_{1t},\ldots,h_{i-1,t}\}\\
|\Omega|=s,\ \Omega\cap h_{it}\neq\emptyset}}
\sum_{q\in\Omega}
\left\langle e_je_q^\top,\widehat\Delta_1\right\rangle
\left\langle e_je_q^\top,\mathcal P_M(Q)\right\rangle
\right|\\
&\leq
\frac{1}{C_{d_2-m_{i-1,t}}^s}
\sum_{g=1}^{\min(|h_{it}|,s-1)}
\sum_{\substack{\widetilde\Omega_1\subset h_{it}\\
|\widetilde\Omega_1|=g}}
\sum_{q\in\widetilde\Omega_1}
\left|
\left\langle e_je_q^\top,\widehat\Delta_1\right\rangle
\left\langle e_je_q^\top,\mathcal P_M(Q)\right\rangle
\right|\\
&\quad+
\sum_{\substack{
\widetilde\Omega_2\subset[d_2]/\{h_{1t},\ldots,h_{it}\}\\
|\widetilde\Omega_2|=s-g}}
\sum_{q\in\widetilde\Omega_2}
\left|
\left\langle e_je_q^\top,\widehat\Delta_1\right\rangle
\left\langle e_je_q^\top,\mathcal P_M(Q)\right\rangle
\right|\\
&\quad+
\frac{1}{C_{d_2-m_{i-1,t}}^s}
\sum_{\substack{\widetilde\Omega\subset h_{it}\\
|\widetilde\Omega|=s}}
\mathbf 1(|h_{it}|\geq s)
\sum_{q\in\widetilde\Omega}
\left|
\left\langle e_je_q^\top,\widehat\Delta_1\right\rangle
\left\langle e_je_q^\top,\mathcal P_M(Q)\right\rangle
\right|.
\end{aligned}
\]Using Cauchy Schwarz inequality similarly,
\[
\begin{aligned}
&\frac{1}{C_{d_2-m_{i-1,t}}^s}
\sum_{g=1}^{\min(|h_{it}|,s-1)}
\sum_{\substack{
\widetilde\Omega_2\subset[d_2]/\{h_{1t},\ldots,h_{it}\}\\
|\widetilde\Omega_2|=s-g}}
\sum_{q\in\widetilde\Omega_2}
\left|
\left\langle e_je_q^\top,\widehat\Delta_1\right\rangle
\left\langle e_je_q^\top,\mathcal P_M(Q)\right\rangle
\right|\\
&=
\frac{1}{C_{d_2-m_{i-1,t}}^s}
\sum_{g=1}^{\min(|h_{it}|,s-1)}
C_{d_2-m_{it}-1}^{s-g-1}
\sum_{q\in[d_2]/\{h_{1t},\ldots,h_{it}\}}
\left|
\left\langle e_je_q^\top,\widehat\Delta_1\right\rangle
\left\langle e_je_q^\top,\mathcal P_M(Q)\right\rangle
\right|\\
&\lesssim
\frac{K}{(d_2-m_{it})^2}
\sqrt{d_2-m_{it}}\,
\|\widehat\Delta_1\|_{\max}
\|\mathcal P_M(Q)(j,)\|,
\end{aligned}
\]where the last inequality comes from
\[
C_{d_2-m_{it}-1}^{s-g-1}/C_{d_2-m_{i-1,t}}^s
\lesssim 1/(d_2-m_{it})^2
\]when $d_1\gg K+s$. Moreover,
\[
\begin{aligned}
&\frac{1}{C_{d_2-m_{i-1,t}}^s}
\left(
\sum_{g=1}^{\min(|h_{it}|,s-1)}
\sum_{\substack{\widetilde\Omega_1\subset h_{it}\\
|\widetilde\Omega_1|=g}}
\sum_{q\in\widetilde\Omega_1}
\left|
\left\langle e_je_q^\top,\widehat\Delta_1\right\rangle
\left\langle e_je_q^\top,\mathcal P_M(Q)\right\rangle
\right| \right.\\ &\left.\quad\quad +
\sum_{\substack{\widetilde\Omega\subset h_{it}\\
|\widetilde\Omega|=s}}
\mathbf 1(|h_{it}|\geq s)
\sum_{q\in\widetilde\Omega}
\left|
\left\langle e_je_q^\top,\widehat\Delta_1\right\rangle
\left\langle e_je_q^\top,\mathcal P_M(Q)\right\rangle
\right|
\right)\\
&=
\frac{1}{C_{d_2-m_{i-1,t}}^s}
\sum_{g=1}^{\min(|h_{it}|,s)}
C_{|h_{it}|-1}^{g-1}
\sum_{q\in h_{it}}
\left|
\left\langle e_je_q^\top,\widehat\Delta_1\right\rangle
\left\langle e_je_q^\top,\mathcal P_M(Q)\right\rangle
\right|\\
&\lesssim
\frac{1}{d_2-m_{i-1,t}}
\sum_{q\in h_{it}}
\left|
\left\langle e_je_q^\top,\widehat\Delta_1\right\rangle
\left\langle e_je_q^\top,\mathcal P_M(Q)\right\rangle
\right|.
\end{aligned}
\]Combine both terms, we have
\[
\begin{aligned}
&\left|
\sum_{s=0}^{K}
\sum_{\substack{
\Omega\subset[d_2]/\{h_{1t},\ldots,h_{i-1,t}\}\\
|\Omega|=s,\ \Omega\cap h_{it}\neq\emptyset}}
\sum_{q\in\Omega}
\left\langle e_je_q^\top,\widehat\Delta_1\right\rangle
\left\langle e_je_q^\top,\mathcal P_M(Q)\right\rangle
C_K^sp_0^s(1-p_0)^{K-s}
\frac{1}{C_{d_2-m_{i-1,t}}^s}
\right|\\
&\lesssim
\frac{K}{(d_2-m_{it})^{3/2}}
\|\widehat\Delta_1\|_{\max}
\|\mathcal P_M(Q)(j,)\|\\
&\quad+
\frac{1}{d_2-m_{i-1,t}}
\sum_{q\in h_{it}}
\left|
\left\langle e_je_q^\top,\widehat\Delta_1\right\rangle
\left\langle e_je_q^\top,\mathcal P_M(Q)\right\rangle
\right|.
\end{aligned}
\]By Cauchy Schwarz inequality,
\[
\begin{aligned}
&\mathbb E\!\left[
\left(
\sum_{j=i+1}^{d_1}\sum_{q\in h_{jt}}
\mathbb E\!\left[
\left\langle\widehat\Delta_1,X_t^{jq}\right\rangle
\left\langle X_t^{jq},\mathcal P_M(Q)\right\rangle
\,\middle|\,\mathcal F_i
\right]-
\mathbb E\!\left[
\left\langle\widehat\Delta_1,X_t^{jq}\right\rangle
\left\langle X_t^{jq},\mathcal P_M(Q)\right\rangle
\,\middle|\,\mathcal F_{i-1}
\right]
\right)^2
\,\middle|\,\mathcal F_{i-1}
\right]\\
&\lesssim
\frac{3K^2d_1}{\gamma^2d_1^2(d_2-m_{it})}
\|\widehat\Delta_1\|_{\max}^2
\sum_{j=i+1}^{d_1}\|\mathcal P_M(Q)(j,)\|^2+
\frac{3K^2d_1}{(d_2-m_{it})^3}
\|\widehat\Delta_1\|_{\max}^2
\sum_{j=i+1}^{d_1}\|\mathcal P_M(Q)(j,)\|^2\\
&\quad+
\frac{3d_1}{(d_2-m_{i-1,t})^2}
\sum_{j=i+1}^{d_1}
\mathbb E\!\left[
\left(
\sum_{q\in h_{it}}
\left\langle e_je_q^\top,\widehat\Delta_1\right\rangle
\left\langle e_je_q^\top,\mathcal P_M(Q)\right\rangle
\right)^2
\,\middle|\,\mathcal F_{i-1}
\right].
\end{aligned}
\]Since condition on $\mathcal F_{i-1}$, $|h_{it}|\sim B(K,p)$ and
$
h_{it}\subset[d_2]/\{h_{1t},\ldots,h_{i-1,t}\},$ then by Cauchy Schwarz inequality,
\[
\begin{aligned}
&\mathbb E\!\left[
\left(
\sum_{q\in h_{it}}
\left\langle e_je_q^\top,\widehat\Delta_1\right\rangle
\left\langle e_je_q^\top,\mathcal P_M(Q)\right\rangle
\right)^2
\,\middle|\,\mathcal F_{i-1}
\right]\\
&\leq
K\mathbb E\!\left[
\sum_{q\in h_{it}}
\left\langle e_je_q^\top,\widehat\Delta_1\right\rangle^2
\left\langle e_je_q^\top,\mathcal P_M(Q)\right\rangle^2
\,\middle|\,\mathcal F_{i-1}
\right]\\
&\leq
K\|\widehat\Delta_1\|_{\max}^2
\sum_{s=0}^{K}
\sum_{\substack{
\Omega\subset[d_2]/\{h_{1t},\ldots,h_{i-1,t}\}\\
|\Omega|=s}}
\sum_{q\in\Omega}
\left\langle e_je_q^\top,\mathcal P_M(Q)\right\rangle^2
C_K^sp_0^s(1-p_0)^{K-s}
\frac{1}{C_{d_2-m_{i-1,t}}^s}\\
&\leq
K\|\widehat\Delta_1\|_{\max}^2
\sum_{s=0}^{K}
C_{d_2-m_{i-1,t}-1}^{s-1}
\frac{1}{C_{d_2-m_{i-1,t}}^s}
C_K^sp_0^s(1-p_0)^{K-s}
\|\mathcal P_M(Q)(j,)\|^2\\
&\leq
\frac{K^2p_0}{d_2-m_{i-1,t}}
\|\widehat\Delta_1\|_{\max}^2
\|\mathcal P_M(Q)(j,)\|^2.
\end{aligned}
\]By $d_2\geq\gamma(1+K)d_1$, and recall
$
\|\mathcal P_M(Q)\|_{\max}
\lesssim
\frac{1}{\sqrt{d_1}}\|\mathcal P_M(Q)\|_{\mathrm F}
$ and
$
\sum_{j=i+1}^{d_1}\|\mathcal P_M(Q)(j,)\|^2
\leq
\|\mathcal P_M(Q)\|_{\mathrm F}^2,
$ we have
\[
\begin{aligned}
&\mathbb E\!\left[
\left(
\sum_{j=i+1}^{d_1}\sum_{q\in h_{jt}}
\mathbb E\!\left[
\left\langle\widehat\Delta_1,X_t^{jq}\right\rangle
\left\langle X_t^{jq},\mathcal P_M(Q)\right\rangle
\,\middle|\,\mathcal F_i
\right]-
\mathbb E\!\left[
\left\langle\widehat\Delta_1,X_t^{jq}\right\rangle
\left\langle X_t^{jq},\mathcal P_M(Q)\right\rangle
\,\middle|\,\mathcal F_{i-1}
\right]
\right)^2
\,\middle|\,\mathcal F_{i-1}
\right]\\
&\lesssim
\frac{K^2}{d_1^2}
\|\widehat\Delta_1\|_{\max}^2
\|\mathcal P_M(Q)\|_{\mathrm F}^2.
\end{aligned}
\]Similarly, we can derive
\[
\begin{aligned}
\mathbb E\!\left[
\left(
\sum_{q\in h_{it}}
\left\langle X_t^{iq},\widehat\Delta_1\right\rangle
\left\langle X_t^{iq},\mathcal P_M(Q)\right\rangle
\right)^2
\,\middle|\,\mathcal F_{i-1}
\right]&\leq
\frac{K^2p_0}{d_2-m_{i-1,t}}
\|\widehat\Delta_1\|_{\max}^2
\|\mathcal P_M(Q)(i,)\|^2\\ &\lesssim
\frac{K^2}{d_1}
\|\widehat\Delta_1\|_{\max}^2
\|\mathcal P_M(Q)(i,)\|^2.
\end{aligned}
\]Combine all the terms above, we have
\[
\sum_{i=1}^{d_1}
\operatorname{Var}[R_i\mid\mathcal F_{i-1}]
\lesssim
\frac{K^2}{d_1}
\|\widehat\Delta_1\|_{\max}^2
\|\mathcal P_M(Q)\|_{\mathrm F}^2.
\]Moreover,
$
\mathbb E[R]
=
\frac{Kp_0}{d_2}
\left\langle\widehat\Delta_1,\mathcal P_M(Q)\right\rangle
$, and for any $i$, following the previous analysis,
\[
\begin{aligned}
&|R_i-R_{i-1}|
\leq
2\left|
\sum_{q\in h_{it}}
\left\langle\widehat\Delta_1,X_t^{iq}\right\rangle
\left\langle X_t^{iq},\mathcal P_M(Q)\right\rangle
\right|\\
&\quad+
\left|
\sum_{j=i+1}^{d_1}\sum_{q\in h_{jt}}
\mathbb E\!\left[
\left\langle\widehat\Delta_1,X_t^{jq}\right\rangle
\left\langle X_t^{jq},\mathcal P_M(Q)\right\rangle
\,\middle|\,\mathcal F_i
\right] -
\mathbb E\!\left[
\left\langle\widehat\Delta_1,X_t^{jq}\right\rangle
\left\langle X_t^{jq},\mathcal P_M(Q)\right\rangle
\,\middle|\,\mathcal F_{i-1}
\right]
\right|\\
&\leq
2K\|\widehat\Delta_1\|_{\max}
\|\mathcal P_M(Q)\|_{\max}\\
&\quad+
\sum_{j=i+1}^{d_1}\sum_{s=0}^{K}
\left|
\sum_{\substack{
\Omega\subset[d_2]/\{h_{1t},\ldots,h_{it}\}\\
|\Omega|=s}}
\sum_{q\in\Omega}
\left\langle e_je_q^\top,\widehat\Delta_1\right\rangle
\left\langle e_je_q^\top,\mathcal P_M(Q)\right\rangle
C_K^sp_0^s(1-p_0)^{K-s}
\left(
\frac{1}{C_{d_2-m_{it}}^s}
-\frac{1}{C_{d_2-m_{i-1,t}}^s}
\right)
\right|\\
&\quad+
\sum_{s=0}^{K}
\left|
\sum_{\substack{
\Omega\subset[d_2]/\{h_{1t},\ldots,h_{i-1,t}\}\\
|\Omega|=s,\ \Omega\cap h_{it}\neq\emptyset}}
\sum_{q\in\Omega}
\left\langle e_je_q^\top,\widehat\Delta_1\right\rangle
\left\langle e_je_q^\top,\mathcal P_M(Q)\right\rangle
C_K^sp_0^s(1-p_0)^{K-s}
\frac{1}{C_{d_2-m_{i-1,t}}^s}
\right|\\
&\lesssim
2K\|\widehat\Delta_1\|_{\max}
\|\mathcal P_M(Q)\|_{\max}\\
&\quad+
\sum_{j=i+1}^{d_1}\sum_{s=0}^{K}
s\|\widehat\Delta_1\|_{\max}
\|\mathcal P_M(Q)\|_{\max}
C_{d_2-m_{it}-1}^{s-1}
C_K^sp_0^s(1-p_0)^{K-s}
\left(
\frac{1}{C_{d_2-m_{it}}^s}
-\frac{1}{C_{d_2-m_{i-1,t}}^s}
\right)\\
&\quad+
\frac{K}{d_2-m_{i-1,t}}
\|\widehat\Delta_1\|_{\max}
\|\mathcal P_M(Q)\|_{\max}\\
&\lesssim
\frac{K}{\sqrt{d_1}}
\|\widehat\Delta_1\|_{\max}
\|\mathcal P_M(Q)\|_{\mathrm F}.
\end{aligned}
\]Finally, by martingale Bernstein inequality (Theorem 18 in \cite{chung2006concentration}), with probability at least $1-d_1^{-10}$,
\[
\left|
\sum_{i=1}^{d_1}\sum_{q\in h_{it}}
\left\langle\widehat\Delta_1,X_t^{iq}\right\rangle
\left\langle X_t^{iq},\mathcal P_M(Q)\right\rangle
-
\frac{Kp_0}{d_2}
\left\langle\widehat\Delta_1,\mathcal P_M(Q)\right\rangle
\right|
\lesssim
\frac{K}{\sqrt{d_1}}
\|\widehat\Delta_1\|_{\max}
\|\mathcal P_M(Q)\|_{\mathrm F}
\log d_1.
\]Then
\[
\begin{aligned}
&\frac{1}{T_0}
\left|
\frac{d_2}{Kp_0}
\sum_{i=1}^{d_1}\sum_{q\in h_{it}}
\left\langle\widehat\Delta_1,X_t^{iq}\right\rangle
\left\langle X_t^{iq},\mathcal P_M(Q)\right\rangle
-
\left\langle\widehat\Delta_1,\mathcal P_M(Q)\right\rangle
\right|\\
&\lesssim
\frac{d_2K}{TKp_0\sqrt{d_1}}
\log d_1
\|\widehat\Delta_1\|_{\max}
\|\mathcal P_M(Q)\|_{\mathrm F}\\
&\lesssim
\sqrt{\frac{r^2d_2^3}{T^3d_1^2Kp_0^3}}
\log^2d_1\,
\sigma\|\mathcal P_M(Q)\|_{\mathrm F},
\end{aligned}
\]where the last inequality comes from the result in Proposition \ref{thm:estimation2}. 

    Next we calculate the variance of $\frac{1}{T_0}\bigg|\frac{d_2}{Kp}\sum_{i=1}^{d_1}\sum_{q\in h_{it}}\inp{\widehat{\Delta}_1}{X_t^{iq}}\inp{X_t^{iq}}{\PMQ} -\inp{\widehat{\Delta}_1}{\PMQ}\bigg|$. Notice that $\widehat{\Delta}_1$ is independent of $X_t^i$, then
    \begin{align*}
        &\frac{1}{T_0^2}\EE\left[\bigg(\frac{d_2}{Kp_0}\sum_{i=1}^{d_1}\sum_{q\in h_{it}} \inp{\widehat{\Delta}_1}{X_t^{iq}}\inp{X_t^{iq}}{\PMQ} -\inp{\widehat{\Delta}_1}{\PMQ} \bigg)^2\right] \\
        &\quad = \frac{1}{T_0^2}\EE\left[\bigg(\frac{d_2}{Kp_0}\sum_{i=1}^{d_1}\sum_{q\in h_{it}}\inp{\widehat{\Delta}_1}{X_t^{iq}}\inp{X_t^{iq}}{\PMQ} \bigg)^2\right]-\frac{1}{T_0^2}\inp{\widehat{\Delta}_1}{\PMQ}^2 \\
        &\quad = \frac{d_2^2}{T_0^2K^2p_0^2}\underbrace{\EE\left[\sum_{i=1}^{d_1}\bigg(\sum_{q\in h_{it}}\inp{\widehat{\Delta}_1}{X_t^{iq}}\inp{X_t^{iq}}{\PMQ}\bigg)^2 \right]}_{\mathcal{I}_1}  - \frac{1}{T_0^2}\inp{\widehat{\Delta}_1}{\PMQ}^2 \\ &\quad\quad + \frac{d_2^2}{T_0^2K^2p_0^2}\underbrace{\EE\left[\sum_{i=1}^{d_1}\sum_{j\neq i} \bigg(\sum_{q\in h_{it}} \inp{\widehat{\Delta}_1}{X_t^{iq}}\inp{X_t^{iq}}{\PMQ}\bigg) \bigg(\sum_{r\in h_{jt}}\inp{\widehat{\Delta}_1}{X_t^{jr}}\inp{X_t^{jr}}{\PMQ}\bigg) \right]}_{\mathcal{I}_2}.
    \end{align*}
    Denote $A=\widehat{\Delta}_1\odot \PMQ$ the Hadamard product of $\widehat{\Delta}_1$ and $\PMQ$, then 
    \begin{align*}
        \mathcal{I}_1\leq K\EE[\sum_{i=1}^{d_1}\sum_{q\in h_{it}}\inp{X_t^i}{A}^2]=\frac{K^2p_0}{d_2}\fro{A}^2\leq \frac{K^2p_0}{d_2}\|\widehat{\Delta}_1\|_{\max}^2\fro{\PMQ}^2.
    \end{align*}
    It is obvious that 
    \begin{align*}
        |\mathcal{I}_2|\leq \frac{K^2p_0^2}{d_2(d_2-K)}\bigg|\sum_{i\neq j}\sum_{p\neq q} A(i,p)A(j,q)\bigg|.
    \end{align*}
    As shown in the proof of Theorem \ref{thm:inference-main},
    \begin{align*}
        \frac{d_2^2}{K^2p_0^2}|\mathcal{I}_2|\leq \frac{d_2}{d_2-K}\inp{\widehat{\Delta}_1}{\PMQ}^2 + d_2\|\widehat{\Delta}_1\|_{\max}^2\fro{\PMQ}^2.
    \end{align*}
   Overall, the variance can be upper bounded by
   \begin{align*}
        &\frac{1}{T_0^2}\EE\left[\bigg(\frac{d_2}{Kp_0}\sum_{i=1}^{d_1}\sum_{q\in h_{it}}\inp{\widehat{\Delta}_1}{X_t^i}\inp{X_t^i}{\PMQ} -\inp{\widehat{\Delta}_1}{\PMQ} \bigg)^2\right]\leq \frac{d_2}{T_0^2p_0}\|\widehat{\Delta}_1\|_{\max}^2\fro{\PMQ}^2 \\ &\quad\quad + \bigg(\frac{d_2}{T_0^2(d_2-K)}-\frac{1}{T_0^2}\bigg)\inp{\widehat{\Delta}_1}{\PMQ}^2 + \frac{d_2^2}{T_0^2(d_2-K)}\|\widehat{\Delta}_1\|_{\max}^2\fro{\PMQ}^2 \\ 
        &\quad\leq \frac{3d_2}{T_0^2p_0}\|\widehat{\Delta}_1\|_{\max}^2\fro{\PMQ}^2 + \frac{2K}{T_0^2d_2}d_1d_2\|\widehat{\Delta}_1\|_{\max}^2\fro{\PMQ}^2\leq \frac{5d_2}{T_0^2p_0}\|\widehat{\Delta}_1\|_{\max}^2\fro{\PMQ}^2
   \end{align*}
   where the last inequality comes from $d_2\geq (1+\gamma)Kd_1$. Combine with Proposition \ref{thm:estimation2} and the previous uniform bound, by Bernstein inequality, with probability at least $1-2d_1^{-10}$, 
   \begin{align*}
        &\left|\frac{1}{T_0}\sum_{t=1}^{T_0} \sum_{i=1}^{d_1}\sum_{q\in h_{it}} \frac{d_2}{Kp_0}\inp{\widehat{\Delta}_1}{X_t^i}\inp{X_t^i}{\PMQ} -\inp{\widehat{\Delta}_1}{\PMQ}\right|\\ &\quad \lesssim \sqrt{\frac{r^2d_2^3}{T^3d_1^2Kp_0^3}}\log^3d_1\sigma\fro{\PMQ} + \sqrt{\frac{r^2d_2^2\log^3 d_1}{T^2d_1Kp_0^2}}\sigma\fro{\PMQ}.
   \end{align*}
    As a result, 
    \begin{align*}
        &\frac{|\frac{1}{2}\sum_{l=1}^{2}\inp{UU^{\top}\widehat{Z}_2^{(l)}V_{\perp}V_{\perp}^{\top}}{Q} + \inp{U_{\perp}U_{\perp}^{\top}\widehat{Z}_2^{(l)}VV^{\top}}{Q} + \inp{UU^{\top}\widehat{Z}_2^{(l)}VV^{\top}}{Q}|}{\sigma\fro{\PMQ}\sqrt{d_2/TKp_0}} \\
        &\quad \lesssim \sqrt{\frac{r^2d_2^2}{T^2d_1^2p_0^2}}\log^3d_1 + \sqrt{\frac{r^2d_2\log^3d_1}{Td_1p_0}}\lesssim \sqrt{\frac{r^2\alpha_d^2 \log^3d_1}{TKp_0}}, 
    \end{align*}
    where in the last inequality we use the fact that $\alpha_d\geq (1+\gamma)K$. 
    
    Finally, following the proof of Lemma \ref{lemmaneg2} and Lemma \ref{lemmaneg3}, as long as we can derive the upper bound of $\|\widehat{Z}^{(l)}\|$, $\|\widehat{U}_l\widehat{U}_l^{\top}-UU^{\top}\|_{2,\max}$ and $\|\widehat{V}_l\widehat{V}_l^{\top}-VV^{\top}\|_{2,\max}$, then we can use them to bound (\ref{neg2-otm}) and (\ref{neg3-otm}). The next two Lemmas show the upper bound of $\|\widehat{Z}^{(l)}\|$, $\|\widehat{U}_l\widehat{U}_l^{\top}-UU^{\top}\|_{2,\max}$ and $\|\widehat{V}_l\widehat{V}_l^{\top}-VV^{\top}\|_{2,\max}$. 

    \begin{Lemma}
        \label{Zbound-otm}
        Under the conditions in Theorem \ref{thm:inference-main-2}, with probability at least $1-2d_1^{-10}$, 
        \begin{align*}
        \|\widehat{Z}^{(l)}\|\lesssim \sqrt{\frac{d_2^2\log d_1}{TKp_0}}\sigma
        \end{align*}
        for $l=1,2$. 
    \end{Lemma}

    \begin{Lemma}
        \label{2maxbound-otm}
        Under the conditions in Theorem \ref{thm:inference-main-2}, with probability at least $1-5d_1^{-5}$,
        \begin{align*}
            &\|\widehat{U}_l\widehat{U}_l^{\top}-UU^{\top}\|_{2,\max}\lesssim \frac{\sigma}{\lambda_{\min}}\sqrt{\frac{d_2^2\log d_1}{TKp_0}}\mu\sqrt{\frac{r}{d_1}} \\
            &\|\widehat{V}_l\widehat{V}_l^{\top}-VV^{\top}\|_{2,\max}\lesssim \frac{\sigma}{\lambda_{\min}}\sqrt{\frac{d_2^2\log d_1}{TKp_0}}\mu\sqrt{\frac{r}{d_2}}. 
        \end{align*}
        for $l=1,2$. 
    \end{Lemma}

    As a result, we can show that, 
    \begin{align*}
        \frac{\big|\frac{1}{2}\sum_{l=1}^{2}\inp{\widehat{U}_l\widehat{U}_l^{\top}\widehat{Z}^{(l)}\widehat{V}_l\widehat{V}_l^{\top}}{Q} - \inp{UU^{\top}\widehat{Z}^{(l)}VV^{\top}}{Q}\big|}{\sigma\fro{\mathcal{P}_M(Q)}\sqrt{d_2/TKp_0}}\lesssim \frac{\sigma\mu^2}{\lambda_{\min}}\sqrt{\frac{r^2d_2^2\log d_1}{Td_1Kp_0}}\frac{\|Q\|_{\ell_1}}{\fro{\PMQ}}. 
    \end{align*}
    and 
    \begin{align*}
        &\frac{\big|\sum_{k=2}^{\infty} \inp{(\mathcal{S}_{A,k}(\widehat{E})A\Theta\Theta^{\top}+\Theta\Theta^{\top}A\mathcal{S}_{A,k}(\widehat{E}))}{\widetilde{Q}}\big|}{\sigma\fro{\mathcal{P}_M(Q)}\sqrt{d_2/TKp_0}}\lesssim \frac{\sigma\mu^2}{\lambda_{\min}}\sqrt{\frac{r^2d_2^2\log d_1}{Td_1Kp_0}}\frac{\|Q\|_{\ell_1}}{\fro{\PMQ}}, \\
        &\frac{\big|\inp{(\widehat{\Theta}\widehat{\Theta}^{\top}-\Theta\Theta^{\top})A(\widehat{\Theta}\widehat{\Theta}^{\top}-\Theta\Theta^{\top})}{\widetilde{Q}}\big|}{\sigma\fro{\mathcal{P}_M(Q)}\sqrt{d_2/TKp_0}} \lesssim \frac{\sigma\mu^2\kappa}{\lambda_{\min}}\sqrt{\frac{r^2d_2^2\log d_1}{Td_1Kp_0}}\frac{\|Q\|_{\ell_1}}{\fro{\PMQ}}.
    \end{align*}
    Then we conclude the proof. 
\end{proof}

\subsection{Proof of Theorem \ref{thm:inference-main-3}}
\begin{proof}
    We start by recalling that $\widehat{M}_l^{\ubs}$ for $l=1,2$ can be decomposed into 
    \begin{align*}
        \widehat{M}_l^{\ubs}= M + \underbrace{\frac{1}{T_0\nu}\sum_{t\in \widetilde{\mathcal{D}}_{l}}\sum_{q\in h_{it}} \xi_t^{iq} X_t^{iq}}_{=:\widehat{Z}_1^{(l)}} + \underbrace{\frac{1}{T_0}\sum_{t\in \widetilde{\mathcal{D}}_{l}} \left(\sum_{(i,q)\in h_{it}} \frac{1}{\nu}\inp{\widehat{\Delta}_{l}}{X_t^{iq}}X_t^{iq}-\widehat{\Delta}_{l}\right)}_{=:\widehat{Z}_2^{(l)}},
    \end{align*}
    where $\widehat{\Delta}_{l}=\widehat{M}_l^{\init}-M$. Following the proof of Theorem \ref{thm:inference-main} and define $\Theta$, $\widehat{\Theta}_l$, $\widehat{E}^{(l)}$, $\mathfrak{P}^{-s}$ and all other variables in the same way, we again can obtain
    \begin{align}
        \inp{\widehat{M}}{Q}-&\inp{M}{Q}= \frac{1}{2}\sum_{l=1}^{2} \inp{UU^{\top}\widehat{Z}_1^{(l)}V_{\perp}V_{\perp}^{\top}}{Q} + \inp{U_{\perp}U_{\perp}^{\top}\widehat{Z}_1^{(l)}VV^{\top}}{Q} + \inp{UU^{\top}\widehat{Z}_1^{(l)}VV^{\top}}{Q} \label{main-tside} \\ &\quad + \frac{1}{2}\sum_{l=1}^{2}\inp{UU^{\top}\widehat{Z}_2^{(l)}V_{\perp}V_{\perp}^{\top}}{Q} + \inp{U_{\perp}U_{\perp}^{\top}\widehat{Z}_2^{(l)}VV^{\top}}{Q} + \inp{UU^{\top}\widehat{Z}_2^{(l)}VV^{\top}}{Q} \label{neg1-tside}\\ &\quad + \frac{1}{2}\sum_{l=1}^{2}\inp{\widehat{U}_l\widehat{U}_l^{\top}\widehat{Z}^{(l)}\widehat{V}_l\widehat{V}_l^{\top}}{Q} - \inp{UU^{\top}\widehat{Z}^{(l)}VV^{\top}}{Q} \label{neg2-tside} \\ &\quad + \frac{1}{2}\sum_{l=1}^{2}\inp{\sum_{k=2}^{\infty} (\mathcal{S}_{A,k}(\widehat{E}^{(l)})A\Theta\Theta^{\top}+\Theta\Theta^{\top}A\mathcal{S}_{A,k}(\widehat{E}^{(l)}))}{\widetilde{Q}} + \inp{(\widehat{\Theta}\widehat{\Theta}^{\top}-\Theta\Theta^{\top})A(\widehat{\Theta}_l\widehat{\Theta}_l^{\top}-\Theta\Theta^{\top})}{\widetilde{Q}} \label{neg3-tside}.
    \end{align}
    In the following Lemmas, we still first apply Berry Esseen Theorem to prove the asymptotic normality of the main term, while showing all the other terms are negligible. Then we can conclude the proof following Theorem \ref{thm:inference-main}.

    \noindent\textbf{Asymptotic normality of (\ref{main-tside})}

    By definition, \eqref{main-tside} can be written as
    \begin{align*}
        L_T^{\tside}:=\frac{1}{2}\sum_{l=1}^{2}\Big(
        &\inp{UU^{\top}\widehat{Z}_1^{(l)}V_{\perp}V_{\perp}^{\top}}{Q}
        +\inp{U_{\perp}U_{\perp}^{\top}\widehat{Z}_1^{(l)}VV^{\top}}{Q}
        +\inp{UU^{\top}\widehat{Z}_1^{(l)}VV^{\top}}{Q}\Big)\\
        &=\frac{1}{T\nu}\sum_{t=1}^{T}\sum_{(i,j)\in g_t}
        \xi_t^{ij}\inp{X_t^{ij}}{B}
        =\frac{1}{T\nu}\sum_{t=1}^T W_t .
    \end{align*}

    For ease of notation, we denote $m_t=\min(B_r,B_s)$ and $c=\min(c_r,c_s)$. Condition on $m_t$, 
    \begin{align*}
        \mathbb P\!\left((i,j),(k,l)\in\Omega_t\mid m_t\right)
&=\frac{m_t}{d_1d_2}\frac{m_t-1}{(d_1-1)(d_2-1)}, \qquad i\ne k,\ j\ne l,
    \end{align*}
    thus
    \begin{align*}
        \mathbb P\!\left((i,j),(k,l)\in\Omega_t\right)
&=\frac{\mathbb E[m_t(m_t-1)]}{d_1(d_1-1)d_2(d_2-1)}.
    \end{align*}
    The truncation in
    \eqref{eq:two-random-complete} and $d_2\geq d_1$ imply
    $c d_1\leq m_t\leq d_1$ almost surely.  Therefore, for
    $d_1,d_2\geq2$,
    \begin{align*}
        \frac{\PP((i,j),(k,l)\in\Omega_t)}{\nu^2}
        &=
        \frac{\EE[m_t(m_t-1)]}{(\EE m_t)^2}
        \frac{d_1d_2}{(d_1-1)(d_2-1)}\\
        &\leq \frac{d_1}{\EE m_t}
        \frac{d_1d_2}{(d_1-1)(d_2-1)}
        \leq \frac{4}{c}.
    \end{align*}
    The apply Lemma \ref{lemma: pairwise} with $\kappa_1=4/c$ and $\rho_Q\lesssim \mu^2r$ gives
    \begin{align*}
        \sup_{x\in\RR}\left|
        \PP\left(
        \frac{L_T^{\tside}}
        {\sigma\fro{\mathcal{P}_M(Q)}\sqrt{1/(T\nu)}}\leq x
        \right)-\Phi(x)\right|
        \lesssim
        \sqrt{\frac{\mu^2r/(d_1\nu)+4/c}{T}}.
    \end{align*}

    \noindent\textbf{Bounding (\ref{neg1-tside})}

    Without loss of generality, we only prove the upper bound for $|\inp{UU^{\top}\widehat{Z}_2^{(1)}V_{\perp}V_{\perp}^{\top}}{Q} + \inp{U_{\perp}U_{\perp}^{\top}\widehat{Z}_2^{(1)}VV^{\top}}{Q} + \inp{UU^{\top}\widehat{Z}_2^{(1)}VV^{\top}}{Q}|$ \emph{under any give $B_r$ and $B_s$}. The the upper bound for $|\inp{UU^{\top}\widehat{Z}_2^{(2)}V_{\perp}V_{\perp}^{\top}}{Q} + \inp{U_{\perp}U_{\perp}^{\top}\widehat{Z}_2^{(2)}VV^{\top}}{Q} + \inp{UU^{\top}\widehat{Z}_2^{(2)}VV^{\top}}{Q}|$ can be derived following the same arguments. Moreover, we assume $B_s\geq (1+\gamma)B_r\geq (1+\gamma)c_rd_1$, while the $B_r\geq (1+\gamma)B_s$ case can be proved using the transpose of the matrix. For ease of notation, we denote 
    \begin{align*}
        X_t^{i}= \begin{cases}
        X_t^{ij} \quad & \text{there exists $j\in[d_2]$ s.t. $(i,j)\in g_t$} \\
        0 \quad &\text{otherwise} 
    \end{cases}.
    \end{align*}
    By definition, 
    \begin{align*}
        &|\inp{UU^{\top}\widehat{Z}_2^{(1)}V_{\perp}V_{\perp}^{\top}}{Q} + \inp{U_{\perp}U_{\perp}^{\top}\widehat{Z}_2^{(1)}VV^{\top}}{Q} + \inp{UU^{\top}\widehat{Z}_2^{(1)}VV^{\top}}{Q}|\\ 
        &\quad =\frac{1}{T_0}\sum_{t=1}^{T_0} \sum_{(i,j)\in g_t} \frac{1}{\nu}\inp{\widehat{\Delta}_1}{X_t^{ij}}\inp{X_t^{ij}}{\PMQ} -\inp{\widehat{\Delta}_1}{\PMQ}.
    \end{align*}
    We first derive the uniform bound for $(1/T_0)|(\nu)^{-1}\sum_{(i,j)\in g_t}\inp{\widehat{\Delta}_1}{X_t^{ij}}\inp{X_t^{ij}}{\PMQ} -\inp{\widehat{\Delta}_1}{\PMQ}|$ under any choices of $B_r$ and $B_s$. We have
\[
\begin{aligned}
&\sum_{(i,j)\in g_t}
\left\langle\widehat{\Delta}_1,X_t^{ij}\right\rangle
\left\langle X_t^{ij},\mathcal P_M(Q)\right\rangle=\sum_{i=1}^{d_1}
\left\langle\widehat{\Delta}_1,X_t^i\right\rangle
\left\langle X_t^i,\mathcal P_M(Q)\right\rangle=\sum_{\substack{\Omega_1\subset[d_1]\\|\Omega_1|=B_r}}
\frac{1}{C_{d_1-1}^{B_r-1}}
\sum_{i\in\Omega_1}
\left\langle\widehat{\Delta}_1,X_t^i\right\rangle
\left\langle X_t^i,\mathcal P_M(Q)\right\rangle .
\end{aligned}
\]Here we denote \(\Omega_1=\{r_1,\ldots,r_{B_r}\}\) to be any possible subset of \([d_1]\). Let \(\mathcal F_{r_l}\) denote the \(\sigma\)-algebra generated by \(X_t^{r_1},\ldots,X_t^{r_l}\), i.e., \(\mathcal F_i=\sigma(X_t^{r_1},\ldots,X_t^{r_l})\) and \(\mathcal F_0=\emptyset\). Then
\[
R:=
\sum_{i\in\Omega_1}
\left\langle\widehat{\Delta}_1,X_t^i\right\rangle
\left\langle X_t^i,\mathcal P_M(Q)\right\rangle
\]is a martingale associated with filter
$
\mathcal F:\quad
\mathcal F_0\subset\mathcal F_{r_1}\subset\cdots\subset
\mathcal F_{r_{B_r}}.
$. Let
\[
\begin{aligned}
R_{r_l}
&:=\mathbb E\!\left[R\mid\mathcal F_{r_{l-1}}\right]=\sum_{j\in\{r_1,\ldots,r_l\}}
\left\langle\widehat{\Delta}_1,X_t^j\right\rangle
\left\langle X_t^j,\mathcal P_M(Q)\right\rangle+
\sum_{j\in\{r_{l+1},\ldots,r_{B_r}\}}
\mathbb E\!\left[
\left\langle\widehat{\Delta}_1,X_t^{jq}\right\rangle
\left\langle X_t^{jq},\mathcal P_M(Q)\right\rangle
\mid\mathcal F_i
\right].
\end{aligned}
\]In particular, \(R_0=\mathbb E[R]\), \(R_{B_r}=R\). Notice that,
\[
\begin{aligned}
\mathbb E\!\left[R_{r_l}\mid\mathcal F_{r_{l-1}}\right]
&=\sum_{j\in\{r_1,\ldots,r_{l-1}\}}
\left\langle\widehat{\Delta}_1,X_t^j\right\rangle
\left\langle X_t^j,\mathcal P_M(Q)\right\rangle +
\sum_{j\in\{r_l,\ldots,r_{B_r}\}}
\mathbb E\!\left[
\left\langle\widehat{\Delta}_1,X_t^j\right\rangle
\left\langle X_t^j,\mathcal P_M(Q)\right\rangle
\mid\mathcal F_{r_{l-1}}
\right] \\ &=R_{r_{l-1}},
\end{aligned}
\]then
\[
\begin{aligned}
R_{r_l}-R_{r_{l-1}}
&=
\left\langle\widehat{\Delta}_1,X_t^{r_l}\right\rangle
\left\langle X_t^{r_l},\mathcal P_M(Q)\right\rangle-
\mathbb E\!\left[
\left\langle\widehat{\Delta}_1,X_t^{r_l}\right\rangle
\left\langle X_t^{r_l},\mathcal P_M(Q)\right\rangle
\mid\mathcal F_{r_{l-1}}
\right]\\
&\quad+
\left(
\sum_{j\in\{r_{l+1},\ldots,r_{B_r}\}}
\mathbb E\!\left[
\left\langle\widehat{\Delta}_1,X_t^j\right\rangle
\left\langle X_t^j,\mathcal P_M(Q)\right\rangle
\mid\mathcal F_{r_l}
\right]-
\mathbb E\!\left[
\left\langle\widehat{\Delta}_1,X_t^j\right\rangle
\left\langle X_t^j,\mathcal P_M(Q)\right\rangle
\mid\mathcal F_{r_{l-1}}
\right]
\right).
\end{aligned}
\]Obviously,
$
\mathbb E\!\left[
R_{r_l}-R_{r_{l-1}}
\mid\mathcal F_{r_{l-1}}
\right]=0,
$ and by \((a+b)^2\leq2(a^2+b^2)\),
\[
\begin{aligned}
&\operatorname{Var}\!\left[R_{r_l}\mid\mathcal F_{r_{l-1}}\right]
=
\mathbb E\!\left[
\left(
R_{r_l}-\mathbb E[R_{r_l}\mid\mathcal F_{r_{l-1}}]
\right)^2
\mid\mathcal F_{r_{l-1}}
\right]=
\mathbb E\!\left[
(R_{r_l}-R_{r_{l-1}})^2
\mid\mathcal F_{r_{l-1}}
\right]\\
&\leq
2\mathbb E\!\left[
\left\langle\widehat{\Delta}_1,X_t^{r_l}\right\rangle^2
\left\langle X_t^{r_l},\mathcal P_M(Q)\right\rangle^2
\mid\mathcal F_{r_{l-1}}
\right]\\
&\quad+
2\mathbb E\!\left[
\left(
\sum_{j\in\{r_{l+1},\ldots,r_{B_r}\}}
\mathbb E\!\left[
\left\langle\widehat{\Delta}_1,X_t^j\right\rangle
\left\langle X_t^j,\mathcal P_M(Q)\right\rangle
\mid\mathcal F_{r_l}
\right]-
\mathbb E\!\left[
\left\langle\widehat{\Delta}_1,X_t^j\right\rangle
\left\langle X_t^j,\mathcal P_M(Q)\right\rangle
\mid\mathcal F_{r_{l-1}}
\right]
\right)^2
\Biggm|\mathcal F_{r_{l-1}}
\right].
\end{aligned}
\]By definition of \(\mathcal F_{r_l}\) and \(\mathcal M_{p_1,p_2}\), \(X_t^j\) with \(j\in\{r_{l+1},\ldots,r_{B_r}\}\) is sampled with the following scheme: given the column set \(\Omega_2\in[d_2]\) which includes all the column indices of \(X_t^{r_1},\ldots,X_t^{r_{B_r}}\), \(X_t^j\) is uniformly sampled from
$
\left\{
e_je_k^\top:
k\in\Omega_2\setminus
\{j_t(r_1),\ldots,j_t(r_l)\}
\right\}.
$ Therefore,
\[
\begin{aligned}
&\left|
\mathbb E\!\left[
\left\langle\widehat{\Delta}_1,X_t^j\right\rangle
\left\langle X_t^j,\mathcal P_M(Q)\right\rangle
\mid\mathcal F_{r_l}
\right]
-
\mathbb E\!\left[
\left\langle\widehat{\Delta}_1,X_t^j\right\rangle
\left\langle X_t^j,\mathcal P_M(Q)\right\rangle
\mid\mathcal F_{r_{l-1}}
\right]
\right|\\
&=
\left|
\sum_{\substack{\Omega_2\subset[d_2]\\|\Omega_2|=B_s}}
\frac{1}{C_{d_2}^{B_s}}\frac{1}{B_s-l}
\sum_{\substack{k\in\Omega_2\setminus
\{j_t(r_1),\ldots,j_t(r_l)\}}}
\left\langle\widehat{\Delta}_1,e_je_k^\top\right\rangle
\left\langle e_je_k^\top,\mathcal P_M(Q)\right\rangle\right.\\
&\qquad\left.-
\sum_{\substack{\Omega_2\subset[d_2]\\|\Omega_2|=B_s}}
\frac{1}{C_{d_2}^{B_s}}\frac{1}{B_s-l+1}
\sum_{\substack{k\in\Omega_2\setminus
\{j_t(r_1),\ldots,j_t(r_{l-1})\}}}
\left\langle\widehat{\Delta}_1,e_je_k^\top\right\rangle
\left\langle e_je_k^\top,\mathcal P_M(Q)\right\rangle
\right|\\
&\leq
\sum_{\substack{\Omega_2\subset[d_2]\\|\Omega_2|=B_s}}
\frac{1}{C_{d_2}^{B_s}}
\frac{1}{(B_s-l)(B_s-l+1)}
\sum_{\substack{k\in\Omega_2\setminus
\{j_t(r_1),\ldots,j_t(r_l)\}}}
\|\widehat{\Delta}_1\|_{\max}
\left|
\left\langle e_je_k^\top,\mathcal P_M(Q)\right\rangle
\right|\\
&\quad+
\sum_{\substack{\Omega_2\subset[d_2]\\|\Omega_2|=B_s}}
\frac{1}{C_{d_2}^{B_s}}
\frac{1}{B_s-l+1}
\|\widehat{\Delta}_1\|_{\max}
\left|
\left\langle e_je_{j_t(r_l)}^\top,\mathcal P_M(Q)\right\rangle
\right|.
\end{aligned}
\]By Cauchy Schwarz inequality
\[
\begin{aligned}
&\sum_{\substack{\Omega_2\subset[d_2]\\|\Omega_2|=B_s}}
\frac{1}{C_{d_2}^{B_s}}
\frac{1}{(B_s-l)(B_s-l+1)}
\sum_{\substack{k\in[d_2]\setminus
\{j_t(r_1),\ldots,j_t(r_l)\}}}
\|\widehat{\Delta}_1\|_{\max}
\left|
\left\langle e_je_k^\top,\mathcal P_M(Q)\right\rangle
\right|\\
&\quad\leq
\sum_{\substack{\Omega_2\subset[d_2]\\|\Omega_2|=B_s}}
\frac{1}{C_{d_2}^{B_s}}
\frac{1}{\sqrt{B_s-l}(B_s-l+1)}
\|\widehat{\Delta}_1\|_{\max}
\|\mathcal P_M(Q)(j,\Omega_2)\|,
\end{aligned}
\]where
$
\|\mathcal P_M(Q)(j,\Omega_2)\|
=
\sqrt{
\sum_{k\in\Omega_2}
\left\langle e_je_k^\top,\mathcal P_M(Q)\right\rangle^2
}.
$ Again by Cauchy Schwarz inequality,
\[
\begin{aligned}
&\left(
\sum_{j\in\{r_{l+1},\ldots,r_{B_r}\}}
\mathbb E\!\left[
\left\langle\widehat{\Delta}_1,X_t^j\right\rangle
\left\langle X_t^j,\mathcal P_M(Q)\right\rangle
\mid\mathcal F_{r_l}
\right] -
\mathbb E\!\left[
\left\langle\widehat{\Delta}_1,X_t^j\right\rangle
\left\langle X_t^j,\mathcal P_M(Q)\right\rangle
\mid\mathcal F_{r_{l-1}}
\right]
\right)^2\\
&\leq
2B_r
\sum_{j\in\{r_{l+1},\ldots,r_{B_r}\}}
\sum_{\substack{\Omega_2\subset[d_2]\\|\Omega_2|=B_s}}
\frac{C_{d_2}^{B_s}}{(C_{d_2}^{B_s})^2}
\frac{
\|\widehat{\Delta}_1\|_{\max}^2
\|\mathcal P_M(Q)(j,\Omega_2)\|^2
}{
(B_s-l)(B_s-l+1)^2
}\\
&\quad+
2B_r
\sum_{j\in\{r_{l+1},\ldots,r_{B_r}\}}
\sum_{\substack{\Omega_2\subset[d_2]\\|\Omega_2|=B_s}}
\frac{C_{d_2}^{B_s}}{(C_{d_2}^{B_s})^2}
\frac{
\left\langle
\widehat{\Delta}_1,e_je_{j_t(r_l)}^\top
\right\rangle^2
\left\langle
e_je_{j_t(r_l)}^\top,\mathcal P_M(Q)
\right\rangle^2
}{
(B_s-l+1)^2
}\\
&\leq
2B_r
\sum_{j\in\{r_{l+1},\ldots,r_{B_r}\}}
\frac{B_s}{d_2}
\frac{
\|\widehat{\Delta}_1\|_{\max}^2
\|\mathcal P_M(Q)(j,\cdot)\|^2
}{
(B_s-l)(B_s-l+1)^2
}\\
&\quad+
2B_r
\sum_{j\in\{r_{l+1},\ldots,r_{B_r}\}}
\sum_{\substack{\Omega_2\subset[d_2]\\|\Omega_2|=B_s}}
\frac{1}{C_{d_2}^{B_s}}
\frac{
\left\langle
\widehat{\Delta}_1,e_je_{j_t(r_l)}^\top
\right\rangle^2
\left\langle
e_je_{j_t(r_l)}^\top,\mathcal P_M(Q)
\right\rangle^2
}{
(B_s-l+1)^2
}\\
&\lesssim
\sum_{i\in\Omega_1}
\frac{1}{d_1^2}
\|\widehat{\Delta}_1\|_{\max}^2
\|\mathcal P_M(Q)(i,\cdot)\|^2+
2B_r
\sum_{j\in\{r_{l+1},\ldots,r_{B_r}\}}
\sum_{\substack{\Omega_2\subset[d_2]\\|\Omega_2|=B_s}}
\frac{1}{C_{d_2}^{B_s}}
\frac{
\left\langle
\widehat{\Delta}_1,e_je_{j_t(r_l)}^\top
\right\rangle^2
\left\langle
e_je_{j_t(r_l)}^\top,\mathcal P_M(Q)
\right\rangle^2
}{
(B_s-l+1)^2
}.
\end{aligned}
\]where the last inequality comes from \(B_s\leq d_2\) and \(B_s\geq(1+\gamma)B_r\geq(1+\gamma)c_rd_1\). Similarly, given the column set \(\Omega_2\) and condition on \(\mathcal F_{r_{l-1}}\), \(j_t(r_l)\) is uniformly sampled from
$
\Omega_2\setminus
\{j_t(r_1),\ldots,j_t(r_{l-1})\},
$ then
\[
\begin{aligned}
&\mathbb E\!\left[
2B_r
\sum_{j\in\{r_{l+1},\ldots,r_{B_r}\}}
\sum_{\substack{\Omega_2\subset[d_2]\\|\Omega_2|=B_s}}
\frac{1}{C_{d_2}^{B_s}}
\frac{
\left\langle
\widehat{\Delta}_1,e_je_{j_t(r_l)}^\top
\right\rangle^2
\left\langle
e_je_{j_t(r_l)}^\top,\mathcal P_M(Q)
\right\rangle^2
}{
(B_s-l+1)^2
}
\Biggm|\mathcal F_{r_{l-1}}
\right]\\
&\leq
2B_r
\sum_{j\in\{r_{l+1},\ldots,r_{B_r}\}}
\sum_{\substack{\Omega_2\subset[d_2]\\|\Omega_2|=B_s}}
\frac{1}{C_{d_2}^{B_s}}
\frac{1}{(B_s-l+1)^3}
\sum_{\substack{k\in\Omega_2\setminus
\{j_t(r_1),\ldots,j_t(r_{l-1})\}}}
\left\langle\widehat{\Delta}_1,e_je_k^\top\right\rangle^2
\left\langle e_je_k^\top,\mathcal P_M(Q)\right\rangle^2\\
&\leq
2B_r
\sum_{j\in\{r_{l+1},\ldots,r_{B_r}\}}
\sum_{\substack{\Omega_2\subset[d_2]\\|\Omega_2|=B_s}}
\frac{1}{C_{d_2}^{B_s}}
\frac{
\|\widehat{\Delta}_1\|_{\max}^2
\|\mathcal P_M(Q)(j,\Omega_2)\|^2
}{
(B_s-l+1)^3
}\\
&\lesssim
\sum_{i\in\Omega_1}
\frac{1}{d_1^2}
\|\widehat{\Delta}_1\|_{\max}^2
\|\mathcal P_M(Q)(i,\cdot)\|^2.
\end{aligned}
\]Following the same argument,
\[
\begin{aligned}
&\quad \mathbb E\!\left[
\left\langle\widehat{\Delta}_1,X_t^{r_l}\right\rangle^2
\left\langle X_t^{r_l},\mathcal P_M(Q)\right\rangle^2
\mid\mathcal F_{r_{l-1}}
\right]
\\ &=
\sum_{\substack{\Omega_2\subset[d_2]\\|\Omega_2|=B_s}}
\frac{1}{C_{d_2}^{B_s}}
\frac{1}{B_s-l+1}
\sum_{\substack{k\in\Omega_2\setminus
\{j_t(r_1),\ldots,j_t(r_{l-1})\}}}
\left\langle\widehat{\Delta}_1,e_{r_l}e_k^\top\right\rangle^2
\left\langle e_{r_l}e_k^\top,\mathcal P_M(Q)\right\rangle^2\lesssim
\frac{1}{d_1}
\|\widehat{\Delta}_1\|_{\max}^2
\|\mathcal P_M(Q)(r_l,\cdot)\|^2.
\end{aligned}
\]Combine the above terms, we have
\[
\sum_{r_l\in\Omega_1}
\operatorname{Var}\!\left[
R_{r_l}\mid\mathcal F_{r_{l-1}}
\right]
\lesssim
\frac{1}{d_1}
\|\widehat{\Delta}_1\|_{\max}^2
\sum_{i\in\Omega_1}
\|\mathcal P_M(Q)(i,\cdot)\|^2.
\]Moreover,
\[
\mathbb E[R]
=
\frac{C_{d_1-1}^{B_r-1}}{C_{d_1}^{B_r}}
\nu
\left\langle
\widehat{\Delta}_1,\mathcal P_M(Q)
\right\rangle,
\]and for any \(r_l\), following the above analysis,
\[
\left|R_{r_l}-R_{r_{l-1}}\right|
\lesssim
\frac{1}{\sqrt{d_1}}
\|\widehat{\Delta}_1\|_{\max}
\|\mathcal P_M(Q)\|_F.
\]Finally, by martingale Bernstein inequality (Theorem 18 in \citep{chung2006concentration}), with probability at least \(1-d_1^{-10}\),
\[
\begin{aligned}
&\left|
\sum_{i\in\Omega_1}
\left\langle\widehat{\Delta}_1,X_t^i\right\rangle
\left\langle X_t^i,\mathcal P_M(Q)\right\rangle
-
\frac{C_{d_1-1}^{B_r-1}}{C_{d_1}^{B_r}}
\nu
\left\langle\widehat{\Delta}_1,\mathcal P_M(Q)\right\rangle
\right|\\
&\lesssim
\frac{1}{\sqrt{d_1}}
\|\widehat{\Delta}_1\|_{\max}
\|\mathcal P_M(Q)\|_F\log d_1+
\frac{1}{\sqrt{d_1}}
\|\widehat{\Delta}_1\|_{\max}
\sqrt{
\sum_{i\in\Omega_1}
\|\mathcal P_M(Q)(i,\cdot)\|^2
}
\log^{1/2}d_1,
\end{aligned}
\]which directly implies
\[
\begin{aligned}
&\left|
\sum_{i=1}^{d_1}
\left\langle\widehat{\Delta}_1,X_t^i\right\rangle
\left\langle X_t^i,\mathcal P_M(Q)\right\rangle
-
\nu
\left\langle\widehat{\Delta}_1,\mathcal P_M(Q)\right\rangle
\right|\\
&=
\left|
\sum_{\substack{\Omega_1\subset[d_1]\\|\Omega_1|=B_r}}
\frac{1}{C_{d_1-1}^{B_r-1}}
\left(
\sum_{i\in\Omega_1}
\left\langle\widehat{\Delta}_1,X_t^i\right\rangle
\left\langle X_t^i,\mathcal P_M(Q)\right\rangle
-
\frac{C_{d_1-1}^{B_r-1}}{C_{d_1}^{B_r}}
\nu
\left\langle\widehat{\Delta}_1,\mathcal P_M(Q)\right\rangle
\right)
\right|\\
&\lesssim
\sum_{\substack{\Omega_1\subset[d_1]\\|\Omega_1|=B_r}}
\frac{1}{C_{d_1-1}^{B_r-1}}
\left(
\frac{1}{\sqrt{d_1}}
\|\widehat{\Delta}_1\|_{\max}
\|\mathcal P_M(Q)\|_F\log d_1 +
\frac{1}{\sqrt{d_1}}
\|\widehat{\Delta}_1\|_{\max}
\sqrt{
\sum_{i\in\Omega_1}
\|\mathcal P_M(Q)(i,\cdot)\|^2
}
\log^{1/2}d_1
\right)\\
&\lesssim
\frac{C_{d_1}^{B_r}}{C_{d_1-1}^{B_r-1}}
\frac{\log d_1}{\sqrt{d_1}}
\|\widehat{\Delta}_1\|_{\max}
\|\mathcal P_M(Q)\|_F +
\frac{\log^{1/2}d_1}
{C_{d_1-1}^{B_r-1}\sqrt{d_1}}
\|\widehat{\Delta}_1\|_{\max}
\sqrt{C_{d_1}^{B_r}}
\sqrt{
\sum_{\substack{\Omega_1\subset[d_1]\\|\Omega_1|=B_r}}
\sum_{i\in\Omega_1}
\|\mathcal P_M(Q)(i,\cdot)\|^2
}\\
&\leq
\frac{C_{d_1}^{B_r}}{C_{d_1-1}^{B_r-1}}
\frac{\log d_1}{\sqrt{d_1}}
\|\widehat{\Delta}_1\|_{\max}
\|\mathcal P_M(Q)\|_F +
\frac{\log^{1/2}d_1}
{C_{d_1-1}^{B_r-1}\sqrt{d_1}}
\|\widehat{\Delta}_1\|_{\max}
\sqrt{C_{d_1}^{B_r}}
\sqrt{
C_{d_1-1}^{B_r-1}
\|\mathcal P_M(Q)\|_F^2
}\\
&\lesssim
\frac{1}{\sqrt{d_1}}
\|\widehat{\Delta}_1\|_{\max}
\|\mathcal P_M(Q)\|_F\log d_1,
\end{aligned}
\]where the second inequality comes from applying Cauchy Schwarz inequality to the last term. Notice that this is a uniform upper bound for any \(B_r\) and \(B_s\). Finally, by Proposition \ref{thm:estimation3}, we can obtain
\[
\begin{aligned}
&\frac{1}{T_0}
\left|
\frac{1}{\nu}
\sum_{(i,j)\in g_t}
\left\langle\widehat{\Delta}_1,X_t^{ij}\right\rangle
\left\langle X_t^{ij},\mathcal P_M(Q)\right\rangle
-
\left\langle\widehat{\Delta}_1,\mathcal P_M(Q)\right\rangle
\right|\\
&\lesssim
\frac{\log d_1}{T\nu\sqrt{d_1}}
\|\widehat{\Delta}_1\|_{\max}
\|\mathcal P_M(Q)\|_F\lesssim
\sqrt{\frac{r^2}{T^3\nu^3d_1^2}}
\log^2d_1\,
\sigma\|\mathcal P_M(Q)\|_F.
\end{aligned}
\]

    Next we calculate the variance of $\frac{1}{T_0}\bigg|(\nu)^{-1}\sum_{(i,j)\in g_t} \inp{\widehat{\Delta}_1}{X_t^{ij}}\inp{X_t^{ij}}{\PMQ} -\inp{\widehat{\Delta}_1}{\PMQ}\bigg|$. Notice that $\widehat{\Delta}_1$ is independent of $X_t^i$, then
    \begin{align*}
        &\frac{1}{T_0^2}\EE\left[\bigg(\frac{1}{\nu}\sum_{(i,j)\in g_t} \inp{\widehat{\Delta}_1}{X_t^{ij}}\inp{X_t^{ij}}{\PMQ} -\inp{\widehat{\Delta}_1}{\PMQ} \bigg)^2\right] \\
        &\quad = \frac{1}{T_0^2}\EE\left[\bigg(\frac{1}{\nu}\sum_{(i,j)\in g_t} \inp{\widehat{\Delta}_1}{X_t^{ij}}\inp{X_t^{ij}}{\PMQ} \bigg)^2\right]-\frac{1}{T_0^2}\inp{\widehat{\Delta}_1}{\PMQ}^2 \\
        &\quad = \frac{1}{T_0^2\nu^2}\underbrace{\EE\left[\sum_{(i,j)\in g_t}\inp{\widehat{\Delta}_1}{X_t^{ij}}^2\inp{X_t^i}{\PMQ}^2 \right]}_{\mathcal{I}_1} \\ &\quad\quad + \frac{1}{T_0^2\nu^2}\underbrace{\EE\left[\sum_{(i,p)\in g_t}\sum_{(j,q)\neq (i,p)} \inp{\widehat{\Delta}_1}{X_t^{ip}}\inp{X_t^{ip}}{\PMQ}\inp{\widehat{\Delta}_1}{X_t^{jq}}\inp{X_t^{jq}}{\PMQ} \right]}_{\mathcal{I}_2} - \frac{1}{T_0^2}\inp{\widehat{\Delta}_1}{\PMQ}^2.
    \end{align*}
    Denote $A=\widehat{\Delta}_1\odot \PMQ$ the Hadamard product of $\widehat{\Delta}_1$ and $\PMQ$, then 
    \begin{align*}
        \mathcal{I}_1=\EE[\sum_{(i,j)\in g_t}\inp{X_t^{ij}}{A}^2]=\nu\fro{A}^2\leq \nu\|\widehat{\Delta}_1\|_{\max}^2\fro{\PMQ}^2.
    \end{align*}
    Since 
    \begin{align*}
        \mathcal{I}_2=\EE\left[\EE\bigg[\sum_{(i,p)\in g_t}\sum_{(j,q)\neq (i,p)} \inp{\widehat{\Delta}_1}{X_t^{ip}}\inp{X_t^{ip}}{\PMQ}\inp{\widehat{\Delta}_1}{X_t^{jq}}\inp{X_t^{jq}}{\PMQ}\bigg|B_r,B_s \bigg]\right].
    \end{align*}
    Then we aim to derive a uniform bound for \\ $\EE\bigg[\sum_{(i,p)\in g_t}\sum_{(j,q)\neq (i,p)} \inp{\widehat{\Delta}_1}{X_t^{ip}}\inp{X_t^{ip}}{\PMQ}\inp{\widehat{\Delta}_1}{X_t^{jq}}\inp{X_t^{jq}}{\PMQ} \bigg]$ under any given fixed $B_r$ and $B_s$. Without loss of generality, we assume $B_r\leq B_s$, then we have  \begin{align*}
        &\EE\bigg[\sum_{(i,p)\in g_t}\sum_{(j,q)\neq (i,p)} \inp{\widehat{\Delta}_1}{X_t^{ip}}\inp{X_t^{ip}}{\PMQ}\inp{\widehat{\Delta}_1}{X_t^{jq}}\inp{X_t^{jq}}{\PMQ}\bigg|B_r,B_s \bigg] \\ &\quad = \frac{1}{C_{d_1}^{B_r}C_{d_2}^{B_s}}\sum_{\substack{\Omega_1 \subset [d_1] \\ |\Omega_1|=B_r}}\sum_{\substack{\Omega_2 \subset [d_2] \\ |\Omega_2|=B_s}} \frac{1}{B_s(B_s-1)}\sum_{\substack{i\neq j \\ i,j \in \Omega_1}}\sum_{\substack{p\neq q \\ p,q \in \Omega_2}} A(i,p)A(j,q) \\
        &\quad = \frac{C_{d_1-2}^{B_r-2}C_{d_2-2}^{B_s-2}}{C_{d_1}^{B_r}C_{d_2}^{B_s}}\frac{1}{B_s(B_s-1)} \sum_{\substack{i\neq j \\ i,j \in [d_1]}}\sum_{\substack{p\neq q \\ p,q \in [d_2]}}  A(i,p)A(j,q) \\
        &\quad = \frac{B_r(B_r-1)}{d_1d_2(d_1-1)(d_2-1)}\sum_{\substack{i\neq j \\ i,j \in [d_1]}}\sum_{\substack{p\neq q \\ p,q \in [d_2]}}  A(i,p)A(j,q).
    \end{align*}
    By proof of Theorem \ref{thm:inference-main}, 
    \begin{align*}
        &\frac{B_r(B_r-1)}{d_1d_2(d_1-1)(d_2-1)}\sum_{\substack{i\neq j \\ i,j \in [d_1]}}\sum_{\substack{p\neq q \\ p,q \in [d_2]}}  A(i,p)A(j,q) \\
        &\quad \leq \frac{B_r(B_r-1)}{d_1d_2(d_1-1)(d_2-1)} \left(\inp{\widehat{\Delta}_1}{\PMQ}^2 + d_2\|\widehat{\Delta}_1\|_{\max}^2\fro{\PMQ}^2\right).
    \end{align*}
    By symmetry, if $B_s\leq B_r$, the above upper bound will be $\frac{B_s(B_s-1)}{d_1d_2(d_1-1)(d_2-1)} (\inp{\widehat{\Delta}_1}{\PMQ}^2 + d_2\|\widehat{\Delta}_1\|_{\max}^2\fro{\PMQ}^2)$. As a result, take expectation over $B_r$ and $B_s$, denote $\widetilde{X}=\min(B_r,B_s)$
   \begin{align*}
        \mathcal{I}_2&\leq  \EE\left[\frac{\widetilde{X}(\widetilde{X}-1)}{d_1d_2(d_1-1)(d_2-1)}\right]\left(\inp{\widehat{\Delta}_1}{\PMQ}^2 + d_2\|\widehat{\Delta}_1\|_{\max}^2\fro{\PMQ}^2\right) \\
        &\leq \frac{\nu d_1d_2(d_1-1)}{d_1d_2(d_1-1)(d_2-1)}\left(\inp{\widehat{\Delta}_1}{\PMQ}^2 + d_2\|\widehat{\Delta}_1\|_{\max}^2\fro{\PMQ}^2\right) \\
        &= \frac{\nu}{d_2-1}\left(\inp{\widehat{\Delta}_1}{\PMQ}^2 + d_2\|\widehat{\Delta}_1\|_{\max}^2\fro{\PMQ}^2\right). 
   \end{align*}
   Overall, the variance can be upper bounded by
   \begin{align*}
        &\frac{1}{T_0^2}\EE\left[\bigg(\frac{1}{\nu}\sum_{i=1}^{d_1}\inp{\widehat{\Delta}_1}{X_t^i}\inp{X_t^i}{\PMQ} -\inp{\widehat{\Delta}_1}{\PMQ} \bigg)^2\right]\leq \frac{1}{T_0^2\nu}\|\widehat{\Delta}_1\|_{\max}^2\fro{\PMQ}^2 \\ &\quad\quad + \bigg(\frac{1}{T_0^2(d_2-1)\nu}-\frac{1}{T_0^2}\bigg)\inp{\widehat{\Delta}_1}{\PMQ}^2 + \frac{d_2}{T_0^2(d_2-1)\nu}\|\widehat{\Delta}_1\|_{\max}^2\fro{\PMQ}^2 \\ 
        &\quad\leq \frac{3}{T_0^2\nu}\|\widehat{\Delta}_1\|_{\max}^2\fro{\PMQ}^2 + \frac{2}{T_0^2\nu}d_1d_2\|\widehat{\Delta}_1\|_{\max}^2\fro{\PMQ}^2\leq \frac{5}{T_0^2\nu}\|\widehat{\Delta}_1\|_{\max}^2\fro{\PMQ}^2.
   \end{align*}
   Combine with Proposition \ref{thm:estimation3} and the previous uniform bound, by Bernstein inequality, with probability at least $1-2d_1^{-10}$, 
   \begin{align*}
        &\left|\frac{1}{T_0}\sum_{t=1}^{T_0} \sum_{i=1}^{d_1} d_2\inp{\widehat{\Delta}_1}{X_t^i}\inp{X_t^i}{\PMQ} -\inp{\widehat{\Delta}_1}{\PMQ}\right|\\ &\quad \lesssim \sqrt{\frac{r^2}{T^3\nu^3d_1^2}}\log^3d_1\sigma\fro{\PMQ} + \sqrt{\frac{r^2\log^3 d_1}{T^2d_1\nu^2}}\sigma\fro{\PMQ}.
   \end{align*}

    Finally, following the proof of Lemma \ref{lemmaneg2} and Lemma \ref{lemmaneg3}, as long as we can derive the upper bound of $\|\widehat{Z}^{(l)}\|$, $\|\widehat{U}_l\widehat{U}_l^{\top}-UU^{\top}\|_{2,\max}$ and $\|\widehat{V}_l\widehat{V}_l^{\top}-VV^{\top}\|_{2,\max}$, then we can use them to bound (\ref{neg2-otm}) and (\ref{neg3-otm}). The next two Lemmas show the upper bound of $\|\widehat{Z}^{(l)}\|$, $\|\widehat{U}_l\widehat{U}_l^{\top}-UU^{\top}\|_{2,\max}$ and $\|\widehat{V}_l\widehat{V}_l^{\top}-VV^{\top}\|_{2,\max}$. 

    \begin{Lemma}
        \label{Zbound-tside}
        Under the conditions in Theorem \ref{thm:inference-main-3}, with probability at least $1-2d_1^{-10}$, 
        \begin{align*}
        \|\widehat{Z}^{(l)}\|\lesssim \sqrt{\frac{d_2\log d_1}{T\nu}}\sigma
        \end{align*}
        for $l=1,2$. 
    \end{Lemma}

    \begin{Lemma}
        \label{2maxbound-tside}
        Under the conditions in Theorem \ref{thm:inference-main-3}, with probability at least $1-5d_1^{-5}$,
        \begin{align*}
            &\|\widehat{U}_l\widehat{U}_l^{\top}-UU^{\top}\|_{2,\max}\lesssim \frac{\sigma}{\lambda_{\min}}\sqrt{\frac{d_2\log d_1}{Td_1\nu}}\mu\sqrt{\frac{r}{d_1}} \\
            &\|\widehat{V}_l\widehat{V}_l^{\top}-VV^{\top}\|_{2,\max}\lesssim \frac{\sigma}{\lambda_{\min}}\sqrt{\frac{d_2\log d_1}{Td_1\nu}}\mu\sqrt{\frac{r}{d_2}}. 
        \end{align*}
        for $l=1,2$. 
    \end{Lemma}

    As a result, we can show that, 
    \begin{align*}
        \frac{\big|\frac{1}{2}\sum_{l=1}^{2}\inp{\widehat{U}_l\widehat{U}_l^{\top}\widehat{Z}^{(l)}\widehat{V}_l\widehat{V}_l^{\top}}{Q} - \inp{UU^{\top}\widehat{Z}^{(l)}VV^{\top}}{Q}\big|}{\sigma\fro{\mathcal{P}_M(Q)}\sqrt{1/T\nu}}\lesssim \frac{\sigma\mu^2}{\lambda_{\min}}\sqrt{\frac{r^2d_2\log d_1}{Td_1\nu}}\frac{\|Q\|_{\ell_1}}{\fro{\PMQ}}. 
    \end{align*}
    and 
    \begin{align*}
        &\frac{\big|\sum_{k=2}^{\infty} \inp{(\mathcal{S}_{A,k}(\widehat{E})A\Theta\Theta^{\top}+\Theta\Theta^{\top}A\mathcal{S}_{A,k}(\widehat{E}))}{\widetilde{Q}}\big|}{\sigma\fro{\mathcal{P}_M(Q)}\sqrt{1/T\nu}}\lesssim \frac{\sigma\mu^2}{\lambda_{\min}}\sqrt{\frac{r^2d_2\log d_1}{Td_1\nu}}\frac{\|Q\|_{\ell_1}}{\fro{\PMQ}}, \\
        &\frac{\big|\inp{(\widehat{\Theta}\widehat{\Theta}^{\top}-\Theta\Theta^{\top})A(\widehat{\Theta}\widehat{\Theta}^{\top}-\Theta\Theta^{\top})}{\widetilde{Q}}\big|}{\sigma\fro{\mathcal{P}_M(Q)}\sqrt{1/T\nu}} \lesssim\frac{\sigma\mu^2\kappa}{\lambda_{\min}}\sqrt{\frac{r^2d_2\log d_1}{Td_1\nu}}\frac{\|Q\|_{\ell_1}}{\fro{\PMQ}}.
    \end{align*}
    Then we conclude the proof.

\end{proof}

\subsection{Proof of Theorem \ref{thm:variance}}
\begin{proof}
     Without loss of generality, we only prove the one-to-one matching case, while the other two cases can be proved following the same arguments. \\
    $\mathbf{\widehat{\sigma}^2\overset{p}{\rightarrow} \sigma^2}$ \quad Notice that, $\text{sum}(X_t)=d_1$ for all $t$. Therefore, $\widehat{\sigma}^2$ can be written as
    \begin{align*}
        \widehat{\sigma}^2&= \underbrace{\frac{1}{Td_1}\sum_{t=1}^{T_0}\sum_{i=1}^{d_1}\inp{M-\widehat{M}_1^{\init}}{X_t^i}^2 + \frac{1}{Td_1}\sum_{t=T_0+1}^{T}\sum_{i=1}^{d_1}\inp{M-\widehat{M}_2^{\init}}{X_t^i}^2}_{\uppercase\expandafter{\romannumeral1}} \\ 
        &\quad + \underbrace{\frac{2}{Td_1}\sum_{t=1}^{T_0}\sum_{i=1}^{d_1} \inp{M-\widehat{M}_1^{\init}}{X_t^i}\xi_t^i + \frac{2}{Td_1}\sum_{t=T_0+1}^{T}\sum_{i=1}^{d_1} \inp{M-\widehat{M}_2^{\init}}{X_t}\xi_t^i}_{\uppercase\expandafter{\romannumeral2}} + \underbrace{\frac{1}{Td_1}\sum_{t=1}^{T}\sum_{i=1}^{d_1} \xi_t^{i2}}_{\uppercase\expandafter{\romannumeral3}}.
    \end{align*}
    Denote $\widehat{\Delta}_{1}=\widehat{M}_1^{\init}-M$, for the first term of \uppercase\expandafter{\romannumeral1}, 
    \begin{align*}
        \big|\frac{1}{Td_1} \sum_{i=1}^{d_1}\inp{M-\widehat{M}_1^{\init}}{X_t^i}^2\big|&\leq \frac{1}{T}\|\widehat{\Delta}_{1}\|_{\max}^2 \lesssim \frac{1}{T}\sigma^2, 
    \end{align*}
    and 
    \begin{align*}
        \EE\left[\frac{1}{T^2d_1^2} \bigg(\sum_{i=1}^{d_1}\inp{M-\widehat{M}_1^{\init}}{X_t^i}^2\bigg)^2\right]&\leq \frac{1}{T^2d_1}d_1^2\|\widehat{\Delta}_{t-1}\|_{\max}^4 \lesssim \frac{1}{T^2}\sigma^4,
    \end{align*}
    where we use the fact that $\|\widehat{\Delta}_{1}\|_{\max}\lesssim \sigma$. By Bernstein inequality, with probability at least $1-d_1^{-2}$,
    \begin{align*}
        \uppercase\expandafter{\romannumeral1}&\lesssim \frac{\log d_1}{T}\sigma^2 + \sqrt{\frac{\log d_1}{T}}\sigma^2,
    \end{align*}
    which indicates $\uppercase\expandafter{\romannumeral1}\overset{p}{\rightarrow} 0$ as long as $\frac{\log d_1}{T}\rightarrow 0$. 
    
    Similarly, for the first term of $\uppercase\expandafter{\romannumeral2}$, note that $|\xi_t^i|$ is bounded with $\sigma\log^{1/2} d_1$ with high probability, 
    \begin{align*}
        \bigg|\frac{2}{Td_1} \sum_{i=1}^{d_1} \inp{M-\widehat{M}_1^{\init}}{X_t^i}\xi_t^i\bigg|\leq \frac{2}{Td_1}\|d_1\widehat{\Delta}_{1}\|_{\max}\sigma\log^{1/2} d_1 \lesssim \frac{\sigma^2\log^{1/2} d_1}{T},
    \end{align*}
    and
    \begin{align*}
        \EE\left[\frac{4}{T^2d_1^2} \bigg(\sum_{i=1}^{d_1}\inp{M-\widehat{M}_1^{\init}}{X_t^i}\xi_t^i\bigg)^2\right] \leq \frac{\sigma^2}{T^2d_1^2}d_1\|\widehat{\Delta}_{1}\|_{\max}^2 \lesssim  \frac{\sigma^4}{T^2d_1}.
    \end{align*}
    where we use $\EE[\xi_t^{i}\xi_t^j]=0$ for $i\neq j$. By Bernstein inequality, with probability at least $1-d_1^{-2}$, 
    \begin{align*}
        &\bigg|\uppercase\expandafter{\romannumeral2} - \frac{2}{Td_1} \sum_{t=1}^{T_0}\sum_{i=1}^{d_1} \EE\left[\inp{M-\widehat{M}_{1}^{\init}}{X_t^i} \xi_t^i\right] - \frac{2}{Td_1} \sum_{t=T_0+1}^{T}\sum_{i=1}^{d_1} \EE\left[\inp{M-\widehat{M}_{2}^{\init}}{X_t^i} \xi_t^i\right]\bigg| \\
        &\quad \lesssim \sqrt{\frac{\log d_1}{Td_1}}\sigma^2 + \frac{\log^{3/2} d_1}{T}\sigma^2,
    \end{align*}
    which indicates $\uppercase\expandafter{\romannumeral2}\overset{p}{\rightarrow} 0$ as long as $\frac{\log^{3/2} d_1}{T} \rightarrow 0$. 
    
    Lastly, for term $\uppercase\expandafter{\romannumeral3}$, directly by Weak Law of Large Numbers, $\uppercase\expandafter{\romannumeral3}\overset{p}{\rightarrow} \sigma^2$. 

    For proof of the one-to-many and two sided random cases, although there is randomness with $\text{sum}(X_t)$, we can first condition on $\text{sum}(X_t)$ then apply concentration and central limit theorem following the same arguments, then show $1/\text{sum}(X_t)$ would concentrate around $1/\EE[\text{sum}(X_t)]$. 
    
    $\mathbf{\fro{\mathcal{P}_{\widehat{M}}(Q)}^2\overset{p}{\rightarrow} \fro{\mathcal{P}_M(Q)}^2}$ \quad  Define $\widehat{U}$ and $\widehat{V}$ to be the left and right singular vectors of $\widehat{M}$, we have
    \begin{align*}
        \left|\fro{\mathcal{P}_M(Q)}^2 - \fro{\mathcal{P}_{\widehat{M}}(Q)}^2\right|&=\left|\fro{U^{\top}Q}^2 + \fro{U_{\perp}^{\top}Q V}^2 - \fro{\widehat{U}^{\top}Q}^2 - \fro{\widehat{U}_{\perp}^{\top}Q \widehat{V}}^2\right| \\
        &\leq \left|\fro{U^{\top}Q}^2 - \fro{\widehat{U}^{\top}Q}^2\right| + \left|\fro{U_{\perp}^{\top}Q V}^2 - \fro{\widehat{U}_{\perp}^{\top}Q \widehat{V}}^2\right|. 
    \end{align*}
    Observing that $U$ and $\widehat{U}$ both have orthonormal columns, then
    \begin{align*}
        \left|\fro{U^{\top}Q}^2 - \fro{\widehat{U}^{\top}Q}^2\right|&=\left|\fro{UU^{\top}Q}^2 - \fro{\widehat{U}\widehat{U}^{\top}Q}^2\right| \\
        &\leq \fro{(UU^{\top} - \widehat{U}\widehat{U}^{\top})Q}^2 + 2\left|\inp{(UU^{\top} - \widehat{U}\widehat{U}^{\top})Q}{UU^{\top}Q}\right|.
    \end{align*}
    Following the same arguments of Lemma \ref{2maxbound}, we can derive 
    \begin{align*}
        &\fro{(UU^{\top} - \widehat{U}\widehat{U}^{\top})Q}^2 \leq \left(\sum_{j_1,j_2} |Q(j_1,j_2)| \|(UU^{\top} - \widehat{U}\widehat{U}^{\top})e_{j_1}\|\right)^2 \\
        &\quad \leq \|Q\|_{\ell_1}^2 \|UU^{\top} - \widehat{U}\widehat{U}^{\top}\|_{2,\max}^2 \lesssim \|Q\|_{\ell_1}^2\frac{\sigma^2}{\lambda_{\min}^2}\frac{\mu^2 r d_2^2\log d_1}{d_1T}.
    \end{align*}
    Similarly,
    \begin{align*}
        &\left|\inp{(UU^{\top} - \widehat{U}\widehat{U}^{\top})Q}{UU^{\top}Q}\right|\leq \fro{U^{\top}Q}\fro{U^{\top}(UU^{\top} - \widehat{U}\widehat{U}^{\top})Q} \\
        &\quad \leq \fro{U^{\top}Q}\|Q\|_{\ell_1}\|\|UU^{\top} - \widehat{U}\widehat{U}^{\top}\|_{2,\max} \lesssim \fro{Q}\|Q\|_{\ell_1}\frac{\sigma}{\lambda_{\min}}\sqrt{\frac{\mu^2 r d_2^2\log d_1}{d_1T}}.
    \end{align*}
    Therefore, as long as $\fro{Q}\|Q\|_{\ell_1}\frac{\sigma}{\lambda_{\min}}\sqrt{\frac{\mu r d_2^2\log d_1}{d_1T}}\rightarrow 0$, the above term converges to 0 in probability.
    
    Next we bound $\left|\fro{U_{\perp}^{\top}Q R}^2 - \fro{\widehat{U}_{\perp}^{\top}Q \widehat{V}}^2\right|$. It can be decomposed into
    \begin{align*}
        \left|\fro{U_{\perp}^{\top}Q V}^2 - \fro{\widehat{U}_{\perp}^{\top}Q \widehat{V}}^2\right|\leq \left|\fro{U_{\perp}^{\top}Q V}^2 - \fro{{U}_{\perp}^{\top}Q \widehat{V}}^2\right| + \left|\fro{U_{\perp}^{\top}Q \widehat{V}}^2 - \fro{\widehat{U}_{\perp}^{\top}Q \widehat{V}}^2\right|.
    \end{align*}
    It follows that
    \begin{align*}
        &\left|\fro{U_{\perp}^{\top}Q V}^2 - \fro{{U}_{\perp}^{\top}Q \widehat{V}}^2\right|\leq \fro{{U}_{\perp}^{\top}Q(VV^{\top}-\widehat{V}\widehat{V}^{\top})}^2 + 2\left|\inp{{U}_{\perp}^{\top}Q(VV^{\top}-\widehat{V}\widehat{V}^{\top})}{U_{\perp}^{\top}Q V}\right| \\
        &\quad \lesssim  \|Q\|_{\ell_1}^2 \frac{\sigma^2}{\lambda_{\min}^2}\frac{\mu rd_2\log d_1}{T} + \fro{L_{\perp}^{\top}Q R}\|Q\|_{\ell_1}\frac{\sigma}{\lambda_{\min}}\sqrt{\frac{\mu rd_2\log d_1}{T}},
    \end{align*}
    and
    \begin{align*}
        &\quad\left|\fro{U_{\perp}^{\top}Q \widehat{V}}^2 - \fro{\widehat{U}_{\perp}^{\top}Q \widehat{V}}^2\right|\leq \fro{(\widehat{U}\widehat{U}^{\top}-UU^{\top})Q\widehat{V}\widehat{V}^{\top}}^2 + 2\left|\inp{{U}_{\perp}^{\top}Q(VV^{\top}-\widehat{V}\widehat{V}^{\top})}{U_{\perp}^{\top}Q \widehat{V}}\right| \\
        & \lesssim \|Q\|_{\ell_1}^2 \frac{\sigma^2}{\lambda_{\min}^2}\frac{\mu r d_2^2\log d_1}{d_1T}  + (\fro{U_{\perp}^{\top}Q {V}}+\fro{{U}_{\perp}^{\top}Q(VV^{\top}-\widehat{V}\widehat{V}^{\top})})\|Q\|_{\ell_1} \frac{\sigma}{\lambda_{\min}}\sqrt{\frac{\mu r d_2\log d_1}{T}} \\
        & \lesssim \|Q\|_{\ell_1}^2 \frac{\sigma^2}{\lambda_{\min}^2}\frac{\mu r d_2^2\log d_1}{d_1T} + \fro{U_{\perp}^{\top}Q {V}}\|Q\|_{\ell_1} \frac{\sigma}{\lambda_{\min}}\sqrt{\frac{\mu r d_2\log d_1}{T}}  + \|Q\|_{\ell_1}^2 \frac{\sigma^2}{\lambda_{\min}^2}\frac{\mu r d_2\log d_1}{T}.
    \end{align*}
    We can show $\left|\fro{U_{\perp}^{\top}Q \widehat{V}}^2 - \fro{\widehat{U}_{\perp}^{\top}Q \widehat{V}}^2\right|$ converges to 0 in probability as long as $\|Q\|_{\ell_1}^2 \frac{\sigma^2}{\lambda_{\min}^2}\frac{d_2^2\log d_1}{d_1T}\rightarrow 0$. As a result, $\fro{\mathcal{P}_{\widehat{M}}(Q)}^2\overset{p}{\rightarrow} \fro{\mathcal{P}_M(Q)}^2$.
\end{proof}

\subsection{Proof of Theorem \ref{thm:inference-main-non}}
\begin{proof}
    We start by recalling that $\widehat{M}_l^{\ubs}$ for $l=1,2$ can be decomposed into 
    \begin{align*}
        \widehat{M}_l^{\ubs}= M + \underbrace{\frac{1}{T_0}\sum_{t\in \widetilde{\mathcal{D}}_{l}}\sum_{i=1}^{d_1} \frac{1}{p_{ij_t(i)}}\xi_t^i X_t^i}_{=:\widehat{Z}_1^{(l)}} + \underbrace{\frac{1}{T_0}\sum_{t\in \widetilde{\mathcal{D}}_{l}} \left(\sum_{i=1}^{d_1}\frac{1}{p_{ij_t(i)}}\inp{\widehat{\Delta}_{l}}{X_t^i}X_t^i-\widehat{\Delta}_{l}\right)}_{=:\widehat{Z}_2^{(l)}},
    \end{align*}
    where $\widehat{\Delta}_{l}=M-\widehat{M}_l^{\init}$. Following the proof of Theorem \ref{thm:inference-main} and define $\Theta$, $\widehat{\Theta}_l$, $\widehat{E}^{(l)}$, $\mathfrak{P}^{-s}$ and all other variables in the same way, we again can obtain the decomposition
    \begin{align}
        \inp{\widehat{M}}{Q}-&\inp{M}{Q}= \frac{1}{2}\sum_{l=1}^{2} \inp{UU^{\top}\widehat{Z}_1^{(l)}V_{\perp}V_{\perp}^{\top}}{Q} + \inp{U_{\perp}U_{\perp}^{\top}\widehat{Z}_1^{(l)}VV^{\top}}{Q} + \inp{UU^{\top}\widehat{Z}_1^{(l)}VV^{\top}}{Q} \label{main1-non} \\ &\quad + \frac{1}{2}\sum_{l=1}^{2}\inp{UU^{\top}\widehat{Z}_2^{(l)}V_{\perp}V_{\perp}^{\top}}{Q} + \inp{U_{\perp}U_{\perp}^{\top}\widehat{Z}_2^{(l)}VV^{\top}}{Q} + \inp{UU^{\top}\widehat{Z}_2^{(l)}VV^{\top}}{Q} \label{neg1-non}\\ &\quad + \frac{1}{2}\sum_{l=1}^{2}\inp{\widehat{U}_l\widehat{U}_l^{\top}\widehat{Z}^{(l)}\widehat{V}_l\widehat{V}_l^{\top}}{Q} - \inp{UU^{\top}\widehat{Z}^{(l)}VV^{\top}}{Q} \label{neg2-non}\\ &\quad + \frac{1}{2}\sum_{l=1}^{2}\inp{\sum_{k=2}^{\infty} (\mathcal{S}_{A,k}(\widehat{E}^{(l)})A\Theta\Theta^{\top}+\Theta\Theta^{\top}A\mathcal{S}_{A,k}(\widehat{E}^{(l)}))}{\widetilde{Q}} + \inp{(\widehat{\Theta}\widehat{\Theta}^{\top}-\Theta\Theta^{\top})A(\widehat{\Theta}_l\widehat{\Theta}_l^{\top}-\Theta\Theta^{\top})}{\widetilde{Q}} \label{neg3-non}.
    \end{align}
    In the following Lemmas, we still first apply Berry Esseen Theorem to prove the asymptotic normality of the main term, while showing all the other terms are negligible. Then we can conclude the proof following Theorem \ref{thm:inference-main}.

    \noindent\textbf{Asymptotic normality of (\ref{main1-non})}
    
     By definition, 
    \begin{align*}
        &\frac{1}{2}\sum_{l=1}^{2}\inp{UU^{\top}\widehat{Z}_1^{(l)}V_{\perp}V_{\perp}^{\top}}{Q} + \inp{U_{\perp}U_{\perp}^{\top}\widehat{Z}_1^{(l)}VV^{\top}}{Q} + \inp{UU^{\top}\widehat{Z}_1^{(l)}VV^{\top}}{Q} \\ &\quad = \frac{1}{T}\sum_{t=1}^{T} \sum_{i=1}^{d_1} \frac{1}{p_{ij_t(i)}}\xi_t^i\inp{X_t^i}{\mathcal{P}_M(Q)},
    \end{align*}
    which is a summation of i.i.d. zero mean random variables $\sum_{i=1}^{d_1}\frac{1}{p_{ij_t(i)}} \xi_t^i\inp{X_t^i}{\mathcal{P}_M(Q)}$. To apply Berry-Essen Theorem \citep{berry1941accuracy,esseen1956moment}, we calculate its variance and third moments. It is easy to verify its variance as
    \begin{align*}
        \EE\left[\bigg(\sum_{i=1}^{d_1} \frac{1}{p_{ij_t(i)}}\xi_t^i\inp{X_t^i}{\mathcal{P}_M(Q)}\bigg)^2\right] = \sigma^2\sum_{i=1}^{d_1}\sum_{j=1}^{d_2}\frac{1}{p_{ij}}\PMQ(i,j)^2 =\sigma^2\fro{P_{-1}^{1/2}\circ\PMQ}^2.
    \end{align*}  
    Next, we bound its third order moment. Notice that condition on $(X_t^1,\cdots,X_t^{d_1})$, $\sum_{i=1}^{d_1}\frac{1}{p_{ij_t(i)}}\xi_t^i\inp{X_t^i}{\mathcal{P}_M(Q)}$ is subGaussian with variance $C_1\sigma^2 \sum_{i=1}^{d_1}\frac{1}{p_{ij_t(i)}^2} \inp{X_t^i}{\mathcal{P}_M(Q)}^2$ for some constant $C_1>0$. Therefore, 
    \begin{align*}
        &\EE\left[\bigg|\sum_{i=1}^{d_1} \frac{1}{p_{ij_t(i)}}\xi_t^i\inp{X_t^i}{\mathcal{P}_M(Q)}\bigg|^3\right]=\EE\left[\EE\bigg[\bigg|\sum_{i=1}^{d_1} \frac{1}{p_{ij_t(i)}}\xi_t^i\inp{X_t^i}{\mathcal{P}_M(Q)}\bigg|^3  \bigg|X_t^1,\cdots,X_t^{d_1} \bigg]\right] \\
        &\quad \lesssim \sigma^3\EE\left[\bigg( \sum_{i=1}^{d_1} \frac{1}{p_{ij_t(i)}^2}\inp{X_t^i}{\mathcal{P}_M(Q)}^2 \bigg)^{3/2}\right]:= \sigma^3\EE\left[S_t^{3/2}\right]. 
    \end{align*}
    Obviously, $\EE[S_t]=\fro{P^{-1/2}\circ \PMQ}^2$, and expanding $S_t^2$ gives
    \begin{align*}
    \EE[S_t^2]
    &= \sum_{i,j}\frac{\PMQ(i,j)^4}{p_{ij}^3}
      +\sum_{(i,j)\neq(k,l)} \frac{\pi_{ij,kl}}{p_{ij}^2p_{kl}^2} \PMQ(i,j)^2\PMQ(k,l)^2 \\
    &\leq \frac{1}{p_{\min}}\|P^{-1/2}\circ \PMQ\|_{\max}^2\fro{P^{-1/2}\circ \PMQ}^2
      +(1+\kappa_2)\fro{P^{-1/2}\circ \PMQ}^4 \\
    &=\fro{P^{-1/2}\circ \PMQ}^4
      \left(\frac{\|P^{-1/2}\circ \PMQ\|_{\max}^2}{p_{\min}\fro{P^{-1/2}\circ \PMQ}^2}+(1+\kappa_2)\right).
\end{align*}
Recall that $\rho_Q:= d_1\|\PMQ\|_{\max}^2/\fro{\PMQ}^2\lesssim \mu^2r$, then  
\begin{align*}
    \frac{\|P^{-1/2}\circ \PMQ\|_{\max}^2}{\fro{P^{-1/2}\circ \PMQ}^2}\leq \frac{p_{\max}\|\PMQ\|_{\max}^2}{p_{\min}\fro{\PMQ}^2}\lesssim \frac{p_{\max}\mu^2r}{p_{\min}d_1}
\end{align*}
Then by Cauchy Schwarz inequality,
\begin{align*}
    \EE S_t^{3/2}
    \leq (\EE S_t)^{1/2}(\EE S_t^2)^{1/2}
    \lesssim  \fro{P^{-1/2}\circ \PMQ}^3
    \left(\frac{p_{\max}\mu^2r}{p_{\min}^2d_1}+(1+\kappa_2)\right)^{1/2},
\end{align*}
which directly leads to 
\begin{align*}
    \EE\left[\bigg|\sum_{i=1}^{d_1} \frac{1}{p_{ij_t(i)}}\xi_t^i\inp{X_t^i}{\mathcal{P}_M(Q)}\bigg|^3\right]\lesssim \fro{P^{-1/2}\circ \PMQ}^3
    \left(\frac{p_{\max}\mu^2r}{p_{\min}^2d_1}+1+\kappa_2\right)^{1/2}.
\end{align*}
Finally, applying the classical Berry-Esseen theorem \citep{berry1941accuracy,esseen1956moment}, we can obtain
    \begin{align*}
        &\sup_{t\in \RR}\left|\PP\left(\frac{\frac{1}{2}\sum_{l=1}^{2} \inp{UU^{\top}\widehat{Z}_1^{(l)}V_{\perp}V_{\perp}^{\top}}{Q} + \inp{U_{\perp}U_{\perp}^{\top}\widehat{Z}_1^{(l)}VV^{\top}}{Q} + \inp{UU^{\top}\widehat{Z}_1^{(l)}VV^{\top}}{Q}}{\sigma\fro{P^{-1/2}\circ \mathcal{P}_M(Q)}\sqrt{1/T}}\leq t\right) - \Phi(t)\right| \\ &\quad \lesssim \sqrt{\frac{p_{\max}\mu^2r /(p_{\min}^2d_1) +1 + \kappa_2}{T}}.
    \end{align*}
Then as long as $\frac{p_{\max}\mu^2r /(p_{\min}^2d_1) +1+\kappa_2}{T}\rightarrow 0$, we have 
\begin{align*}
    \frac{\frac{1}{2}\sum_{l=1}^{2} \inp{UU^{\top}\widehat{Z}_1^{(l)}V_{\perp}V_{\perp}^{\top}}{Q} + \inp{U_{\perp}U_{\perp}^{\top}\widehat{Z}_1^{(l)}VV^{\top}}{Q} + \inp{UU^{\top}\widehat{Z}_1^{(l)}VV^{\top}}{Q}}{\sigma\fro{P^{-1/2}\circ \mathcal{P}_M(Q)}\sqrt{1/T}} \overset{d.}{\rightarrow} N(0,1). 
\end{align*}

    \noindent\textbf{Bounding (\ref{neg1-non})}

     Without loss of generality, we only prove the upper bound for $|\inp{UU^{\top}\widehat{Z}_2^{(1)}V_{\perp}V_{\perp}^{\top}}{Q} + \inp{U_{\perp}U_{\perp}^{\top}\widehat{Z}_2^{(1)}VV^{\top}}{Q} + \inp{UU^{\top}\widehat{Z}_2^{(1)}VV^{\top}}{Q}|$. The the upper bound for $|\inp{UU^{\top}\widehat{Z}_2^{(2)}V_{\perp}V_{\perp}^{\top}}{Q} + \inp{U_{\perp}U_{\perp}^{\top}\widehat{Z}_2^{(2)}VV^{\top}}{Q} + \inp{UU^{\top}\widehat{Z}_2^{(2)}VV^{\top}}{Q}|$ can be derived following the same arguments. By definition, 
    \begin{align*}
        &|\inp{UU^{\top}\widehat{Z}_2^{(1)}V_{\perp}V_{\perp}^{\top}}{Q} + \inp{U_{\perp}U_{\perp}^{\top}\widehat{Z}_2^{(1)}VV^{\top}}{Q} + \inp{UU^{\top}\widehat{Z}_2^{(1)}VV^{\top}}{Q}|\\ 
        &\quad =\frac{1}{T_0}\sum_{t=1}^{T_0} \sum_{i=1}^{d_1} \frac{1}{p_{ij_t(i)}}\inp{\widehat{\Delta}_1}{X_t^i}\inp{X_t^i}{\PMQ} -\inp{\widehat{\Delta}_1}{\PMQ}.
    \end{align*}

    Next we calculate the variance of $\frac{1}{T_0}\bigg|\frac{1}{p_{ij_t(i)}}\sum_{i=1}^{d_1}\inp{\widehat{\Delta}_1}{X_t^i}\inp{X_t^i}{\PMQ} -\inp{\widehat{\Delta}_1}{\PMQ}\bigg|$. Notice that $\widehat{\Delta}_1$ is independent of $X_t^i$, then
    \begin{align*}
        &\frac{1}{T_0^2}\EE\left[\bigg(\sum_{i=1}^{d_1}\frac{1}{p_{ij_t(i)}}\inp{\widehat{\Delta}_1}{X_t^i}\inp{X_t^i}{\PMQ} -\inp{\widehat{\Delta}_1}{\PMQ} \bigg)^2\right] \\
        &\quad = \frac{1}{T_0^2}\EE\left[\bigg(\sum_{i=1}^{d_1}\frac{1}{p_{ij_t(i)}}\inp{\widehat{\Delta}_1}{X_t^i}\inp{X_t^i}{\PMQ} \bigg)^2\right]-\frac{1}{T_0^2}\inp{\widehat{\Delta}_1}{\PMQ}^2 \\
        &\quad = \frac{1}{T_0^2}\underbrace{\EE\left[\sum_{i=1}^{d_1}\frac{1}{p_{ij_t(i)}^2}\inp{\widehat{\Delta}_1}{X_t^i}^2\inp{X_t^i}{\PMQ}^2 \right]}_{\mathcal{I}_1} \\ &\quad\quad + \frac{1}{T_0^2}\underbrace{\EE\left[\sum_{i=1}^{d_1}\sum_{j\neq i} \frac{1}{p_{ij_t(i)}p_{jj_t(j)} }\inp{\widehat{\Delta}_1}{X_t^i}\inp{X_t^i}{\PMQ}\inp{\widehat{\Delta}_1}{X_t^j}\inp{X_t^j}{\PMQ} \right]}_{\mathcal{I}_2} - \frac{1}{T_0^2}\inp{\widehat{\Delta}_1}{\PMQ}^2.
    \end{align*}
    Denote $A=\widehat{\Delta}_1\odot \PMQ$ the Hadamard product of $\widehat{\Delta}_1$ and $\PMQ$, then 
    \begin{align*}
        \mathcal{I}_1=\EE[\sum_{i=1}^{d_1}\frac{1}{p_{ij_t(i)}^2}\inp{X_t^i}{A}^2]= \fro{P^{-1/2} \circ A}^2.
    \end{align*}
    It is obvious that 
    \begin{align*}
        \mathcal{I}_2=\sum_{i\neq j}\sum_{p\neq q} \frac{\pi_{ip, jq}}{p_{ip}p_{jq}} A(i,p)A(j,q).
    \end{align*}
    Since 
    \begin{align*}
        \inp{\widehat{\Delta}_1}{\PMQ}^2&=\big(\sum_{i,p} A(i,p) \big)^2= \sum_{i,p} A(i,p)^2 + \sum_i \sum_{p\neq q} A(i,p)A(i,q) \\ &\quad + \sum_{i\neq j}\sum_p A(i,p)A(j,p) + \sum_{i\neq j}\sum_{p\neq q} A(i,p)A(j,q) \\
        &\geq \sum_i \sum_{p\neq q} A(i,p)A(i,q) + \sum_{i\neq j}\sum_p A(i,p)A(j,p) + \sum_{i\neq j}\sum_{p\neq q} A(i,p)A(j,q).
    \end{align*}
    Notice that by Cauchy Schwarz inequality,
    \begin{align*}
        |\sum_i \sum_{p\neq q} A(i,p)A(i,q)|&\leq \sum_{i} \left(\sum_{p} A(i,p)\right)^2 \leq \sum_{i}\left(\sum_{p} p_{ip}\right) \left(\sum_{p} \frac{A(i,p)^2}{p_{ip}} \right)=\fro{P^{-1/2}\circ A}^2,  
    \end{align*}
    where the last inequality comes from $\sum_{p} p_{ip}=1$ since every row is matched exactly once. Similarly, since each column can be matched to at most one row, $\sum_{i} p_{ip}\leq 1$ for $\forall p\in [d_2]$,  
    \begin{align*}
        |\sum_{i\neq j}\sum_p A(i,p)A(j,p)|\leq \fro{P^{-1/2}\circ A}^2.
    \end{align*}
    Moreover, by the condition $|\pi_{ip,jq} - p_{ip}p_{jq}|\leq \kappa_2p_{ip}p_{jq}$, 
    \begin{align*}
        \mathcal{I}_2 - \sum_{i\neq j}\sum_{p\neq q} A(i,p)A(j,q) &= \sum_{i\neq j}\sum_{p\neq q}(\frac{\pi_{ip, jq}}{p_{ip}p_{jq}}-1) A(i,p)A(j,q) \leq \kappa_2\sum_{i\neq j}\sum_{p\neq q} |A(i,p)A(j,q)| \\
        &\leq \kappa_2\left(\sum_{i,p} A(i,p)\right)^2\leq \kappa_2\left(\sum_{i,p} p_{ip}\right) \left(\sum_{p} \frac{A(i,p)^2}{p_{ip}} \right)=\kappa_2d_1\fro{P^{-1/2}\circ A}^2.
    \end{align*}
   Overall, the variance can be upper bounded by
   \begin{align*}
        &\frac{1}{T_0^2}\EE\left[\bigg(\sum_{i=1}^{d_1}\frac{1}{p_{ij_t(i)}}\inp{\widehat{\Delta}_1}{X_t^i}\inp{X_t^i}{\PMQ} -\inp{\widehat{\Delta}_1}{\PMQ} \bigg)^2\right] \\ &\quad \leq (3 + \kappa_2d_1)\fro{P^{-1/2} \circ A}^2\leq (3 + \kappa_2d_1)\|\widehat{\Delta}_1\|_{\max}^2\fro{P^{-1/2} \circ \PMQ}^2.
   \end{align*}
   By Chebyshev inequality, for any $\delta>0$, 
   \begin{align*}
    &\PP\left(\frac{|\inp{UU^{\top}\widehat{Z}_2^{(1)}V_{\perp}V_{\perp}^{\top}}{Q} + \inp{U_{\perp}U_{\perp}^{\top}\widehat{Z}_2^{(1)}VV^{\top}}{Q} + \inp{UU^{\top}\widehat{Z}_2^{(1)}VV^{\top}}{Q}|}{\sigma\fro{P^{-1/2}\circ \PMQ}\sqrt{1/T}}>\delta\right) \\
    &\quad \leq \frac{(3 + \kappa_2d_1)\|\widehat{\Delta}_1\|_{\max}^2\fro{P^{-1/2} \circ \PMQ}^2/T}{\delta^2\sigma^2\fro{P^{-1/2}\circ \PMQ}^2/T}
   \end{align*}
   Then as long as $\|\widehat{\Delta}_1\|_{\max}=o_p(\sigma)$ and $\kappa_2=O(d_1^{-1})$, \\ $\frac{|\inp{UU^{\top}\widehat{Z}_2^{(1)}V_{\perp}V_{\perp}^{\top}}{Q} + \inp{U_{\perp}U_{\perp}^{\top}\widehat{Z}_2^{(1)}VV^{\top}}{Q} + \inp{UU^{\top}\widehat{Z}_2^{(1)}VV^{\top}}{Q}|}{\sigma\fro{P^{-1/2}\circ \PMQ}\sqrt{1/T}}=o_p(1)$.

   Finally, following the proof of Lemma \ref{lemmaneg2} and Lemma \ref{lemmaneg3}, as long as we can derive the upper bound of $\|\widehat{Z}^{(l)}\|$, $\|\widehat{U}_l\widehat{U}_l^{\top}-UU^{\top}\|_{2,\max}$ and $\|\widehat{V}_l\widehat{V}_l^{\top}-VV^{\top}\|_{2,\max}$, then we can use them to bound (\ref{neg2-otm}) and (\ref{neg3-otm}). The next two Lemmas show the upper bound of $\|\widehat{Z}^{(l)}\|$, $\|\widehat{U}_l\widehat{U}_l^{\top}-UU^{\top}\|_{2,\max}$ and $\|\widehat{V}_l\widehat{V}_l^{\top}-VV^{\top}\|_{2,\max}$. 

    \begin{Lemma}
        \label{Zbound-non}
        Under the conditions in Theorem \ref{thm:inference-main-non}, with probability at least $1-2d_1^{-10}$, 
        \begin{align*}
        \|\widehat{Z}^{(l)}\|\lesssim \sqrt{\frac{\log d_1}{d_1T}\left(\sum_{i=1}^{d_1}\sum_{j=1}^{d_2}\frac{1}{p_{ij}}\right)}\sigma
        \end{align*}
        for $l=1,2$. 
    \end{Lemma}

    \begin{Lemma}
        \label{2maxbound-non}
        Under the conditions in Theorem \ref{thm:inference-main-non}, with probability at least $1-5d_1^{-5}$,
        \begin{align*}
            &\|\widehat{U}_l\widehat{U}_l^{\top}-UU^{\top}\|_{2,\max}\lesssim \frac{\sigma}{\lambda_{\min}}\sqrt{\frac{\log d_1}{d_1T}\left(\sum_{i=1}^{d_1}\sum_{j=1}^{d_2}\frac{1}{p_{ij}}\right)}\mu\sqrt{\frac{r}{d_1}} \\
            &\|\widehat{V}_l\widehat{V}_l^{\top}-VV^{\top}\|_{2,\max}\lesssim \frac{\sigma}{\lambda_{\min}}\sqrt{\frac{\log d_1}{d_1T}\left(\sum_{i=1}^{d_1}\sum_{j=1}^{d_2}\frac{1}{p_{ij}}\right)}\mu\sqrt{\frac{r}{d_2}}. 
        \end{align*}
        for $l=1,2$. 
    \end{Lemma}

    As a result, we can show that, 
    \begin{align*}
        \frac{\big|\frac{1}{2}\sum_{l=1}^{2}\inp{\widehat{U}_l\widehat{U}_l^{\top}\widehat{Z}^{(l)}\widehat{V}_l\widehat{V}_l^{\top}}{Q} - \inp{UU^{\top}\widehat{Z}^{(l)}VV^{\top}}{Q}\big|}{\sigma\fro{P^{-1/2}\circ \mathcal{P}_M(Q)}\sqrt{1/T}}\lesssim \frac{\sigma\mu^2}{\lambda_{\min}}\sqrt{\frac{r^2d_2\log d_1}{Td_1^2}\left(\sum_{i=1}^{d_1}\sum_{j=1}^{d_2}\frac{1}{p_{ij}}\right)}\frac{\|Q\|_{\ell_1}}{\fro{\PMQ}}. 
    \end{align*}
    and 
    \begin{align*}
        &\frac{\big|\sum_{k=2}^{\infty} \inp{(\mathcal{S}_{A,k}(\widehat{E})A\Theta\Theta^{\top}+\Theta\Theta^{\top}A\mathcal{S}_{A,k}(\widehat{E}))}{\widetilde{Q}}\big|}{\sigma\fro{P^{-1/2}\circ \mathcal{P}_M(Q)}\sqrt{1/T}}\lesssim \frac{\sigma\mu^2}{\lambda_{\min}}\sqrt{\frac{r^2\log d_1}{Td_1^2}\left(\sum_{i=1}^{d_1}\sum_{j=1}^{d_2}\frac{1}{p_{ij}}\right)}\frac{\|Q\|_{\ell_1}}{\fro{\PMQ}}, \\
        &\frac{\big|\inp{(\widehat{\Theta}\widehat{\Theta}^{\top}-\Theta\Theta^{\top})A(\widehat{\Theta}\widehat{\Theta}^{\top}-\Theta\Theta^{\top})}{\widetilde{Q}}\big|}{\sigma\fro{P^{-1/2}\circ \mathcal{P}_M(Q)}\sqrt{1/T}} \lesssim \frac{\sigma\mu^2\kappa}{\lambda_{\min}}\sqrt{\frac{r^2\log d_1}{Td_1^2}\left(\sum_{i=1}^{d_1}\sum_{j=1}^{d_2}\frac{1}{p_{ij}}\right)}\frac{\|Q\|_{\ell_1}}{\fro{\PMQ}}.
    \end{align*}
    Then we conclude the proof.

\end{proof}

\section{Proof of Propositions}\label{sec: proof-estimation}

\subsection{Proof of Proposition \ref{thm:estimation}}

We introduce several more notations to facilitate the proof. We first define
\begin{align*}
    \widehat{O}_U^{(1)}=\argmin_{O\in \OO^{r\times r}} \|\widehat{U}^{(1)} - UO\| \quad \text{and} \quad \widehat{O}_V^{(1)}=\argmin_{O\in \OO^{r\times r}} \|\widehat{V}^{(1)}- VO\|,
\end{align*}
and for all $p=1,\cdots,m-1$, denote the SVDs
\begin{align*}
    \widehat{U}^{(p+0.5)}=\widehat{U}^{(p+1)}\widehat{\Sigma}_U^{(p+1)}\widehat{K}_U^{(p+1)\top} \quad \text{and} \quad \widehat{V}^{(p+0.5)}=\widehat{V}^{(p+1)}\widehat{\Sigma}_V^{(p+1)}\widehat{K}_V^{(p+1)\top}.
\end{align*}
and the orthogonal matrices
\begin{align*}
    \widehat{O}_U^{(p+1)}=\widehat{O}_U^{(p)}\widehat{L}_G^{(p)}\widehat{K}_U^{(p+1)} \quad \text{and} \quad \widehat{O}_V^{(p+1)}=\widehat{O}_V^{(p)}\widehat{R}_G^{(p)}\widehat{K}_V^{(p+1)}.
\end{align*}

\begin{proof}
    We first prove the following statement: for $p=1,\cdots,m-1$, with probability at least $1-4md_1^{-10}$,
    \begin{align*}
         &\sqrt{\frac{1}{d_2}}\|\widehat{U}^{(p+1)}-U\widehat{O}_U^{(p+1)}\|_{2,\max} + \sqrt{\frac{1}{d_1}}\|\widehat{V}^{(p+1)}-V\widehat{O}_V^{(p+1)}\|_{2,\max}\leq 2C\eta\frac{\sigma}{\lambda_{\min}}\sqrt{\frac{rd_2\log d_1}{Td_1}} \\ &\quad\quad + (1-\frac{2}{3}\eta)\left(\sqrt{\frac{1}{d_2}}\|\widehat{U}^{(p)}-U\widehat{O}_U^{(p)}\|_{2,\max} + \sqrt{\frac{1}{d_1}}\|\widehat{V}^{(p)}-V\widehat{O}_V^{(p)}\|_{2,\max}\right)
    \end{align*}
    for some constant $C$.

    We begin with the accuracy of $\widehat{G}^{(p)}$. By the definition of $\widehat{G}^{(p)}$, we have
    \begin{align*}
        \frac{d_2}{N_0}\sum_{t\in \mathcal{D}_{2p}}\sum_{i=1}^{d_1}(\inp{X_t^i}{\widehat{U}^{(p)}\widehat{G}^{(p)}\widehat{V}^{(p)\top}}-y_t^{i})\widehat{U}^{(p)\top}X_t^i\widehat{V}^{(p)}=0.
    \end{align*}
    Let $\widehat{O}_U^{(p)}$ and $\widehat{O}_V^{(p)}$ be any orthogonal matrices such that 
    \begin{align*}
        \max(\|\widehat{U}^{(p)}-U\widehat{O}_U^{(p)}\|, \|\widehat{V}^{(p)}-V\widehat{O}_U^{(p)}\|)\leq \frac{1}{C_1\mu\kappa^2\sqrt{r}}
    \end{align*}
    for some large constant $C_1>0$.
    \begin{Lemma}\label{lemma:estimation1}
        Suppose that $\|\widehat{U}^{(p)}\|_{2,\max}\leq 2\mu\sqrt{r/d_1}$, $\|\widehat{V}^{(p)}\|_{2,\max}\leq 2\mu\sqrt{r/d_2}$ and $T\geq C_2\mu^4r^3\log d_1$, then with probability at least $1-3d_1^{-10}$,
        \begin{align*}
            &\|\widehat{G}^{(p)}-\widehat{O}_U^{(p)\top}\Lambda\widehat{O}_V^{(p)}\|\leq C_3(\frac{r\sigma}{N_0}\sqrt{d_1d_2}\log d_1 + \sqrt{\frac{\sigma^2 rd_2\log d_1}{N_0}}) \\ &\quad + 2\lambda_{\max}(\|\widehat{U}^{(p)}-U\widehat{O}_U^{(p)}\|^2+\|\widehat{V}^{(p)}-V\widehat{O}_U^{(p)}\|^2) \\ &\quad + C_4\lambda_{\max}\left(\sqrt{\frac{1}{d_2}}\|\widehat{U}^{(p)}-U\widehat{O}_U^{(p)}\|_{2,\max} + \sqrt{\frac{1}{d_1}}\|\widehat{V}^{(p)}-V\widehat{O}_V^{(p)}\|_{2,\max}\right)\mu\sqrt{\frac{r^3d_1d_2\log d_1}{N_0}}
        \end{align*}
        for some constant $C_2, C_3, C_4>0$.
    \end{Lemma}
    Recall $\widehat{G}^{(p)}=\widehat{L}_G^{(p)}\widehat{\Lambda}^{(p)}\widehat{R}_G^{(p)\top}$ denotes $\widehat{G}^{(p)}$'s SVD. By the gradient descent step of Algorithm \ref{alg:gd},
    \begin{align*}
        \widehat{U}^{(p+0.5)}&=\widehat{U}^{(p)}\widehat{L}_G^{(p)} - \eta\frac{d_2}{N_0}\sum_{t\in \mathcal{D}_{2p+1}} \sum_{i=1}^{d_1} \inp{\widehat{U}^{(p)}\widehat{G}^{(p)}\widehat{V}^{(p)\top} - U\Lambda V^{\top}}{X_t^i}X_t^{i}\widehat{V}^{(p)}\widehat{R}_G^{(p)}(\widehat{\Lambda}^{(p)})^{-1} \\ &\quad - \eta\frac{d_2}{N_0}\sum_{t\in \mathcal{D}_{2p+1}} \sum_{i=1}^{d_1}\xi_t^i X_t^{i}\widehat{V}^{(p)}\widehat{R}_G^{(p)}(\widehat{\Lambda}^{(p)})^{-1}.
    \end{align*}
    Observe that $(\widehat{U}^{(p)}, \widehat{G}^{(p)}, \widehat{V}^{(p)})$ are independent with $\mathcal{D}_{2p+1}$. Then we write
    \begin{align*}
        \widehat{U}^{(p+0.5)}=\widehat{U}^{(p)}\widehat{L}_G^{(p)} - \eta(\widehat{U}^{(p)}\widehat{G}^{(p)}\widehat{V}^{(p)\top} - U\Lambda V^{\top})\widehat{V}^{(p)}\widehat{R}_G^{(p)}(\widehat{\Lambda}^{(p)})^{-1} + \widehat{E}_V^{(p)} + \widehat{E}_{\xi,V}^{(p)},
    \end{align*} 
    where
    \begin{align*}
        &\widehat{E}_V^{(p)}=\eta(\widehat{U}^{(p)}\widehat{G}^{(p)}\widehat{V}^{(p)\top} - U\Lambda V^{\top})\widehat{V}^{(p)}\widehat{R}_G^{(p)}(\widehat{\Lambda}^{(p)})^{-1} \\ &\quad - \eta\frac{d_2}{N_0}\sum_{t\in \mathcal{D}_{2p+1}} \sum_{i=1}^{d_1} \inp{\widehat{U}^{(p)}\widehat{G}^{(p)}\widehat{V}^{(p)\top} - U\Lambda V^{\top}}{X_t^i}X_t^{i}\widehat{V}^{(p)}\widehat{R}_G^{(p)}(\widehat{\Lambda}^{(p)})^{-1}
    \end{align*}
    and
    \begin{align*}
        \widehat{E}_{\xi,V}^{(p)}=- \eta\frac{d_2}{N_0}\sum_{t\in \mathcal{D}_{2p+1}} \sum_{i=1}^{d_1}\xi_t^i X_t^{i}\widehat{V}^{(p)}\widehat{R}_G^{(p)}(\widehat{\Lambda}^{(p)})^{-1}.
    \end{align*}
    Note that 
    \begin{align*}
        &(\widehat{U}^{(p)}\widehat{G}^{(p)}\widehat{V}^{(p)\top} - U\Lambda V^{\top})\widehat{V}^{(p)}\widehat{R}_G^{(p)}(\widehat{\Lambda}^{(p)})^{-1} = \widehat{U}^{(p)}(\widehat{G}^{(p)}-\widehat{O}_U^{(p)\top}\Lambda\widehat{O}_V^{(p)})\widehat{R}_G^{(p)}(\widehat{\Lambda}^{(p)})^{-1} \\
        &\quad + (\widehat{U}^{(p)}\widehat{L}_G^{(p)}-U\widehat{O}_U^{(p)}\widehat{L}_G^{(p)})\widehat{L}_G^{(p)\top}\widehat{O}_U^{(p)\top}\Lambda\widehat{O}_V^{(p)}\widehat{R}_G^{(p)}(\widehat{\Lambda}^{(p)})^{-1} + U\Lambda(\widehat{V}^{(p)}\widehat{O}_V^{(p)\top}-V)^{\top}\widehat{V}^{(p)}\widehat{R}_G^{(p)}(\widehat{\Lambda}^{(p)})^{-1}.
    \end{align*}
    Therefore,
    \begin{align*}
        &\widehat{U}^{(p+0.5)}=U\widehat{O}_U^{(p)}\widehat{L}_G^{(p)} + (\widehat{U}^{(p)}\widehat{L}_G^{(p)}-U\widehat{O}_U^{(p)}\widehat{L}_G^{(p)})(I-\eta\widehat{L}_G^{(p)\top}\widehat{O}_U^{(p)\top}\Lambda\widehat{O}_V^{(p)}\widehat{R}_G^{(p)}(\widehat{\Lambda}^{(p)})^{-1}) \\
        &\quad -\eta \widehat{U}^{(p)}(\widehat{G}^{(p)}-\widehat{O}_U^{(p)\top}\Lambda\widehat{O}_V^{(p)})\widehat{R}_G^{(p)}(\widehat{\Lambda}^{(p)})^{-1} - \eta U\Lambda(\widehat{V}^{(p)}\widehat{O}_V^{(p)\top}-V)^{\top}\widehat{V}^{(p)}\widehat{R}_G^{(p)}(\widehat{\Lambda}^{(p)})^{-1} + \widehat{E}_V^{(p)} + \widehat{E}_{\xi,V}^{(p)}.
    \end{align*}
    \begin{Lemma}\label{lemma:estimation2}
        Under the conditions in Proposition \ref{thm:estimation} and Lemma \ref{lemma:estimation1}, and 
        \begin{align*}
            \max\{\|\widehat{U}^{(p)}-U\widehat{O}_U^{(p)}\|, \|\widehat{V}^{(p)}-V\widehat{O}_V^{(p)}\|\}\leq 1/(C_1\mu\kappa r\sqrt{d_2/d_1})
        \end{align*}
        for some large constant $C_1$. If $\eta\in [0.25, 0.75]$, then with probability at least $1-2d_1^{-10}$,
        \begin{align*}
            &\|\widehat{U}^{(p+0.5)}-U\widehat{O}_U^{(p)}\widehat{L}_G^{(p)}\|_{2,\max}\leq (1-\frac{9\eta}{10})\|\widehat{U}^{(p)}-U\widehat{O}_U^{(p)}\|_{2,\max} + C_2\eta\frac{\sigma}{\lambda_{\min}}\sqrt{\frac{d_2r\log d_1}{N_0}} \\ &\quad + \frac{\eta}{8}(\sqrt{\frac{d_1}{d_2}}\|\widehat{U}^{(p)}-U\widehat{O}_U^{(p)}\|_{2,\max}+\|\widehat{V}^{(p)}-V\widehat{O}_V^{(p)}\|_{2,\max}),
        \end{align*}
        for some constants $C_2,>0$. Moreover, with probability at least $1-2d_1^{-10}$,
        \begin{align*}
            &\left\{|1-\lambda_{\min}(\widehat{U}^{(p+0.5)})|, |1-\lambda_{\max}(\widehat{U}^{(p+0.5)})|\right\}\lesssim \eta\frac{\sigma}{\lambda_{\min}}\sqrt{\frac{d_2^2\log d_1}{N_0}} \\ &\quad + (\kappa^2\eta^2+\kappa\eta)(\|\widehat{U}^{(p)}-U\widehat{O}_U^{(p)}\|^2 + \|\widehat{V}^{(p)}-V\widehat{O}_V^{(p)}\|^2) \\
            &\quad +  \eta\sqrt{\frac{d_2^2\log d_1}{N_0}}\left(\sqrt{\frac{r}{d_2}}\|\widehat{U}^{(p)}-U\widehat{O}_U^{(p)}\|_{2,\max} + \sqrt{\frac{r}{d_1}}\|\widehat{V}^{(p)}-V\widehat{O}_V^{(p)}\|_{2,\max}\right).
        \end{align*}
    \end{Lemma} 
    By Lemma \ref{lemma:estimation2}, we denote the SVD of $\widehat{U}^{(p+0.5)}$ by $\widehat{U}^{(p+0.5)}=\widehat{U}^{(p+1)}\widehat{\Sigma}_U^{(p+1)}\widehat{K}_U^{(p+1)\top}$ where $\widehat{\Sigma}_U^{(p+1)}$ is diagonal and
    \begin{align*}
        &\|\widehat{\Sigma}_U^{(p+1)}-I\|\leq C_3\eta\frac{\sigma}{\lambda_{\min}}\sqrt{\frac{d_2^2\log d_1}{N_0}} \\ &\quad + C_4(\kappa^2\eta^2+\kappa\eta)(\|\widehat{U}^{(p)}-U\widehat{O}_U^{(p)}\|^2 + \|\widehat{V}^{(p)}-V\widehat{O}_V^{(p)}\|^2) \\
        &\quad +  C_5\eta\sqrt{\frac{d_2^2\log d_1}{N_0}}\left(\sqrt{\frac{r}{d_2}}\|\widehat{U}^{(p)}-U\widehat{O}_U^{(p)}\|_{2,\max} + \sqrt{\frac{r}{d_1}}\|\widehat{V}^{(p)}-V\widehat{O}_V^{(p)}\|_{2,\max}\right).
    \end{align*}
    By $\widehat{U}^{(p+1)}\widehat{\Sigma}_U^{(p+1)}\widehat{K}_U^{(p+1)\top}=U\widehat{O}_U^{(p)}\widehat{L}_G^{(p)} + (\widehat{\Sigma}_U^{(p+0.5)}-U\widehat{O}_U^{(p)}\widehat{L}_G^{(p)})$, we write
    \begin{align*}
        \widehat{U}^{(p+1)}=U\widehat{O}_U^{(p)}\widehat{L}_G^{(p)}\widehat{K}_U^{(p+1)}(\widehat{\Sigma}_U^{(p+1)})^{-1} + (\widehat{U}^{(p+0.5)}-U\widehat{O}_U^{(p)}\widehat{L}_G^{(p)})\widehat{K}_U^{(p+1)}(\widehat{\Sigma}_U^{(p+1)})^{-1}
    \end{align*}
    and 
    \begin{align*}
        \widehat{U}^{(p+1)} - U\widehat{O}_U^{(p)}\widehat{L}_G^{(p)}\widehat{K}_U^{(p+1)}=U\widehat{O}_U^{(p)}\widehat{L}_G^{(p)}\widehat{K}_U^{(p+1)}((\widehat{\Sigma}_U^{(p+1)})^{-1}-I) + (\widehat{U}^{(p+0.5)}-U\widehat{O}_U^{(p)}\widehat{L}_G^{(p)})\widehat{K}_U^{(p+1)}(\widehat{\Sigma}_U^{(p+1)})^{-1}.
    \end{align*}
    Note that $\widehat{O}_U^{(p)}\widehat{L}_G^{(p)}\widehat{K}_U^{(p+1)}$ is an $r\times r$ orthogonal matrix. The Assumptions of Lemma \ref{lemma:estimation2} can guarantee $\lambda_{\min}(\widehat{\Sigma}_U^{(p+1)})\geq 1-\eta/20$ so that $\|(\widehat{\Sigma}_U^{(p+1)})^{-1}\|\leq 1+\eta/10$. Therefore, 
    \begin{align*}
        &\sqrt{\frac{1}{d_2}}\|\widehat{U}^{(p+1)} - U\widehat{O}_U^{(p)}\widehat{L}_G^{(p)}\widehat{K}_U^{(p+1)}\|_{2,\max}\leq \sqrt{\frac{1}{d_2}}\|U\|_{2,\max} \|(\widehat{\Sigma}_U^{(p+1)})^{-1}-I\| \\ &\quad\quad + \sqrt{\frac{1}{d_2}}(1+\eta/10)\| \widehat{U}^{(p+0.5)} -  U\widehat{O}_U^{(p)}\widehat{L}_G^{(p)}\|_{2,\max} \\
        &\quad \leq C_3\eta\frac{\sigma}{\lambda_{\min}}\sqrt{\frac{rd_2\log d_1}{N_0d_1}} + C_4(\kappa^2\eta^2+\kappa\eta)(\|\widehat{U}^{(p)}-U\widehat{O}_U^{(p)}\|^2 + \|\widehat{V}^{(p)}-V\widehat{O}_V^{(p)}\|^2)\sqrt{\frac{r}{d_1d_2}} \\
        &\quad\quad +  C_5\eta\sqrt{\frac{rd_2\log d_1}{N_0d_1}}\left(\sqrt{\frac{r}{d_2}}\|\widehat{U}^{(p)}-U\widehat{O}_U^{(p)}\|_{2,\max} + \sqrt{\frac{r}{d_1}}\|\widehat{V}^{(p)}-V\widehat{O}_V^{(p)}\|_{2,\max}\right) \\ 
        &\quad\quad + (1-\frac{4\eta}{5})\sqrt{\frac{1}{d_2}}\|\widehat{U}^{(p)}-U\widehat{O}_U^{(p)}\|_{2,\max} + C_2\eta\frac{\sigma}{\lambda_{\min}}\sqrt{\frac{r^2\log d_1}{N_0}} \\ &\quad\quad + \frac{\eta}{7}(\sqrt{\frac{d_1}{d_2}}\|\widehat{U}^{(p)}-U\widehat{O}_U^{(p)}\|_{2,\max}+\|\widehat{V}^{(p)}-V\widehat{O}_V^{(p)}\|_{2,\max})\sqrt{\frac{r}{d_2}} \\
        &\quad \leq (1-\frac{4\eta}{5})\sqrt{\frac{1}{d_2}}\|\widehat{U}^{(p)}-U\widehat{O}_U^{(p)}\|_{2,\max} + C_3\eta\frac{\sigma}{\lambda_{\min}}\sqrt{\frac{rd_2\log d_1}{N_0d_1}} \\ &\quad\quad + \frac{\eta}{6}\left(\sqrt{\frac{r}{d_2}}\|\widehat{U}^{(p)}-U\widehat{O}_U^{(p)}\|_{2,\max} + \sqrt{\frac{r}{d_1}}\|\widehat{V}^{(p)}-V\widehat{O}_V^{(p)}\|_{2,\max}\right)
    \end{align*}
    as long as $T\gtrsim \alpha_dr\log d_1$ and $\|\widehat{U}^{(p)}-U\widehat{O}_U^{(p)}\| + \|\widehat{V}^{(p)}-V\widehat{O}_V^{(p)}\| \lesssim 1/\kappa^2r$. Similarly, we have 
    \begin{align*}
        &\sqrt{\frac{1}{d_1}}\|\widehat{V}^{(p+1)} - V\widehat{O}_V^{(p)}\widehat{R}_G^{(p)}\widehat{K}_V^{(p+1)}\|_{2,\max}\leq (1-\frac{4\eta}{5})\sqrt{\frac{1}{d_1}}\|\widehat{V}^{(p)}-V\widehat{O}_V^{(p)}\|_{2,\max} + C_3\eta\frac{\sigma}{\lambda_{\min}}\sqrt{\frac{rd_2\log d_1}{N_0d_1}} \\ &\quad\quad + \frac{\eta}{6}\left(\sqrt{\frac{r}{d_2}}\|\widehat{U}^{(p)}-U\widehat{O}_U^{(p)}\|_{2,\max} + \sqrt{\frac{r}{d_1}}\|\widehat{V}^{(p)}-V\widehat{O}_V^{(p)}\|_{2,\max}\right).
    \end{align*}
    As a result, 
    \begin{align*}
        &\sqrt{\frac{1}{d_2}}\|\widehat{U}^{(p+1)} - U\widehat{O}_U^{(p)}\widehat{L}_G^{(p)}\widehat{K}_U^{(p+1)}\|_{2,\max} + \sqrt{\frac{1}{d_1}}\|\widehat{V}^{(p+1)} - V\widehat{O}_V^{(p)}\widehat{R}_G^{(p)}\widehat{K}_V^{(p+1)}\|_{2,\max} \\ &\quad \leq C_3\eta\frac{\sigma}{\lambda_{\min}}\sqrt{\frac{rd_2\log d_1}{N_0d_1}}  + (1-\frac{2\eta}{3})\left(\sqrt{\frac{r}{d_2}}\|\widehat{U}^{(p)}-U\widehat{O}_U^{(p)}\|_{2,\max} + \sqrt{\frac{r}{d_1}}\|\widehat{V}^{(p)}-V\widehat{O}_V^{(p)}\|_{2,\max}\right),
    \end{align*}
    where both $\widehat{O}_U^{(p)}\widehat{L}_G^{(p)}\widehat{K}_U^{(p+1)}$ and $\widehat{O}_V^{(p)}\widehat{R}_G^{(p)}\widehat{K}_V^{(p+1)}$ are orthogonal matrices.

    The contraction property in the beginning is then proved if we show 
    \begin{align*}
        \max{\|\widehat{U}^{(p)}-U\widehat{O}_U^{(p)}\|, \|\widehat{V}^{(p)}-V\widehat{O}_V^{(p)}\|}\lesssim \frac{1}{\kappa^2r\sqrt{d_2/d_1}}
    \end{align*}
    and $\|\widehat{U}^{(p)}\|\leq 2\mu\sqrt{r/d_1}$, $\|\widehat{V}^{(p)}\|\leq 2\mu\sqrt{r/d_2}$ for all $1\leq p\leq m$. 

    We first show $\|\widehat{U}^{(p)}\|\leq 2\mu\sqrt{r/d_1}$ and $\|\widehat{V}^{(p)}\|\leq 2\mu\sqrt{r/d_2}$. By the above contraction equation, it suffices to show $\|\widehat{U}^{(1)}-U\widehat{O}_U^{(1)}\|_{2,\max}\leq \mu\sqrt{r/d_1}$ and $\|\widehat{U}^{(1)}-U\widehat{O}_U^{(1)}\|_{2,\max}\leq \mu\sqrt{r/d_1}$ as long as $\lambda_{\min}/\sigma\gtrsim \sqrt{d_2^2\log d_1/T}$. Similarly as the proof of Lemma \ref{2maxbound}, with probability at least $1-5d_1^{-10}$,
    \begin{align*}
        &\|\widehat{U}^{(1)}-U\|_{2,\max}\lesssim \frac{\sigma+\|M\|_{\max}}{\lambda_{\min}}\sqrt{\frac{rd_2^2\log d_1}{N_0d_1}} \\
        &\|\widehat{V}^{(1)}-V\|_{2,\max}\lesssim \frac{\sigma+\|M\|_{\max}}{\lambda_{\min}}\sqrt{\frac{rd_2\log d_1}{N_0}}.
    \end{align*}
    By incoherence of $U$ and $V$, $\|M\|_{\max}\leq \lambda_{\max}\mu^2r/\sqrt{d_1d_2}$. Then as long as $T\gtrsim \kappa^2r^2\alpha_d\log d_1$ and $\lambda_{\min}/\sigma\gtrsim \sqrt{rd_2^2\log d_1/T}$.

    We then prove $ \max{\|\widehat{U}^{(p)}-U\widehat{O}_U^{(p)}\|, \|\widehat{V}^{(p)}-V\widehat{O}_V^{(p)}\|}\lesssim 1/\kappa^2r\sqrt{d_2/d_1}$. Recall that
    \begin{align*}
        &\widehat{U}^{(p+1)}\widehat{\Sigma}_U^{(p+1)}-U\widehat{O}_U^{(p)}\widehat{L}_G^{(p)}\widehat{K}_U^{(p+1)} \\ 
        &\quad = (\widehat{U}^{(p)}\widehat{L}_G^{(p)}-U\widehat{O}_U^{(p)}\widehat{L}_G^{(p)})(I-\eta\widehat{L}_G^{(p)\top}\widehat{O}_U^{(p)\top}\Lambda\widehat{O}_V^{(p)}\widehat{R}_G^{(p)}(\widehat{\Lambda}^{(p)})^{-1})\widehat{K}_U^{(p+1)}\\
        &\quad\quad -\eta \widehat{U}^{(p)}(\widehat{G}^{(p)}-\widehat{O}_U^{(p)\top}\Lambda\widehat{O}_V^{(p)})\widehat{R}_G^{(p)}(\widehat{\Lambda}^{(p)})^{-1}\widehat{K}_U^{(p+1)} \\ &\quad\quad - \eta U\Lambda(\widehat{V}^{(p)}\widehat{O}_V^{(p)\top}-V)^{\top}\widehat{V}^{(p)}\widehat{R}_G^{(p)}(\widehat{\Lambda}^{(p)})^{-1} + \widehat{E}_V^{(p)}\widehat{K}_U^{(p+1)} + \widehat{E}_{\xi,V}^{(p)}\widehat{K}_U^{(p+1)}.
    \end{align*}
    Similar as the proof of Lemma \ref{lemma:estimation2}, we can write
    \begin{align*}
        &\|\widehat{U}^{(p+1)}\widehat{\Sigma}_U^{(p+1)}-U\widehat{O}_U^{(p)}\widehat{L}_G^{(p)}\widehat{K}_U^{(p+1)}\|\leq (1-\frac{9\eta}{10})\|\widehat{U}^{(p)}-U\widehat{O}_U^{(p)}\| + 2\eta\frac{\|\widehat{G}^{(p)} - \widehat{O}_U^{(p)\top}\Lambda\widehat{O}_V^{(p)}\|}{\lambda_{\min}} \\
        &\quad\quad + 2\eta\kappa\|\widehat{V}^{(p)}-V\widehat{O}_V^{(p)}\|^2 + \|\widehat{E}_V^{(p)}\widehat{K}_U^{(p+1)} + \widehat{E}_{\xi,V}^{(p)}\widehat{K}_U^{(p+1)}\|,
    \end{align*}
    and as a result
    \begin{align*}
        &\|\widehat{U}^{(p+1)}-U\widehat{O}_U^{(p)}\widehat{L}_G^{(p)}\widehat{K}_U^{(p+1)}\|\leq (1-\frac{9\eta}{10})\|\widehat{U}^{(p)}-U\widehat{O}_U^{(p)}\| + 2\eta\frac{\|\widehat{G}^{(p)} - \widehat{O}_U^{(p)\top}\Lambda\widehat{O}_V^{(p)}\|}{\lambda_{\min}} \\
        &\quad\quad + 2\eta\kappa\|\widehat{V}^{(p)}-V\widehat{O}_V^{(p)}\|^2 + \|\widehat{E}_V^{(p)}\widehat{K}_U^{(p+1)} + \widehat{E}_{\xi,V}^{(p)}\widehat{K}_U^{(p+1)}\| + \|\widehat{\Sigma}_U^{(p+1)}-I\|.
    \end{align*}
    Then, by Lemma \ref{lemma:estimation1}-\ref{lemma:estimation2} and the upper bound of $\|\widehat{E}_V^{(p)} + \widehat{E}_{\xi,V}^{(p)}\|$ in the proof of Lemma \ref{lemma:estimation2},
    \begin{align*}
        &\|\widehat{U}^{(p+1)}-U\widehat{O}_U^{(p)}\widehat{L}_G^{(p)}\widehat{K}_U^{(p+1)}\|\leq (1-0.8\eta)\|\widehat{U}^{(p)}-U\widehat{O}_U^{(p)}\| + C_3\eta\frac{\sigma}{\lambda_{\min}}\sqrt{\frac{d_2^2\log d_1}{N_0}} \\ &\quad + C_4(\kappa^2\eta^2+\kappa\eta)(\|\widehat{U}^{(p)}-U\widehat{O}_U^{(p)}\|^2 + \|\widehat{V}^{(p)}-V\widehat{O}_V^{(p)}\|^2) \\
        &\quad +  C_5\eta\sqrt{\frac{d_2^2\log d_1}{N_0}}\left(\sqrt{\frac{r}{d_2}}\|\widehat{U}^{(p)}-U\widehat{O}_U^{(p)}\|_{2,\max} + \sqrt{\frac{r}{d_1}}\|\widehat{V}^{(p)}-V\widehat{O}_V^{(p)}\|_{2,\max}\right).
    \end{align*}
    Similarly, we can get the bound for $\|\widehat{V}^{(p+1)}-V\widehat{O}_V^{(p)}\widehat{R}_G^{(p)}\widehat{K}_V^{(p+1)}\|$ and as a result, if $\|\widehat{U}^{(p)}-U\widehat{O}_U^{(p)}\| + \|\widehat{V}^{(p)}-V\widehat{O}_V^{(p)}\|\lesssim 1/\kappa^2r\sqrt{d_2/d_1}$, we have
    \begin{align*}
        &\|\widehat{U}^{(p+1)}-U\widehat{O}_U^{(p)}\widehat{L}_G^{(p)}\widehat{K}_U^{(p+1)}\| + \|\widehat{V}^{(p+1)}-V\widehat{O}_V^{(p)}\widehat{R}_G^{(p)}\widehat{K}_V^{(p+1)}\| \\
        &\quad \leq (1-0.8\eta)(\|\widehat{U}^{(p)}-U\widehat{O}_U^{(p)}\| + \|\widehat{V}^{(p)}-V\widehat{O}_V^{(p)}\|) + C_3\eta\frac{\sigma}{\lambda_{\min}}\sqrt{\frac{d_2^2\log d_1}{N_0}} \\ &\quad\quad + C_4(\kappa^2\eta^2+\kappa\eta)(\|\widehat{U}^{(p)}-U\widehat{O}_U^{(p)}\|^2 + \|\widehat{V}^{(p)}-V\widehat{O}_V^{(p)}\|^2) \\
        &\quad\quad +  C_5\eta\sqrt{\frac{d_2^2\log d_1}{N_0}}\left(\sqrt{\frac{r}{d_2}}\|\widehat{U}^{(p)}-U\widehat{O}_U^{(p)}\|_{2,\max} + \sqrt{\frac{r}{d_1}}\|\widehat{V}^{(p)}-V\widehat{O}_V^{(p)}\|_{2,\max}\right) \\
        &\quad \leq C_3\eta\frac{\sigma}{\lambda_{\min}}\sqrt{\frac{d_2^2\log d_1}{N_0}} + (1-\frac{\eta}{2})(\|\widehat{U}^{(p)}-U\widehat{O}_U^{(p)}\| + \|\widehat{V}^{(p)}-V\widehat{O}_V^{(p)}\|) + C_5\eta\sqrt{\frac{r^2d_2\log d_1}{N_0d_1}} \\
        &\lesssim \frac{1}{\kappa^2r\sqrt{d_2/d_1}}
    \end{align*}
    where the second last inequality holds as long as $\|\widehat{U}^{(p)}-U\widehat{O}_U^{(p)}\|_{2,\max}\lesssim \mu\sqrt{r/d_1}$, $\|\widehat{V}^{(p)}-V\widehat{O}_V^{(p)}\|_{2,\max}\lesssim \mu\sqrt{r/d_2}$, and the last inequality holds as long as $T\gtrsim r^2\alpha_d\log d_1$ and $\lambda_{\min}/\sigma\gtrsim \sqrt{d_2^2\log d_1/T}$. Then it suffices to prove $\|\widehat{U}^{(1)}-U\widehat{O}_U^{(1)}\| + \|\widehat{V}^{(1)}-V\widehat{O}_V^{(1)}\|\lesssim 1/\kappa^2r\sqrt{d_2/d_1}$. By Davis-Kahan Theorem, with probability at least $1-2d_1^{-10}$,
    \begin{align*}
        \|\widehat{U}^{(1)}\widehat{U}^{(1)\top}-UU^{\top}\| + \|\widehat{V}^{(1)}\widehat{V}^{(1)\top}-VV^{\top}\|\lesssim \frac{\sigma+\|M\|_{\max}}{\lambda_{\min}}\sqrt{\frac{d_2^2\log d_1}{N_0}}\lesssim \frac{1}{\kappa^2r\sqrt{d_2/d_1}}
    \end{align*}
    as long as $T\gtrsim \kappa^4\alpha_d^2\log d_1$ and $\lambda_{\min}/\sigma\gtrsim \sqrt{\kappa^4r\alpha_dd_2^2\log d_1/T}$. Then we conclude the proof of the statement in the beginning. 
    
    Recall that $N_0\asymp T/\log d_1$, by the beginning statement with $\eta=0.75$, we get with probability at least $1-4md_1^{-10}$,
    \begin{align*}
        &\sqrt{\frac{1}{d_2}}\|\widehat{U}^{(m)}-U\widehat{O}_U^{(m)}\|_{2,\max} + \sqrt{\frac{1}{d_1}}\|\widehat{V}^{(m)}-V\widehat{O}_V^{(m)}\|_{2,\max} - 2C_3\eta\frac{\sigma}{\lambda_{\min}}\sqrt{\frac{rd_2\log d_1}{N_0d_1}} \\ 
        &\quad \leq (\frac{1}{2})^m \left(\sqrt{\frac{1}{d_2}}\|\widehat{U}^{(1)}-U\widehat{O}_U^{(1)}\|_{2,\max} + \sqrt{\frac{1}{d_1}}\|\widehat{V}^{(1)}-V\widehat{O}_V^{(1)}\|_{2,\max} - 2C_3\eta\frac{\sigma}{\lambda_{\min}}\sqrt{\frac{rd_2\log d_1}{N_0d_1}}\right).
    \end{align*}
    Similar as the proof of Lemma \ref{2maxbound}, with probability at least $1-d_1^{-10}$,
    \begin{align*}
        \sqrt{\frac{1}{d_2}}\|\widehat{U}^{(1)}-U\widehat{O}_U^{(1)}\|_{2,\max} + \sqrt{\frac{1}{d_1}}\|\widehat{V}^{(1)}-V\widehat{O}_V^{(1)}\|_{2,\max}\lesssim \frac{\sigma+\|M\|_{\max}}{\lambda_{\min}}\sqrt{\frac{rd_2\log d_1}{d_1N_0}}.
    \end{align*}
    Therefore, if $m=2\lceil \log (\alpha_d\|M\|_{\max}/\sigma)$, we get
    \begin{align*}
        \sqrt{\frac{1}{d_2}}\|\widehat{U}^{(m)}-U\widehat{O}_U^{(m)}\|_{2,\max} + \sqrt{\frac{1}{d_1}}\|\widehat{V}^{(m)}-V\widehat{O}_V^{(m)}\|_{2,\max}\lesssim \frac{\sigma}{\lambda_{\min}}\sqrt{\frac{rd_2\log d_1}{d_1N_0}}.
    \end{align*}
    Then by proof of Lemma \ref{lemma:estimation2} and the result of Lemma \ref{lemma:estimation1}, 
    \begin{align*}
        &\|\widehat{M}^{(m)}-M\|_{\max}\leq 2\mu\lambda_{\max}\left(\sqrt{\frac{r}{d_2}}\|\widehat{U}^{(m)}-U\widehat{O}_U^{(m)}\|_{2,\max} + \sqrt{\frac{r}{d_1}}\|\widehat{V}^{(m)}-V\widehat{O}_V^{(m)}\|_{2,\max}\right)  \\ &\quad + \mu^2\sqrt{\frac{r^2}{d_1d_2}}\|\widehat{G}^{(m)}-\widehat{O}_U^{(m)}\Lambda\widehat{O}_V^{(m)}\|\lesssim \mu\kappa\sigma\sqrt{\frac{r^2d_2\log^2 d_1}{Td_1}}
    \end{align*}
\end{proof}

\subsection{Proof of Proposition \ref{thm:estimation2}}

\begin{proof}
    The proof follows the similar arguments as Proposition \ref{thm:estimation}. We first prove the following statement: for $p=1,\cdots,m-1$, with probability at least $1-4md_1^{-10}$,
    \begin{align*}
         &\sqrt{\frac{1}{d_2}}\|\widehat{U}^{(p+1)}-U\widehat{O}_U^{(p+1)}\|_{2,\max} + \sqrt{\frac{1}{d_1}}\|\widehat{V}^{(p+1)}-V\widehat{O}_V^{(p+1)}\|_{2,\max}\leq 2C\eta\frac{\sigma}{\lambda_{\min}}\sqrt{\frac{rd_2\log d_1}{Td_1Kp_0}} \\ &\quad\quad + (1-\frac{2}{3}\eta)\left(\sqrt{\frac{1}{d_2}}\|\widehat{U}^{(p)}-U\widehat{O}_U^{(p+1)}\|_{2,\max} + \sqrt{\frac{1}{d_1}}\|\widehat{V}^{(p)}-V\widehat{O}_V^{(p)}\|_{2,\max}\right)
    \end{align*}
    for some constant $C$.
    
    We begin with the accuracy of $\widehat{G}^{(p)}$. By the definition of $\widehat{G}^{(p)}$, we have
    \begin{align*}
        \frac{d_2}{N_0Kp_0}\sum_{t\in \mathcal{D}_{2p}} \sum_{(i,q)\in h_{t}}(\inp{X_t^{iq}}{\widehat{U}^{(p)}\widehat{G}^{(p)}\widehat{V}^{(p)\top}}-y_t^{iq})\widehat{U}^{(p)\top}X_t^{iq}\widehat{V}^{(p)}=0.
    \end{align*}
    Let $\widehat{O}_U^{(p)}$ and $\widehat{O}_V^{(p)}$ be any orthogonal matrices such that 
    \begin{align*}
        \max(\|\widehat{U}^{(p)}-U\widehat{O}_U^{(p)}\|, \|\widehat{V}^{(p)}-V\widehat{O}_U^{(p)}\|)\leq \frac{1}{C_1\mu\kappa^2\sqrt{r}}
    \end{align*}
    for some large constant $C_1>0$.
    \begin{Lemma}\label{lemma:estimation2-1}
        Suppose that $\|\widehat{U}^{(p)}\|_{2,\max}\leq 2\mu\sqrt{r/d_1}$, $\|\widehat{V}^{(p)}\|_{2,\max}\leq 2\mu\sqrt{r/d_2}$ and $T\geq C_2p_0^{-1}\mu^4r^3\log d_1$, then with probability at least $1-3d_1^{-10}$,
        \begin{align*}
            &\|\widehat{G}^{(p)}-\widehat{O}_U^{(p)\top}\Lambda\widehat{O}_V^{(p)}\|\leq C_3(\frac{r\sigma}{N_0p_0}\sqrt{d_1d_2}\log d_1 + \sqrt{\frac{\sigma^2 rd_2\log d_1}{N_0Kp_0}}) \\ &\quad + 2\lambda_{\max}(\|\widehat{U}^{(p)}-U\widehat{O}_U^{(p)}\|^2+\|\widehat{V}^{(p)}-V\widehat{O}_U^{(p)}\|^2) \\ &\quad + C_4\lambda_{\max}\left(\sqrt{\frac{1}{d_2}}\|\widehat{U}^{(p)}-U\widehat{O}_U^{(p)}\|_{2,\max} + \sqrt{\frac{1}{d_1}}\|\widehat{V}^{(p)}-V\widehat{O}_V^{(p)}\|_{2,\max}\right)\mu\sqrt{\frac{r^3d_1d_2\log d_1}{N_0p_0}}
        \end{align*}
        for some constant $C_2, C_3, C_4>0$.
    \end{Lemma}
    Recall $\widehat{G}^{(p)}=\widehat{L}_G^{(p)}\widehat{\Lambda}^{(p)}\widehat{R}_G^{(p)\top}$ denotes $\widehat{G}^{(p)}$'s SVD. By the gradient descent step and following proof of Proposition \ref{thm:estimation},
    \begin{align*}
        \widehat{U}^{(p+0.5)}=\widehat{U}^{(p)}\widehat{L}_G^{(p)} - \eta(\widehat{U}^{(p)}\widehat{G}^{(p)}\widehat{V}^{(p)\top} - U\Lambda V^{\top})\widehat{V}^{(p)}\widehat{R}_G^{(p)}(\widehat{\Lambda}^{(p)})^{-1} + \widehat{E}_V^{(p)} + \widehat{E}_{\xi,V}^{(p)},
    \end{align*} 
    where
    \begin{align*}
        &\widehat{E}_V^{(p)}=\eta(\widehat{U}^{(p)}\widehat{G}^{(p)}\widehat{V}^{(p)\top} - U\Lambda V^{\top})\widehat{V}^{(p)}\widehat{R}_G^{(p)}(\widehat{\Lambda}^{(p)})^{-1} \\ &\quad - \eta\frac{d_2}{N_0Kp_0}\sum_{t\in \mathcal{D}_{2p+1}} \sum_{i=1}^{d_1}\sum_{q\in h_{it}} \inp{\widehat{U}^{(p)}\widehat{G}^{(p)}\widehat{V}^{(p)\top} - U\Lambda V^{\top}}{X_t^{iq}}X_t^{iq}\widehat{V}^{(p)}\widehat{R}_G^{(p)}(\widehat{\Lambda}^{(p)})^{-1}
    \end{align*}
    and
    \begin{align*}
        \widehat{E}_{\xi,V}^{(p)}=- \eta\frac{d_2}{N_0Kp_0}\sum_{t\in \mathcal{D}_{2p+1}} \sum_{i=1}^{d_1}\sum_{q\in h_{it}}  \xi_t^{iq} X_t^{iq}\widehat{V}^{(p)}\widehat{R}_G^{(p)}(\widehat{\Lambda}^{(p)})^{-1}.
    \end{align*}
    This directly leads to
    \begin{align*}
        &\widehat{U}^{(p+0.5)}=U\widehat{O}_U^{(p)}\widehat{L}_G^{(p)} + (\widehat{U}^{(p)}\widehat{L}_G^{(p)}-U\widehat{O}_U^{(p)}\widehat{L}_G^{(p)})(I-\eta\widehat{L}_G^{(p)\top}\widehat{O}_U^{(p)\top}\Lambda\widehat{O}_V^{(p)}\widehat{R}_G^{(p)}(\widehat{\Lambda}^{(p)})^{-1}) \\
        &\quad -\eta \widehat{U}^{(p)}(\widehat{G}^{(p)}-\widehat{O}_U^{(p)\top}\Lambda\widehat{O}_V^{(p)})\widehat{R}_G^{(p)}(\widehat{\Lambda}^{(p)})^{-1} - \eta U\Lambda(\widehat{V}^{(p)}\widehat{O}_V^{(p)\top}-V)^{\top}\widehat{V}^{(p)}\widehat{R}_G^{(p)}(\widehat{\Lambda}^{(p)})^{-1} + \widehat{E}_V^{(p)} + \widehat{E}_{\xi,V}^{(p)}.
    \end{align*}
    \begin{Lemma}\label{lemma:estimation2-2}
        Under the conditions in Proposition \ref{thm:estimation2} and Lemma \ref{lemma:estimation2-1}, and 
        \begin{align*}
            \max\{\|\widehat{U}^{(p)}-U\widehat{O}_U^{(p)}\|, \|\widehat{V}^{(p)}-V\widehat{O}_V^{(p)}\|\}\leq 1/(C_1\mu\kappa r\sqrt{d_2/d_1})
        \end{align*}
        for some large constant $C_1$. If $\eta\in [0.25, 0.75]$, then with probability at least $1-2d_1^{-10}$,
        \begin{align*}
            &\|\widehat{U}^{(p+0.5)}-U\widehat{O}_U^{(p)}\widehat{L}_G^{(p)}\|_{2,\max}\leq (1-\frac{9\eta}{10})\|\widehat{U}^{(p)}-U\widehat{O}_U^{(p)}\|_{2,\max} + C_2\eta\frac{\sigma}{\lambda_{\min}}\sqrt{\frac{d_2r\log d_1}{N_0Kp_0}} \\ &\quad + \frac{\eta}{8}(\sqrt{\frac{d_1}{d_2}}\|\widehat{U}^{(p)}-U\widehat{O}_U^{(p)}\|_{2,\max}+\|\widehat{V}^{(p)}-V\widehat{O}_V^{(p)}\|_{2,\max}),
        \end{align*}
        for some constants $C_2,>0$. Moreover, with probability at least $1-2d_1^{-10}$,
        \begin{align*}
            &\left\{|1-\lambda_{\min}(\widehat{U}^{(p+0.5)})|, |1-\lambda_{\max}(\widehat{U}^{(p+0.5)})|\right\}\lesssim \eta\frac{\sigma}{\lambda_{\min}}\sqrt{\frac{d_2^2\log^2 d_1}{N_0Kp_0}} \\ &\quad + (\kappa^2\eta^2+\kappa\eta)(\|\widehat{U}^{(p)}-U\widehat{O}_U^{(p)}\|^2 + \|\widehat{V}^{(p)}-V\widehat{O}_V^{(p)}\|^2) \\
            &\quad +  \eta\sqrt{\frac{d_2^2\log^2 d_1}{N_0Kp_0}}\left(\sqrt{\frac{r}{d_2}}\|\widehat{U}^{(p)}-U\widehat{O}_U^{(p)}\|_{2,\max} + \sqrt{\frac{r}{d_1}}\|\widehat{V}^{(p)}-V\widehat{O}_V^{(p)}\|_{2,\max}\right).
        \end{align*}
    \end{Lemma} 
    Following the proof of Proposition \ref{thm:estimation},  
    \begin{align*}
        &\sqrt{\frac{1}{d_2}}\|\widehat{U}^{(p+1)} - U\widehat{O}_U^{(p)}\widehat{L}_G^{(p)}\widehat{K}_U^{(p+1)}\|_{2,\max}\leq \sqrt{\frac{1}{d_2}}\|U\|_{2,\max} \|(\widehat{\Sigma}_U^{(p+1)})^{-1}-I\| \\ &\quad\quad + \sqrt{\frac{1}{d_2}}(1+\eta/10)\| \widehat{U}^{(p+0.5)} -  U\widehat{O}_U^{(p)}\widehat{L}_G^{(p)}\|_{2,\max} \\
        &\quad \leq C_3\eta\frac{\sigma}{\lambda_{\min}}\sqrt{\frac{rd_2\log d_1}{N_0d_1Kp_0}} + C_4(\kappa^2\eta^2+\kappa\eta)(\|\widehat{U}^{(p)}-U\widehat{O}_U^{(p)}\|^2 + \|\widehat{V}^{(p)}-V\widehat{O}_V^{(p)}\|^2)\sqrt{\frac{r}{d_1d_2}} \\
        &\quad\quad +  C_5\eta\sqrt{\frac{rd_2\log d_1}{N_0d_1Kp_0}}\left(\sqrt{\frac{r}{d_2}}\|\widehat{U}^{(p)}-U\widehat{O}_U^{(p)}\|_{2,\max} + \sqrt{\frac{r}{d_1}}\|\widehat{V}^{(p)}-V\widehat{O}_V^{(p)}\|_{2,\max}\right) \\ 
        &\quad\quad + (1-\frac{4\eta}{5})\sqrt{\frac{1}{d_2}}\|\widehat{U}^{(p)}-U\widehat{O}_U^{(p)}\|_{2,\max} + C_2\eta\frac{\sigma}{\lambda_{\min}}\sqrt{\frac{r^2\log d_1}{N_0Kp_0}} \\ &\quad\quad + \frac{\eta}{7}(\sqrt{\frac{d_1}{d_2}}\|\widehat{U}^{(p)}-U\widehat{O}_U^{(p)}\|_{2,\max}+\|\widehat{V}^{(p)}-V\widehat{O}_V^{(p)}\|_{2,\max})\sqrt{\frac{r}{d_2}} \\
        &\quad \leq (1-\frac{4\eta}{5})\sqrt{\frac{1}{d_2}}\|\widehat{U}^{(p)}-U\widehat{O}_U^{(p)}\|_{2,\max} + C_3\eta\frac{\sigma}{\lambda_{\min}}\sqrt{\frac{rd_2\log d_1}{N_0d_1Kp_0}} \\ &\quad\quad + \frac{\eta}{6}\left(\sqrt{\frac{r}{d_2}}\|\widehat{U}^{(p)}-U\widehat{O}_U^{(p)}\|_{2,\max} + \sqrt{\frac{r}{d_1}}\|\widehat{V}^{(p)}-V\widehat{O}_V^{(p)}\|_{2,\max}\right)
    \end{align*}
    as long as $T\gtrsim (Kp_0)^{-1}\alpha_dr\log d_1$ and $\|\widehat{U}^{(p)}-U\widehat{O}_U^{(p)}\| + \|\widehat{V}^{(p)}-V\widehat{O}_V^{(p)}\| \lesssim 1/\kappa^2r$. Similarly, we have 
    \begin{align*}
        &\sqrt{\frac{1}{d_1}}\|\widehat{V}^{(p+1)} - V\widehat{O}_V^{(p)}\widehat{R}_G^{(p)}\widehat{K}_V^{(p+1)}\|_{2,\max}\leq (1-\frac{4\eta}{5})\sqrt{\frac{1}{d_1}}\|\widehat{V}^{(p)}-V\widehat{O}_V^{(p)}\|_{2,\max} + C_3\eta\frac{\sigma}{\lambda_{\min}}\sqrt{\frac{rd_2\log d_1}{N_0d_1Kp_0}} \\ &\quad\quad + \frac{\eta}{6}\left(\sqrt{\frac{r}{d_2}}\|\widehat{U}^{(p)}-U\widehat{O}_U^{(p)}\|_{2,\max} + \sqrt{\frac{r}{d_1}}\|\widehat{V}^{(p)}-V\widehat{O}_V^{(p)}\|_{2,\max}\right).
    \end{align*}
    As a result, 
    \begin{align*}
        &\sqrt{\frac{1}{d_2}}\|\widehat{U}^{(p+1)} - U\widehat{O}_U^{(p)}\widehat{L}_G^{(p)}\widehat{K}_U^{(p+1)}\|_{2,\max} + \sqrt{\frac{1}{d_1}}\|\widehat{V}^{(p+1)} - V\widehat{O}_V^{(p)}\widehat{R}_G^{(p)}\widehat{K}_V^{(p+1)}\|_{2,\max} \\ &\quad \leq C_3\eta\frac{\sigma}{\lambda_{\min}}\sqrt{\frac{rd_2\log d_1}{N_0d_1Kp_0}}  + (1-\frac{2\eta}{3})\left(\sqrt{\frac{r}{d_2}}\|\widehat{U}^{(p)}-U\widehat{O}_U^{(p)}\|_{2,\max} + \sqrt{\frac{r}{d_1}}\|\widehat{V}^{(p)}-V\widehat{O}_V^{(p)}\|_{2,\max}\right),
    \end{align*}
    where both $\widehat{O}_U^{(p)}\widehat{L}_G^{(p)}\widehat{K}_U^{(p+1)}$ and $\widehat{O}_V^{(p)}\widehat{R}_G^{(p)}\widehat{K}_V^{(p+1)}$ are orthogonal matrices.

    The contraction property in the beginning of Proposition \ref{thm:estimation2} is then proved if we show 
    \begin{align*}
        \max{\|\widehat{U}^{(p)}-U\widehat{O}_U^{(p)}\|, \|\widehat{V}^{(p)}-V\widehat{O}_V^{(p)}\|}\lesssim \frac{1}{\kappa^2r\sqrt{d_2/d_1}}
    \end{align*}
    and $\|\widehat{U}^{(p)}\|\leq 2\mu\sqrt{r/d_1}$, $\|\widehat{V}^{(p)}\|\leq 2\mu\sqrt{r/d_2}$ for all $1\leq p\leq m$. They can be proved in the same way following the proof of Proposition \ref{thm:estimation}, under the sample size condition $T\gtrsim (Kp_0)^{-1}\alpha_d^2\log^2d_2$.

    As a result, recall that $N_0\asymp T/\log d_1$, by the contraction property with $\eta=0.75$, we get with probability at least $1-4md_1^{-10}$,
    \begin{align*}
        &\sqrt{\frac{1}{d_2}}\|\widehat{U}^{(m)}-U\widehat{O}_U^{(m)}\|_{2,\max} + \sqrt{\frac{1}{d_1}}\|\widehat{V}^{(m)}-V\widehat{O}_V^{(m)}\|_{2,\max} - 2C_3\eta\frac{\sigma}{\lambda_{\min}}\sqrt{\frac{rd_2\log d_1}{N_0d_1Kp_0}} \\ 
        &\quad \leq (\frac{1}{2})^m \left(\sqrt{\frac{1}{d_2}}\|\widehat{U}^{(1)}-U\widehat{O}_U^{(1)}\|_{2,\max} + \sqrt{\frac{1}{d_1}}\|\widehat{V}^{(1)}-V\widehat{O}_V^{(1)}\|_{2,\max} - 2C_3\eta\frac{\sigma}{\lambda_{\min}}\sqrt{\frac{rd_2\log d_1}{N_0d_1Kp_0}}\right).
    \end{align*}
    Similar as the proof of Lemma \ref{2maxbound-otm}, with probability at least $1-d_1^{-10}$,
    \begin{align*}
        \sqrt{\frac{1}{d_2}}\|\widehat{U}^{(1)}-U\widehat{O}_U^{(1)}\|_{2,\max} + \sqrt{\frac{1}{d_1}}\|\widehat{V}^{(1)}-V\widehat{O}_V^{(1)}\|_{2,\max}\lesssim \frac{\sigma/\sqrt{Kp_0}+\|M\|_{\max}}{\lambda_{\min}}\sqrt{\frac{rd_2\log d_1}{d_1N_0}}.
    \end{align*}
    Therefore, if $m=2\lceil \log (\alpha_d\|M\|_{\max}\sqrt{Kp_0}/\sigma)$, we get
    \begin{align*}
        \sqrt{\frac{1}{d_2}}\|\widehat{U}^{(m)}-U\widehat{O}_U^{(m)}\|_{2,\max} + \sqrt{\frac{1}{d_1}}\|\widehat{V}^{(m)}-V\widehat{O}_V^{(m)}\|_{2,\max}\lesssim \frac{\sigma}{\lambda_{\min}}\sqrt{\frac{rd_2\log d_1}{d_1N_0Kp_0}}.
    \end{align*}
    Then by proof of Lemma \ref{lemma:estimation2-2} and the result of Lemma \ref{lemma:estimation2-1}, 
    \begin{align*}
        &\|\widehat{M}^{(m)}-M\|_{\max}\leq 2\mu\lambda_{\max}\left(\sqrt{\frac{r}{d_2}}\|\widehat{U}^{(m)}-U\widehat{O}_U^{(m)}\|_{2,\max} + \sqrt{\frac{r}{d_1}}\|\widehat{V}^{(m)}-V\widehat{O}_V^{(m)}\|_{2,\max}\right)  \\ &\quad + \mu^2\sqrt{\frac{r^2}{d_1d_2}}\|\widehat{G}^{(m)}-\widehat{O}_U^{(m)}\Lambda\widehat{O}_V^{(m)}\|\lesssim \mu\kappa\sigma\sqrt{\frac{r^2d_2\log^2 d_1}{Td_1Kp_0}}
    \end{align*}
\end{proof}

\subsection{Proof of Proposition \ref{thm:estimation3}}
\begin{proof}
    The proof follows the similar arguments as Proposition \ref{thm:estimation}. We first prove the following statement: for $p=1,\cdots,m-1$, with probability at least $1-4md_1^{-10}$,
    \begin{align*}
         &\sqrt{\frac{1}{d_2}}\|\widehat{U}^{(p+1)}-U\widehat{O}_U^{(p+1)}\|_{2,\max} + \sqrt{\frac{1}{d_1}}\|\widehat{V}^{(p+1)}-V\widehat{O}_V^{(p+1)}\|_{2,\max}\leq 2C\eta\frac{\sigma}{\lambda_{\min}}\sqrt{\frac{r\log d_1}{Td_1\nu}} \\ &\quad\quad + (1-\frac{2}{3}\eta)\left(\sqrt{\frac{1}{d_2}}\|\widehat{U}^{(p)}-U\widehat{O}_U^{(p)}\|_{2,\max} + \sqrt{\frac{1}{d_1}}\|\widehat{V}^{(p)}-V\widehat{O}_V^{(p)}\|_{2,\max}\right)
    \end{align*}
    for some constant $C$. 
    We begin with the accuracy of $\widehat{G}^{(p)}$. By the definition of $\widehat{G}^{(p)}$, we have
    \begin{align*}
        \frac{1}{N_0\nu}\sum_{t\in \mathcal{D}_{2p}} \sum_{(i,j)\in g_{t}}(\inp{X_t^{ij}}{\widehat{U}^{(p)}\widehat{G}^{(p)}\widehat{V}^{(p)\top}}-y_t^{ij})\widehat{U}^{(p)\top}X_t^{ij}\widehat{V}^{(p)}=0.
    \end{align*}
    Let $\widehat{O}_U^{(p)}$ and $\widehat{O}_V^{(p)}$ be any orthogonal matrices such that 
    \begin{align*}
        \max(\|\widehat{U}^{(p)}-U\widehat{O}_U^{(p)}\|, \|\widehat{V}^{(p)}-V\widehat{O}_U^{(p)}\|)\leq \frac{1}{C_1\mu\kappa^2\sqrt{r}}
    \end{align*}
    for some large constant $C_1>0$.
    \begin{Lemma}\label{lemma:estimation3-1}
        Suppose that $\|\widehat{U}^{(p)}\|_{2,\max}\leq 2\mu\sqrt{r/d_1}$, $\|\widehat{V}^{(p)}\|_{2,\max}\leq 2\mu\sqrt{r/d_2}$ and $T\geq C_2(d_2\nu)^{-1}\mu^4r^3\log d_1$, then with probability at least $1-3d_1^{-10}$,
        \begin{align*}
            &\|\widehat{G}^{(p)}-\widehat{O}_U^{(p)\top}\Lambda\widehat{O}_V^{(p)}\|\leq C_3(\frac{r\sigma}{N_0\nu}\sqrt{\frac{d_1}{d_2}}\log d_1 + \sqrt{\frac{\sigma^2 rd_2\log d_1}{N_0\nu}}) \\ &\quad + 2\lambda_{\max}(\|\widehat{U}^{(p)}-U\widehat{O}_U^{(p)}\|^2+\|\widehat{V}^{(p)}-V\widehat{O}_U^{(p)}\|^2) \\ &\quad + C_4\lambda_{\max}\left(\sqrt{\frac{1}{d_2}}\|\widehat{U}^{(p)}-U\widehat{O}_U^{(p)}\|_{2,\max} + \sqrt{\frac{1}{d_1}}\|\widehat{V}^{(p)}-V\widehat{O}_V^{(p)}\|_{2,\max}\right)\mu\sqrt{\frac{r^3d_1\log d_1}{N_0d_2\nu^2}}
        \end{align*}
        for some constant $C_2, C_3, C_4>0$.
    \end{Lemma}
    Recall $\widehat{G}^{(p)}=\widehat{L}_G^{(p)}\widehat{\Lambda}^{(p)}\widehat{R}_G^{(p)\top}$ denotes $\widehat{G}^{(p)}$'s SVD. By the gradient descent step and following proof of Proposition \ref{thm:estimation},
    \begin{align*}
        \widehat{U}^{(p+0.5)}=\widehat{U}^{(p)}\widehat{L}_G^{(p)} - \eta(\widehat{U}^{(p)}\widehat{G}^{(p)}\widehat{V}^{(p)\top} - U\Lambda V^{\top})\widehat{V}^{(p)}\widehat{R}_G^{(p)}(\widehat{\Lambda}^{(p)})^{-1} + \widehat{E}_V^{(p)} + \widehat{E}_{\xi,V}^{(p)},
    \end{align*} 
    where
    \begin{align*}
        &\widehat{E}_V^{(p)}=\eta(\widehat{U}^{(p)}\widehat{G}^{(p)}\widehat{V}^{(p)\top} - U\Lambda V^{\top})\widehat{V}^{(p)}\widehat{R}_G^{(p)}(\widehat{\Lambda}^{(p)})^{-1} \\ &\quad - \eta\frac{1}{N_0\nu}\sum_{t\in \mathcal{D}_{2p+1}} \sum_{(i,j)\in g_{t}} \inp{\widehat{U}^{(p)}\widehat{G}^{(p)}\widehat{V}^{(p)\top} - U\Lambda V^{\top}}{X_t^{ij}}X_t^{ij}\widehat{V}^{(p)}\widehat{R}_G^{(p)}(\widehat{\Lambda}^{(p)})^{-1}
    \end{align*}
    and
    \begin{align*}
        \widehat{E}_{\xi,V}^{(p)}=- \eta\frac{1}{N_0\nu}\sum_{t\in \mathcal{D}_{2p+1}} \sum_{(i,j)\in g_{t}}  \xi_t^{ij} X_t^{ij}\widehat{V}^{(p)}\widehat{R}_G^{(p)}(\widehat{\Lambda}^{(p)})^{-1}.
    \end{align*}
    This directly leads to
    \begin{align*}
        &\widehat{U}^{(p+0.5)}=U\widehat{O}_U^{(p)}\widehat{L}_G^{(p)} + (\widehat{U}^{(p)}\widehat{L}_G^{(p)}-U\widehat{O}_U^{(p)}\widehat{L}_G^{(p)})(I-\eta\widehat{L}_G^{(p)\top}\widehat{O}_U^{(p)\top}\Lambda\widehat{O}_V^{(p)}\widehat{R}_G^{(p)}(\widehat{\Lambda}^{(p)})^{-1}) \\
        &\quad -\eta \widehat{U}^{(p)}(\widehat{G}^{(p)}-\widehat{O}_U^{(p)\top}\Lambda\widehat{O}_V^{(p)})\widehat{R}_G^{(p)}(\widehat{\Lambda}^{(p)})^{-1} - \eta U\Lambda(\widehat{V}^{(p)}\widehat{O}_V^{(p)\top}-V)^{\top}\widehat{V}^{(p)}\widehat{R}_G^{(p)}(\widehat{\Lambda}^{(p)})^{-1} + \widehat{E}_V^{(p)} + \widehat{E}_{\xi,V}^{(p)}.
    \end{align*}
    \begin{Lemma}\label{lemma:estimation3-2}
        Under the conditions in Proposition \ref{thm:estimation3} and Lemma \ref{lemma:estimation3-1}, and 
        \begin{align*}
            \max\{\|\widehat{U}^{(p)}-U\widehat{O}_U^{(p)}\|, \|\widehat{V}^{(p)}-V\widehat{O}_V^{(p)}\|\}\leq 1/(C_1\mu\kappa r\sqrt{d_2/d_1})
        \end{align*}
        for some large constant $C_1$. If $\eta\in [0.25, 0.75]$, then with probability at least $1-2d_1^{-10}$,
        \begin{align*}
            &\|\widehat{U}^{(p+0.5)}-U\widehat{O}_U^{(p)}\widehat{L}_G^{(p)}\|_{2,\max}\leq (1-\frac{9\eta}{10})\|\widehat{U}^{(p)}-U\widehat{O}_U^{(p)}\|_{2,\max} + C_2\eta\frac{\sigma}{\lambda_{\min}}\sqrt{\frac{r\log d_1}{N_0\nu}} \\ &\quad + \frac{\eta}{8}(\sqrt{\frac{d_1}{d_2}}\|\widehat{U}^{(p)}-U\widehat{O}_U^{(p)}\|_{2,\max}+\|\widehat{V}^{(p)}-V\widehat{O}_V^{(p)}\|_{2,\max}),
        \end{align*}
        for some constants $C_2,>0$. Moreover, with probability at least $1-2d_1^{-10}$,
        \begin{align*}
            &\left\{|1-\lambda_{\min}(\widehat{U}^{(p+0.5)})|, |1-\lambda_{\max}(\widehat{U}^{(p+0.5)})|\right\}\lesssim \eta\frac{\sigma}{\lambda_{\min}}\sqrt{\frac{d_2\log^2 d_1}{N_0\nu}} \\ &\quad + (\kappa^2\eta^2+\kappa\eta)(\|\widehat{U}^{(p)}-U\widehat{O}_U^{(p)}\|^2 + \|\widehat{V}^{(p)}-V\widehat{O}_V^{(p)}\|^2) \\
            &\quad +  \eta\sqrt{\frac{d_2\log^2 d_1}{N_0\nu}}\left(\sqrt{\frac{r}{d_2}}\|\widehat{U}^{(p)}-U\widehat{O}_U^{(p)}\|_{2,\max} + \sqrt{\frac{r}{d_1}}\|\widehat{V}^{(p)}-V\widehat{O}_V^{(p)}\|_{2,\max}\right).
        \end{align*}
    \end{Lemma} 
    Following the proof of Proposition \ref{thm:estimation},  
    \begin{align*}
        &\sqrt{\frac{1}{d_2}}\|\widehat{U}^{(p+1)} - U\widehat{O}_U^{(p)}\widehat{L}_G^{(p)}\widehat{K}_U^{(p+1)}\|_{2,\max}\leq \sqrt{\frac{1}{d_2}}\|U\|_{2,\max} \|(\widehat{\Sigma}_U^{(p+1)})^{-1}-I\| \\ &\quad\quad + \sqrt{\frac{1}{d_2}}(1+\eta/10)\| \widehat{U}^{(p+0.5)} -  U\widehat{O}_U^{(p)}\widehat{L}_G^{(p)}\|_{2,\max} \\
        &\quad \leq C_3\eta\frac{\sigma}{\lambda_{\min}}\sqrt{\frac{r\log d_1}{N_0d_1\nu}} + C_4(\kappa^2\eta^2+\kappa\eta)(\|\widehat{U}^{(p)}-U\widehat{O}_U^{(p)}\|^2 + \|\widehat{V}^{(p)}-V\widehat{O}_V^{(p)}\|^2)\sqrt{\frac{r}{d_1d_2}} \\
        &\quad\quad +  C_5\eta\sqrt{\frac{r\log d_1}{N_0d_1\nu}}\left(\sqrt{\frac{r}{d_2}}\|\widehat{U}^{(p)}-U\widehat{O}_U^{(p)}\|_{2,\max} + \sqrt{\frac{r}{d_1}}\|\widehat{V}^{(p)}-V\widehat{O}_V^{(p)}\|_{2,\max}\right) \\ 
        &\quad\quad + (1-\frac{4\eta}{5})\sqrt{\frac{1}{d_2}}\|\widehat{U}^{(p)}-U\widehat{O}_U^{(p)}\|_{2,\max} + C_2\eta\frac{\sigma}{\lambda_{\min}}\sqrt{\frac{r^2\log d_1}{N_0d_2\nu}} \\ &\quad\quad + \frac{\eta}{7}(\sqrt{\frac{d_1}{d_2}}\|\widehat{U}^{(p)}-U\widehat{O}_U^{(p)}\|_{2,\max}+\|\widehat{V}^{(p)}-V\widehat{O}_V^{(p)}\|_{2,\max})\sqrt{\frac{r}{d_2}} \\
        &\quad \leq (1-\frac{4\eta}{5})\sqrt{\frac{1}{d_2}}\|\widehat{U}^{(p)}-U\widehat{O}_U^{(p)}\|_{2,\max} + C_3\eta\frac{\sigma}{\lambda_{\min}}\sqrt{\frac{r\log d_1}{N_0d_1\nu}} \\ &\quad\quad + \frac{\eta}{6}\left(\sqrt{\frac{r}{d_2}}\|\widehat{U}^{(p)}-U\widehat{O}_U^{(p)}\|_{2,\max} + \sqrt{\frac{r}{d_1}}\|\widehat{V}^{(p)}-V\widehat{O}_V^{(p)}\|_{2,\max}\right)
    \end{align*}
    as long as $T\gtrsim (1/d_2\nu)\alpha_dr\log d_1$ and $\|\widehat{U}^{(p)}-U\widehat{O}_U^{(p)}\| + \|\widehat{V}^{(p)}-V\widehat{O}_V^{(p)}\| \lesssim 1/\kappa^2r$. Similarly, we have 
    \begin{align*}
        &\sqrt{\frac{1}{d_1}}\|\widehat{V}^{(p+1)} - V\widehat{O}_V^{(p)}\widehat{R}_G^{(p)}\widehat{K}_V^{(p+1)}\|_{2,\max}\leq (1-\frac{4\eta}{5})\sqrt{\frac{1}{d_1}}\|\widehat{V}^{(p)}-V\widehat{O}_V^{(p)}\|_{2,\max} + C_3\eta\frac{\sigma}{\lambda_{\min}}\sqrt{\frac{r\log d_1}{N_0d_1\nu}} \\ &\quad\quad + \frac{\eta}{6}\left(\sqrt{\frac{r}{d_2}}\|\widehat{U}^{(p)}-U\widehat{O}_U^{(p)}\|_{2,\max} + \sqrt{\frac{r}{d_1}}\|\widehat{V}^{(p)}-V\widehat{O}_V^{(p)}\|_{2,\max}\right).
    \end{align*}
    As a result, 
    \begin{align*}
        &\sqrt{\frac{1}{d_2}}\|\widehat{U}^{(p+1)} - U\widehat{O}_U^{(p)}\widehat{L}_G^{(p)}\widehat{K}_U^{(p+1)}\|_{2,\max} + \sqrt{\frac{1}{d_1}}\|\widehat{V}^{(p+1)} - V\widehat{O}_V^{(p)}\widehat{R}_G^{(p)}\widehat{K}_V^{(p+1)}\|_{2,\max} \\ &\quad \leq C_3\eta\frac{\sigma}{\lambda_{\min}}\sqrt{\frac{r\log d_1}{N_0d_1\nu}}  + (1-\frac{2\eta}{3})\left(\sqrt{\frac{r}{d_2}}\|\widehat{U}^{(p)}-U\widehat{O}_U^{(p)}\|_{2,\max} + \sqrt{\frac{r}{d_1}}\|\widehat{V}^{(p)}-V\widehat{O}_V^{(p)}\|_{2,\max}\right),
    \end{align*}
    where both $\widehat{O}_U^{(p)}\widehat{L}_G^{(p)}\widehat{K}_U^{(p+1)}$ and $\widehat{O}_V^{(p)}\widehat{R}_G^{(p)}\widehat{K}_V^{(p+1)}$ are orthogonal matrices.

    The contraction property in the beginning is then proved if we show 
    \begin{align*}
        \max{\|\widehat{U}^{(p)}-U\widehat{O}_U^{(p)}\|, \|\widehat{V}^{(p)}-V\widehat{O}_V^{(p)}\|}\lesssim \frac{1}{\kappa^2r\sqrt{d_2/d_1}}
    \end{align*}
    and $\|\widehat{U}^{(p)}\|\leq 2\mu\sqrt{r/d_1}$, $\|\widehat{V}^{(p)}\|\leq 2\mu\sqrt{r/d_2}$ for all $1\leq p\leq m$. They can be proved in the same way following the proof of Proposition \ref{thm:estimation}, with sample size condition $T\geq C_1(d_2\nu)^{-1}\alpha_d^2r^2\log^2 d_2$.  

    As a result, recall that $N_0\asymp T/\log d_1$, by the beginning statement with $\eta=0.75$, we get with probability at least $1-4md_1^{-10}$,
    \begin{align*}
        &\sqrt{\frac{1}{d_2}}\|\widehat{U}^{(m)}-U\widehat{O}_U^{(m)}\|_{2,\max} + \sqrt{\frac{1}{d_1}}\|\widehat{V}^{(m)}-V\widehat{O}_V^{(m)}\|_{2,\max} - 2C_3\eta\frac{\sigma}{\lambda_{\min}}\sqrt{\frac{r\log d_1}{N_0d_1\nu}} \\ 
        &\quad \leq (\frac{1}{2})^m \left(\sqrt{\frac{1}{d_2}}\|\widehat{U}^{(1)}-U\widehat{O}_U^{(1)}\|_{2,\max} + \sqrt{\frac{1}{d_1}}\|\widehat{V}^{(1)}-V\widehat{O}_V^{(1)}\|_{2,\max} - 2C_3\eta\frac{\sigma}{\lambda_{\min}}\sqrt{\frac{r\log d_1}{N_0d_1\nu}}\right).
    \end{align*}
    Similar as the proof of Lemma \ref{2maxbound-tside}, with probability at least $1-d_1^{-10}$,
    \begin{align*}
        \sqrt{\frac{1}{d_2}}\|\widehat{U}^{(1)}-U\widehat{O}_U^{(1)}\|_{2,\max} + \sqrt{\frac{1}{d_1}}\|\widehat{V}^{(1)}-V\widehat{O}_V^{(1)}\|_{2,\max}\lesssim \frac{\sigma+\|M\|_{\max}}{\lambda_{\min}}\sqrt{\frac{r\log d_1}{N_0d_1\nu}}.
    \end{align*}
    Therefore, if $m=2\lceil \log (\alpha_d\|M\|_{\max}/\sigma)$, we get
    \begin{align*}
        \sqrt{\frac{1}{d_2}}\|\widehat{U}^{(m)}-U\widehat{O}_U^{(m)}\|_{2,\max} + \sqrt{\frac{1}{d_1}}\|\widehat{V}^{(m)}-V\widehat{O}_V^{(m)}\|_{2,\max}\lesssim \frac{\sigma}{\lambda_{\min}}\sqrt{\frac{r\log d_1}{d_1N_0\nu}}.
    \end{align*}
    Then by proof of Lemma \ref{lemma:estimation3-2} and the result of Lemma \ref{lemma:estimation3-1}, 
    \begin{align*}
        &\|\widehat{M}^{(m)}-M\|_{\max}\leq 2\mu\lambda_{\max}\left(\sqrt{\frac{r}{d_2}}\|\widehat{U}^{(m)}-U\widehat{O}_U^{(m)}\|_{2,\max} + \sqrt{\frac{r}{d_1}}\|\widehat{V}^{(m)}-V\widehat{O}_V^{(m)}\|_{2,\max}\right)  \\ &\quad + \mu^2\sqrt{\frac{r^2}{d_1d_2}}\|\widehat{G}^{(m)}-\widehat{O}_U^{(m)}\Lambda\widehat{O}_V^{(m)}\|\lesssim \mu\kappa\sigma\sqrt{\frac{r^2\log^2 d_1}{Td_1\nu}}
    \end{align*}
\end{proof}

\subsection{Proof of Proposition \ref{thm:estimation-non}}

\begin{proof}
    The proof follows the similar arguments as Proposition \ref{thm:estimation}. We first prove the following statement: for $p=1,\cdots,m-1$, with probability at least $1-4md_1^{-10}$,
    \begin{align*}
         &\sqrt{\frac{1}{d_2}}\|\widehat{U}^{(p+1)}-U\widehat{O}_U^{(p+1)}\|_{2,\max} + \sqrt{\frac{1}{d_1}}\|\widehat{V}^{(p+1)}-V\widehat{O}_V^{(p+1)}\|_{2,\max}\leq 2C\eta\frac{\sigma}{\lambda_{\min}}\sqrt{\frac{\log d_1}{d_1^2d_2N_0}\left(\sum_{i=1}^{d_1}\sum_{j=1}^{d_2}\frac{1}{p_{ij}}\right)} \\ &\quad\quad + (1-\frac{2}{3}\eta)\left(\sqrt{\frac{1}{d_2}}\|\widehat{U}^{(p)}-U\widehat{O}_U^{(p)}\|_{2,\max} + \sqrt{\frac{1}{d_1}}\|\widehat{V}^{(p)}-V\widehat{O}_V^{(p)}\|_{2,\max}\right)
    \end{align*}
    for some constant $C$.

    We begin with the accuracy of $\widehat{G}^{(p)}$. By the definition of $\widehat{G}^{(p)}$, we have
    \begin{align*}
        \frac{1}{N_0}\sum_{t\in \mathcal{D}_{2p}}\sum_{i=1}^{d_1}\frac{1}{p_{ij_t(i)}}(\inp{X_t^i}{\widehat{U}^{(p)}\widehat{G}^{(p)}\widehat{V}^{(p)\top}}-y_t^{i})\widehat{U}^{(p)\top}X_t^i\widehat{V}^{(p)}=0.
    \end{align*}
    Let $\widehat{O}_U^{(p)}$ and $\widehat{O}_V^{(p)}$ be any orthogonal matrices such that 
    \begin{align*}
        \max(\|\widehat{U}^{(p)}-U\widehat{O}_U^{(p)}\|, \|\widehat{V}^{(p)}-V\widehat{O}_U^{(p)}\|)\leq \frac{1}{C_1\mu\kappa^2\sqrt{r}}
    \end{align*}
    for some large constant $C_1>0$.
    \begin{Lemma}\label{lemma:estimation1-nonuniform}
        Suppose that $\|\widehat{U}^{(p)}\|_{2,\max}\leq 2\mu\sqrt{r/d_1}$, $\|\widehat{V}^{(p)}\|_{2,\max}\leq 2\mu\sqrt{r/d_2}$ and $p_{\min}^{-1}\lesssim \sqrt{N_0d_1^{-1}\sum_{i,j}p_{ij}^{-1}/\log d_1}$, then with probability at least $1-3d_1^{-10}$,
        \begin{align*}
            &\|\widehat{G}^{(p)}-\widehat{O}_U^{(p)\top}\Lambda\widehat{O}_V^{(p)}\|\leq C_3\left(\frac{\sqrt{d_1}\sigma r}{N_0p_{\min}\sqrt{d_2}}\log d_1 + \sqrt{\frac{\sigma^2r^2}{d_1d_2N_0}\left(\sum_{i=1}^{d_1}\sum_{j=1}^{d_2} \frac{1}{p_{ij}}\right)\log d_1}\right) \\ &\quad + 2\lambda_{\max}(\|\widehat{U}^{(p)}-U\widehat{O}_U^{(p)}\|^2 + \|\widehat{V}^{(p)}-V\widehat{O}_V^{(p)}\|^2) \\ &\quad + C_4\lambda_{\max}\left(\sqrt{\frac{r}{d_2}}\|\widehat{U}^{(p)}-U\widehat{O}_U^{(p)}\|_{2,\max} + \sqrt{\frac{r}{d_1}}\|\widehat{V}^{(p)}-V\widehat{O}_V^{(p)}\|_{2,\max}\right)\sqrt{\frac{r^2\log d_1}{d_2N_0}\left(\sum_{i=1}^{d_1}\sum_{j=1}^{d_2}\frac{1}{p_{ij}}\right)}.
        \end{align*}
        for some constant $C_2, C_3, C_4>0$.
    \end{Lemma}
    Recall $\widehat{G}^{(p)}=\widehat{L}_G^{(p)}\widehat{\Lambda}^{(p)}\widehat{R}_G^{(p)\top}$ denotes $\widehat{G}^{(p)}$'s SVD. By the gradient descent step of Algorithm \ref{alg:gd},
    \begin{align*}
        \widehat{U}^{(p+0.5)}&=\widehat{U}^{(p)}\widehat{L}_G^{(p)} - \eta\frac{1}{N_0}\sum_{t\in \mathcal{D}_{2p+1}} \sum_{i=1}^{d_1} \frac{1}{p_{ij_t(i)}}\inp{\widehat{U}^{(p)}\widehat{G}^{(p)}\widehat{V}^{(p)\top} - U\Lambda V^{\top}}{X_t^i}X_t^{i}\widehat{V}^{(p)}\widehat{R}_G^{(p)}(\widehat{\Lambda}^{(p)})^{-1} \\ &\quad - \eta\frac{1}{N_0}\sum_{t\in \mathcal{D}_{2p+1}} \sum_{i=1}^{d_1}\frac{1}{p_{ij_t(i)}}\xi_t^i X_t^{i}\widehat{V}^{(p)}\widehat{R}_G^{(p)}(\widehat{\Lambda}^{(p)})^{-1}.
    \end{align*}
    Observe that $(\widehat{U}^{(p)}, \widehat{G}^{(p)}, \widehat{V}^{(p)})$ are independent with $\mathcal{D}_{2p+1}$. Then we write
    \begin{align*}
        \widehat{U}^{(p+0.5)}=\widehat{U}^{(p)}\widehat{L}_G^{(p)} - \eta(\widehat{U}^{(p)}\widehat{G}^{(p)}\widehat{V}^{(p)\top} - U\Lambda V^{\top})\widehat{V}^{(p)}\widehat{R}_G^{(p)}(\widehat{\Lambda}^{(p)})^{-1} + \widehat{E}_V^{(p)} + \widehat{E}_{\xi,V}^{(p)},
    \end{align*} 
    where
    \begin{align*}
        &\widehat{E}_V^{(p)}=\eta(\widehat{U}^{(p)}\widehat{G}^{(p)}\widehat{V}^{(p)\top} - U\Lambda V^{\top})\widehat{V}^{(p)}\widehat{R}_G^{(p)}(\widehat{\Lambda}^{(p)})^{-1} \\ &\quad - \eta\frac{1}{N_0}\sum_{t\in \mathcal{D}_{2p+1}} \sum_{i=1}^{d_1} \frac{1}{p_{ij_t(i)}}\inp{\widehat{U}^{(p)}\widehat{G}^{(p)}\widehat{V}^{(p)\top} - U\Lambda V^{\top}}{X_t^i}X_t^{i}\widehat{V}^{(p)}\widehat{R}_G^{(p)}(\widehat{\Lambda}^{(p)})^{-1}
    \end{align*}
    and
    \begin{align*}
        \widehat{E}_{\xi,V}^{(p)}=- \eta\frac{1}{N_0}\sum_{t\in \mathcal{D}_{2p+1}} \sum_{i=1}^{d_1}\frac{1}{p_{ij_t(i)}}\xi_t^i X_t^{i}\widehat{V}^{(p)}\widehat{R}_G^{(p)}(\widehat{\Lambda}^{(p)})^{-1}.
    \end{align*}
    Note that 
    \begin{align*}
        &(\widehat{U}^{(p)}\widehat{G}^{(p)}\widehat{V}^{(p)\top} - U\Lambda V^{\top})\widehat{V}^{(p)}\widehat{R}_G^{(p)}(\widehat{\Lambda}^{(p)})^{-1} = \widehat{U}^{(p)}(\widehat{G}^{(p)}-\widehat{O}_U^{(p)\top}\Lambda\widehat{O}_V^{(p)})\widehat{R}_G^{(p)}(\widehat{\Lambda}^{(p)})^{-1} \\
        &\quad + (\widehat{U}^{(p)}\widehat{L}_G^{(p)}-U\widehat{O}_U^{(p)}\widehat{L}_G^{(p)})\widehat{L}_G^{(p)\top}\widehat{O}_U^{(p)\top}\Lambda\widehat{O}_V^{(p)}\widehat{R}_G^{(p)}(\widehat{\Lambda}^{(p)})^{-1} + U\Lambda(\widehat{V}^{(p)}\widehat{O}_V^{(p)\top}-V)^{\top}\widehat{V}^{(p)}\widehat{R}_G^{(p)}(\widehat{\Lambda}^{(p)})^{-1}.
    \end{align*}
    Therefore,
    \begin{align*}
        &\widehat{U}^{(p+0.5)}=U\widehat{O}_U^{(p)}\widehat{L}_G^{(p)} + (\widehat{U}^{(p)}\widehat{L}_G^{(p)}-U\widehat{O}_U^{(p)}\widehat{L}_G^{(p)})(I-\eta\widehat{L}_G^{(p)\top}\widehat{O}_U^{(p)\top}\Lambda\widehat{O}_V^{(p)}\widehat{R}_G^{(p)}(\widehat{\Lambda}^{(p)})^{-1}) \\
        &\quad -\eta \widehat{U}^{(p)}(\widehat{G}^{(p)}-\widehat{O}_U^{(p)\top}\Lambda\widehat{O}_V^{(p)})\widehat{R}_G^{(p)}(\widehat{\Lambda}^{(p)})^{-1} - \eta U\Lambda(\widehat{V}^{(p)}\widehat{O}_V^{(p)\top}-V)^{\top}\widehat{V}^{(p)}\widehat{R}_G^{(p)}(\widehat{\Lambda}^{(p)})^{-1} + \widehat{E}_V^{(p)} + \widehat{E}_{\xi,V}^{(p)}.
    \end{align*}
    \begin{Lemma}\label{lemma:estimation2-nonuniform}
        Under the conditions in Proposition \ref{thm:estimation-non} and Lemma \ref{lemma:estimation1-nonuniform}, and 
        \begin{align*}
            \max\{\|\widehat{U}^{(p)}-U\widehat{O}_U^{(p)}\|, \|\widehat{V}^{(p)}-V\widehat{O}_V^{(p)}\|\}\leq 1/(C_1\mu\kappa r\sqrt{d_2/d_1})
        \end{align*}
        for some large constant $C_1$. If $\eta\in [0.25, 0.75]$, then with probability at least $1-2d_1^{-10}$,
        \begin{align*}
            &\|\widehat{U}^{(p+0.5)}-U\widehat{O}_U^{(p)}\widehat{L}_G^{(p)}\|_{2,\max}\leq (1-\frac{9\eta}{10})\|\widehat{U}^{(p)}-U\widehat{O}_U^{(p)}\|_{2,\max} + C_2\eta\frac{\sigma}{\lambda_{\min}}\sqrt{\frac{r\log d_1}{p_{\min}N_0}} \\ &\quad + \frac{\eta}{8}(\sqrt{\frac{d_1}{d_2}}\|\widehat{U}^{(p)}-U\widehat{O}_U^{(p)}\|_{2,\max}+\|\widehat{V}^{(p)}-V\widehat{O}_V^{(p)}\|_{2,\max}),
        \end{align*}
        for some constants $C_2,>0$. Moreover, with probability at least $1-2d_1^{-10}$,
        \begin{align*}
            &\left\{|1-\lambda_{\min}(\widehat{U}^{(p+0.5)})|, |1-\lambda_{\max}(\widehat{U}^{(p+0.5)})|\right\}\lesssim \eta\frac{\sigma}{\lambda_{\min}}\sqrt{\frac{\log d_1}{d_2N_0}\left(\sum_{i=1}^{d_1}\sum_{j=1}^{d_2}\frac{1}{p_{ij}}\right)} \\ &\quad + (\kappa^2\eta^2+\kappa\eta)(\|\widehat{U}^{(p)}-U\widehat{O}_U^{(p)}\|^2 + \|\widehat{V}^{(p)}-V\widehat{O}_V^{(p)}\|^2) \\
            &\quad +  \eta\sqrt{\frac{\log d_1}{d_2N_0}\left(\sum_{i=1}^{d_1}\sum_{j=1}^{d_2}\frac{1}{p_{ij}}\right)}\left(\sqrt{\frac{r}{d_2}}\|\widehat{U}^{(p)}-U\widehat{O}_U^{(p)}\|_{2,\max} + \sqrt{\frac{r}{d_1}}\|\widehat{V}^{(p)}-V\widehat{O}_V^{(p)}\|_{2,\max}\right).
        \end{align*}
    \end{Lemma} 
    By Lemma \ref{lemma:estimation2-nonuniform}, we denote the SVD of $\widehat{U}^{(p+0.5)}$ by $\widehat{U}^{(p+0.5)}=\widehat{U}^{(p+1)}\widehat{\Sigma}_U^{(p+1)}\widehat{K}_U^{(p+1)\top}$ where $\widehat{\Sigma}_U^{(p+1)}$ is diagonal and
    \begin{align*}
        &\|\widehat{\Sigma}_U^{(p+1)}-I\|\leq C_3\eta\frac{\sigma}{\lambda_{\min}}\sqrt{\frac{\log d_1}{d_2N_0}\left(\sum_{i=1}^{d_1}\sum_{j=1}^{d_2}\frac{1}{p_{ij}}\right)} \\ &\quad + C_4(\kappa^2\eta^2+\kappa\eta)(\|\widehat{U}^{(p)}-U\widehat{O}_U^{(p)}\|^2 + \|\widehat{V}^{(p)}-V\widehat{O}_V^{(p)}\|^2) \\
        &\quad +  C_5\eta\sqrt{\frac{\log d_1}{d_2N_0}\left(\sum_{i=1}^{d_1}\sum_{j=1}^{d_2}\frac{1}{p_{ij}}\right)}\left(\sqrt{\frac{r}{d_2}}\|\widehat{U}^{(p)}-U\widehat{O}_U^{(p)}\|_{2,\max} + \sqrt{\frac{r}{d_1}}\|\widehat{V}^{(p)}-V\widehat{O}_V^{(p)}\|_{2,\max}\right).
    \end{align*}
    By $\widehat{U}^{(p+1)}\widehat{\Sigma}_U^{(p+1)}\widehat{K}_U^{(p+1)\top}=U\widehat{O}_U^{(p)}\widehat{L}_G^{(p)} + (\widehat{\Sigma}_U^{(p+0.5)}-U\widehat{O}_U^{(p)}\widehat{L}_G^{(p)})$, we write
    \begin{align*}
        \widehat{U}^{(p+1)}=U\widehat{O}_U^{(p)}\widehat{L}_G^{(p)}\widehat{K}_U^{(p+1)}(\widehat{\Sigma}_U^{(p+1)})^{-1} + (\widehat{U}^{(p+0.5)}-U\widehat{O}_U^{(p)}\widehat{L}_G^{(p)})\widehat{K}_U^{(p+1)}(\widehat{\Sigma}_U^{(p+1)})^{-1}
    \end{align*}
    and 
    \begin{align*}
        \widehat{U}^{(p+1)} - U\widehat{O}_U^{(p)}\widehat{L}_G^{(p)}\widehat{K}_U^{(p+1)}=U\widehat{O}_U^{(p)}\widehat{L}_G^{(p)}\widehat{K}_U^{(p+1)}((\widehat{\Sigma}_U^{(p+1)})^{-1}-I) + (\widehat{U}^{(p+0.5)}-U\widehat{O}_U^{(p)}\widehat{L}_G^{(p)})\widehat{K}_U^{(p+1)}(\widehat{\Sigma}_U^{(p+1)})^{-1}.
    \end{align*}
    Note that $\widehat{O}_U^{(p)}\widehat{L}_G^{(p)}\widehat{K}_U^{(p+1)}$ is an $r\times r$ orthogonal matrix. The Assumptions of Lemma \ref{lemma:estimation2-nonuniform} can guarantee $\lambda_{\min}(\widehat{\Sigma}_U^{(p+1)})\geq 1-\eta/20$ so that $\|(\widehat{\Sigma}_U^{(p+1)})^{-1}\|\leq 1+\eta/10$. Therefore,
    \begin{align*}
        &\sqrt{\frac{1}{d_2}}\|\widehat{U}^{(p+1)} - U\widehat{O}_U^{(p)}\widehat{L}_G^{(p)}\widehat{K}_U^{(p+1)}\|_{2,\max}\leq \sqrt{\frac{1}{d_2}}\|U\|_{2,\max} \|(\widehat{\Sigma}_U^{(p+1)})^{-1}-I\| \\ &\quad\quad + \sqrt{\frac{1}{d_2}}(1+\eta/10)\| \widehat{U}^{(p+0.5)} -  U\widehat{O}_U^{(p)}\widehat{L}_G^{(p)}\|_{2,\max} \\
        &\quad \leq C_3\eta\frac{\sigma}{\lambda_{\min}}\sqrt{\frac{\log d_1}{d_1d_2^2N_0}\left(\sum_{i=1}^{d_1}\sum_{j=1}^{d_2}\frac{1}{p_{ij}}\right)} + C_4(\kappa^2\eta^2+\kappa\eta)(\|\widehat{U}^{(p)}-U\widehat{O}_U^{(p)}\|^2 + \|\widehat{V}^{(p)}-V\widehat{O}_V^{(p)}\|^2)\sqrt{\frac{r}{d_1d_2}} \\
        &\quad\quad +  C_5\eta\sqrt{\frac{\log d_1}{d_1d_2^2N_0}\left(\sum_{i=1}^{d_1}\sum_{j=1}^{d_2}\frac{1}{p_{ij}}\right)}\left(\sqrt{\frac{r}{d_2}}\|\widehat{U}^{(p)}-U\widehat{O}_U^{(p)}\|_{2,\max} + \sqrt{\frac{r}{d_1}}\|\widehat{V}^{(p)}-V\widehat{O}_V^{(p)}\|_{2,\max}\right) \\ 
        &\quad\quad + (1-\frac{4\eta}{5})\sqrt{\frac{1}{d_2}}\|\widehat{U}^{(p)}-U\widehat{O}_U^{(p)}\|_{2,\max} + C_2\eta\frac{\sigma}{\lambda_{\min}}\sqrt{\frac{r\log d_1}{d_2p_{\min}N_0}} \\ &\quad\quad + \frac{\eta}{7}(\sqrt{\frac{d_1}{d_2}}\|\widehat{U}^{(p)}-U\widehat{O}_U^{(p)}\|_{2,\max}+\|\widehat{V}^{(p)}-V\widehat{O}_V^{(p)}\|_{2,\max})\sqrt{\frac{r}{d_2}} \\
        &\quad \leq (1-\frac{4\eta}{5})\sqrt{\frac{1}{d_2}}\|\widehat{U}^{(p)}-U\widehat{O}_U^{(p)}\|_{2,\max} + C_3\eta\frac{\sigma}{\lambda_{\min}}\left(\sqrt{\frac{\log d_1}{d_1d_2^2N_0}\left(\sum_{i=1}^{d_1}\sum_{j=1}^{d_2}\frac{1}{p_{ij}}\right)} + \sqrt{\frac{r\log d_1}{d_2p_{\min}N_0}}\right) \\ &\quad\quad + \frac{\eta}{6}\left(\sqrt{\frac{r}{d_2}}\|\widehat{U}^{(p)}-U\widehat{O}_U^{(p)}\|_{2,\max} + \sqrt{\frac{r}{d_1}}\|\widehat{V}^{(p)}-V\widehat{O}_V^{(p)}\|_{2,\max}\right)
    \end{align*}
    as long as $N_0\gtrsim (\sum_i\sum_j (p_{ij})^{-1})/(d_1d_2^2) \log d_1$ $\|\widehat{U}^{(p)}-U\widehat{O}_U^{(p)}\| + \|\widehat{V}^{(p)}-V\widehat{O}_V^{(p)}\| \lesssim 1/\kappa^2r$. Similarly, we have 
    \begin{align*}
        &\sqrt{\frac{1}{d_1}}\|\widehat{V}^{(p+1)} - V\widehat{O}_V^{(p)}\widehat{R}_G^{(p)}\widehat{K}_V^{(p+1)}\|_{2,\max}\leq (1-\frac{4\eta}{5})\sqrt{\frac{1}{d_1}}\|\widehat{V}^{(p)}-V\widehat{O}_V^{(p)}\|_{2,\max} \\ &\quad + C_3\eta\frac{\sigma}{\lambda_{\min}}\left(\sqrt{\frac{\log d_1}{d_1^2d_2N_0}\left(\sum_{i=1}^{d_1}\sum_{j=1}^{d_2}\frac{1}{p_{ij}}\right)} + \sqrt{\frac{r\log d_1}{d_1p_{\min}N_0}}\right) \\ &\quad\quad + \frac{\eta}{6}\left(\sqrt{\frac{r}{d_2}}\|\widehat{U}^{(p)}-U\widehat{O}_U^{(p)}\|_{2,\max} + \sqrt{\frac{r}{d_1}}\|\widehat{V}^{(p)}-V\widehat{O}_V^{(p)}\|_{2,\max}\right).
    \end{align*}
    As a result, as long as $p_{\min}^{-1}\lesssim (1/d_1d_2)(\sum_i\sum_j (p_{ij})^{-1})$,
    \begin{align*}
        &\sqrt{\frac{1}{d_2}}\|\widehat{U}^{(p+1)} - U\widehat{O}_U^{(p)}\widehat{L}_G^{(p)}\widehat{K}_U^{(p+1)}\|_{2,\max} + \sqrt{\frac{1}{d_1}}\|\widehat{V}^{(p+1)} - V\widehat{O}_V^{(p)}\widehat{R}_G^{(p)}\widehat{K}_V^{(p+1)}\|_{2,\max} \\ &\quad \leq C_3\eta\frac{\sigma}{\lambda_{\min}}\sqrt{\frac{\log d_1}{d_1^2d_2N_0}\left(\sum_{i=1}^{d_1}\sum_{j=1}^{d_2}\frac{1}{p_{ij}}\right)}   \\ &\quad\quad + (1-\frac{2\eta}{3})\left(\sqrt{\frac{r}{d_2}}\|\widehat{U}^{(p)}-U\widehat{O}_U^{(p)}\|_{2,\max} + \sqrt{\frac{r}{d_1}}\|\widehat{V}^{(p)}-V\widehat{O}_V^{(p)}\|_{2,\max}\right),
    \end{align*}
    where both $\widehat{O}_U^{(p)}\widehat{L}_G^{(p)}\widehat{K}_U^{(p+1)}$ and $\widehat{O}_V^{(p)}\widehat{R}_G^{(p)}\widehat{K}_V^{(p+1)}$ are orthogonal matrices. 

    The contraction property in the beginning is then proved if we show 
    \begin{align*}
        \max{\|\widehat{U}^{(p)}-U\widehat{O}_U^{(p)}\|, \|\widehat{V}^{(p)}-V\widehat{O}_V^{(p)}\|}\lesssim \frac{1}{\kappa^2r\sqrt{d_2/d_1}}
    \end{align*}
    and $\|\widehat{U}^{(p)}\|\leq 2\mu\sqrt{r/d_1}$, $\|\widehat{V}^{(p)}\|\leq 2\mu\sqrt{r/d_2}$ for all $1\leq p\leq m$. They can be proved in the same way following the proof of Proposition \ref{thm:estimation}, with sample size condition $N_0\gtrsim \alpha_d^2(d_1d_2^2)^{-1}(\sum_{i,j} p_{ij}^{-1})\log d_1$ and SNR $\lambda_{\min}/\sigma\gtrsim \sqrt{\alpha_d (\sum_{i,j} p_{ij}^{-1})\log d_1/(d_1N_0)}$.

    As a result, with $\eta=0.75$, we get with probability at least $1-4md_1^{-10}$,
    \begin{align*}
        &\sqrt{\frac{1}{d_2}}\|\widehat{U}^{(m)}-U\widehat{O}_U^{(m)}\|_{2,\max} + \sqrt{\frac{1}{d_1}}\|\widehat{V}^{(m)}-V\widehat{O}_V^{(m)}\|_{2,\max} - 2C_3\eta\frac{\sigma}{\lambda_{\min}}\sqrt{\frac{\log d_1}{d_1^2d_2N_0}\left(\sum_{i=1}^{d_1}\sum_{j=1}^{d_2}\frac{1}{p_{ij}}\right)} \\ 
        &\quad \leq (\frac{1}{2})^m \left(\sqrt{\frac{1}{d_2}}\|\widehat{U}^{(1)}-U\widehat{O}_U^{(1)}\|_{2,\max} + \sqrt{\frac{1}{d_1}}\|\widehat{V}^{(1)}-V\widehat{O}_V^{(1)}\|_{2,\max} - 2C_3\eta\frac{\sigma}{\lambda_{\min}}\sqrt{\frac{\log d_1}{d_1^2d_2N_0}\left(\sum_{i=1}^{d_1}\sum_{j=1}^{d_2}\frac{1}{p_{ij}}\right)}\right).
    \end{align*}
    Similar as the proof of Lemma \ref{2maxbound-non}, with probability at least $1-d_1^{-10}$,
    \begin{align*}
        \sqrt{\frac{1}{d_2}}\|\widehat{U}^{(1)}-U\widehat{O}_U^{(1)}\|_{2,\max} + \sqrt{\frac{1}{d_1}}\|\widehat{V}^{(1)}-V\widehat{O}_V^{(1)}\|_{2,\max}\lesssim \frac{\sigma+\|M\|_{\max}}{\lambda_{\min}}\sqrt{\frac{\log d_1}{d_1^2d_2N_0}\left(\sum_{i=1}^{d_1}\sum_{j=1}^{d_2}\frac{1}{p_{ij}}\right)}.
    \end{align*}
    Therefore, if $m=2\lceil \log (\alpha_d\|M\|_{\max}/\sigma)$, we get
    \begin{align*}
        \sqrt{\frac{1}{d_2}}\|\widehat{U}^{(m)}-U\widehat{O}_U^{(m)}\|_{2,\max} + \sqrt{\frac{1}{d_1}}\|\widehat{V}^{(m)}-V\widehat{O}_V^{(m)}\|_{2,\max}\lesssim \frac{\sigma}{\lambda_{\min}}\sqrt{\frac{\log d_1}{d_1^2d_2N_0}\left(\sum_{i=1}^{d_1}\sum_{j=1}^{d_2}\frac{1}{p_{ij}}\right)}.
    \end{align*}
    Then by proof of Lemma \ref{lemma:estimation2-nonuniform} and the result of Lemma \ref{lemma:estimation1-nonuniform}, as long as $p_{\min}^{-1}\lesssim \sqrt{N_0d_2(d_1)^{-2}\sum_{i,j}p_{ij}^{-1}}$,
    \begin{align*}
        &\|\widehat{M}^{(m)}-M\|_{\max}\leq 2\mu\lambda_{\max}\left(\sqrt{\frac{r}{d_2}}\|\widehat{U}^{(m)}-U\widehat{O}_U^{(m)}\|_{2,\max} + \sqrt{\frac{r}{d_1}}\|\widehat{V}^{(m)}-V\widehat{O}_V^{(m)}\|_{2,\max}\right)  \\ &\quad + \mu^2\sqrt{\frac{r^2}{d_1d_2}}\|\widehat{G}^{(m)}-\widehat{O}_U^{(m)}\Lambda\widehat{O}_V^{(m)}\|\lesssim \mu\kappa\sigma\sqrt{\frac{r\log^2 d_1}{d_1^2d_2N_0}\left(\sum_{i=1}^{d_1}\sum_{j=1}^{d_2}\frac{1}{p_{ij}}\right)}.
    \end{align*}
\end{proof}

\section{Proofs of Technical Lemmas in Appendix \ref{sec: proof-inference}}

\subsection{Proof of Lemma \ref{lemmaneg2}} 
\begin{proof}
    We first state the following Lemmas.
    \begin{Lemma}
        \label{Zbound}
        Under the conditions in Theorem \ref{thm:inference-main}, with probability at least $1-2d_1^{-10}$, 
        \begin{align*}
        \|\widehat{Z}^{(l)}\|\lesssim \sqrt{\frac{d_2^2\log d_1}{T}}\sigma
        \end{align*}
        for $l=1,2$. 
    \end{Lemma}

    \begin{Lemma}
        \label{2maxbound}
        Under the conditions in Theorem \ref{thm:inference-main}, with probability at least $1-5d_1^{-5}$,
        \begin{align*}
            &\|\widehat{U}_l\widehat{U}_l^{\top}-UU^{\top}\|_{2,\max}\lesssim \frac{\sigma}{\lambda_{\min}}\sqrt{\frac{d_2^2\log d_1}{T}}\mu\sqrt{\frac{r}{d_1}} \\
            &\|\widehat{V}_l\widehat{V}_l^{\top}-VV^{\top}\|_{2,\max}\lesssim \frac{\sigma}{\lambda_{\min}}\sqrt{\frac{d_2^2\log d_1}{T}}\mu\sqrt{\frac{r}{d_2}}. 
        \end{align*}
        for $l=1,2$. 
    \end{Lemma}

    Without loss of generality, we only prove the upper bound for $|\inp{\widehat{U}_1\widehat{U}_1^{\top}\widehat{Z}^{(1)}\widehat{V}_1\widehat{V}_1^{\top}}{Q} - \inp{UU^{\top}\widehat{Z}^{(1)}VV^{\top}}{Q}|$ and omit the subscript and superscript $l$ for ease of notation. Notice that 
    \begin{align*}
        |\inp{\widehat{U}\widehat{U}^\top \widehat{Z} \widehat{V}\widehat{V}^\top - UU^\top \widehat{Z}VV^\top} {Q}| \le \|Q\|_{\ell_1} \|\widehat{U}\widehat{U}^\top \widehat{Z} \widehat{V}\widehat{V}^\top - UU^\top \widehat{Z}VV^\top\|_{\max}.
    \end{align*}
    Now by triangular inequality, under the conditions in Theorem \ref{thm:inference-main},
    \begin{align*}
        \|\widehat{U}\widehat{U}^\top \widehat{Z} \widehat{V}\widehat{V}^\top - UU^\top \widehat{Z}VV^\top\|_{\max} \le  &   \|(\widehat{U}\widehat{U}^\top-UU^\top) \widehat{Z}VV^\top\|_{\max} + \|UU^\top \widehat{Z}( \widehat{V}\widehat{V}^\top -VV^\top)\|_{\max} \\
    &+  \|(\widehat{U}\widehat{U}^\top-UU^\top)  \widehat{Z} ( \widehat{V}\widehat{V}^\top- VV^\top)\|_{\max}\\
    \le & \|\widehat{Z}\| \left(\|\widehat{U} \widehat{U}^{\top}-UU^{\top}\|_{2,  { \max }}\|V\|_{2, { \max }} +\|\widehat{V}\widehat{V}^\top- VV^\top\|_{2,\max} \|U\|_{2, { \max }} \right) \\
    & + \|\widehat{Z}\| \|\widehat{U} \widehat{U}^{\top}-UU^{\top}\|_{2,  { \max }}\|\widehat{V}\widehat{V}^\top- VV^\top\|_{2,\max} \\
    \lesssim & \frac{\sigma^2}{\lambda_{\min}}\frac{\mu^2 rd_2^{3/2}\log d_1}{Td_1^{1/2}}.
    \end{align*}
\end{proof}

\subsection{Proof of Lemma \ref{lemmaneg3}}
\begin{proof}
    By Lemma 5 and Lemma 6 in \cite{xia2021statistical}, under the the result of Lemma \ref{Zbound} and Lemma \ref{2maxbound}, 
    \begin{align*}
        \big|\sum_{k=2}^{\infty} \inp{(\mathcal{S}_{A,k}(\widehat{E})A\Theta\Theta^{\top}+\Theta\Theta^{\top}A\mathcal{S}_{A,k}(\widehat{E}))}{\widetilde{Q}}\big|&\lesssim \|Q\|_{\ell_1}\frac{\mu^2 r}{\lambda_{\min}\sqrt{d_1d_2}}\delta^2 \\
        &\lesssim \|Q\|_{\ell_1}\frac{\sigma^2}{\lambda_{\min}}\frac{\mu^2 rd_2^{3/2}\log d_1}{Td_1^{1/2}},
    \end{align*}
    and
    \begin{align*}
        \big|\inp{(\widehat{\Theta}\widehat{\Theta}^{\top}-\Theta\Theta^{\top})A(\widehat{\Theta}\widehat{\Theta}^{\top}-\Theta\Theta^{\top})}{\widetilde{Q}}\big|&\lesssim \|Q\|_{\ell_1}\|\Lambda\|\|\widehat{U} \widehat{U}^{\top}-UU^{\top}\|_{2,  { \max }}\|\widehat{V}\widehat{V}^\top- VV^\top\|_{2,\max} \\
        &\lesssim \|Q\|_{\ell_1}\frac{\sigma^2}{\lambda_{\min}}\frac{\kappa\mu^2 rd_2^{3/2}\log d_1}{Td_1^{1/2}},
    \end{align*}
    where $\delta$ is defined as the upper bound $\|\widehat{E}\|\leq C_1\sqrt{d_2^2\log d_1/T}\sigma:= \delta$ by result of Lemma \ref{Zbound}.
\end{proof}

\subsection{Proof of Lemma \ref{Zbound}}
\begin{proof}
    Without loss of generality, we only prove the upper bound for $\|\widehat{Z}^{(1)}\|$ and omit the subscript and superscript $l$ for ease of notation. We analysis $\|\widehat{Z}_1\|$ and $\|\widehat{Z}_2\|$ separately. Since $\xi_t^i$ is subGaussian, for any $t$,
    \begin{align*}
        \bigg\|\big\|\frac{d_2}{T_0}\sum_{i=1}^{d_1}\xi_t^iX_t^i\big\|\bigg\|_{\psi_2}\lesssim \frac{d_2\sigma}{T_0}\bigg\|\sum_{i=1}^{d_1}X_t^i\bigg\|\lesssim \frac{d_2\sigma}{T_0}.
    \end{align*} 
    We also calculate the variance
    \begin{align*}
        \frac{d_2^2}{T_0^2}\EE\left[\bigg(\sum_{i=1}^{d_1}\xi_t^iX_t^i\bigg)\bigg(\sum_{i=1}^{d_1}\xi_t^iX_t^i\bigg)^{\top}\right]=\frac{d_2^2}{T_0^2}\EE\left[\sum_{i=1}^{d_1}\xi_t^{i2}X_t^iX_t^{i\top} \right]=\frac{d_2^2\sigma^2}{T_0^2}I_{d_1}.
    \end{align*}
    since $X_t^iX_t^{j\top}=0$ for all $i\neq j$. Similarly,
    \begin{align*}
        \frac{d_2^2}{T_0^2}\EE\left[\bigg(\sum_{i=1}^{d_1}\xi_t^iX_t^i\bigg)^{\top}\bigg(\sum_{i=1}^{d_1}\xi_t^iX_t^i\bigg)\right]=\frac{d_1d_2\sigma^2}{T_0^2}I_{d_2}.
    \end{align*}
    As a result,
    \begin{align*}
        &\max\left\{\bigg\|\sum_{t=1}^{T_0}  \frac{d_2^2}{T_0^2}\EE\left[\bigg(\sum_{i=1}^{d_1}\xi_t^iX_t^i\bigg)\bigg(\sum_{i=1}^{d_1}\xi_t^iX_t^i\bigg)^{\top}\right] \bigg\|, \right.\\ & \left.\quad\quad\quad \bigg\|\sum_{t=1}^{T_0} \frac{d_2^2}{T_0^2}\EE\left[\bigg(\sum_{i=1}^{d_1}\xi_t^iX_t^i\bigg)^{\top}\bigg(\sum_{i=1}^{d_1}\xi_t^iX_t^i\bigg)\right]\bigg\|\right\} \leq \frac{d_2^2\sigma^2}{T_0}.
    \end{align*}
    By matrix Bernstein inequality, with probability at least $1-d_1^{-10}$,
    \begin{align*}
        \|\widehat{Z}_1\|\lesssim \frac{d_2\sigma}{T}\log d_1 + \frac{d_2\sigma}{\sqrt{T}}\log^{1/2}d_1\lesssim  \frac{d_2\sigma}{\sqrt{T}}\log^{1/2}d_1
    \end{align*}
    as long as $T\gtrsim \log d_1$. Similarly, we have
    \begin{align*}
        \frac{1}{T_0}\bigg\|\sum_{i=1}^{d_1} d_2\inp{\widehat{\Delta}_1}{X_t^i}X_t^i -\widehat{\Delta}_1\bigg\|\leq \frac{d_2}{T_0}\|\widehat{\Delta}_1\|_{\max} + \frac{\sqrt{d_1d_2}}{T_0}\|\widehat{\Delta}_1\|_{\max}\leq  \frac{2d_2}{T_0}\|\widehat{\Delta}_1\|_{\max}
    \end{align*}
    and 
    \begin{align*}
        &\left\|\frac{1}{T_0^2}\EE\bigg[\bigg(\sum_{i=1}^{d_1} d_2\inp{\widehat{\Delta}_1}{X_t^i}X_t^i -\widehat{\Delta}_1\bigg)\bigg(\sum_{i=1}^{d_1} d_2\inp{\widehat{\Delta}_1}{X_t^i}X_t^i -\widehat{\Delta}_1\bigg)^{\top}\bigg]\right\| \\ 
        &\quad \leq  \frac{d_2^2}{T_0^2}\|\widehat{\Delta}_1\|_{\max}^2\left\|\EE\bigg[\bigg(\sum_{i=1}^{d_1} X_t^i\bigg)\bigg(\sum_{i=1}^{d_1} X_t^i\bigg)^{\top} \bigg]\right\|  + \frac{1}{T_0^2}\|\widehat{\Delta}_1\|^2 \\ 
        &\quad \leq \frac{d_2^2}{T_0^2}\|\widehat{\Delta}_1\|_{\max}^2+ \frac{d_1d_2}{T_0^2}\|\widehat{\Delta}_1\|_{\max}^2 \lesssim \frac{d_2^2\sigma^2}{T_0^2}
    \end{align*}
    where the last inequality comes from $\|\widehat{\Delta}_1\|_{\max}^2\lesssim \sigma^2$ under the conditions in Theorem \ref{thm:inference-main}. In the same way, by matrix Bernstein inequality, with probability at least $1-d_1^{-10}$,
    \begin{align*}
        \|\widehat{Z}_2\|\lesssim \frac{d_2\sigma}{T}\log d_1 + \frac{d_2\sigma}{\sqrt{T}}\log^{1/2}d_1\lesssim  \frac{d_2\sigma}{\sqrt{T}}\log^{1/2}d_1
    \end{align*}
    as long as $T\gtrsim \log d_1$. Then we conclude the proof. 
\end{proof}

\subsection{Proof of Lemma \ref{2maxbound}}
\begin{proof}
    Without loss loss of generality, we only prove $l=1$ case and omit all the subscripts and superscripts. We first denote the upper bound of $\|\widehat{E}\|$ defined in the proof of Theorem \ref{thm:inference-main} to be $\|\widehat{E}\|\leq C_1\sqrt{d_2^2\log d_1/T}\sigma:= \delta$ by result of Lemma \ref{Zbound}. According to Theorem 4 in \cite{xia2021statistical}, it suffices to prove that, there exist constants $C_1,C_2>0$ such that, for all $k\geq 0$, with probability at least $1-2(k+1)d_1^{-10}$,
    \begin{align*}
        &\max_{j \in [d_1]} \|e_{j}^{\top}\mathfrak{P}^{\perp}(\mathfrak{P}^{\perp}\widehat{E}\mathfrak{P}^{\perp})^k\widehat{E}\Theta\|\leq C_1(C_2\delta)^{k+1}\mu\sqrt{\frac{r}{d_1}}, \\
        &\max_{j \in [d_2]} \|e_{j+d_1}^{\top}\mathfrak{P}^{\perp}(\mathfrak{P}^{\perp}\widehat{E}\mathfrak{P}^{\perp})^k\widehat{E}\Theta\|\leq C_1(C_2\delta)^{k+1}\mu\sqrt{\frac{ r}{d_2}}.
    \end{align*}
    where $\Theta$ and $\mathfrak{P}^{\perp}$ are also defined in the proof of Theorem \ref{thm:inference-main}. \\
    \noindent\textbf{Case 0: k=0}  \\
    For any $j\in [d_1]$, clearly,
    \begin{align*}
        \|e_j^{\top}\mathfrak{P}^{\perp}\widehat{E}\Theta\|\leq \|e_j^{\top}\Theta\Theta^{\top}\widehat{E}\Theta\| + \|e_j^{\top}\widehat{E}\Theta\|\leq \delta\mu\sqrt{\frac{r}{d_1}} + \|e_j^{\top}\widehat{E}\Theta\|= \delta\mu\sqrt{\frac{r}{d_1}} + \|e_j^{\top}\widehat{Z}V\|,
    \end{align*}
    where in the last inequality, we transfer $e_j$ to be the canonical basis vector from $\RR^{d_1+d_2}$ to $\RR^{d_1}$.  We write
    \begin{align*}
        e_{j}^{\top}\widehat{Z}V= \frac{d_2}{T_0}\sum_{t=1}^{T_0} \sum_{i=1}^{d_1}\xi_t^ie_j^{\top}X_t^i V  + \frac{1}{T_0}\sum_{t=1}^{T_0}\sum_{i=1}^{d_1} d_2\inp{\widehat{\Delta}_{1}}{X_t^{i}}e_j^{\top}X_t^iV - e_j\widehat{\Delta}_{1}V.
    \end{align*} 
    Since $e_j^{\top}X_t^i\neq 0$ only when $i=j$, we have
    \begin{align*}
        \left\|\bigg\|\frac{d_2}{T_0} \sum_{i=1}^{d_1}\xi_t^ie_j^{\top}X_t^i V \bigg\|\right\|_{\psi_2}\lesssim \frac{d_2\sigma}{T_0}\|V\|_{2,\max}\leq \frac{d_2\sigma\mu}{T_0}\sqrt{\frac{r}{d_2}}
    \end{align*}
    and
    \begin{align*}
        &\frac{d_2^2}{T_0^2}\EE\left[e_j^{\top}\bigg(\sum_{i=1}^{d_1}\xi_t^iX_t^i \bigg)VV^{\top}\bigg(\sum_{i=1}^{d_1}\xi_t^iX_t^i \bigg)^{\top} e_j \right] = \frac{d_2^2\sigma^2}{T_0^2}\EE\left[e_j^{\top}X_t^jVV^{\top}X_t^{j\top}e_j\right] \\
        &\quad = \frac{d_2\sigma^2}{T_0^2}\text{tr}(VV^{\top})=\frac{rd_2\sigma^2}{T_0^2}.
    \end{align*}
    By Bernstein inequality, with probability at least $1-d_1^{-10}$,
    \begin{align*}
        \|e_j^{\top}\widehat{Z}_1V\|\lesssim \mu\sigma\sqrt{\frac{rd_2\log d_1}{T}}
    \end{align*}
    as long as $T\gtrsim \log d_1$. Similarly, 
    \begin{align*}
        \frac{1}{T_0}\bigg\|e_j^{\top}\big(\sum_{i=1}^{d_1} d_2\inp{\widehat{\Delta}_1}{X_t^i}X_t^i -\widehat{\Delta}_1\big)V\bigg\|&\leq \frac{d_2}{T_0}\bigg\|e_j^{\top}\inp{\widehat{\Delta}_1}{X_t^j}X_t^jV \bigg\| + \frac{\sqrt{d_2}}{T_0}\|\widehat{\Delta}_1\|_{\max} \\
        &\leq \frac{\mu\sqrt{rd_2}}{T_0}\|\widehat{\Delta}_1\|_{\max}\lesssim \frac{\mu\sqrt{rd_2}}{T_0}\sigma
    \end{align*}
    and 
    \begin{align*}
        &\frac{d_2^2}{T_0^2} \EE\left[e_j^{\top}\bigg(\sum_{i=1}^{d_1} \inp{\widehat{\Delta}_1}{X_t^i}X_t^iV\bigg)\bigg(\sum_{i=1}^{d_1} \inp{\widehat{\Delta}_1}{X_t^i}X_t^iV\bigg)^{\top}e_j \right] \\ 
        &\quad \leq \frac{d_2^2}{T_0^2}\|\widehat{\Delta}_1\|_{\max}^2 \EE\left[e_j^{\top}X_t^jVV^{\top}X_t^{j\top}e_j\right] \lesssim \frac{rd_2\sigma^2}{T_0^2}.
    \end{align*}
    where we again use $\|\widehat{\Delta}_1\|_{\max}\lesssim \sigma$. By Bernstein inequality, with probability at least $1-d_1^{-10}$,
    \begin{align*}
        \|e_j^{\top}\widehat{Z}_2V\|\lesssim \mu\sigma\sqrt{\frac{rd_2\log d_1}{T}}
    \end{align*}
    as long as $T\gtrsim \log d_1$. Then this directly implies
    \begin{align*}
        \max_{j\in [d_1]}\|e_j^{\top}\mathfrak{P}^{\perp}\widehat{E}\Theta\|\lesssim \delta\mu\sqrt{\frac{r}{d_1}}.
    \end{align*}
    Following the same arguments, we can show 
    \begin{align*}
        \max_{j\in [d_2]}\|e_{j+d_1}^{\top}\mathfrak{P}^{\perp}\widehat{E}\Theta\|\lesssim \delta\mu\sqrt{\frac{r}{d_2}}.
    \end{align*}
    \noindent\textbf{Case 1: k=1}  \\
    Observe that for any $j\in [d_1]$, 
    \begin{align*}
        \|e_j^{\top}\mathfrak{P}^{\perp}\widehat{E}\mathfrak{P}^{\perp}\widehat{E}\Theta\|\leq \|e_j^{\top}\Theta\Theta^{\top} \widehat{E}\mathfrak{P}^{\perp}\widehat{E}\Theta\| + \|e_j^{\top}\widehat{E}\mathfrak{P}^{\perp}\widehat{E}\Theta\|\leq \delta^2\mu\sqrt{\frac{r}{d_1}} + \|e_j^{\top}\widehat{E}\mathfrak{P}^{\perp}\widehat{E}\Theta\|.
    \end{align*}
    By definition of $\widehat{E}$ and $\mathfrak{P}^{\perp}$, we have 
    \begin{align*}
        \widehat{E}\mathfrak{P}^{\perp}\widehat{E}\Theta = \begin{pmatrix}
            \widehat{Z}V_{\perp}V_{\perp}^{\top}\widehat{Z}^{\top}U & 0 \\
            0 & \widehat{Z}^{\top}U_{\perp}U_{\perp}^{\top}\widehat{Z}V 
        \end{pmatrix}.
    \end{align*}
    It suffices to prove the upper bound for $\|e_j^{\top}\widehat{Z}V_{\perp}V_{\perp}^{\top}\widehat{Z}^{\top}U\|$. Define $\mathfrak{J}_j=e_je_j^{\top}\in \RR^{d_1\times d_1}$ and $\mathfrak{J}^{\perp}=I-\mathfrak{J}$. Then write $\widehat{Z}=\mathfrak{J}_j\widehat{Z} + \mathfrak{J}_j^{\perp}\widehat{Z}$ and as a result,
    \begin{align*}
        \|e_j^{\top}\widehat{Z}V_{\perp}V_{\perp}^{\top}\widehat{Z}^{\top}U\|&\leq \|e_j^{\top}\widehat{Z}V_{\perp}V_{\perp}^{\top}(\mathfrak{J}_j\widehat{Z})^{\top}U\| + \|e_j^{\top}\widehat{Z}V_{\perp}V_{\perp}^{\top}(\mathfrak{J}_j^{\perp}\widehat{Z})^{\top}U\|  \\
        &\leq \|e_j^{\top}\widehat{Z}V_{\perp}V_{\perp}^{\top}\widehat{Z}^{\top}e_j\|\|e_j^{\top}U\| + \|e_j^{\top}\widehat{Z}V_{\perp}V_{\perp}^{\top}(\mathfrak{J}_j^{\perp}\widehat{Z})^{\top}U\| \\
        &\leq \delta^2\mu\sqrt{\frac{r}{d_1}} + \|e_j^{\top}\widehat{Z}_1V_{\perp}V_{\perp}^{\top}(\mathfrak{J}_j^{\perp}\widehat{Z})^{\top}U\| + \|e_j^{\top}\widehat{Z}_2V_{\perp}V_{\perp}^{\top}(\mathfrak{J}_j^{\perp}\widehat{Z})^{\top}U\|.
    \end{align*}
    We first prove the upper bound of $\|e_j^{\top}\widehat{Z}_1V_{\perp}V_{\perp}^{\top}(\mathfrak{J}_j^{\perp}\widehat{Z})^{\top}U\|$. Since $\mathfrak{J}_j^{\perp}\widehat{Z}$ does not include $X_t^j$ for any $t$, condition on $\mathfrak{J}_j^{\perp}\widehat{Z}$, we have
    \begin{align*}
        \left\|\bigg\|\frac{d_2}{T_0} \sum_{i=1}^{d_1}\xi_t^ie_j^{\top}X_t^iV_{\perp}V_{\perp}^{\top}(\mathfrak{J}_j^{\perp}\widehat{Z})^{\top}U  \bigg\|\right\|_{\psi_2} \lesssim \frac{d_2\sigma}{T_0}\|V_{\perp}V_{\perp}^{\top}(\mathfrak{J}_j^{\perp}\widehat{Z})^{\top}U\|_{2,\max}.  
    \end{align*}
    since only $e_j^{\top}X_t^{j}\neq 0$. Under the result of \textbf{Case 0}, we have
    \begin{align*}
        \|V_{\perp}V_{\perp}^{\top}(\mathfrak{J}_j^{\perp}\widehat{Z})^{\top}U\|_{2,\max}&\leq \|V_{\perp}V_{\perp}^{\top}(\mathfrak{J}_j\widehat{Z})^{\top}U\|_{2,\max} + \|V_{\perp}V_{\perp}^{\top}\widehat{Z}^{\top}U\|_{2,\max} \\
        &\lesssim \delta\mu\sqrt{\frac{r}{d_1}} + \delta\mu\sqrt{\frac{r}{d_2}}\lesssim \delta\mu\sqrt{\frac{r}{d_1}}.
    \end{align*}
    We also calculate the variance, 
     \begin{align*}
        &\frac{d_2^2}{T_0^2}\EE\left[ \big(\sum_{i=1}^{d_1}\xi_t^ie_j^{\top}X_t^iV_{\perp}V_{\perp}^{\top}(\mathfrak{J}_j^{\perp}\widehat{Z})^{\top}U\big)\big(\sum_{i=1}^{d_1}\xi_t^ie_j^{\top}X_t^iV_{\perp}V_{\perp}^{\top}(\mathfrak{J}_j^{\perp}\widehat{Z})^{\top}U\big)^{\top} \bigg| \mathfrak{J}_j^{\perp}\widehat{Z} \right] \\
        &\leq \frac{d_2^2\sigma^2}{T_0^2}\EE\left[\|e_j^{\top}X_t^jV_{\perp}V_{\perp}^{\top}(\mathfrak{J}_j^{\perp}\widehat{Z})^{\top}U \|^2 \bigg| \mathfrak{J}_j^{\perp}\widehat{Z} \right]\leq \frac{d_2^2\sigma^2}{T_0^2}\sum_{s=1}^{d_2} \PP\left(j_t(j)=s\bigg| \mathfrak{J}_j^{\perp}\widehat{Z}\right)\|V_{\perp}V_{\perp}^{\top}(\mathfrak{J}_j^{\perp}\widehat{Z})^{\top}U\|_{2,\max}  \\
        &\leq \frac{d_2^2\sigma^2}{T_0^2}\|V_{\perp}V_{\perp}^{\top}(\mathfrak{J}_j^{\perp}\widehat{Z})^{\top}U\|_{2,\max}\lesssim \frac{d_2^2\mu^2r\sigma^2}{d_1T^2}\delta^2. 
    \end{align*}
    Then by Bernstein inequality, with probability at least $1-d_1^{-10}$,
    \begin{align*}
        \|e_j^{\top}\widehat{Z}_1V_{\perp}V_{\perp}^{\top}(\mathfrak{J}_j^{\perp}\widehat{Z})^{\top}U\|\lesssim \frac{d_2\sigma}{T}\delta\mu\sqrt{\frac{r}{d_1}}\log d_1 + \mu\sqrt{\frac{rd_2^2\sigma^2\log^{1/2}d_1}{d_1T}}\delta\lesssim \delta^2\mu\sqrt{\frac{r}{d_1}}
    \end{align*}
    as long as $T\gtrsim \log d_1$. Similarly, by the fact that $\|\widehat{\Delta}_1\|_{\max}\lesssim \sigma$ and $e_j^{\top}X_t^{i}= 0$ for $i\neq j$, under the result of \textbf{Case 0}, we have with probability at least $1-d_1^{-10}$,
    \begin{align*}
        \|e_j^{\top}\widehat{Z}_2V_{\perp}V_{\perp}^{\top}(\mathfrak{J}_j^{\perp}\widehat{Z})^{\top}U\|\lesssim \delta^2\mu\sqrt{\frac{r}{d_1}}.
    \end{align*}
    The upper bound for $\max_{j\in [d_2]}\|e_{j+d_1}^{\top} \mathfrak{P}^{\perp}\widehat{E}\mathfrak{P}^{\perp}\widehat{E}\Theta\|$ can be proved in the same way. Taking a union bound, with probability at least $1-4d_1^{-10}$,
    \begin{align*}
        \max_{j\in [d_1]}\|e_{j}^{\top} \mathfrak{P}^{\perp}\widehat{E}\mathfrak{P}^{\perp}\widehat{E}\Theta\|\lesssim \delta^2\mu\sqrt{\frac{r}{d_1}} \quad\text{and}\quad \max_{j\in [d_2]}\|e_{j+d_1}^{\top} \mathfrak{P}^{\perp}\widehat{E}\mathfrak{P}^{\perp}\widehat{E}\Theta\|\lesssim \delta^2\mu\sqrt{\frac{r}{d_2}}.
    \end{align*}
    Finally, the general case $k\geq 2$ can be proved by induction, following Lemma 9 in \cite{xia2021statistical}. We omit the details here. 
\end{proof}

\subsection{Proof of Lemma \ref{Zbound-otm}}
\begin{proof}
    Without loss of generality, we only prove the upper bound for $\|\widehat{Z}^{(1)}\|$ and omit the subscript and superscript $l$ for ease of notation. We analysis $\|\widehat{Z}_1\|$ and $\|\widehat{Z}_2\|$ separately. Since $\xi_t^i$ is subGaussian, for any $t$,
    \begin{align*}
        \bigg\|\big\|\frac{d_2}{T_0Kp_0}\sum_{i=1}^{d_1} \sum_{q\in h_{it}} \xi_t^{iq}X_t^{iq}\big\|\bigg\|_{\psi_2}\lesssim \frac{d_2\sigma}{T_0Kp_0}\bigg\|\sum_{i=1}^{d_1} \sum_{q\in h_{it}} X_t^{iq}\bigg\|\lesssim \frac{d_2\sigma}{T_0\sqrt{K}p_0}
    \end{align*} 
    since by definition of $\mathcal{M}_{K,p_0}$, $\big\|\sum_{i=1}^{d_1} \sum_{q\in h_{it}} X_t^{iq}\big\|\leq \sqrt{K}$. We also calculate the variance
    \begin{align*}
        \frac{d_2^2}{T_0^2K^2p_0^2}\EE\left[\bigg(\sum_{i=1}^{d_1}\sum_{q\in h_{it}} \xi_t^{iq}X_t^{iq}\bigg)\bigg(\sum_{i=1}^{d_1} \sum_{q\in h_{it}} \xi_t^{iq}X_t^{iq}\bigg)^{\top}\right]=\frac{d_2^2}{T_0^2K^2p_0^2}\EE\left[\sum_{i=1}^{d_1}\sum_{q\in h_{it}} \xi_t^{iq}X_t^{iq}X_t^{iq\top} \right]=\frac{d_2^2\sigma^2}{T_0^2Kp_0}I_{d_1}.
    \end{align*}
    since $X_t^{iq}X_t^{jr\top}\neq 0$ only when $i=j, q=r$. Similarly,
    \begin{align*}
        \frac{d_2^2}{T_0^2K^2p_0^2}\EE\left[\bigg(\sum_{i=1}^{d_1}\sum_{q\in h_{it}} \xi_t^{iq}X_t^{iq}\bigg)^{\top}\bigg(\sum_{i=1}^{d_1}\sum_{q\in h_{it}} \xi_t^{iq}X_t^{iq}\bigg)\right]=\frac{d_1d_2\sigma^2}{T_0^2Kp_0}I_{d_2}.
    \end{align*}
    As a result,
    \begin{align*}
        &\max\left\{\bigg\|\sum_{t=1}^{T_0}  \frac{d_2^2}{T_0^2K^2p_0^2}\EE\left[\bigg(\sum_{i=1}^{d_1}\sum_{q\in h_{it}} \xi_t^{iq}X_t^{iq}\bigg)\bigg(\sum_{i=1}^{d_1}\sum_{q\in h_{it}}  \xi_t^{iq}X_t^{iq}\bigg)^{\top}\right] \bigg\|, \right.\\ & \left.\quad\quad\quad \bigg\|\sum_{t=1}^{T_0} \frac{d_2^2}{T_0^2K^2p_0^2}\EE\left[\bigg(\sum_{i=1}^{d_1} \sum_{q\in h_{it}} \xi_t^{iq}X_t^{iq}\bigg)^{\top}\bigg(\sum_{i=1}^{d_1}\sum_{q\in h_{it}} \xi_t^{iq}X_t^{iq}\bigg)\right]\bigg\|\right\} \leq \frac{d_2^2\sigma^2}{T_0Kp_0}.
    \end{align*}
    By matrix Bernstein inequality, with probability at least $1-d_1^{-10}$,
    \begin{align*}
        \|\widehat{Z}_1\|\lesssim \frac{d_2\sigma}{T\sqrt{K}p_0}\log d_1 + \frac{d_2\sigma}{\sqrt{TKp_0}}\log^{1/2}d_1\lesssim  \frac{d_2\sigma}{\sqrt{TKp_0}}\log^{1/2}d_1
    \end{align*}
    as long as $T\gtrsim p_0^{-1}\log d_1$. Similarly, we have
    \begin{align*}
        \frac{1}{T_0}\bigg\|\frac{d_2}{Kp_0}\sum_{i=1}^{d_1} \sum_{q\in h_{it}}\inp{\widehat{\Delta}_1}{X_t^{iq}}X_t^{iq} -\widehat{\Delta}_1\bigg\|\leq \frac{d_2}{T_0\sqrt{K}p_0}\|\widehat{\Delta}_1\|_{\max} + \frac{\sqrt{d_1d_2}}{T_0}\|\widehat{\Delta}_1\|_{\max}\leq  \frac{2d_2}{T_0\sqrt{K}p_0}\|\widehat{\Delta}_1\|_{\max}
    \end{align*}
    since $d_2\geq K(1+\gamma)d_1$, and 
    \begin{align*}
        &\left\|\frac{1}{T_0^2}\EE\bigg[\bigg(\sum_{i=1}^{d_1}\sum_{q\in h_{it}} \frac{d_2}{Kp_0}\inp{\widehat{\Delta}_1}{X_t^{iq}}X_t^{iq} -\widehat{\Delta}_1\bigg)\bigg(\sum_{i=1}^{d_1}\sum_{q\in h_{it}}  \frac{d_2}{Kp_0}\inp{\widehat{\Delta}_1}{X_t^{iq}}X_t^{iq} -\widehat{\Delta}_1\bigg)^{\top}\bigg]\right\| \\ 
        &\quad \leq  \frac{d_2^2}{T_0^2K^2p_0^2}\|\widehat{\Delta}_1\|_{\max}^2\left\|\EE\bigg[\bigg(\sum_{i=1}^{d_1}\sum_{q\in h_{it}} X_t^{iq}\bigg)\bigg(\sum_{i=1}^{d_1}\sum_{q\in h_{it}} X_t^{iq}\bigg)^{\top} \bigg]\right\|  + \frac{1}{T_0^2}\|\widehat{\Delta}_1\|^2 \\ 
        &\quad \leq \frac{d_2^2}{T_0^2Kp_0}\|\widehat{\Delta}_1\|_{\max}^2+ \frac{d_1d_2}{T_0^2}\|\widehat{\Delta}_1\|_{\max}^2 \lesssim \frac{d_2^2\sigma^2}{T_0^2Kp_0}
    \end{align*}
    where the last inequality comes from $\|\widehat{\Delta}_1\|_{\max}^2\lesssim \sigma^2$ under the conditions in Theorem \ref{thm:inference-main}. In the same way, by matrix Bernstein inequality, with probability at least $1-d_1^{-10}$,
    \begin{align*}
        \|\widehat{Z}_2\|\lesssim \frac{d_2\sigma}{T\sqrt{K}p_0}\log d_1 + \frac{d_2\sigma}{\sqrt{TKp_0}}\log^{1/2}d_1\lesssim  \frac{d_2\sigma}{\sqrt{TKp_0}}\log^{1/2}d_1
    \end{align*}
    as long as $T\gtrsim p_0^{-1}\log d_1$. Then we conclude the proof. 
\end{proof}

\subsection{Proof of Lemma \ref{2maxbound-otm}}
\begin{proof}
    Without loss loss of generality, we only prove $l=1$ case and omit all the subscripts and superscripts. Recall that we have defined 
    \begin{align*}
         \widehat{E}^{(l)}= \begin{pmatrix}
            0 & \widehat{Z}^{(l)} \\
            \widehat{Z}^{(l)\top} & 0
        \end{pmatrix}.
    \end{align*}
    By Lemma \ref{Zbound-otm}, we have $\|\widehat{E}\|\leq C_1\sqrt{d_2^2\log d_1/TKp_0}\sigma:= \delta_1$. Following proof of Theorem \ref{thm:inference-main} and Lemma \ref{2maxbound}, it suffices to prove that, there exist constants $C_1,C_2>0$ such that, for all $k\geq 0$, with probability at least $1-2(k+1)d_1^{-10}$,
    \begin{align*}
        &\max_{j \in [d_1]} \|e_{j}^{\top}\mathfrak{P}^{\perp}(\mathfrak{P}^{\perp}\widehat{E}\mathfrak{P}^{\perp})^k\widehat{E}\Theta\|\leq C_1(C_2\delta_1)^{k+1}\mu\sqrt{\frac{r}{d_1}}, \\
        &\max_{j \in [d_2]} \|e_{j+d_1}^{\top}\mathfrak{P}^{\perp}(\mathfrak{P}^{\perp}\widehat{E}\mathfrak{P}^{\perp})^k\widehat{E}\Theta\|\leq C_1(C_2\delta_1)^{k+1}\mu\sqrt{\frac{r}{d_2}}.
    \end{align*} 
    The technical differences from Lemma \ref{2maxbound} are mainly from the several concentrations due to the different sampling pattern. \\
    \noindent\textbf{Case 0: k=0}  \\
    For any $j\in [d_1]$, clearly,
    \begin{align*}
        \|e_j^{\top}\mathfrak{P}^{\perp}\widehat{E}\Theta\|\leq \|e_j^{\top}\Theta\Theta^{\top}\widehat{E}\Theta\| + \|e_j^{\top}\widehat{E}\Theta\|\leq \delta_1\mu\sqrt{\frac{r}{d_1}} + \|e_j^{\top}\widehat{E}\Theta\|= \delta_1\mu\sqrt{\frac{r}{d_1}} + \|e_j^{\top}\widehat{Z}V\|,
    \end{align*}
    where in the last inequality, we transfer $e_j$ to be the canonical basis vector from $\RR^{d_1+d_2}$ to $\RR^{d_1}$.  We write
    \begin{align*}
        e_{j}^{\top}\widehat{Z}V= \frac{d_2}{T_0Kp_0}\sum_{t=1}^{T_0} \sum_{i=1}^{d_1}\sum_{q\in h_{it}}\xi_t^{iq}e_j^{\top}X_t^{iq} V  + \frac{1}{T_0}\sum_{t=1}^{T_0}\sum_{i=1}^{d_1}\sum_{q\in h_{it}} \frac{d_2}{Kp_0}\inp{\widehat{\Delta}_{1}}{X_t^{iq}}e_j^{\top}X_t^iV - e_j\widehat{\Delta}_{1}V.
    \end{align*} 
    Since $e_j^{\top}X_t^{iq}\neq 0$ only when $i=j$, we have
    \begin{align*}
        \left\|\bigg\|\frac{d_2}{T_0Kp_0} \sum_{i=1}^{d_1}\sum_{q\in h_{it}} \xi_t^{iq}e_j^{\top}X_t^{iq} V \bigg\|\right\|_{\psi_2}\lesssim \frac{d_2\sigma}{T_0\sqrt{K}p_0}\|V\|_{2,\max}\leq \frac{d_2\sigma\mu}{T_0\sqrt{K}p_0}\sqrt{\frac{r}{d_2}}
    \end{align*}
    and
    \begin{align*}
        &\frac{d_2^2}{T_0^2K^2p_0^2}\EE\left[e_j^{\top}\bigg(\sum_{i=1}^{d_1}\sum_{q\in h_{it}}\xi_t^{iq}X_t^{iq} \bigg)VV^{\top}\bigg(\sum_{i=1}^{d_1}\sum_{q\in h_{it}} \xi_t^{iq}X_t^{iq} \bigg)^{\top} e_j \right] = \frac{d_2^2\sigma^2}{T_0^2K^2p_0^2}\sum_{q\in h_{jt}}\EE\left[e_j^{\top}X_t^jVV^{\top}X_t^{j\top}e_j\right] \\
        &\quad = \frac{d_2\sigma^2}{T_0^2Kp_0}\text{tr}(VV^{\top})=\frac{rd_2\sigma^2}{T_0^2Kp_0}.
    \end{align*}
    By Bernstein inequality, with probability at least $1-d_1^{-10}$,
    \begin{align*}
        \|e_j^{\top}\widehat{Z}_1V\|\lesssim \mu\sigma\sqrt{\frac{rd_2\log d_1}{TKp_0}}
    \end{align*}
    as long as $T\gtrsim p_0^{-1}\log d_1$. Similarly, 
    \begin{align*}
        \frac{1}{T_0}\bigg\|e_j^{\top}\big(\sum_{i=1}^{d_1}\sum_{q\in h_{it}} \frac{d_2}{Kp_0}\inp{\widehat{\Delta}_1}{X_t^{iq}}X_t^{iq} -\widehat{\Delta}_1\big)V\bigg\|&\leq \frac{d_2}{T_0Kp}\bigg\|\sum_{q\in h_{jt}} e_j^{\top}\inp{\widehat{\Delta}_1}{X_t^{jq}}X_t^{jq}V \bigg\| + \frac{\sqrt{d_2}}{T_0}\|\widehat{\Delta}_1\|_{\max} \\
        &\leq \frac{\mu\sqrt{rd_2}}{T_0p_0}\|\widehat{\Delta}_1\|_{\max} + \frac{\sqrt{d_2}}{T_0}\|\widehat{\Delta}_1\|_{\max} \lesssim \frac{\mu\sqrt{rd_2}}{T_0p_0}\sigma
    \end{align*}
    and 
    \begin{align*}
        &\frac{d_2^2}{T_0^2K^2p_0^2} \EE\left[e_j^{\top}\bigg(\sum_{i=1}^{d_1}\sum_{q\in h_{it}} \inp{\widehat{\Delta}_1}{X_t^{iq}}X_t^{iq}V\bigg)\bigg(\sum_{i=1}^{d_1}\sum_{q\in h_{it}} \inp{\widehat{\Delta}_1}{X_t^{iq}}X_t^{iq}V\bigg)^{\top}e_j \right] \\ 
        &\quad \leq \frac{d_2^2}{T_0^2K^2p_0^2}\|\widehat{\Delta}_1\|_{\max}^2 \EE\left[\bigg\|e_{j}^{\top}\sum_{q\in h_{jt}}X_t^{jq}V\bigg\|^2\right] \lesssim \frac{\mu rd_2\sigma^2}{T_0^2Kp_0}.
    \end{align*}
    where we again use $\|\widehat{\Delta}_1\|_{\max}\lesssim (1/\sqrt{Kp_0})\sigma$. The last inequality comes from the incoherence of $V$ and $|h_{jt}|\sim B(K,p_0)$. By Bernstein inequality, with probability at least $1-d_1^{-10}$,
    \begin{align*}
        \|e_j^{\top}\widehat{Z}_2V\|\lesssim \mu\sigma\sqrt{\frac{rd_2\log d_1}{TKp_0}}
    \end{align*}
    as long as $T\gtrsim Kp_0^{-1}\log d_1$. As before, this condition can be written as $T\gtrsim \alpha_dp_0^{-1}\log d_1$. Then this directly implies
    \begin{align*}
        \max_{j\in [d_1]}\|e_j^{\top}\mathfrak{P}^{\perp}\widehat{E}\Theta\|\lesssim \delta_1\mu\sqrt{\frac{r}{d_1}}.
    \end{align*}
    Following the same arguments, we can show 
    \begin{align*}
        \max_{j\in [d_2]}\|e_{j+d_1}^{\top}\mathfrak{P}^{\perp}\widehat{E}\Theta\|\lesssim \delta_1\mu\sqrt{\frac{r}{d_2}}.
    \end{align*}
    \noindent\textbf{Case 1: k=1}  \\
    Observe that for any $j\in [d_1]$, 
    \begin{align*}
        \|e_j^{\top}\mathfrak{P}^{\perp}\widehat{E}\mathfrak{P}^{\perp}\widehat{E}\Theta\|\leq \|e_j^{\top}\Theta\Theta^{\top} \widehat{E}\mathfrak{P}^{\perp}\widehat{E}\Theta\| + \|e_j^{\top}\widehat{E}\mathfrak{P}^{\perp}\widehat{E}\Theta\|\leq \delta_1^2\mu\sqrt{\frac{r}{d_1}} + \|e_j^{\top}\widehat{E}\mathfrak{P}^{\perp}\widehat{E}\Theta\|.
    \end{align*}
    By definition of $\widehat{E}$ and $\mathfrak{P}^{\perp}$, we have 
    \begin{align*}
        \widehat{E}\mathfrak{P}^{\perp}\widehat{E}\Theta = \begin{pmatrix}
            \widehat{Z}V_{\perp}V_{\perp}^{\top}\widehat{Z}^{\top}U & 0 \\
            0 & \widehat{Z}^{\top}U_{\perp}U_{\perp}^{\top}\widehat{Z}V 
        \end{pmatrix}.
    \end{align*}
    It suffices to prove the upper bound for $\|e_j^{\top}\widehat{Z}V_{\perp}V_{\perp}^{\top}\widehat{Z}^{\top}U\|$. Define $\mathfrak{J}_j=e_je_j^{\top}\in \RR^{d_1\times d_1}$ and $\mathfrak{J}^{\perp}=I-\mathfrak{J}$. Then write $\widehat{Z}=\mathfrak{J}_j\widehat{Z} + \mathfrak{J}_j^{\perp}\widehat{Z}$ and as a result,
    \begin{align*}
        \|e_j^{\top}\widehat{Z}V_{\perp}V_{\perp}^{\top}\widehat{Z}^{\top}U\|&\leq \delta_1^2\mu\sqrt{\frac{r}{d_1}} + \|e_j^{\top}\widehat{Z}_1V_{\perp}V_{\perp}^{\top}(\mathfrak{J}_j^{\perp}\widehat{Z})^{\top}U\| + \|e_j^{\top}\widehat{Z}_2V_{\perp}V_{\perp}^{\top}(\mathfrak{J}_j^{\perp}\widehat{Z})^{\top}U\|.
    \end{align*}
    We first prove the upper bound of $\|e_j^{\top}\widehat{Z}_1V_{\perp}V_{\perp}^{\top}(\mathfrak{J}_j^{\perp}\widehat{Z})^{\top}U\|$. Since $\mathfrak{J}_j^{\perp}\widehat{Z}$ does not include $X_t^{jq}$ for any $t$, condition on $\mathfrak{J}_j^{\perp}\widehat{Z}$, we have
    \begin{align*}
        \left\|\bigg\|\frac{d_2}{T_0Kp_0} \sum_{i=1}^{d_1}\sum_{q\in h_{it}} \xi_t^{iq} e_j^{\top}X_t^{iq}V_{\perp}V_{\perp}^{\top}(\mathfrak{J}_j^{\perp}\widehat{Z})^{\top}U  \bigg\|\right\|_{\psi_2} \lesssim \frac{d_2\sigma}{T_0p_0}\|V_{\perp}V_{\perp}^{\top}(\mathfrak{J}_j^{\perp}\widehat{Z})^{\top}U\|_{2,\max}.  
    \end{align*}
    since $e_j^{\top}X_t^{iq}\neq 0$ only when $i=j$. Under the result of \textbf{Case 0}, we have
    \begin{align*}
        \|V_{\perp}V_{\perp}^{\top}(\mathfrak{J}_j^{\perp}\widehat{Z})^{\top}U\|_{2,\max}&\leq \|V_{\perp}V_{\perp}^{\top}(\mathfrak{J}_j\widehat{Z})^{\top}U\|_{2,\max} + \|V_{\perp}V_{\perp}^{\top}\widehat{Z}^{\top}U\|_{2,\max} \\
        &\lesssim \delta_1\mu\sqrt{\frac{r}{d_1}} + \delta_1\mu\sqrt{\frac{r}{d_2}}\lesssim \delta_1\mu\sqrt{\frac{r}{d_1}}.
    \end{align*}
    We also calculate the variance, 
    \begin{align*}
        &\frac{d_2^2}{T_0^2K^2p_0^2}\EE\left[ \big(\sum_{i=1}^{d_1}\sum_{q\in h_{it}}\xi_t^{iq}e_j^{\top}X_t^{iq}V_{\perp}V_{\perp}^{\top}(\mathfrak{J}_j^{\perp}\widehat{Z})^{\top}U\big)\big(\sum_{i=1}^{d_1}\sum_{q\in h_{it}}\xi_t^{iq}e_j^{\top}X_t^{iq}V_{\perp}V_{\perp}^{\top}(\mathfrak{J}_j^{\perp}\widehat{Z})^{\top}U\big)^{\top} \bigg| \mathfrak{J}_j^{\perp}\widehat{Z} \right] \\
        &\leq \frac{d_2^2\sigma^2}{T_0^2K^2p_0^2}\EE\left[\sum_{q\in h_{jt}} \|e_j^{\top}X_t^{jq}V_{\perp}V_{\perp}^{\top}(\mathfrak{J}_j^{\perp}\widehat{Z})^{\top}U \|^2 \bigg| \mathfrak{J}_j^{\perp}\widehat{Z} \right] \\ 
        &\leq \frac{d_2^2\sigma^2}{T_0^2K^2p_0^2}\sum_{s=1}^{d_2} \PP\left(j_t(j)=s\bigg| \mathfrak{J}_j^{\perp}\widehat{Z}\right)\|V_{\perp}V_{\perp}^{\top}(\mathfrak{J}_j^{\perp}\widehat{Z})^{\top}U\|_{2,\max}  \\
        &\leq \frac{d_2^2\sigma^2}{T_0^2K^2p_0^2}\|V_{\perp}V_{\perp}^{\top}(\mathfrak{J}_j^{\perp}\widehat{Z})^{\top}U\|_{2,\max}\lesssim \frac{d_2^2\mu^2r\sigma^2}{d_1T^2K^2p_0^2}\delta_1^2.
    \end{align*}
    Then by Bernstein inequality, with probability at least $1-d_1^{-10}$,
    \begin{align*}
        \|e_j^{\top}\widehat{Z}_1V_{\perp}V_{\perp}^{\top}(\mathfrak{J}_j^{\perp}\widehat{Z})^{\top}U\|\lesssim \frac{d_2\sigma}{Tp_0}\delta_1\mu\sqrt{\frac{r}{d_1}}\log d_1 + \mu\sqrt{\frac{rd_2^2\sigma^2\log d_1}{d_1TK^2p_0^2}}\delta_1\lesssim \delta_1^2\mu\sqrt{\frac{r}{d_1}}
    \end{align*}
    as long as $Kp_0=\Omega(1)$. Similarly, by the fact that $\|\widehat{\Delta}_1\|_{\max}\lesssim (1/\sqrt{Kp})\sigma$, $|h_{jt}|\sim B(K,p_0)$ and the upper bound of $ \|V_{\perp}V_{\perp}^{\top}(\mathfrak{J}_j^{\perp}\widehat{Z})^{\top}U\|_{2,\max}$ derived above, we have with probability at least $1-d_1^{-10}$,
    \begin{align*}
        \|e_j^{\top}\widehat{Z}_2V_{\perp}V_{\perp}^{\top}(\mathfrak{J}_j^{\perp}\widehat{Z})^{\top}U\|\lesssim \delta_1^2\mu\sqrt{\frac{r}{d_1}}.
    \end{align*}
    The upper bound for $\max_{j\in [d_2]}\|e_{j+d_1}^{\top} \mathfrak{P}^{\perp}\widehat{E}\mathfrak{P}^{\perp}\widehat{E}\Theta\|$ can be proved in the same way. Taking a union bound, with probability at least $1-4d_1^{-10}$,
    \begin{align*}
        \max_{j\in [d_1]}\|e_{j}^{\top} \mathfrak{P}^{\perp}\widehat{E}\mathfrak{P}^{\perp}\widehat{E}\Theta\|\lesssim \delta_1^2\mu\sqrt{\frac{r}{d_1}} \quad\text{and}\quad \max_{j\in [d_2]}\|e_{j+d_1}^{\top} \mathfrak{P}^{\perp}\widehat{E}\mathfrak{P}^{\perp}\widehat{E}\Theta\|\lesssim \delta_1^2\mu\sqrt{\frac{r}{d_2}}.
    \end{align*}
    Finally, the general case $k\geq 2$ can be proved by induction, following Lemma 9 in \cite{xia2021statistical}. We omit the details here. 
\end{proof}

\subsection{Proof of Lemma \ref{Zbound-tside}}
\begin{proof}
    Without loss of generality, we only prove the upper bound for $\|\widehat{Z}^{(1)}\|$ and omit the subscript and superscript $l$ for ease of notation. We analyze $\|\widehat{Z}_1\|$ and $\|\widehat{Z}_2\|$ separately. Since $\xi_t^i$ is subGaussian, for any $t$,
    \begin{align*}
        \bigg\|\big\|\frac{1}{T_0\nu}\sum_{(i,q)\in g_t} \xi_t^{iq} X_t^{iq}\big\|\bigg\|_{\psi_2}\lesssim \frac{\sigma}{T_0\nu}\bigg\|\sum_{(i,q)\in g_t} X_t^{iq} \bigg\|\lesssim \frac{\sigma}{T_0\nu}.
    \end{align*} 
    We also calculate the variance
    \begin{align*}
        \frac{1}{T_0^2\nu^2}\EE\left[\bigg(\sum_{(i,q)\in g_t} \xi_t^{iq} X_t^{iq}\bigg)\bigg(\sum_{(i,q)\in g_t} \xi_t^{iq} X_t^{iq}\bigg)^{\top}\right]=\frac{1}{T_0^2\nu^2}\EE\left[\sum_{(i,q)\in g_t} \xi_t^{iq2}X_t^{iq}X_t^{iq\top} \right]=\frac{d_2\sigma^2}{T_0^2\nu}I_{d_1}.
    \end{align*}
    since $X_t^iX_t^{j\top}=0$ for all $i\neq j$. Similarly,
    \begin{align*}
        \frac{1}{T_0^2\nu^2}\EE\left[\bigg(\sum_{i=1}^{d_1}\xi_t^iX_t^i\bigg)^{\top}\bigg(\sum_{i=1}^{d_1}\xi_t^iX_t^i\bigg)\right]=\frac{d_1\sigma^2}{T_0^2\nu}I_{d_2}.
    \end{align*}
    As a result,
    \begin{align*}
        &\max\left\{\bigg\|\sum_{t=1}^{T_0}  \frac{1}{T_0^2\nu^2}\EE\left[\bigg(\sum_{(i,q)\in g_t} \xi_t^{iq} X_t^{iq}\bigg)\bigg(\sum_{(i,q)\in g_t} \xi_t^{iq} X_t^{iq}\bigg)^{\top}\right] \bigg\|, \right.\\ & \left.\quad\quad\quad \bigg\|\sum_{t=1}^{T_0} \frac{1}{T_0^2\nu^2}\EE\left[\bigg(\sum_{(i,q)\in g_t} \xi_t^{iq} X_t^{iq}\bigg)^{\top}\bigg(\sum_{(i,q)\in g_t} \xi_t^{iq} X_t^{iq}\bigg)\right]\bigg\|\right\} \leq \frac{d_2\sigma^2}{T_0\nu}.
    \end{align*}
    By matrix Bernstein inequality, with probability at least $1-d_1^{-10}$,
    \begin{align*}
        \|\widehat{Z}_1\|\lesssim \frac{\sigma}{T\nu}\log d_1 + \sqrt{\frac{d_2\sigma^2}{T\nu}\log d_1}\lesssim  \sqrt{\frac{d_2\sigma}{T\nu}\log d_1}
    \end{align*}
    as long as $T\gtrsim (1/d_2\nu)\log d_1$. Similarly, we have
    \begin{align*}
        \frac{1}{T_0}\bigg\|\sum_{(i,q)\in g_t} \frac{1}{\nu}\inp{\widehat{\Delta}_1}{X_t^{iq}}X_t^{iq} -\widehat{\Delta}_1\bigg\|\leq \frac{1}{T_0\nu}\|\widehat{\Delta}_1\|_{\max} + \frac{\sqrt{d_1d_2}}{T_0}\|\widehat{\Delta}_1\|_{\max}\leq  \frac{2}{T_0\nu}\|\widehat{\Delta}_1\|_{\max}
    \end{align*}
    and 
    \begin{align*}
        &\left\|\frac{1}{T_0^2}\EE\bigg[\bigg(\sum_{(i,q)\in g_t} \frac{1}{\nu}\inp{\widehat{\Delta}_1}{X_t^{iq}}X_t^{iq} -\widehat{\Delta}_1\bigg)\bigg(\sum_{(i,q)\in g_t} \frac{1}{\nu}\inp{\widehat{\Delta}_1}{X_t^{iq}}X_t^{iq} -\widehat{\Delta}_1\bigg)^{\top}\bigg]\right\| \\ 
        &\quad \leq  \frac{1}{T_0^2\nu^2}\|\widehat{\Delta}_1\|_{\max}^2\left\|\EE\bigg[\bigg(\sum_{(i,q)\in g_t} X_t^{iq}\bigg)\bigg(\sum_{(i,q)\in g_t} X_t^{iq}\bigg)^{\top} \bigg]\right\|  + \frac{1}{T_0^2}\|\widehat{\Delta}_1\|^2 \\ 
        &\quad \leq \frac{d_2}{T_0^2\nu}\|\widehat{\Delta}_1\|_{\max}^2+ \frac{d_1d_2}{T_0^2}\|\widehat{\Delta}_1\|_{\max}^2 \lesssim \frac{d_2\sigma^2}{T_0^2\nu}
    \end{align*}
    where the last inequality comes from $\|\widehat{\Delta}_1\|_{\max}^2\lesssim \sigma^2$ under the conditions in Proposition \ref{thm:estimation3}. In the same way, by matrix Bernstein inequality, with probability at least $1-d_1^{-10}$,
    \begin{align*}
        \|\widehat{Z}_2\|\lesssim \frac{\sigma}{T\nu}\log d_1 + \sqrt{\frac{d_2\sigma^2}{T\nu}\log d_1}\lesssim  \sqrt{\frac{d_2\sigma^2}{T\nu}\log d_1}
    \end{align*}
    as long as $T\gtrsim (1/d_2\nu)\log d_1$. Then we conclude the proof. 
\end{proof}

\subsection{Proof of Lemma \ref{2maxbound-tside}}
\begin{proof}
    Without loss loss of generality, we only prove $l=1$ case and omit all the subscripts and superscripts. Recall that we have defined 
    \begin{align*}
         \widehat{E}^{(l)}= \begin{pmatrix}
            0 & \widehat{Z}^{(l)} \\
            \widehat{Z}^{(l)\top} & 0
        \end{pmatrix}.
    \end{align*}
    By Lemma \ref{Zbound-tside}, we have $\|\widehat{E}\|\leq C_1\sqrt{d_2\log d_1/T\nu}\sigma:= \delta_2$. Following proof of Theorem \ref{thm:inference-main} and Lemma \ref{2maxbound}, it suffices to prove that, there exist constants $C_1,C_2>0$ such that, for all $k\geq 0$, with probability at least $1-2(k+1)d_1^{-10}$,
    \begin{align*}
        &\max_{j \in [d_1]} \|e_{j}^{\top}\mathfrak{P}^{\perp}(\mathfrak{P}^{\perp}\widehat{E}\mathfrak{P}^{\perp})^k\widehat{E}\Theta\|\leq C_1(C_2\delta_2)^{k+1}\mu\sqrt{\frac{r}{d_1}}, \\
        &\max_{j \in [d_2]} \|e_{j+d_1}^{\top}\mathfrak{P}^{\perp}(\mathfrak{P}^{\perp}\widehat{E}\mathfrak{P}^{\perp})^k\widehat{E}\Theta\|\leq C_1(C_2\delta_2)^{k+1}\mu\sqrt{\frac{r}{d_2}}.
    \end{align*} 
    The technical differences from Lemma \ref{2maxbound} are mainly from the several concentrations due to the different sampling pattern. \\
    \noindent\textbf{Case 0: k=0}  \\
    For any $j\in [d_1]$, clearly,
    \begin{align*}
        \|e_j^{\top}\mathfrak{P}^{\perp}\widehat{E}\Theta\|\leq \|e_j^{\top}\Theta\Theta^{\top}\widehat{E}\Theta\| + \|e_j^{\top}\widehat{E}\Theta\|\leq \delta_2\mu\sqrt{\frac{r}{d_1}} + \|e_j^{\top}\widehat{E}\Theta\|= \delta_2\mu\sqrt{\frac{r}{d_1}} + \|e_j^{\top}\widehat{Z}V\|,
    \end{align*}
    where in the last inequality, we transfer $e_j$ to be the canonical basis vector from $\RR^{d_1+d_2}$ to $\RR^{d_1}$.  We write
    \begin{align*}
        e_{j}^{\top}\widehat{Z}V= \frac{1}{T_0\nu}\sum_{t=1}^{T_0} \sum_{(i,q)\in h_{t}}\xi_t^{iq}e_j^{\top}X_t^{iq} V  + \frac{1}{T_0}\sum_{t=1}^{T_0} \sum_{(i,q)\in h_{t}} \frac{1}{\nu}\inp{\widehat{\Delta}_{1}}{X_t^{iq}}e_j^{\top}X_t^iV - e_j\widehat{\Delta}_{1}V.
    \end{align*} 
    Since $e_j^{\top}X_t^{iq}\neq 0$ only when $i=j$, we have
    \begin{align*}
        \left\|\bigg\|\frac{1}{T_0\nu} \sum_{(i,q)\in h_{t}} \xi_t^{iq} e_j^{\top}X_t^{iq} V \bigg\|\right\|_{\psi_2}\lesssim \frac{\sigma}{T_0\nu}\|V\|_{2,\max}\leq \frac{\sigma\mu}{T_0\nu}\sqrt{\frac{r}{d_2}}
    \end{align*}
    and
    \begin{align*}
        &\frac{1}{T_0^2\nu^2}\EE\left[e_j^{\top}\bigg(\sum_{(i,q)\in h_{t}}\xi_t^{iq}X_t^{iq} \bigg)VV^{\top}\bigg(\sum_{(i,q)\in h_{t}} \xi_t^{iq}X_t^{iq} \bigg)^{\top} e_j \right] = \frac{\sigma^2}{T_0^2\nu^2} \EE\left[\sum_{q=1}^{d_2} e_j^{\top}X_t^{jq}VV^{\top}X_t^{jq\top}e_j\right] \\
        &\quad = \frac{\sigma^2}{T_0^2\nu}\text{tr}(VV^{\top})=\frac{r\sigma^2}{T_0^2\nu}.
    \end{align*}
    By Bernstein inequality, with probability at least $1-d_1^{-10}$,
    \begin{align*}
        \|e_j^{\top}\widehat{Z}_1V\|\lesssim \sigma\sqrt{\frac{r\log d_1}{T\nu}}
    \end{align*}
    as long as $T\gtrsim (1/d_2\nu)\log d_1$. Similarly, 
    \begin{align*}
        \frac{1}{T_0}\bigg\|e_j^{\top}\big(\sum_{(i,q)\in h_{t}} \frac{1}{\nu}\inp{\widehat{\Delta}_1}{X_t^{iq}}X_t^{iq} -\widehat{\Delta}_1\big)V\bigg\|&\leq \frac{1}{T_0\nu}\bigg\|\sum_{(i,q)\in h_{t}} e_j^{\top}\inp{\widehat{\Delta}_1}{X_t^{jq}}X_t^{jq}V \bigg\| + \frac{\sqrt{d_2}}{T_0}\|\widehat{\Delta}_1\|_{\max} \\
        &\leq \frac{\mu\sqrt{r}}{T_0\nu\sqrt{d_2}}\|\widehat{\Delta}_1\|_{\max} + \frac{\sqrt{d_2}}{T_0}\|\widehat{\Delta}_1\|_{\max} \lesssim \frac{\mu\sqrt{rd_2}}{T_0\nu}\sigma
    \end{align*}
    and 
    \begin{align*}
        &\frac{1}{T_0^2\nu^2} \EE\left[e_j^{\top}\bigg(\sum_{(i,q)\in h_{t}} \inp{\widehat{\Delta}_1}{X_t^{iq}}X_t^{iq}V\bigg)\bigg(\sum_{(i,q)\in h_{t}} \inp{\widehat{\Delta}_1}{X_t^{iq}}X_t^{iq}V\bigg)^{\top}e_j \right] \\ 
        &\quad \leq \frac{1}{T_0^2\nu}\|\widehat{\Delta}_1\|_{\max}^2\text{tr}(VV^{\top})  \lesssim \frac{r\sigma^2}{T_0^2\nu}.
    \end{align*}
   By Bernstein inequality, with probability at least $1-d_1^{-10}$,
    \begin{align*}
        \|e_j^{\top}\widehat{Z}_2V\|\lesssim \sigma\sqrt{\frac{r\log d_1}{T\sigma}}
    \end{align*}
    as long as $T\gtrsim (1/d_2\nu)\log d_1$. Then this directly implies
    \begin{align*}
        \max_{j\in [d_1]}\|e_j^{\top}\mathfrak{P}^{\perp}\widehat{E}\Theta\|\lesssim \delta_2\mu\sqrt{\frac{r}{d_1}}.
    \end{align*}
    Following the same arguments, we can show 
    \begin{align*}
        \max_{j\in [d_2]}\|e_{j+d_1}^{\top}\mathfrak{P}^{\perp}\widehat{E}\Theta\|\lesssim \delta_2\mu\sqrt{\frac{r}{d_2}}.
    \end{align*}
    \noindent\textbf{Case 1: k=1}  \\
    Observe that for any $j\in [d_1]$, 
    \begin{align*}
        \|e_j^{\top}\mathfrak{P}^{\perp}\widehat{E}\mathfrak{P}^{\perp}\widehat{E}\Theta\|\leq \|e_j^{\top}\Theta\Theta^{\top} \widehat{E}\mathfrak{P}^{\perp}\widehat{E}\Theta\| + \|e_j^{\top}\widehat{E}\mathfrak{P}^{\perp}\widehat{E}\Theta\|\leq \delta_2^2\mu\sqrt{\frac{r}{d_1}} + \|e_j^{\top}\widehat{E}\mathfrak{P}^{\perp}\widehat{E}\Theta\|.
    \end{align*}
    By definition of $\widehat{E}$ and $\mathfrak{P}^{\perp}$, we have 
    \begin{align*}
        \widehat{E}\mathfrak{P}^{\perp}\widehat{E}\Theta = \begin{pmatrix}
            \widehat{Z}V_{\perp}V_{\perp}^{\top}\widehat{Z}^{\top}U & 0 \\
            0 & \widehat{Z}^{\top}U_{\perp}U_{\perp}^{\top}\widehat{Z}V 
        \end{pmatrix}.
    \end{align*}
    It suffices to prove the upper bound for $\|e_j^{\top}\widehat{Z}V_{\perp}V_{\perp}^{\top}\widehat{Z}^{\top}U\|$. Define $\mathfrak{J}_j=e_je_j^{\top}\in \RR^{d_1\times d_1}$ and $\mathfrak{J}^{\perp}=I-\mathfrak{J}$. Then write $\widehat{Z}=\mathfrak{J}_j\widehat{Z} + \mathfrak{J}_j^{\perp}\widehat{Z}$ and as a result,
    \begin{align*}
        \|e_j^{\top}\widehat{Z}V_{\perp}V_{\perp}^{\top}\widehat{Z}^{\top}U\|&\leq \delta_2^2\mu\sqrt{\frac{r}{d_1}} + \|e_j^{\top}\widehat{Z}_1V_{\perp}V_{\perp}^{\top}(\mathfrak{J}_j^{\perp}\widehat{Z})^{\top}U\| + \|e_j^{\top}\widehat{Z}_2V_{\perp}V_{\perp}^{\top}(\mathfrak{J}_j^{\perp}\widehat{Z})^{\top}U\|.
    \end{align*}
    We first prove the upper bound of $\|e_j^{\top}\widehat{Z}_1V_{\perp}V_{\perp}^{\top}(\mathfrak{J}_j^{\perp}\widehat{Z})^{\top}U\|$. Since $\mathfrak{J}_j^{\perp}\widehat{Z}$ does not include $X_t^{jq}$ for any $t$, condition on $\mathfrak{J}_j^{\perp}\widehat{Z}$, we have
    \begin{align*}
        \left\|\bigg\|\frac{1}{T_0\nu} \sum_{(i,q)\in h_{t}} \xi_t^{iq} e_j^{\top}X_t^{iq}V_{\perp}V_{\perp}^{\top}(\mathfrak{J}_j^{\perp}\widehat{Z})^{\top}U  \bigg\|\right\|_{\psi_2} \lesssim \frac{\sigma}{T_0\nu}\|V_{\perp}V_{\perp}^{\top}(\mathfrak{J}_j^{\perp}\widehat{Z})^{\top}U\|_{2,\max}
    \end{align*}
    since $e_j^{\top}X_t^{iq}\neq 0$ only when $i=j$. Under the result of \textbf{Case 0}, we have
    \begin{align*}
        \|V_{\perp}V_{\perp}^{\top}(\mathfrak{J}_j^{\perp}\widehat{Z})^{\top}U\|_{2,\max}&\leq \|V_{\perp}V_{\perp}^{\top}(\mathfrak{J}_j\widehat{Z})^{\top}U\|_{2,\max} + \|V_{\perp}V_{\perp}^{\top}\widehat{Z}^{\top}U\|_{2,\max} \\
        &\lesssim \delta_2\mu\sqrt{\frac{r}{d_1}} + \delta_2\mu\sqrt{\frac{r}{d_2}}\lesssim \delta_2\mu\sqrt{\frac{r}{d_1}}.
    \end{align*}
    We also calculate the variance, 
     \begin{align*}
        &\frac{1}{T_0^2\nu^2}\EE\left[ \big(\sum_{(i,q)\in h_{t}}\xi_t^{iq}e_j^{\top}X_t^{iq}V_{\perp}V_{\perp}^{\top}(\mathfrak{J}_j^{\perp}\widehat{Z})^{\top}U\big)\big(\sum_{(i,q)\in h_{t}}\xi_t^{iq}e_j^{\top}X_t^{iq}V_{\perp}V_{\perp}^{\top}(\mathfrak{J}_j^{\perp}\widehat{Z})^{\top}U\big)^{\top} \bigg| \mathfrak{J}_j^{\perp}\widehat{Z} \right] \\
        &\leq \frac{\sigma^2}{T_0^2\nu^2}\EE\left[\sum_{(i,q)\in h_{t}, i=j} \|e_j^{\top}X_t^{jq}V_{\perp}V_{\perp}^{\top}(\mathfrak{J}_j^{\perp}\widehat{Z})^{\top}U \|^2 \bigg| \mathfrak{J}_j^{\perp}\widehat{Z} \right]\leq \frac{\sigma^2}{T_0^2\nu^2}\sum_{s=1}^{d_2} \PP\left(j_t(j)=s\bigg| \mathfrak{J}_j^{\perp}\widehat{Z}\right)\|V_{\perp}V_{\perp}^{\top}(\mathfrak{J}_j^{\perp}\widehat{Z})^{\top}U\|_{2,\max}  \\
        &\leq \frac{\sigma^2}{T_0^2\nu^2}\|V_{\perp}V_{\perp}^{\top}(\mathfrak{J}_j^{\perp}\widehat{Z})^{\top}U\|_{2,\max}\lesssim \frac{\mu^2r\sigma^2}{d_1T^2\nu^2}\delta_2^2,
    \end{align*}
    Then by Bernstein inequality, with probability at least $1-d_1^{-10}$,
    \begin{align*}
        \|e_j^{\top}\widehat{Z}_1V_{\perp}V_{\perp}^{\top}(\mathfrak{J}_j^{\perp}\widehat{Z})^{\top}U\|\lesssim \frac{\sigma}{T\nu}\delta_2\mu\sqrt{\frac{r}{d_1}}\log d_1 + \mu\sqrt{\frac{r\sigma^2\log d_1}{Td_1\nu^2}}\delta_2\lesssim \delta_2^2\mu\sqrt{\frac{r}{d_1}}
    \end{align*}
    since $\nu\asymp d_2^{-1}$ if $d_2\geq d_1$. Similarly, with the same probability,
    \begin{align*}
        \|e_j^{\top}\widehat{Z}_2V_{\perp}V_{\perp}^{\top}(\mathfrak{J}_j^{\perp}\widehat{Z})^{\top}U\|\lesssim \delta_2^2\mu\sqrt{\frac{r}{d_1}}.
    \end{align*}
    The upper bound for $\max_{j\in [d_2]}\|e_{j+d_1}^{\top} \mathfrak{P}^{\perp}\widehat{E}\mathfrak{P}^{\perp}\widehat{E}\Theta\|$ can be proved in the same way. Taking a union bound, with probability at least $1-4d_1^{-10}$,
    \begin{align*}
        \max_{j\in [d_1]}\|e_{j}^{\top} \mathfrak{P}^{\perp}\widehat{E}\mathfrak{P}^{\perp}\widehat{E}\Theta\|\lesssim \delta_2^2\mu\sqrt{\frac{r}{d_1}} \quad\text{and}\quad \max_{j\in [d_2]}\|e_{j+d_1}^{\top} \mathfrak{P}^{\perp}\widehat{E}\mathfrak{P}^{\perp}\widehat{E}\Theta\|\lesssim \delta_2^2\mu\sqrt{\frac{r}{d_2}}.
    \end{align*}
    Finally, the general case $k\geq 2$ can be proved by induction, following Lemma 9 in \cite{xia2021statistical}. We omit the details here. 
\end{proof}

\subsection{Proof of Lemma \ref{Zbound-non}}
\begin{proof}
    Without loss of generality, we only prove the upper bound for $\|\widehat{Z}^{(1)}\|$ and omit the subscript and superscript $l$ for ease of notation. We analysis $\|\widehat{Z}_1\|$ and $\|\widehat{Z}_2\|$ separately. Since $\xi_t^i$ is subGaussian, for any $t$,
    \begin{align*}
        \bigg\|\big\|\frac{1}{T_0}\sum_{i=1}^{d_1}\frac{1}{p_{ij_t(i)}}\xi_t^iX_t^i\big\|\bigg\|_{\psi_2}\lesssim \frac{\sigma}{p_{\min}T_0}\bigg\|\sum_{i=1}^{d_1}X_t^i\bigg\|= \frac{\sigma}{p_{\min}T_0}.
    \end{align*} 
    We also calculate the variance
    \begin{align*}
        \frac{1}{T_0^2}\EE\left[\bigg(\sum_{i=1}^{d_1}\frac{1}{p_{ij_t(i)}}\xi_t^iX_t^i\bigg)\bigg(\sum_{i=1}^{d_1}\frac{1}{p_{ij_t(i)}}\xi_t^iX_t^i\bigg)^{\top}\right]=\frac{1}{T_0^2}\EE\left[\sum_{i=1}^{d_1}\xi_t^{i2}\frac{1}{p_{ij_t(i)}^2}X_t^iX_t^{i\top} \right]\leq \frac{d_2\sigma^2}{p_{\min}T_0^2}I_{d_1}.
    \end{align*}
    since $X_t^iX_t^{j\top}=0$ for all $i\neq j$. Similarly,
    \begin{align*}
        \frac{1}{T_0^2}\EE\left[\bigg(\sum_{i=1}^{d_1}\frac{1}{p_{ij_t(i)}}\xi_t^iX_t^i\bigg)^{\top}\bigg(\sum_{i=1}^{d_1}\frac{1}{p_{ij_t(i)}}\xi_t^iX_t^i\bigg)\right]\leq \frac{d_1\sigma^2}{p_{\min}T_0^2}I_{d_2}.
    \end{align*}
    As a result, 
    \begin{align*}
        &\max\left\{\bigg\|\sum_{t=1}^{T_0}  \frac{d_2^2}{T_0^2}\EE\left[\bigg(\sum_{i=1}^{d_1}\xi_t^iX_t^i\bigg)\bigg(\sum_{i=1}^{d_1}\xi_t^iX_t^i\bigg)^{\top}\right] \bigg\|, \right.\\ & \left.\quad\quad\quad \bigg\|\sum_{t=1}^{T_0} \frac{d_2^2}{T_0^2}\EE\left[\bigg(\sum_{i=1}^{d_1}\xi_t^iX_t^i\bigg)^{\top}\bigg(\sum_{i=1}^{d_1}\xi_t^iX_t^i\bigg)\right]\bigg\|\right\} \leq \frac{d_2\sigma^2}{p_{\min}T_0}.
    \end{align*}
    By matrix Bernstein inequality, with probability at least $1-d_1^{-10}$,
    \begin{align*}
        \|\widehat{Z}_1\|\lesssim \frac{\sigma}{p_{\min}T}\log d_1 + \frac{\sqrt{d_2}\sigma}{\sqrt{p_{\min}T}}\log^{1/2}d_1\lesssim  \frac{d_2\sigma}{\sqrt{p_{\min}T}}\log^{1/2}d_1
    \end{align*}
    as long as $(p_{\min})^{-1}\log d_1\lesssim Td_2$. Similarly, we have
    \begin{align*}
        \frac{1}{T_0}\bigg\|\sum_{i=1}^{d_1} \frac{1}{p_{ij_t(i)}}\inp{\widehat{\Delta}_1}{X_t^i}X_t^i -\widehat{\Delta}_1\bigg\|\leq \frac{1}{p_{\min}T_0}\|\widehat{\Delta}_1\|_{\max} + \frac{\sqrt{d_1d_2}}{T_0}\|\widehat{\Delta}_1\|_{\max}\lesssim  \frac{1}{p_{\min}T_0}\|\widehat{\Delta}_1\|_{\max}
    \end{align*}
    since $p_{\min}^{-1}\leq (d_2)^{-1}$, and 
    \begin{align*}
        &\left\|\frac{1}{T_0^2}\EE\bigg[\bigg(\sum_{i=1}^{d_1}  \frac{1}{p_{ij_t(i)}}\inp{\widehat{\Delta}_1}{X_t^i}X_t^i -\widehat{\Delta}_1\bigg)\bigg(\sum_{i=1}^{d_1}  \frac{1}{p_{ij_t(i)}}\inp{\widehat{\Delta}_1}{X_t^i}X_t^i -\widehat{\Delta}_1\bigg)^{\top}\bigg]\right\| \\ 
        &\quad \leq  \frac{1}{T_0^2}\|\widehat{\Delta}_1\|_{\max}^2\left\|\EE\bigg[\bigg(\sum_{i=1}^{d_1}\frac{1}{p_{ij_t(i)}} X_t^i\bigg)\bigg(\sum_{i=1}^{d_1} \frac{1}{p_{ij_t(i)}}X_t^i\bigg)^{\top} \bigg]\right\|\leq \frac{d_2}{p_{\min}T_0^2}\|\widehat{\Delta}_1\|_{\max}^2\lesssim \frac{d_2\sigma^2}{p_{\min}T_0^2}. 
    \end{align*}
    where the last inequality comes from $\|\widehat{\Delta}_1\|_{\max}^2\lesssim \sigma^2$ under the conditions in Theorem \ref{thm:inference-main}. In the same way, by matrix Bernstein inequality, with probability at least $1-d_1^{-10}$,
    \begin{align*}
        \|\widehat{Z}_2\|\lesssim \frac{\sigma}{p_{\min}T}\log d_1 + \frac{\sqrt{d_2}\sigma}{\sqrt{p_{\min}T}}\log^{1/2}d_1\lesssim  \frac{\sqrt{d_2}\sigma}{\sqrt{p_{\min}T}}\log^{1/2}d_1,
    \end{align*}
    as long as $(p_{\min})^{-1}\log d_1\lesssim Td_2$. Finally, combine with the conditions that $p_{\min}^{-1}\lesssim (1/d_1d_2)\sum_i\sum_j p_{ij}^{-1}$, we can conclude that 
    \begin{align*}
        \|\widehat{Z}\|\lesssim \sqrt{\frac{\sigma^2\log d_1}{d_1T}\left(\sum_{i=1}^{d_1}\sum_{j=1}^{d_2} \frac{1}{p_{ij}}\right)}. 
    \end{align*}
\end{proof}

\subsection{Proof of Lemma \ref{2maxbound-non}}
\begin{proof}
    Without loss loss of generality, we only prove $l=1$ case and omit all the subscripts and superscripts. We first denote the upper bound of $\|\widehat{E}\|$ defined in the proof of Theorem \ref{thm:inference-main-non} to be $\|\widehat{E}\|\leq C_1\sqrt{ (\sum_i\sum_j p_{ij}^{-1})\log d_1/d_1T}\sigma:= \delta_3$ by result of Lemma \ref{Zbound-non}. According to Theorem 4 in \cite{xia2021statistical}, it suffices to prove that, there exist constants $C_1,C_2>0$ such that, for all $k\geq 0$, with probability at least $1-2(k+1)d_1^{-10}$,
    \begin{align*}
        &\max_{j \in [d_1]} \|e_{j}^{\top}\mathfrak{P}^{\perp}(\mathfrak{P}^{\perp}\widehat{E}\mathfrak{P}^{\perp})^k\widehat{E}\Theta\|\leq C_1(C_2\delta_3)^{k+1}\mu\sqrt{\frac{r}{d_1}}, \\
        &\max_{j \in [d_2]} \|e_{j+d_1}^{\top}\mathfrak{P}^{\perp}(\mathfrak{P}^{\perp}\widehat{E}\mathfrak{P}^{\perp})^k\widehat{E}\Theta\|\leq C_1(C_2\delta_3)^{k+1}\mu\sqrt{\frac{ r}{d_2}}.
    \end{align*}
    where $\Theta$ and $\mathfrak{P}^{\perp}$ are also defined in the proof of Theorem \ref{thm:inference-main}. \\
    \noindent\textbf{Case 0: k=0}  \\
    For any $j\in [d_1]$, clearly,
    \begin{align*}
        \|e_j^{\top}\mathfrak{P}^{\perp}\widehat{E}\Theta\|\leq \|e_j^{\top}\Theta\Theta^{\top}\widehat{E}\Theta\| + \|e_j^{\top}\widehat{E}\Theta\|\leq \delta_3\mu\sqrt{\frac{r}{d_1}} + \|e_j^{\top}\widehat{E}\Theta\|= \delta_3\mu\sqrt{\frac{r}{d_1}} + \|e_j^{\top}\widehat{Z}V\|,
    \end{align*}
    where in the last inequality, we transfer $e_j$ to be the canonical basis vector from $\RR^{d_1+d_2}$ to $\RR^{d_1}$.  We write
    \begin{align*}
        e_{j}^{\top}\widehat{Z}V= \frac{1}{T_0}\sum_{t=1}^{T_0} \sum_{i=1}^{d_1}\frac{1}{p_{ij_t(i)}}\xi_t^ie_j^{\top}X_t^i V  + \frac{1}{T_0}\sum_{t=1}^{T_0}\sum_{i=1}^{d_1} \frac{1}{p_{ij_t(i)}}\inp{\widehat{\Delta}_{1}}{X_t^{i}}e_j^{\top}X_t^iV - e_j\widehat{\Delta}_{1}V.
    \end{align*} 
    Since $e_j^{\top}X_t^i\neq 0$ only when $i=j$, we have
    \begin{align*}
        \left\|\bigg\|\frac{1}{T_0} \sum_{i=1}^{d_1}\frac{1}{p_{ij_t(i)}}\xi_t^ie_j^{\top}X_t^i V \bigg\|\right\|_{\psi_2}\lesssim \frac{\sigma}{p_{\min}T_0}\|V\|_{2,\max}\leq \frac{\sigma\mu}{p_{\min}T_0}\sqrt{\frac{r}{d_2}}
    \end{align*}
    and
    \begin{align*}
        &\frac{1}{T_0^2}\EE\left[e_j^{\top}\bigg(\sum_{i=1}^{d_1}\frac{1}{p_{ij_t(i)}}\xi_t^iX_t^i \bigg)VV^{\top}\bigg(\sum_{i=1}^{d_1}\frac{1}{p_{ij_t(i)}}\xi_t^iX_t^i \bigg)^{\top} e_j \right] = \frac{\sigma^2}{T_0^2}\EE\left[\frac{1}{p_{jj_t(j)}^2} \left\|V^{\top}e_{j_t(j)}e_{j_t(j)}^{\top}V\right\| \right]  \\
        &\quad \leq \sum_{s=1}^{d_2}\frac{1}{p_{js}}\|V\|_{2,\max}^2 \leq \frac{\sigma^2\mu^2r}{T_0^2p_{\min}}. 
    \end{align*}
    By Bernstein inequality, with probability at least $1-d_1^{-10}$,
    \begin{align*}
        \|e_j^{\top}\widehat{Z}_1V\|\lesssim \mu\sigma\sqrt{\frac{r\log d_1}{p_{\min}T}}
    \end{align*}
    as long as $p_{\min}^{-1}\log d_1\lesssim Td_2$. Similarly, since $p_{\min}^{-1}\geq d_2$, 
    \begin{align*}
        \frac{1}{T_0}\bigg\|e_j^{\top}\big(\sum_{i=1}^{d_1} \frac{1}{p_{ij_t(i)}}\inp{\widehat{\Delta}_1}{X_t^i}X_t^i -\widehat{\Delta}_1\big)V\bigg\|&\leq \frac{1}{p_{\min}T_0}\bigg\|e_j^{\top}\inp{\widehat{\Delta}_1}{X_t^j}X_t^jV \bigg\| + \frac{\sqrt{d_2}}{T_0}\|\widehat{\Delta}_1\|_{\max} \\
        &\leq \frac{\mu\sqrt{r}}{\sqrt{d_2}p_{\min}T_0}\|\widehat{\Delta}_1\|_{\max}\lesssim \frac{\mu\sqrt{r}}{\sqrt{d_2}p_{\min}T_0}\sigma
    \end{align*}
    and 
    \begin{align*}
        &\frac{1}{T_0^2} \EE\left[e_j^{\top}\bigg(\sum_{i=1}^{d_1} \frac{1}{p_{ij_t(i)}}\inp{\widehat{\Delta}_1}{X_t^i}X_t^iV\bigg)\bigg(\sum_{i=1}^{d_1} \frac{1}{p_{ij_t(i)}}\inp{\widehat{\Delta}_1}{X_t^i}X_t^iV\bigg)^{\top}e_j \right] \\ 
        &\quad \leq \frac{1}{T_0^2}\|\widehat{\Delta}_1\|_{\max}^2 \EE\left[\frac{1}{p_{ij_t(i)}^2}e_j^{\top}X_t^jVV^{\top}X_t^{j\top}e_j\right] \lesssim \frac{\sigma^2\mu^2r}{T_0^2p_{\min}}.
    \end{align*}
    where we again use $\|\widehat{\Delta}_1\|_{\max}\lesssim \sigma$. By Bernstein inequality, with probability at least $1-d_1^{-10}$,
    \begin{align*}
        \|e_j^{\top}\widehat{Z}_2V\|\lesssim \mu\sigma\sqrt{\frac{r\log d_1}{p_{\min}T}}
    \end{align*}
    as long as $p_{\min}^{-1}\log d_1\lesssim Td_2$. Combine with $p_{\min}^{-1}\lesssim (1/d_1^2)(\sum_i\sum_j p_{ij}^{-1})$, this directly implies
    \begin{align*}
        \max_{j\in [d_1]}\|e_j^{\top}\mathfrak{P}^{\perp}\widehat{E}\Theta\|\lesssim \delta_3\mu\sqrt{\frac{r}{d_1}}.
    \end{align*}
    Following the same arguments, we can show under $p_{\min}^{-1}\lesssim (1/d_1d_2)(\sum_i\sum_j p_{ij}^{-1})$,
    \begin{align*}
        \max_{j\in [d_2]}\|e_{j+d_1}^{\top}\mathfrak{P}^{\perp}\widehat{E}\Theta\|\lesssim \delta_3\mu\sqrt{\frac{r}{d_2}}.
    \end{align*}
    \noindent\textbf{Case 1: k=1}  \\
    Observe that for any $j\in [d_1]$, 
    \begin{align*}
        \|e_j^{\top}\mathfrak{P}^{\perp}\widehat{E}\mathfrak{P}^{\perp}\widehat{E}\Theta\|\leq \|e_j^{\top}\Theta\Theta^{\top} \widehat{E}\mathfrak{P}^{\perp}\widehat{E}\Theta\| + \|e_j^{\top}\widehat{E}\mathfrak{P}^{\perp}\widehat{E}\Theta\|\leq \delta_3^2\mu\sqrt{\frac{r}{d_1}} + \|e_j^{\top}\widehat{E}\mathfrak{P}^{\perp}\widehat{E}\Theta\|.
    \end{align*}
    By definition of $\widehat{E}$ and $\mathfrak{P}^{\perp}$, we have 
    \begin{align*}
        \widehat{E}\mathfrak{P}^{\perp}\widehat{E}\Theta = \begin{pmatrix}
            \widehat{Z}V_{\perp}V_{\perp}^{\top}\widehat{Z}^{\top}U & 0 \\
            0 & \widehat{Z}^{\top}U_{\perp}U_{\perp}^{\top}\widehat{Z}V 
        \end{pmatrix}.
    \end{align*}
    It suffices to prove the upper bound for $\|e_j^{\top}\widehat{Z}V_{\perp}V_{\perp}^{\top}\widehat{Z}^{\top}U\|$. Define $\mathfrak{J}_j=e_je_j^{\top}\in \RR^{d_1\times d_1}$ and $\mathfrak{J}^{\perp}=I-\mathfrak{J}$. Then write $\widehat{Z}=\mathfrak{J}_j\widehat{Z} + \mathfrak{J}_j^{\perp}\widehat{Z}$ and as a result,
    \begin{align*}
        \|e_j^{\top}\widehat{Z}V_{\perp}V_{\perp}^{\top}\widehat{Z}^{\top}U\|&\leq \|e_j^{\top}\widehat{Z}V_{\perp}V_{\perp}^{\top}(\mathfrak{J}_j\widehat{Z})^{\top}U\| + \|e_j^{\top}\widehat{Z}V_{\perp}V_{\perp}^{\top}(\mathfrak{J}_j^{\perp}\widehat{Z})^{\top}U\|  \\
        &\leq \|e_j^{\top}\widehat{Z}V_{\perp}V_{\perp}^{\top}\widehat{Z}^{\top}e_j\|\|e_j^{\top}U\| + \|e_j^{\top}\widehat{Z}V_{\perp}V_{\perp}^{\top}(\mathfrak{J}_j^{\perp}\widehat{Z})^{\top}U\| \\
        &\leq \delta_3^2\mu\sqrt{\frac{r}{d_1}} + \|e_j^{\top}\widehat{Z}_1V_{\perp}V_{\perp}^{\top}(\mathfrak{J}_j^{\perp}\widehat{Z})^{\top}U\| + \|e_j^{\top}\widehat{Z}_2V_{\perp}V_{\perp}^{\top}(\mathfrak{J}_j^{\perp}\widehat{Z})^{\top}U\|.
    \end{align*}
    We first prove the upper bound of $\|e_j^{\top}\widehat{Z}_1V_{\perp}V_{\perp}^{\top}(\mathfrak{J}_j^{\perp}\widehat{Z})^{\top}U\|$. Since $\mathfrak{J}_j^{\perp}\widehat{Z}$ does not include $X_t^j$ for any $t$, condition on the $\mathfrak{J}_j^{\perp}\widehat{Z}$, we have
    \begin{align*}
        \left\|\bigg\|\frac{1}{T_0} \sum_{i=1}^{d_1}\frac{1}{p_{ij_t(i)}}\xi_t^ie_j^{\top}X_t^iV_{\perp}V_{\perp}^{\top}(\mathfrak{J}_j^{\perp}\widehat{Z})^{\top}U  \bigg\|\right\|_{\psi_2} \lesssim \frac{\sigma}{p_{\min}T_0}\|V_{\perp}V_{\perp}^{\top}(\mathfrak{J}_j^{\perp}\widehat{Z})^{\top}U\|_{2,\max},  
    \end{align*}
    since only $e_j^{\top}X_t^{j}\neq 0$. Under the result of \textbf{Case 0}, we have
    \begin{align*}
        \|V_{\perp}V_{\perp}^{\top}(\mathfrak{J}_j^{\perp}\widehat{Z})^{\top}U\|_{2,\max}&\leq \|V_{\perp}V_{\perp}^{\top}(\mathfrak{J}_j\widehat{Z})^{\top}U\|_{2,\max} + \|V_{\perp}V_{\perp}^{\top}\widehat{Z}^{\top}U\|_{2,\max} \\
        &\lesssim \delta_3\mu\sqrt{\frac{r}{d_1}} + \delta_3\mu\sqrt{\frac{r}{d_2}}\lesssim \delta_3\mu\sqrt{\frac{r}{d_1}}.
    \end{align*}
    We also calculate the variance,  
    \begin{align*}
        &\frac{1}{T_0^2}\EE\left[ \big(\sum_{i=1}^{d_1}\frac{1}{p_{ij_t(i)}}\xi_t^ie_j^{\top}X_t^iV_{\perp}V_{\perp}^{\top}(\mathfrak{J}_j^{\perp}\widehat{Z})^{\top}U\big)\big(\sum_{i=1}^{d_1}\frac{1}{p_{ij_t(i)}}\xi_t^ie_j^{\top}X_t^iV_{\perp}V_{\perp}^{\top}(\mathfrak{J}_j^{\perp}\widehat{Z})^{\top}U\big)^{\top} \bigg| \mathfrak{J}_j^{\perp}\widehat{Z} \right] \\
        &\leq \frac{\sigma^2}{p_{\min}^2T_0^2}\EE\left[\|e_j^{\top}X_t^jV_{\perp}V_{\perp}^{\top}(\mathfrak{J}_j^{\perp}\widehat{Z})^{\top}U \|^2 \bigg| \mathfrak{J}_j^{\perp}\widehat{Z} \right]\leq \frac{\sigma^2}{p_{\min}^2T_0^2}\sum_{s=1}^{d_2} \PP\left(j_t(j)=s\bigg| \mathfrak{J}_j^{\perp}\widehat{Z}\right)\|V_{\perp}V_{\perp}^{\top}(\mathfrak{J}_j^{\perp}\widehat{Z})^{\top}U\|_{2,\max}  \\
        &\leq \frac{\sigma^2}{p_{\min}^2T_0^2}\|V_{\perp}V_{\perp}^{\top}(\mathfrak{J}_j^{\perp}\widehat{Z})^{\top}U\|_{2,\max}\lesssim \frac{\mu^2r\sigma^2}{p_{\min}^2d_1T^2}\delta_3^2,
    \end{align*}
    Then by Bernstein inequality, with probability at least $1-d_1^{-10}$,
    \begin{align*}
        \|e_j^{\top}\widehat{Z}_1V_{\perp}V_{\perp}^{\top}(\mathfrak{J}_j^{\perp}\widehat{Z})^{\top}U\|\lesssim \frac{\delta_3\mu\sigma}{p_{\min}}\sqrt{\frac{r\log d_1}{d_1T}}.
    \end{align*}
    Then as long as $p_{\min}^{-1}\lesssim \sqrt{(\sum_i\sum_j p_{ij}^{-1})/d_1}$, 
    \begin{align*}
        \|e_j^{\top}\widehat{Z}_1V_{\perp}V_{\perp}^{\top}(\mathfrak{J}_j^{\perp}\widehat{Z})^{\top}U\|\lesssim \delta_3^3\mu\sqrt{\frac{r}{d_1}}.
    \end{align*}
    Notice that the condition $p_{\min}^{-1}\lesssim \sqrt{(\sum_i\sum_j p_{ij}^{-1})/d_1}$ can be satisfied by $p_{\min}\asymp d_2^{-1}$, and can be further implied by $p_{\min}^{-1}\lesssim (1/d_1d_2)(\sum_{i,j} p_{ij}^{-1})$. Similarly, by the fact that $\|\widehat{\Delta}_1\|_{\max}\lesssim \sigma$ and $e_j^{\top}X_t^{i}= 0$ for $i\neq j$, under the result of \textbf{Case 0}, we have with probability at least $1-d_1^{-10}$,
    \begin{align*}
        \|e_j^{\top}\widehat{Z}_2V_{\perp}V_{\perp}^{\top}(\mathfrak{J}_j^{\perp}\widehat{Z})^{\top}U\|\lesssim \delta_3^2\mu\sqrt{\frac{r}{d_1}}.
    \end{align*}
    The upper bound for $\max_{j\in [d_2]}\|e_{j+d_1}^{\top} \mathfrak{P}^{\perp}\widehat{E}\mathfrak{P}^{\perp}\widehat{E}\Theta\|$ can be proved in the same way. Taking a union bound, with probability at least $1-4d_1^{-10}$,
    \begin{align*}
        \max_{j\in [d_1]}\|e_{j}^{\top} \mathfrak{P}^{\perp}\widehat{E}\mathfrak{P}^{\perp}\widehat{E}\Theta\|\lesssim \delta_3^2\mu\sqrt{\frac{r}{d_1}} \quad\text{and}\quad \max_{j\in [d_2]}\|e_{j+d_1}^{\top} \mathfrak{P}^{\perp}\widehat{E}\mathfrak{P}^{\perp}\widehat{E}\Theta\|\lesssim \delta_3^2\mu\sqrt{\frac{r}{d_2}}.
    \end{align*}
    Finally, the general case $k\geq 2$ can be proved by induction, following Lemma 9 in \cite{xia2021statistical}. We omit the details here. 
\end{proof}

\section{Proofs of Technical Lemmas in Appendix \ref{sec: proof-estimation}}

\subsection{Proof of Lemma \ref{lemma:estimation1}}
\begin{proof}
    We first write
    \begin{align*}
        \frac{d_2}{N_0}\sum_{t\in \mathcal{D}_{2p}} \sum_{i=1}^{d_1} \inp{\widehat{U}^{(p)}\widehat{G}^{(p)}\widehat{V}^{(p)\top} - U\Lambda V^{\top}}{X_t^i}\widehat{U}^{(p)\top}X_t^{i}\widehat{V}^{(p)} - \frac{d_2}{N_0}\sum_{t\in \mathcal{D}_{2p}} \sum_{i=1}^{d_1}\xi_t^i\widehat{U}^{(p)\top}X_t^{i}\widehat{V}^{(p)}=0
    \end{align*}
    where due to data splitting, $(\widehat{U}^{(p)},\widehat{V}^{(p)})$ are independent with $\mathcal{D}_{2p}$. Note that
    \begin{align*}
        \widehat{U}^{(p)}\widehat{G}^{(p)}\widehat{V}^{(p)\top} - U\Lambda V^{\top}=\widehat{U}^{(p)}(\widehat{G}^{(p)} -\widehat{O}_U^{(p)\top}\Lambda \widehat{O}_V^{(p)})\widehat{V}^{(p)\top} + (\widehat{U}^{(p)}\widehat{O}_U^{(p)\top}\Lambda(V^{(p)}\widehat{O}_V^{(p)\top})^{\top}- U\Lambda V^{\top}).
    \end{align*}
    Then
    \begin{align*}
        &\widehat{G}^{(p)}-\widehat{O}_U^{(p)\top}\Lambda\widehat{O}_V^{(p)}=(\widehat{G}^{(p)}-\widehat{O}_U^{(p)\top}\Lambda\widehat{O}_V^{(p)}) - \frac{d_2}{N_0}\sum_{t\in \mathcal{D}_{2p}} \sum_{i=1}^{d_1} \inp{\widehat{G}^{(p)} - \widehat{O}_U^{(p)\top}\Lambda\widehat{O}_V^{(p)}}{\widehat{U}^{(p)\top}X_t^i\widehat{V}^{(p)}}\widehat{U}^{(p)\top}X_t^{i}\widehat{V}^{(p)} \\ &\quad - \frac{d_2}{N_0}\sum_{t\in \mathcal{D}_{2p}} \sum_{i=1}^{d_1} \inp{\widehat{U}^{(p)}\widehat{O}_U^{(p)\top}\Lambda(V^{(p)}\widehat{O}_V^{(p)\top})^{\top}- U\Lambda V^{\top}}{X_t^i}\widehat{U}^{(p)\top}X_t^{i}\widehat{V}^{(p)}  + \frac{d_2}{N_0}\sum_{t\in \mathcal{D}_{2p}} \sum_{i=1}^{d_1}\xi_t^i\widehat{U}^{(p)\top}X_t^{i}\widehat{V}^{(p)}.
    \end{align*}
    Since $\|\widehat{U}^{(p)}\|\leq 2\mu\sqrt{r/d_1}$, $\|\widehat{V}^{(p)}\|\leq 2\mu\sqrt{r/d_2}$, then
    \begin{align*}
        \left\|\bigg\|\sum_{i=1}^{d_1}\xi_t^i\widehat{U}^{(p)\top}X_t^{i}\widehat{V}^{(p)}\bigg\|\right\|_{\psi_2}\lesssim \sigma\frac{\mu^2r}{\sqrt{d_1d_2}}d_1=\sigma\mu^2r\sqrt{\frac{d_1}{d_2}}
    \end{align*}
    and 
    \begin{align*}
        &\left\|\EE\left[\bigg(\sum_{i=1}^{d_1}\xi_t^i\widehat{U}^{(p)\top}X_t^{i}\widehat{V}^{(p)}\bigg)\bigg(\sum_{i=1}^{d_1}\xi_t^i\widehat{U}^{(p)\top}X_t^{i}\widehat{V}^{(p)}\bigg)^{\top}\right]\right\| \\ &\quad = \left\|\EE\left[\sum_{i=1}^{d_1}\xi_t^{i2}\widehat{U}^{(p)\top}X_t^{i}\widehat{V}^{(p)}\widehat{V}^{(p)\top}X_t^{i\top}\widehat{U}^{(p)}\right]\right\|=\sigma^2\frac{r}{d_2}
    \end{align*}
    by the independence of $\xi_t^i$. By matrix Bernstein inequality (\cite{tropp2012user}), with probability at least $1-d_1^{-10}$,
    \begin{align*}
        \left\|\frac{d_2}{N_0}\sum_{t\in \mathcal{D}_{2p}} \sum_{i=1}^{d_1}\xi_t^i\widehat{U}^{(p)\top}X_t^{i}\widehat{V}^{(p)}\right\|\lesssim \frac{\sigma r}{N_0}\sqrt{d_1d_2}\log d_1 + \sqrt{\frac{\sigma^2d_2r}{N_0}\log d_1}.
    \end{align*}
    Notice that
    \begin{align*}
        &\left\|(\widehat{G}^{(p)}-\widehat{O}_U^{(p)\top}\Lambda\widehat{O}_V^{(p)}) - \frac{d_2}{N_0}\sum_{t\in \mathcal{D}_{2p}} \sum_{i=1}^{d_1} \inp{\widehat{G}^{(p)} - \widehat{O}_U^{(p)\top}\Lambda\widehat{O}_V^{(p)}}{\widehat{U}^{(p)\top}X_t^i\widehat{V}^{(p)}}\widehat{U}^{(p)\top}X_t^{i}\widehat{V}^{(p)}\right\| \\
        &\quad \leq \|\widehat{G}^{(p)}-\widehat{O}_U^{(p)\top}\Lambda\widehat{O}_V^{(p)}\|\sup_{A\in \RR^{r\times r}, \|A\|\leq 1} \left\|A - \frac{d_2}{N_0}\sum_{t\in \mathcal{D}_{2p}} \sum_{i=1}^{d_1} \inp{A}{\widehat{U}^{(p)\top}X_t^i\widehat{V}^{(p)}}\widehat{U}^{(p)\top}X_t^i\widehat{V}^{(p)}\right\|.
    \end{align*}
    Denote $\mathcal{O}_r=\{A\in \RR^{r\times r},\|A\|\leq 1\}$ and $\mathcal{N}_{1/3}(\mathcal{O}_r)$ the $1/3$-net of $\mathcal{O}_r$, i.e., for any $A\in \mathcal{O}_r$, these exists $A_0\in \mathcal{N}_{1/3}(\mathcal{O}_r)$ so that $\|A-A_0\|\leq 1/3$. It is well-known by (\cite{pajor1998metric, koltchinskii2015optimal}) that $\text{Card}(\mathcal{N}_{1/3}(\mathcal{O}_r))\leq 3^{C_2r^2}$ for some constant $C_2>0$. By definition of $\mathcal{N}_{1/3}(\mathcal{O}_r)$,
    \begin{align*}
        &\left\|(\widehat{G}^{(p)}-\widehat{O}_U^{(p)\top}\Lambda\widehat{O}_V^{(p)}) - \frac{d_2}{N_0}\sum_{t\in \mathcal{D}_{2p}} \sum_{i=1}^{d_1} \inp{\widehat{G}^{(p)} - \widehat{O}_U^{(p)\top}\Lambda\widehat{O}_V^{(p)}}{\widehat{U}^{(p)\top}X_t^i\widehat{V}^{(p)}}\widehat{U}^{(p)\top}X_t^{i}\widehat{V}^{(p)}\right\| \\
        &\quad \leq  \|\widehat{G}^{(p)}-\widehat{O}_U^{(p)\top}\Lambda\widehat{O}_V^{(p)}\|\max_{A\in \mathcal{N}_{1/3}(\mathcal{O}_r)} \left\|A - \frac{d_2}{N_0}\sum_{t\in \mathcal{D}_{2p}} \sum_{i=1}^{d_1} \inp{A}{\widehat{U}^{(p)\top}X_t^i\widehat{V}^{(p)}}\widehat{U}^{(p)\top}X_t^i\widehat{V}^{(p)}\right\|.
    \end{align*}
    For each $A\in \mathcal{N}_{1/3}(\mathcal{O}_r)$,
    \begin{align*}
        \|\sum_{i=1}^{d_1}\inp{A}{\widehat{U}^{(p)\top}X_t^i\widehat{V}^{(p)}}\widehat{U}^{(p)\top}X_t^i\widehat{V}^{(p)}\|&\leq \sum_{i=1}^{d_1}\|\widehat{U}^{(p)\top}X_t^i\widehat{V}^{(p)}\|_*\|\widehat{U}^{(p)\top}X_t^i\widehat{V}^{(p)}\| \\ &\leq d_1 \|\widehat{U}^{(p)}\|_{2,\max}^2\|\widehat{V}^{(p)}\|_{2,\max}^2\leq \frac{\mu^2r^2}{d_2}
    \end{align*}
    where $\|\cdot\|_*$ denotes the matrix nuclear norm. Moreover,
    \begin{align*}
        &\left\|\EE\left[\bigg(\sum_{i=1}^{d_1} \inp{A}{\widehat{U}^{(p)\top}X_t^i\widehat{V}^{(p)}}\widehat{U}^{(p)\top}X_t^i\widehat{V}^{(p)}\bigg)\bigg(\sum_{i=1}^{d_1} \inp{A}{\widehat{U}^{(p)\top}X_t^i\widehat{V}^{(p)}}\widehat{U}^{(p)\top}X_t^i\widehat{V}^{(p)}\bigg)^{\top}\right]\right\| \\
        &\quad \leq \max_{i,j}|\inp{A}{\widehat{U}^{(p)\top}X_t^i\widehat{V}^{(p)}}|^2d_1^2\max_{i,j,k,l}\|(\widehat{U}^{(p)\top}e_ie_j^{\top}\widehat{V}^{(p)})(\widehat{U}^{(p)\top}e_ke_l^{\top}\widehat{V}^{(p)})^{\top}\| \\
        &\quad \leq \fro{A}^2\max_{i,j}\fro{\widehat{U}^{(p)\top}e_ie_j^{\top}\widehat{V}^{(p)}}^2d_1^2\max_{i,j}\|\widehat{U}^{(p)\top}e_ie_j^{\top}\widehat{V}^{(p)}\|^2\leq \frac{\mu^4r^3}{d_2^2}.
    \end{align*}
    Then by matrix Bernstein inequality, with probability at least $1-d_1^{-10}$,
    \begin{align*}
        \left\|A - \frac{d_2}{N_0}\sum_{t\in \mathcal{D}_{2p}} \sum_{i=1}^{d_1} \inp{A}{\widehat{U}^{(p)\top}X_t^i\widehat{V}^{(p)}}\widehat{U}^{(p)\top}X_t^i\widehat{V}^{(p)}\right\|\lesssim \sqrt{\frac{r^3\log d_1}{N_0}}
    \end{align*}
    as long as $T\gtrsim r\log d_1$. This directly implies that
    \begin{align*}
        &\left\|(\widehat{G}^{(p)}-\widehat{O}_U^{(p)\top}\Lambda\widehat{O}_V^{(p)}) - \frac{d_2}{N_0}\sum_{t\in \mathcal{D}_{2p}} \sum_{i=1}^{d_1} \inp{\widehat{G}^{(p)} - \widehat{O}_U^{(p)\top}\Lambda\widehat{O}_V^{(p)}}{\widehat{U}^{(p)\top}X_t^i\widehat{V}^{(p)}}\widehat{U}^{(p)\top}X_t^{i}\widehat{V}^{(p)}\right\| \\
        &\quad \lesssim  \|\widehat{G}^{(p)}-\widehat{O}_U^{(p)\top}\Lambda\widehat{O}_V^{(p)}\|\sqrt{\frac{r^3\log d_1}{N_0}}.
    \end{align*}
    Similarly, we first notice that 
    \begin{align*}
        &\left\|\EE\left[\sum_{i=1}^{d_1} \inp{\widehat{U}^{(p)}\widehat{O}_U^{(p)\top}\Lambda(V^{(p)}\widehat{O}_V^{(p)\top})^{\top}- U\Lambda V^{\top}}{X_t^i}\widehat{U}^{(p)\top}X_t^{i}\widehat{V}^{(p)}\right]\right\| \\ &\quad =\frac{1}{d_2}\left\|\widehat{U}^{(p)\top}(\widehat{U}^{(p)}\widehat{O}_U^{(p)\top}\Lambda(V^{(p)}\widehat{O}_V^{(p)\top})^{\top}- U\Lambda V^{\top})\widehat{V}^{(p)}\right\|.
    \end{align*}
    We also calculate the uniform bound
    \begin{align*}
        &\left\|\sum_{i=1}^{d_1} \inp{\widehat{U}^{(p)}\widehat{O}_U^{(p)\top}\Lambda(V^{(p)}\widehat{O}_V^{(p)\top})^{\top}- U\Lambda V^{\top}}{X_t^i}\widehat{U}^{(p)\top}X_t^{i}\widehat{V}^{(p)}\right\| \\ 
        &\quad \leq \|\widehat{U}^{(p)}\widehat{O}_U^{(p)\top}\Lambda(V^{(p)}\widehat{O}_V^{(p)\top})^{\top}- U\Lambda V^{\top}\|_{\max}^2d_1\max_{i,j}\|\widehat{U}^{(p)\top}e_ie_j^{\top}\widehat{V}^{(p)}\| \\ 
        &\quad \leq \|\widehat{U}^{(p)}\widehat{O}_U^{(p)\top}\Lambda(V^{(p)}\widehat{O}_V^{(p)\top})^{\top}- U\Lambda V^{\top}\|_{\max}\mu^2r\sqrt{\frac{d_1}{d_2}}
    \end{align*}
    and variance
    \begin{align*}
        &\left\|\EE\left[\bigg(\sum_{i=1}^{d_1} \inp{\widehat{U}^{(p)}\widehat{O}_U^{(p)\top}\Lambda(V^{(p)}\widehat{O}_V^{(p)\top})^{\top}- U\Lambda V^{\top}}{X_t^i}\widehat{U}^{(p)\top}X_t^{i}\widehat{V}^{(p)}\bigg) \right.\right.\\ &\left.\left.\quad \bigg(\sum_{i=1}^{d_1} \inp{\widehat{U}^{(p)}\widehat{O}_U^{(p)\top}\Lambda(V^{(p)}\widehat{O}_V^{(p)\top})^{\top}- U\Lambda V^{\top}}{X_t^i}\widehat{U}^{(p)\top}X_t^{i}\widehat{V}^{(p)}\bigg)^{\top}\right]\right\| \\
        &\quad\quad \leq \|\widehat{U}^{(p)}\widehat{O}_U^{(p)\top}\Lambda(V^{(p)}\widehat{O}_V^{(p)\top})^{\top}- U\Lambda V^{\top}\|_{\max}^2d_1^2\max_{i,j}\|\widehat{U}^{(p)\top}e_ie_j^{\top}\widehat{V}^{(p)}\|^2 \\
        &\quad\quad \leq \|\widehat{U}^{(p)}\widehat{O}_U^{(p)\top}\Lambda(V^{(p)}\widehat{O}_V^{(p)\top})^{\top}- U\Lambda V^{\top}\|_{\max}^2\frac{\mu^4r^2d_1}{d_2}
    \end{align*}
    By matrix Bernstein inequality, with probability at least $1-d_1^{-10}$, 
    \begin{align*}
        &\left\|\frac{d_2}{N_0}\sum_{t\in \mathcal{D}_{2p}} \sum_{i=1}^{d_1} \inp{\widehat{U}^{(p)}\widehat{O}_U^{(p)\top}\Lambda(V^{(p)}\widehat{O}_V^{(p)\top})^{\top}- U\Lambda V^{\top}}{X_t^i}\widehat{U}^{(p)\top}X_t^{i}\widehat{V}^{(p)}\right\| \\
        &\quad \leq \left\|\widehat{U}^{(p)\top}(\widehat{U}^{(p)}\widehat{O}_U^{(p)\top}\Lambda(V^{(p)}\widehat{O}_V^{(p)\top})^{\top}- U\Lambda V^{\top})\widehat{V}^{(p)}\right\| \\ &\quad\quad + C_2\|\widehat{U}^{(p)}\widehat{O}_U^{(p)\top}\Lambda(V^{(p)}\widehat{O}_V^{(p)\top})^{\top}- U\Lambda V^{\top}\|_{\max}\sqrt{\frac{d_1d_2r^2\log d_1}{N_0}}.
    \end{align*}
    Note that 
    \begin{align*}
        &\|\widehat{U}^{(p)}\widehat{O}_U^{(p)\top}\Lambda(V^{(p)}\widehat{O}_V^{(p)\top})^{\top}- U\Lambda V^{\top}\|_{\max} \\&\quad \leq 3\lambda_{\max}\mu\left(\sqrt{\frac{r}{d_2}}\|\widehat{U}^{(p)}-U\widehat{O}_U^{(p)}\|_{2,\max} + \sqrt{\frac{r}{d_1}}\|\widehat{V}^{(p)}-V\widehat{O}_V^{(p)}\|_{2,\max}\right).
    \end{align*}
    By the differential preperty of Grassmannians (\cite{edelman1998geometry,keshavan2010matrix,xia2019polynomial}),
    \begin{align*}
        &\left\|\widehat{U}^{(p)\top}(\widehat{U}^{(p)}\widehat{O}_U^{(p)\top}\Lambda(V^{(p)}\widehat{O}_V^{(p)\top})^{\top}- U\Lambda V^{\top})\widehat{V}^{(p)}\right\| \\
        &\quad \leq \lambda_{\max}\|\widehat{U}^{(p)\top}(\widehat{U}^{(p)}\widehat{O}_U^{(p)\top}-U)\| + \lambda_{\max}\|\widehat{V}^{(p)\top}(\widehat{V}^{(p)}\widehat{O}_V^{(p)\top}-V)\| \\
        &\quad \leq 2\lambda_{\max}(\|\widehat{U}^{(p)}-U\widehat{O}_U^{(p)}\|^2 + \|\widehat{V}^{(p)}-V\widehat{O}_V^{(p)}\|^2).
    \end{align*}
    As a result, as long as $T\gtrsim r^3\log d_1$, with probability at least $1-3d_1^{-10}$,
    \begin{align*}
        &\|\widehat{G}^{(p)}-\widehat{O}_U^{(p)\top}\Lambda\widehat{O}_V^{(p)}\|\leq C_3\left( \frac{\sigma r}{N_0}\sqrt{d_1d_2}\log d_1 + \sqrt{\frac{\sigma^2d_2r}{N_0}\log d_1}\right) \\ &\quad + 2\lambda_{\max}(\|\widehat{U}^{(p)}-U\widehat{O}_U^{(p)}\|^2 + \|\widehat{V}^{(p)}-V\widehat{O}_V^{(p)}\|^2) \\ &\quad + C_4\lambda_{\max}\left(\sqrt{\frac{r}{d_2}}\|\widehat{U}^{(p)}-U\widehat{O}_U^{(p)}\|_{2,\max} + \sqrt{\frac{r}{d_1}}\|\widehat{V}^{(p)}-V\widehat{O}_V^{(p)}\|_{2,\max}\right)\sqrt{\frac{d_1d_2r^2\log d_1}{N_0}}.
    \end{align*}
\end{proof}

\subsection{Proof of Lemma \ref{lemma:estimation2}}
\begin{proof}
    Recall that
    \begin{align*}
        &\widehat{U}^{(p+0.5)}-U\widehat{O}_U^{(p)}\widehat{L}_G^{(p)} = (\widehat{U}^{(p)}\widehat{L}_G^{(p)}-U\widehat{O}_U^{(p)}\widehat{L}_G^{(p)})(I-\eta\widehat{L}_G^{(p)\top}\widehat{O}_U^{(p)\top}\Lambda\widehat{O}_V^{(p)}\widehat{R}_G^{(p)}(\widehat{\Lambda}^{(p)})^{-1}) \\
        &\quad -\eta \widehat{U}^{(p)}(\widehat{G}^{(p)}-\widehat{O}_U^{(p)\top}\Lambda\widehat{O}_V^{(p)})\widehat{R}_G^{(p)}(\widehat{\Lambda}^{(p)})^{-1} - \eta U\Lambda(\widehat{V}^{(p)}\widehat{O}_V^{(p)\top}-V)^{\top}\widehat{V}^{(p)}\widehat{R}_G^{(p)}(\widehat{\Lambda}^{(p)})^{-1} + \widehat{E}_V^{(p)} + \widehat{E}_{\xi,V}^{(p)}.
    \end{align*}
    By Lemma \ref{lemma:estimation1}, 
    \begin{align*}
        &\|\widehat{\Lambda}^{(p)}-\widehat{L}_G^{(p)\top}\widehat{O}_U^{(p)\top}\Lambda\widehat{O}_V^{(p)}\widehat{R}_G^{(p)}\|\leq C_3(\frac{r\sigma}{N_0}\sqrt{d_1d_2}\log d_1 + \sqrt{\frac{\sigma^2 rd_2\log d_1}{N_0}}) \\ &\quad + 2\lambda_{\max}(\|\widehat{U}^{(p)}-U\widehat{O}_U^{(p)}\|^2+\|\widehat{V}^{(p)}-V\widehat{O}_U^{(p)}\|^2) \\ &\quad + C_4\lambda_{\max}\left(\sqrt{\frac{1}{d_2}}\|\widehat{U}^{(p)}-U\widehat{O}_U^{(p)}\|_{2,\max} + \sqrt{\frac{1}{d_1}}\|\widehat{V}^{(p)}-V\widehat{O}_V^{(p)}\|_{2,\max}\right)\mu\sqrt{\frac{r^3d_1d_2\log d_1}{N_0}}
    \end{align*}
    which implies $\|\widehat{\Lambda}^{(p)}-\widehat{L}_G^{(p)\top}\widehat{O}_U^{(p)\top}\Lambda\widehat{O}_V^{(p)}\widehat{R}_G^{(p)}\|\leq \lambda_{\min}/20$ and $\lambda_{\min}(\widehat{\Lambda}^{(p)})\geq \lambda_{\min}/2$ as long as $\lambda_{\min}/\sigma \gtrsim \max\{\sqrt{d_1d_2}/T, \sqrt{d_2/T}\}r\log d_1$, $T\gtrsim r^3\log d_1$, $\|\widehat{U}^{(p)}\|_{2,\max}\leq 2\mu\sqrt{r/d_1}$, $\|\widehat{V}^{(p)}\|_{2,\max}\leq 2\mu\sqrt{r/d_2}$ and $\max\{\|\widehat{U}^{(p)}-U\widehat{O}_U^{(p)}\|, \|\widehat{V}^{(p)}-V\widehat{O}_V^{(p)}\|\}\lesssim 1/\sqrt{\kappa}$. 

    Since $\eta\leq 0.75$, we have
    \begin{align*}
        &\|(\widehat{U}^{(p)}\widehat{L}_G^{(p)}-U\widehat{O}_U^{(p)}\widehat{L}_G^{(p)})(I-\eta\widehat{L}_G^{(p)\top}\widehat{O}_U^{(p)\top}\Lambda\widehat{O}_V^{(p)}\widehat{R}_G^{(p)}(\widehat{\Lambda}^{(p)})^{-1})\|_{2,\max} \\
        &\quad \leq \|\widehat{U}^{(p)}\widehat{L}_G^{(p)}-U\widehat{O}_U^{(p)}\widehat{L}_G^{(p)}\|_{2,\max} \|I-\eta\widehat{L}_G^{(p)\top}\widehat{O}_U^{(p)\top}\Lambda\widehat{O}_V^{(p)}\widehat{R}_G^{(p)}(\widehat{\Lambda}^{(p)})^{-1}\| \\
        &\quad \leq \|\widehat{U}^{(p)}-U\widehat{O}_U^{(p)}\|_{2,\max}(1-\eta) + \|\widehat{U}^{(p)}-U\widehat{O}_U^{(p)}\|_{2,\max}\eta\|\widehat{\Lambda}^{(p)}-\widehat{L}_G^{(p)\top}\widehat{O}_U^{(p)\top}\Lambda\widehat{O}_V^{(p)}\widehat{R}_G^{(p)}\|\|(\widehat{\Lambda}^{(p)})^{-1}\| \\
        &\quad \leq (1-\eta)\|\widehat{U}^{(p)}-U\widehat{O}_U^{(p)}\|_{2,\max} + 2\lambda_{\min}^{-1}\|\widehat{U}^{(p)}-U\widehat{O}_U^{(p)}\|_{2,\max}\eta\|\widehat{\Lambda}^{(p)}-\widehat{L}_G^{(p)\top}\widehat{O}_U^{(p)\top}\Lambda\widehat{O}_V^{(p)}\widehat{R}_G^{(p)}\|.
    \end{align*}
    Then we can obtain
    \begin{align*}
        \|(\widehat{U}^{(p)}\widehat{L}_G^{(p)}-U\widehat{O}_U^{(p)}\widehat{L}_G^{(p)})(I-\eta\widehat{L}_G^{(p)\top}\widehat{O}_U^{(p)\top}\Lambda\widehat{O}_V^{(p)}\widehat{R}_G^{(p)}(\widehat{\Lambda}^{(p)})^{-1})\|_{2,\max}\leq (1-\frac{9\eta}{10})\|\widehat{U}^{(p)}-U\widehat{O}_U^{(p)}\|_{2,\max}.
    \end{align*}
    By $\|\widehat{U}^{(p)}\|_{2,\max}\leq 2\mu\sqrt{r/d_1}$, we have
    \begin{align*}
        &\eta\|\widehat{U}^{(p)}(\widehat{G}^{(p)}-\widehat{O}_U^{(p)\top}\Lambda\widehat{O}_V^{(p)})\widehat{R}_G^{(p)}(\widehat{\Lambda}^{(p)})^{-1} \|_{2,\max}\leq 2\eta\|\widehat{U}^{(p)}\|_{2,\max}\|\widehat{G}^{(p)}-\widehat{O}_U^{(p)\top}\Lambda\widehat{O}_V^{(p)}\|\lambda_{\min}^{-1} \\
        &\leq C_3\eta\mu(\frac{r^{3/2}\sigma}{\lambda_{\min}}\frac{\sqrt{d_2}}{N_0}\log d_1 + \frac{\sigma}{\lambda_{\min}}\sqrt{\frac{r^2d_2\log d_1}{N_0d_1}})  + C_4\eta\frac{\mu\kappa r^{1/2}}{\sqrt{d_1}}(\|\widehat{U}^{(p)}-U\widehat{O}_U^{(p)}\|^2+\|\widehat{V}^{(p)}-V\widehat{O}_U^{(p)}\|^2) \\ &\quad + C_5\eta\mu\kappa\left(\sqrt{\frac{1}{d_2}}\|\widehat{U}^{(p)}-U\widehat{O}_U^{(p)}\|_{2,\max} + \sqrt{\frac{1}{d_1}}\|\widehat{V}^{(p)}-V\widehat{O}_V^{(p)}\|_{2,\max}\right)\mu\sqrt{\frac{r^4d_2\log d_1}{N_0}}.
    \end{align*}
    Observe that 
    \begin{align*}
        &C_4\frac{\mu\kappa r^{1/2}}{\sqrt{d_1}}\|\widehat{U}^{(p)}-U\widehat{O}_U^{(p)}\|^2= C_4\mu\kappa r^{1/2} \|\widehat{U}^{(p)}-U\widehat{O}_U^{(p)}\|\frac{\|\widehat{U}^{(p)}-U\widehat{O}_U^{(p)}\|}{\sqrt{d_1}} \\
        &\quad \leq C_4\mu\kappa r^{1/2} \|\widehat{U}^{(p)}-U\widehat{O}_U^{(p)}\|\frac{\fro{\widehat{U}^{(p)}-U\widehat{O}_U^{(p)}}}{\sqrt{d_1}} \leq \frac{\|\widehat{U}^{(p)}-U\widehat{O}_U^{(p)}\|_{2,\max}}{20\sqrt{d_2/d_1}}
    \end{align*}
    if $\|\widehat{U}^{(p)}-U\widehat{O}_U^{(p)}\|\lesssim 1/\kappa\mu\sqrt{rd_2/d_1}$. Similarly, we have
    \begin{align*}
        C_4\frac{\mu\kappa r^{1/2}}{\sqrt{d_1}}\|\widehat{V}^{(p)}-V\widehat{O}_V^{(p)}\|^2\leq \frac{\|\widehat{V}^{(p)}-V\widehat{O}_V^{(p)}\|_{2,\max}}{20}.
    \end{align*}
    Therefore, 
    \begin{align*}
        C_4\frac{\mu\kappa r^{1/2}}{\sqrt{d_1}}(\|\widehat{U}^{(p)}-U\widehat{O}_U^{(p)}\|^2+\|\widehat{V}^{(p)}-V\widehat{O}_U^{(p)}\|^2)\leq \frac{\sqrt{\frac{d_1}{d_2}}\|\widehat{U}^{(p)}-U\widehat{O}_U^{(p)}\|_{2,\max} + \|\widehat{V}^{(p)}-V\widehat{O}_V^{(p)}\|_{2,\max} }{20}.
    \end{align*}
    Moreover, if $T\gtrsim \mu^4\kappa^2r^4\alpha_d\log d_1$,
    \begin{align*}
        &C_5\mu\kappa\left(\sqrt{\frac{1}{d_2}}\|\widehat{U}^{(p)}-U\widehat{O}_U^{(p)}\|_{2,\max} + \sqrt{\frac{1}{d_1}}\|\widehat{V}^{(p)}-V\widehat{O}_V^{(p)}\|_{2,\max}\right)\mu\sqrt{\frac{r^4d_2\log d_1}{N_0}} \\
        &\leq \frac{\sqrt{\frac{d_1}{d_2}}\|\widehat{U}^{(p)}-U\widehat{O}_U^{(p)}\|_{2,\max} + \|\widehat{V}^{(p)}-V\widehat{O}_V^{(p)}\|_{2,\max}}{40}.
    \end{align*}
    Then we have
    \begin{align*}
        &\eta\|\widehat{U}^{(p)}(\widehat{G}^{(p)}-\widehat{O}_U^{(p)\top}\Lambda\widehat{O}_V^{(p)})\widehat{R}_G^{(p)}(\widehat{\Lambda}^{(p)})^{-1} \|_{2,\max}\leq C_3\mu(\frac{r^{3/2}\sigma}{\lambda_{\min}}\frac{\sqrt{d_2}}{N_0}\log d_1 + \frac{\sigma}{\lambda_{\min}}\sqrt{\frac{r^2d_2\log d_1}{N_0d_1}}) \\
        &\quad + \frac{3\eta}{40}\left(\sqrt{\frac{d_1}{d_2}}\|\widehat{U}^{(p)}-U\widehat{O}_U^{(p)}\|_{2,\max} + \|\widehat{V}^{(p)}-V\widehat{O}_V^{(p)}\|_{2,\max}\right).
    \end{align*}
    Since $\|(\widehat{V}^{(p)}\widehat{O}_V^{(p)\top}-V)^{\top}\widehat{V}^{(p)}\|\leq \|\widehat{V}^{(p)}\widehat{O}_V^{(p)\top}-V\|^2$, we get
    \begin{align*}
        \|U\Lambda(\widehat{V}^{(p)}\widehat{O}_V^{(p)\top}-V)^{\top}\widehat{V}^{(p)}\widehat{R}_G^{(p)}(\widehat{\Lambda}^{(p)})^{-1}\|_{2,\max}&\leq 2\kappa\mu\sqrt{\frac{r}{d_1}}\|\widehat{V}^{(p)}\widehat{O}_V^{(p)\top}-V\|^2 \\ &\leq \frac{1}{40}\|\widehat{V}^{(p)}-V\widehat{O}_V^{(p)\top}\|_{2,\max}
    \end{align*}
    as long as $\|\widehat{V}^{(p)}\widehat{O}_V^{(p)\top}-V\|\lesssim 1/\kappa\mu\sqrt{r}\sqrt{d_2/d_1}$. Putting together the above bounds, we obtain
    \begin{align*}
        &\|\widehat{U}^{(p+0.5)}-U\widehat{O}_U^{(p)}\widehat{L}_G^{(p)}\|_{2,\max}\leq (1-\frac{9\eta}{10})\|\widehat{U}^{(p)}-U\widehat{O}_U^{(p)}\|_{2,\max} + C_3\mu(\frac{r^{3/2}\sigma}{\lambda_{\min}}\frac{\sqrt{d_2}}{N_0}\log d_1 + \frac{\sigma}{\lambda_{\min}}\sqrt{\frac{r^2d_2\log d_1}{N_0d_1}}) \\ &\quad + \frac{\eta}{10}\left(\sqrt{\frac{d_1}{d_2}}\|\widehat{U}^{(p)}-U\widehat{O}_U^{(p)}\|_{2,\max} + \|\widehat{V}^{(p)}-V\widehat{O}_V^{(p)}\|_{2,\max}\right) + \|\widehat{E}_V^{(p)}\|_{2,\max} + \|\widehat{E}_{\xi,V}^{(p)}\|_{2,\max}.
    \end{align*}
    Since $(\widehat{U}^{(p)}, \widehat{G}^{(p)}, \widehat{V}^{(p)})$ are independent with $\mathcal{D}_{2p+1}$ and $\widehat{U}^{(p)}, \widehat{V}^{(p)}$ are incoherent, for any $j\in [d_1]$, we have
    \begin{align*}
        \bigg\|\big\|e_j^{\top}\sum_{i=1}^{d_1}\xi_t^i X_t^{i}\widehat{V}^{(p)}\widehat{R}_G^{(p)}(\widehat{\Lambda}^{(p)})^{-1}\big\|\bigg\|_{\psi_2}\lesssim \sigma\|\widehat{V}^{(p)}\|_{2,\max}\lambda_{\min}^{-1}\lesssim \frac{\sigma}{\lambda_{\min}}\sqrt{\frac{r}{d_2}}
    \end{align*}
    and 
    \begin{align*}
        &\EE\left[(e_j^{\top}\sum_{i=1}^{d_1}\xi_t^i X_t^{i}\widehat{V}^{(p)}\widehat{R}_G^{(p)}(\widehat{\Lambda}^{(p)})^{-1})(e_j^{\top}\sum_{i=1}^{d_1}\xi_t^i X_t^{i}\widehat{V}^{(p)}\widehat{R}_G^{(p)}(\widehat{\Lambda}^{(p)})^{-1})^{\top} \right] \\
        &\quad \leq \sigma^2\|e_j^{\top} X_t^{j}\widehat{V}^{(p)}\widehat{R}_G^{(p)}(\widehat{\Lambda}^{(p)})^{-1}\|^2\lesssim \frac{\sigma^2}{\lambda_{\min}^2}\frac{r}{d_2}.
    \end{align*}
    since $e_j^{\top}X_t^i\neq 0$ only when $i=j$. Then by Bernstein inequality, with probability at least $1-d_1^{-10}$,
    \begin{align*}
        \|\widehat{E}_{\xi,V}^{(p)}\|_{2,\max}\lesssim \eta\frac{\sigma}{\lambda_{\min}}\sqrt{\frac{rd_2\log d_1}{N_0}}.
    \end{align*}
    Similarly, with the same probability
    \begin{align*}
        \|\widehat{E}_V^{(p)}\|_{2,\max}\lesssim \eta\frac{\|\widehat{M}^{(p)}-M\|_{\max}}{\lambda_{\min}}\sqrt{\frac{rd_2\log d_1}{N_0}}
    \end{align*}
    where $\widehat{M}^{(p)}=\widehat{U}^{(p)}\widehat{G}^{(p)}\widehat{V}^{(p)\top}$. Note that
    \begin{align*}
        &\|\widehat{M}^{(p)}-M\|_{\max}\leq \|(\widehat{U}^{(p)}-U\widehat{O}_U^{(p)})\widehat{G}^{(p)}\widehat{V}^{(p)\top}\|_{\max} + \|U(\widehat{O}_U^{(p)}\widehat{G}^{(p)}-\Lambda\widehat{O}_V^{(p)})\widehat{V}^{(p)\top}\|_{\max} \\ &\quad\quad  + \|U\Lambda(\widehat{V}^{(p)}\widehat{O}_V^{(p)}-V)^{\top}\|_{2,\max} \\
        &\quad \leq 2\mu\lambda_{\max}\left(\sqrt{\frac{r}{d_2}}\|\widehat{U}^{(p)}-U\widehat{O}_U^{(p)}\|_{2,\max} + \sqrt{\frac{r}{d_1}}\|\widehat{V}^{(p)}-V\widehat{O}_V^{(p)}\|_{2,\max}\right)  + \mu^2\sqrt{\frac{r^2}{d_1d_2}}\|\widehat{G}^{(p)}-\widehat{O}_U^{(p)}\Lambda\widehat{O}_V^{(p)}\|.
    \end{align*}
    Together with Lemma \ref{lemma:estimation1},
    \begin{align*}
        &\mu^2\sqrt{\frac{r^2}{d_1d_2}}\|\widehat{G}^{(p)}-\widehat{O}_U^{(p)}\Lambda\widehat{O}_V^{(p)}\|\leq C_3\mu^2(\frac{r^2\sigma}{N_0}\log d_1 + \sqrt{\frac{\sigma^2 r^3\log d_1}{d_1N_0}}) \\ &\quad + 2\lambda_{\max}\mu^2\sqrt{\frac{r^2}{d_1d_2}}(\|\widehat{U}^{(p)}-U\widehat{O}_U^{(p)}\|^2+\|\widehat{V}^{(p)}-V\widehat{O}_U^{(p)}\|^2) \\ &\quad + C_4\lambda_{\max}\left(\sqrt{\frac{1}{d_2}}\|\widehat{U}^{(p)}-U\widehat{O}_U^{(p)}\|_{2,\max} + \sqrt{\frac{1}{d_1}}\|\widehat{V}^{(p)}-V\widehat{O}_V^{(p)}\|_{2,\max}\right)\mu^3\sqrt{\frac{r^5\log d_1}{N_0}}.
    \end{align*}
    As long as $T\gtrsim \mu^6r^4\log d_1$ and $\|\widehat{U}^{(p)}-U\widehat{O}_U^{(p)}\|+\|\widehat{V}^{(p)}-V\widehat{O}_U^{(p)}\|\lesssim \frac{1}{\mu^2r}$,
    \begin{align*}
        \|\widehat{M}^{(p)}-M\|_{\max}\leq \sigma+2\mu\lambda_{\max}\left(\sqrt{\frac{r}{d_2}}\|\widehat{U}^{(p)}-U\widehat{O}_U^{(p)}\|_{2,\max} + \sqrt{\frac{r}{d_1}}\|\widehat{V}^{(p)}-V\widehat{O}_V^{(p)}\|_{2,\max}\right).
    \end{align*}
    Therefore, as long as $T\gtrsim \kappa^2\mu^6r^4\alpha_d\log d_1$, with probability at least $1-2d_1^{-10}$,
    \begin{align*}
        \|\widehat{E}_V^{(p)}\|_{2,\max} + \|\widehat{E}_{\xi,V}^{(p)}\|_{2,\max}\leq C_6\eta\frac{\sigma}{\lambda_{\min}}\sqrt{\frac{rd_2\log d_1}{N_0}} + \frac{\eta}{40}\left(\sqrt{\frac{d_1}{d_2}}\|\widehat{U}^{(p)}-U\widehat{O}_U^{(p)}\|_{2,\max} + \|\widehat{V}^{(p)}-V\widehat{O}_V^{(p)}\|_{2,\max}\right).
    \end{align*}
    As a result,
    \begin{align*}
        &\|\widehat{U}^{(p+0.5)}-U\widehat{O}_U^{(p)}\widehat{L}_G^{(p)}\|_{2,\max}\leq (1-\frac{9\eta}{10})\|\widehat{U}^{(p)}-U\widehat{O}_U^{(p)}\|_{2,\max} + C_6\eta\frac{\sigma}{\lambda_{\min}}\sqrt{\frac{rd_2\log d_1}{N_0}}\\
        &\quad + \frac{\eta}{8}\left(\sqrt{\frac{d_1}{d_2}}\|\widehat{U}^{(p)}-U\widehat{O}_U^{(p)}\|_{2,\max} + \|\widehat{V}^{(p)}-V\widehat{O}_V^{(p)}\|_{2,\max}\right).
    \end{align*}
    Next, we investigate the singular values of $\widehat{U}^{(p+0.5)}-U\widehat{O}_U^{(p)}\widehat{L}_G^{(p)}$. Recall
    \begin{align*}
        \widehat{U}^{(p+0.5)}= \underbrace{\widehat{U}^{(p)}\widehat{L}_G^{(p)}-\eta(\widehat{U}^{(p)}\widehat{G}^{(p)}\widehat{V}^{(p)\top} - U\Lambda V^{\top})\widehat{V}^{(p)}\widehat{R}_G^{(p)}(\widehat{\Lambda}^{(p)})^{-1}}_{\mathcal{I}_1} + \underbrace{\widehat{E}_V^{(p)} + \widehat{E}_{\xi,V}^{(p)}}_{\mathcal{I}_2}. 
    \end{align*}
    By the independence between $(\widehat{U}^{(p)}, \widehat{G}^{(p)}, \widehat{V}^{(p)}, \widehat{L}_G^{(p)}, \widehat{R}_G^{(p)}, \widehat{\Lambda}^{(p)})$ and $\mathcal{D}_{2p+1}$, and notice that $\|X_t\|=1$, we have
    \begin{align*}
        \bigg\|\big\|\sum_{i=1}^{d_1}\xi_t^i X_t^{i}\widehat{V}^{(p)}\widehat{R}_G^{(p)}(\widehat{\Lambda}^{(p)})^{-1} \big\|\bigg\|_{\psi_2}\lesssim \frac{\sigma}{\lambda_{\min}}
    \end{align*}
    and
    \begin{align*}
        &\bigg\|\EE\left[(\sum_{i=1}^{d_1}\xi_t^i X_t^{i}\widehat{V}^{(p)}\widehat{R}_G^{(p)}(\widehat{\Lambda}^{(p)})^{-1})(\sum_{i=1}^{d_1}\xi_t^i X_t^{i}\widehat{V}^{(p)}\widehat{R}_G^{(p)}(\widehat{\Lambda}^{(p)})^{-1})^{\top} \right]\bigg\| \\ 
        &\quad \leq \sigma^2\|X_t\widehat{V}^{(p)}\widehat{R}_G^{(p)}(\widehat{\Lambda}^{(p)})^{-1}\|^2\lesssim \frac{\sigma^2}{\lambda_{\min}^2}.
    \end{align*}
    By matrix Bernstein inequality, with probability at least $1-d_1^{-10}$,
    \begin{align*}
        \|\widehat{E}_{\xi,V}^{(p)}\|\lesssim \eta\frac{\sigma}{\lambda_{\min}}\sqrt{\frac{d_2^2\log d_1}{N_0}}.
    \end{align*} 
    Similarly, with the same probability
    \begin{align*}
        \|\widehat{E}_V^{(p)}\|\lesssim \eta\frac{\|\widehat{M}^{(p)}-M\|_{\max}}{\lambda_{\min}}\sqrt{\frac{d_2^2\log d_1}{N_0}}.
    \end{align*}
    Combine with the previous result, with probability at least $1-2d_1^{-10}$,
    \begin{align*}
        \|\widehat{E}_V^{(p)} + \widehat{E}_{\xi,V}^{(p)}\|\lesssim \eta\frac{\sigma}{\lambda_{\min}}\sqrt{\frac{d_2^2\log d_1}{N_0}}  + \eta\sqrt{\frac{d_2^2\log d_1}{N_0}}\left(\sqrt{\frac{r}{d_2}}\|\widehat{U}^{(p)}-U\widehat{O}_U^{(p)}\|_{2,\max} + \sqrt{\frac{r}{d_1}}\|\widehat{V}^{(p)}-V\widehat{O}_V^{(p)}\|_{2,\max}\right).
    \end{align*}
    Note that the singular values of $\widehat{U}^{(p+0.5)}$ are the square root of eigenvalues of $\widehat{U}^{(p+0.5)\top}\widehat{U}^{(p+0.5)}$. We write
    \begin{align*}
        \widehat{U}^{(p+0.5)\top}\widehat{U}^{(p+0.5)}=\mathcal{I}_1^{\top}\mathcal{I}_1 + \mathcal{I}_2^{\top}\mathcal{I}_2 + \mathcal{I}_1^{\top}\mathcal{I}_2 + \mathcal{I}_2^{\top}\mathcal{I}_1.
    \end{align*}
    Since $\|\widehat{U}^{(p)\top}(\widehat{U}^{(p)}-U\widehat{O}_U^{(p)})\|\leq 2\|\widehat{U}^{(p)}-U\widehat{O}_U^{(p)}\|^2$ and $\|\widehat{V}^{(p)\top}(\widehat{V}^{(p)}-V\widehat{O}_V^{(p)})\|\leq 2\|\widehat{V}^{(p)}-V\widehat{O}_V^{(p)}\|^2$, by Lemma \ref{lemma:estimation1}, we get
    \begin{align*}
        &\|\mathcal{I}_1^{\top}\mathcal{I}_1-I\|\leq 3\kappa\eta(\|\widehat{U}^{(p)}-U\widehat{O}_U^{(p)}\|^2 + \|\widehat{V}^{(p)}-V\widehat{O}_V^{(p)}\|^2) + 2\frac{\eta}{\lambda_{\min}}\|\widehat{O}_U^{(p)}\widehat{G}^{(p)} - \Lambda\widehat{O}_V^{(p)} \| \\ &\quad\quad + 2\eta^2(\kappa^2\|\widehat{U}^{(p)}-U\widehat{O}_U^{(p)}\|^2 + \kappa^2\|\widehat{V}^{(p)}-V\widehat{O}_V^{(p)}\|^2 + \lambda_{\min}^{-2}\|\widehat{O}_U^{(p)}\widehat{G}^{(p)} - \Lambda\widehat{O}_V^{(p)} \|^2) \\
        &\quad\leq 3(\kappa^2\eta^2+\kappa\eta)(\|\widehat{U}^{(p)}-U\widehat{O}_U^{(p)}\|^2 + \|\widehat{V}^{(p)}-V\widehat{O}_V^{(p)}\|^2) + 4\frac{\eta}{\lambda_{\min}}\|\widehat{O}_U^{(p)}\widehat{G}^{(p)} - \Lambda\widehat{O}_V^{(p)} \| \\
        &\quad\lesssim (\kappa^2\eta^2+\kappa\eta)(\|\widehat{U}^{(p)}-U\widehat{O}_U^{(p)}\|^2 + \|\widehat{V}^{(p)}-V\widehat{O}_V^{(p)}\|^2) + \eta\frac{\sigma}{\lambda_{\min}}\left( \frac{r}{N_0}\sqrt{d_1d_2}\log d_1 + \sqrt{\frac{d_2r}{N_0}\log d_1}\right) \\ &\quad\quad +  \eta\kappa\left(\sqrt{\frac{r}{d_2}}\|\widehat{U}^{(p)}-U\widehat{O}_U^{(p)}\|_{2,\max} + \sqrt{\frac{r}{d_1}}\|\widehat{V}^{(p)}-V\widehat{O}_V^{(p)}\|_{2,\max}\right)\sqrt{\frac{d_1d_2r^2\log d_1}{N_0}}.
    \end{align*}
    Moreover, as long as $\|\widehat{U}^{(p)}-U\widehat{O}_U^{(p)}\| + \|\widehat{V}^{(p)}-V\widehat{O}_V^{(p)}\|\lesssim \frac{1}{\kappa}$, $\|\mathcal{I}_1\|\leq 2$, then $\|\mathcal{I}_1^{\top}\mathcal{I}_2 + \mathcal{I}_2^{\top}\mathcal{I}_1\|\leq 4\|\widehat{E}_V^{(p)} + \widehat{E}_{\xi,V}^{(p)}\|$. Finally, with probability at least $1-2d_1^{-10}$,
    \begin{align*}
        &\|\widehat{U}^{(p+0.5)\top}\widehat{U}^{(p+0.5)}-I\|\lesssim \eta\frac{\sigma}{\lambda_{\min}}\sqrt{\frac{d_2^2\log d_1}{N_0}} + (\kappa^2\eta^2+\kappa\eta)(\|\widehat{U}^{(p)}-U\widehat{O}_U^{(p)}\|^2 + \|\widehat{V}^{(p)}-V\widehat{O}_V^{(p)}\|^2) \\
        &\quad +  \eta\sqrt{\frac{d_2^2\log d_1}{N_0}}\left(\sqrt{\frac{r}{d_2}}\|\widehat{U}^{(p)}-U\widehat{O}_U^{(p)}\|_{2,\max} + \sqrt{\frac{r}{d_1}}\|\widehat{V}^{(p)}-V\widehat{O}_V^{(p)}\|_{2,\max}\right),
    \end{align*}
    implying that $\{|1-\lambda_{\min}(\widehat{U}^{(p+0.5)})|, |1-\lambda_{\max}(\widehat{U}^{(p+0.5)})|\}$ can be upper bounded by the above rate.
\end{proof}

\subsection{Proof of Lemma \ref{lemma:estimation2-1}}
\begin{proof}
    We first write
    \begin{align*}
        &\frac{d_2}{N_0Kp_0}\sum_{t\in \mathcal{D}_{2p}} \sum_{i=1}^{d_1}\sum_{q\in h_{it}} \inp{\widehat{U}^{(p)}\widehat{G}^{(p)}\widehat{V}^{(p)\top} - U\Lambda V^{\top}}{X_t^{iq}}\widehat{U}^{(p)\top}X_t^{iq}\widehat{V}^{(p)} \\ &\quad - \frac{d_2}{N_0Kp_0}\sum_{t\in \mathcal{D}_{2p}} \sum_{i=1}^{d_1}\sum_{q\in h_{it}} \xi_t^{iq}\widehat{U}^{(p)\top}X_t^{iq}\widehat{V}^{(p)}=0
    \end{align*}
    where due to data splitting, $(\widehat{U}^{(p)},\widehat{V}^{(p)})$ are independent with $\mathcal{D}_{2p}$. Note that from proof of Lemma \ref{lemma:estimation1},
    \begin{align*}
        &\widehat{G}^{(p)}-\widehat{O}_U^{(p)\top}\Lambda\widehat{O}_V^{(p)}=(\widehat{G}^{(p)}-\widehat{O}_U^{(p)\top}\Lambda\widehat{O}_V^{(p)})  + \frac{d_2}{N_0Kp_0}\sum_{t\in \mathcal{D}_{2p}} \sum_{i=1}^{d_1}\sum_{q\in h_{it}} \xi_t^{iq}\widehat{U}^{(p)\top}X_t^{iq}\widehat{V}^{(p)}\\ &\quad - \frac{d_2}{N_0Kp_0}\sum_{t\in \mathcal{D}_{2p}} \sum_{i=1}^{d_1}\sum_{q\in h_{it}}  \inp{\widehat{G}^{(p)} - \widehat{O}_U^{(p)\top}\Lambda\widehat{O}_V^{(p)}}{\widehat{U}^{(p)\top}X_t^{iq}\widehat{V}^{(p)}}\widehat{U}^{(p)\top}X_t^{iq}\widehat{V}^{(p)} \\ &\quad - \frac{d_2}{N_0Kp_0}\sum_{t\in \mathcal{D}_{2p}} \sum_{i=1}^{d_1}\sum_{q\in h_{it}}  \inp{\widehat{U}^{(p)}\widehat{O}_U^{(p)\top}\Lambda(V^{(p)}\widehat{O}_V^{(p)\top})^{\top}- U\Lambda V^{\top}}{X_t^{iq}}\widehat{U}^{(p)\top}X_t^{iq}\widehat{V}^{(p)}.
    \end{align*}
    Since $\|\widehat{U}^{(p)}\|\leq 2\mu\sqrt{r/d_1}$, $\|\widehat{V}^{(p)}\|\leq 2\mu\sqrt{r/d_2}$, then
    \begin{align*}
        \left\|\bigg\|\sum_{i=1}^{d_1}\sum_{q\in h_{it}} \xi_t^{iq}\widehat{U}^{(p)\top}X_t^{iq}\widehat{V}^{(p)}\bigg\|\right\|_{\psi_2}\lesssim \sigma\frac{K\mu^2r}{\sqrt{d_1d_2}}d_1=\sigma\mu^2rK\sqrt{\frac{d_1}{d_2}}
    \end{align*}
    and 
    \begin{align*}
        &\left\|\EE\left[\bigg(\sum_{i=1}^{d_1} \sum_{q\in h_{it}} \xi_t^{iq}\widehat{U}^{(p)\top}X_t^{iq}\widehat{V}^{(p)}\bigg)\bigg(\sum_{i=1}^{d_1}\sum_{q\in h_{it}} \xi_t^{iq}\widehat{U}^{(p)\top}X_t^{iq}\widehat{V}^{(p)}\bigg)^{\top}\right]\right\| \\ &\quad = \left\|\EE\left[\sum_{i=1}^{d_1} \sum_{q\in h_{it}}\xi_t^{iq2}\widehat{U}^{(p)\top}X_t^{iq}\widehat{V}^{(p)}\widehat{V}^{(p)\top}X_t^{iq\top}\widehat{U}^{(p)}\right]\right\|=\sigma^2\frac{rKp_0}{d_2}
    \end{align*}
    by the independence of $\xi_t^{iq}$. By matrix Bernstein inequality (\cite{tropp2012user}), with probability at least $1-d_1^{-10}$,
    \begin{align*}
        \left\|\frac{d_2}{N_0Kp_0}\sum_{t\in \mathcal{D}_{2p}} \sum_{i=1}^{d_1}\sum_{q\in h_{it}} \xi_t^{iq}\widehat{U}^{(p)\top}X_t^{iq}\widehat{V}^{(p)}\right\|\lesssim \frac{\sigma r}{N_0p_0}\sqrt{d_1d_2}\log d_1 + \sqrt{\frac{\sigma^2d_2r}{N_0Kp_0}\log d_1}.
    \end{align*}
    Also from the proof of Lemma \ref{lemma:estimation1},
    \begin{align*}
        &\left\|(\widehat{G}^{(p)}-\widehat{O}_U^{(p)\top}\Lambda\widehat{O}_V^{(p)}) - \frac{d_2}{N_0Kp_0}\sum_{t\in \mathcal{D}_{2p}} \sum_{i=1}^{d_1}\sum_{q\in h_{it}} \inp{\widehat{G}^{(p)} - \widehat{O}_U^{(p)\top}\Lambda\widehat{O}_V^{(p)}}{\widehat{U}^{(p)\top}X_t^{iq}\widehat{V}^{(p)}}\widehat{U}^{(p)\top}X_t^{iq}\widehat{V}^{(p)}\right\| \\
        &\quad \leq  \|\widehat{G}^{(p)}-\widehat{O}_U^{(p)\top}\Lambda\widehat{O}_V^{(p)}\|\max_{A\in \mathcal{N}_{1/3}(\mathcal{O}_r)} \left\|A - \frac{d_2}{N_0Kp_0}\sum_{t\in \mathcal{D}_{2p}} \sum_{i=1}^{d_1}\sum_{q\in h_{it}} \inp{A}{\widehat{U}^{(p)\top}X_t^{iq}\widehat{V}^{(p)}}\widehat{U}^{(p)\top}X_t^{iq}\widehat{V}^{(p)}\right\|.
    \end{align*}
    For each $A\in \mathcal{N}_{1/3}(\mathcal{O}_r)$,
    \begin{align*}
        \|\sum_{i=1}^{d_1}\sum_{q\in h_{it}} \inp{A}{\widehat{U}^{(p)\top}X_t^{iq}\widehat{V}^{(p)}}\widehat{U}^{(p)\top}X_t^i\widehat{V}^{(p)}\|&\leq \sum_{i=1}^{d_1}\sum_{q\in h_{it}} \|\widehat{U}^{(p)\top}X_t^{iq}\widehat{V}^{(p)}\|_*\|\widehat{U}^{(p)\top}X_t^{iq}\widehat{V}^{(p)}\| \\ &\leq d_1K \|\widehat{U}^{(p)}\|_{2,\max}^2\|\widehat{V}^{(p)}\|_{2,\max}^2\leq \frac{K\mu^2r^2}{d_2}
    \end{align*}
    where $\|\cdot\|_*$ denotes the matrix nuclear norm. Moreover, since each $|h_{it}|\sim B(K,p)$,
    \begin{align*}
        &\left\|\EE\left[\bigg(\sum_{i=1}^{d_1}\sum_{q\in h_{it}} \inp{A}{\widehat{U}^{(p)\top}X_t^{iq}\widehat{V}^{(p)}}\widehat{U}^{(p)\top}X_t^{iq}\widehat{V}^{(p)}\bigg)\bigg(\sum_{i=1}^{d_1}\sum_{q\in h_{it}} \inp{A}{\widehat{U}^{(p)\top}X_t^{iq}\widehat{V}^{(p)}}\widehat{U}^{(p)\top}X_t^{iq}\widehat{V}^{(p)}\bigg)^{\top}\right]\right\| \\
        &\quad \leq \max_{i,j}|\inp{A}{\widehat{U}^{(p)\top}e_ie_j^{\top}\widehat{V}^{(p)}}|^2(d_1^2K^2p_0^2+d_1Kp_0(1-p_0))\max_{i,j,k,l}\|(\widehat{U}^{(p)\top}e_ie_j^{\top}\widehat{V}^{(p)})(\widehat{U}^{(p)\top}e_ke_l^{\top}\widehat{V}^{(p)})^{\top}\| \\
        &\quad \leq 2\fro{A}^2\max_{i,j}\fro{\widehat{U}^{(p)\top}e_ie_j^{\top}\widehat{V}^{(p)}}^2d_1^2K^2p_0^2\max_{i,j}\|\widehat{U}^{(p)\top}e_ie_j^{\top}\widehat{V}^{(p)}\|^2\leq \frac{2K^2p^2\mu^4r^3}{d_2^2}.
    \end{align*}
    Then by matrix Bernstein inequality, with probability at least $1-d_1^{-10}$,
    \begin{align*}
        \left\|A - \frac{d_2}{N_0Kp_0}\sum_{t\in \mathcal{D}_{2p}} \sum_{i=1}^{d_1}\sum_{q\in h_{it}} \inp{A}{\widehat{U}^{(p)\top}X_t^{iq}\widehat{V}^{(p)}}\widehat{U}^{(p)\top}X_t^{iq}\widehat{V}^{(p)}\right\|\lesssim \sqrt{\frac{r^3\log d_1}{N_0}} + \frac{r^2\log d_1}{N_0p_0}.
    \end{align*}
   This directly implies that
    \begin{align*}
        &\left\|(\widehat{G}^{(p)}-\widehat{O}_U^{(p)\top}\Lambda\widehat{O}_V^{(p)}) - \frac{d_2}{N_0Kp_0}\sum_{t\in \mathcal{D}_{2p}} \sum_{i=1}^{d_1}\sum_{q\in h_{it}}  \inp{\widehat{G}^{(p)} - \widehat{O}_U^{(p)\top}\Lambda\widehat{O}_V^{(p)}}{\widehat{U}^{(p)\top}X_t^{iq}\widehat{V}^{(p)}}\widehat{U}^{(p)\top}X_t^{i}\widehat{V}^{(p)}\right\| \\
        &\quad \lesssim  \|\widehat{G}^{(p)}-\widehat{O}_U^{(p)\top}\Lambda\widehat{O}_V^{(p)}\|\left(\sqrt{\frac{r^3\log d_1}{N_0}} + \frac{r^2\log d_1}{N_0p_0}\right).
    \end{align*}
    Similarly, we first notice that 
    \begin{align*}
        &\left\|\EE\left[\sum_{i=1}^{d_1}\sum_{q\in h_{it}}  \inp{\widehat{U}^{(p)}\widehat{O}_U^{(p)\top}\Lambda(V^{(p)}\widehat{O}_V^{(p)\top})^{\top}- U\Lambda V^{\top}}{X_t^{iq}}\widehat{U}^{(p)\top}X_t^{iq}\widehat{V}^{(p)}\right]\right\| \\ &\quad =\frac{Kp_0}{d_2}\left\|\widehat{U}^{(p)\top}(\widehat{U}^{(p)}\widehat{O}_U^{(p)\top}\Lambda(V^{(p)}\widehat{O}_V^{(p)\top})^{\top}- U\Lambda V^{\top})\widehat{V}^{(p)}\right\|.
    \end{align*}
    We also calculate the uniform bound
    \begin{align*}
        &\left\|\sum_{i=1}^{d_1}\sum_{q\in h_{it}}  \inp{\widehat{U}^{(p)}\widehat{O}_U^{(p)\top}\Lambda(V^{(p)}\widehat{O}_V^{(p)\top})^{\top}- U\Lambda V^{\top}}{X_t^{iq}}\widehat{U}^{(p)\top}X_t^{iq}\widehat{V}^{(p)}\right\| \\ 
        &\quad \leq \|\widehat{U}^{(p)}\widehat{O}_U^{(p)\top}\Lambda(V^{(p)}\widehat{O}_V^{(p)\top})^{\top}- U\Lambda V^{\top}\|_{\max}d_1K\max_{i,j}\|\widehat{U}^{(p)\top}e_ie_j^{\top}\widehat{V}^{(p)}\| \\ 
        &\quad \leq \|\widehat{U}^{(p)}\widehat{O}_U^{(p)\top}\Lambda(V^{(p)}\widehat{O}_V^{(p)\top})^{\top}- U\Lambda V^{\top}\|_{\max}\mu^2rK\sqrt{\frac{d_1}{d_2}}
    \end{align*}
    and variance
    \begin{align*}
        &\left\|\EE\left[\bigg(\sum_{i=1}^{d_1}\sum_{q\in h_{it}}  \inp{\widehat{U}^{(p)}\widehat{O}_U^{(p)\top}\Lambda(V^{(p)}\widehat{O}_V^{(p)\top})^{\top}- U\Lambda V^{\top}}{X_t^{iq}}\widehat{U}^{(p)\top}X_t^{iq}\widehat{V}^{(p)}\bigg) \right.\right.\\ &\left.\left.\quad \bigg(\sum_{i=1}^{d_1}\sum_{q\in h_{it}}  \inp{\widehat{U}^{(p)}\widehat{O}_U^{(p)\top}\Lambda(V^{(p)}\widehat{O}_V^{(p)\top})^{\top}- U\Lambda V^{\top}}{X_t^{iq}}\widehat{U}^{(p)\top}X_t^{iq}\widehat{V}^{(p)}\bigg)^{\top}\right]\right\| \\
        &\quad\quad \leq \|\widehat{U}^{(p)}\widehat{O}_U^{(p)\top}\Lambda(V^{(p)}\widehat{O}_V^{(p)\top})^{\top}- U\Lambda V^{\top}\|_{\max}^2(d_1^2K^2p_0^2+d_1Kp_0(1-p_0))\max_{i,j}\|\widehat{U}^{(p)\top}e_ie_j^{\top}\widehat{V}^{(p)}\|^2 \\
        &\quad\quad \leq \|\widehat{U}^{(p)}\widehat{O}_U^{(p)\top}\Lambda(V^{(p)}\widehat{O}_V^{(p)\top})^{\top}- U\Lambda V^{\top}\|_{\max}^2\frac{2K^2p_0^2\mu^4r^2d_1}{d_2}
    \end{align*}
    By matrix Bernstein inequality, with probability at least $1-d_1^{-10}$, 
    \begin{align*}
        &\left\|\frac{d_2}{N_0Kp_0}\sum_{t\in \mathcal{D}_{2p}} \sum_{i=1}^{d_1}\sum_{q\in h_{it}}  \inp{\widehat{U}^{(p)}\widehat{O}_U^{(p)\top}\Lambda(V^{(p)}\widehat{O}_V^{(p)\top})^{\top}- U\Lambda V^{\top}}{X_t^{iq}}\widehat{U}^{(p)\top}X_t^{iq}\widehat{V}^{(p)}\right\| \\
        &\quad \leq \left\|\widehat{U}^{(p)\top}(\widehat{U}^{(p)}\widehat{O}_U^{(p)\top}\Lambda(V^{(p)}\widehat{O}_V^{(p)\top})^{\top}- U\Lambda V^{\top})\widehat{V}^{(p)}\right\| \\ &\quad\quad + C_2\|\widehat{U}^{(p)}\widehat{O}_U^{(p)\top}\Lambda(V^{(p)}\widehat{O}_V^{(p)\top})^{\top}- U\Lambda V^{\top}\|_{\max}\sqrt{\frac{d_1d_2r^2\log d_1}{N_0p_0}}
    \end{align*}
    as long as $T\gtrsim p_0^{-1}\log d_1$. Then following the last parts of the proof of Lemma \ref{lemma:estimation1}, as long as $T\gtrsim p_0^{-1}r^3\log d_1$, with probability at least $1-3d_1^{-10}$,
    \begin{align*}
        &\|\widehat{G}^{(p)}-\widehat{O}_U^{(p)\top}\Lambda\widehat{O}_V^{(p)}\|\leq C_3\left( \frac{\sigma r}{N_0p_0}\sqrt{d_1d_2}\log d_1 + \sqrt{\frac{\sigma^2d_2r}{N_0Kp_0}\log d_1}\right) \\ &\quad + 2\lambda_{\max}(\|\widehat{U}^{(p)}-U\widehat{O}_U^{(p)}\|^2 + \|\widehat{V}^{(p)}-V\widehat{O}_V^{(p)}\|^2) \\ &\quad + C_4\lambda_{\max}\left(\sqrt{\frac{r}{d_2}}\|\widehat{U}^{(p)}-U\widehat{O}_U^{(p)}\|_{2,\max} + \sqrt{\frac{r}{d_1}}\|\widehat{V}^{(p)}-V\widehat{O}_V^{(p)}\|_{2,\max}\right)\sqrt{\frac{d_1d_2r^2\log d_1}{N_0p_0}}.
    \end{align*}
\end{proof}

\subsection{Proof of Lemma \ref{lemma:estimation2-2}}
\begin{proof}
    Recall that, 
    \begin{align*}
        &\widehat{U}^{(p+0.5)}-U\widehat{O}_U^{(p)}\widehat{L}_G^{(p)} = (\widehat{U}^{(p)}\widehat{L}_G^{(p)}-U\widehat{O}_U^{(p)}\widehat{L}_G^{(p)})(I-\eta\widehat{L}_G^{(p)\top}\widehat{O}_U^{(p)\top}\Lambda\widehat{O}_V^{(p)}\widehat{R}_G^{(p)}(\widehat{\Lambda}^{(p)})^{-1}) \\
        &\quad -\eta \widehat{U}^{(p)}(\widehat{G}^{(p)}-\widehat{O}_U^{(p)\top}\Lambda\widehat{O}_V^{(p)})\widehat{R}_G^{(p)}(\widehat{\Lambda}^{(p)})^{-1} - \eta U\Lambda(\widehat{V}^{(p)}\widehat{O}_V^{(p)\top}-V)^{\top}\widehat{V}^{(p)}\widehat{R}_G^{(p)}(\widehat{\Lambda}^{(p)})^{-1} + \widehat{E}_V^{(p)} + \widehat{E}_{\xi,V}^{(p)}.
    \end{align*}
    By Lemma \ref{lemma:estimation2-1}, 
    \begin{align*}
        &\|\widehat{\Lambda}^{(p)}-\widehat{L}_G^{(p)\top}\widehat{O}_U^{(p)\top}\Lambda\widehat{O}_V^{(p)}\widehat{R}_G^{(p)}\|\leq C_3(\frac{r\sigma}{N_0p_0}\sqrt{d_1d_2}\log d_1 + \sqrt{\frac{\sigma^2 rd_2\log d_1}{N_0Kp_0}}) \\ &\quad + 2\lambda_{\max}(\|\widehat{U}^{(p)}-U\widehat{O}_U^{(p)}\|^2+\|\widehat{V}^{(p)}-V\widehat{O}_U^{(p)}\|^2) \\ &\quad + C_4\lambda_{\max}\left(\sqrt{\frac{1}{d_2}}\|\widehat{U}^{(p)}-U\widehat{O}_U^{(p)}\|_{2,\max} + \sqrt{\frac{1}{d_1}}\|\widehat{V}^{(p)}-V\widehat{O}_V^{(p)}\|_{2,\max}\right)\mu\sqrt{\frac{r^3d_1d_2\log d_1}{N_0p_0}}
    \end{align*}
    which implies $\|\widehat{\Lambda}^{(p)}-\widehat{L}_G^{(p)\top}\widehat{O}_U^{(p)\top}\Lambda\widehat{O}_V^{(p)}\widehat{R}_G^{(p)}\|\leq \lambda_{\min}/20$ and $\lambda_{\min}(\widehat{\Lambda}^{(p)})\geq \lambda_{\min}/2$ as long as $\lambda_{\min}/\sigma \gtrsim \max\{\sqrt{d_1d_2}/Tp_0, \sqrt{d_2/TKp_0}\}r\log d_1$, $T\gtrsim p_0^{-1}r^3\log d_1$, $\|\widehat{U}^{(p)}\|_{2,\max}\leq 2\mu\sqrt{r/d_1}$, $\|\widehat{V}^{(p)}\|_{2,\max}\leq 2\mu\sqrt{r/d_2}$ and $\max\{\|\widehat{U}^{(p)}-U\widehat{O}_U^{(p)}\|, \|\widehat{V}^{(p)}-V\widehat{O}_V^{(p)}\|\}\lesssim 1/\sqrt{\kappa}$. 

    Since $\eta\leq 0.75$, by proof of Lemma \ref{lemma:estimation2}, we have
    \begin{align*}
        \|(\widehat{U}^{(p)}\widehat{L}_G^{(p)}-U\widehat{O}_U^{(p)}\widehat{L}_G^{(p)})(I-\eta\widehat{L}_G^{(p)\top}\widehat{O}_U^{(p)\top}\Lambda\widehat{O}_V^{(p)}\widehat{R}_G^{(p)}(\widehat{\Lambda}^{(p)})^{-1})\|_{2,\max}\leq (1-\frac{9\eta}{10})\|\widehat{U}^{(p)}-U\widehat{O}_U^{(p)}\|_{2,\max}.
    \end{align*}
    By $\|\widehat{U}^{(p)}\|_{2,\max}\leq 2\mu\sqrt{r/d_1}$, we have
    \begin{align*}
        &\eta\|\widehat{U}^{(p)}(\widehat{G}^{(p)}-\widehat{O}_U^{(p)\top}\Lambda\widehat{O}_V^{(p)})\widehat{R}_G^{(p)}(\widehat{\Lambda}^{(p)})^{-1} \|_{2,\max}\leq 2\eta\|\widehat{U}^{(p)}\|_{2,\max}\|\widehat{G}^{(p)}-\widehat{O}_U^{(p)\top}\Lambda\widehat{O}_V^{(p)}\|\lambda_{\min}^{-1} \\
        &\leq C_3\eta\mu(\frac{r^{3/2}\sigma}{\lambda_{\min}}\frac{\sqrt{d_2}}{N_0p_0}\log d_1 + \frac{\sigma}{\lambda_{\min}}\sqrt{\frac{r^2d_2\log d_1}{N_0p_0d_1}})  + C_4\eta\frac{\mu\kappa r^{1/2}}{\sqrt{d_1}}(\|\widehat{U}^{(p)}-U\widehat{O}_U^{(p)}\|^2+\|\widehat{V}^{(p)}-V\widehat{O}_U^{(p)}\|^2) \\ &\quad + C_5\eta\mu\kappa\left(\sqrt{\frac{1}{d_2}}\|\widehat{U}^{(p)}-U\widehat{O}_U^{(p)}\|_{2,\max} + \sqrt{\frac{1}{d_1}}\|\widehat{V}^{(p)}-V\widehat{O}_V^{(p)}\|_{2,\max}\right)\mu\sqrt{\frac{r^4d_2\log d_1}{N_0p_0}}.
    \end{align*}
    Again according to proof of Lemma \ref{lemma:estimation2}, if $T\gtrsim p_0^{-1}\mu^4\kappa^2r^4\alpha_d\log d_1$, 
    \begin{align*}
        &C_5\mu\kappa\left(\sqrt{\frac{1}{d_2}}\|\widehat{U}^{(p)}-U\widehat{O}_U^{(p)}\|_{2,\max} + \sqrt{\frac{1}{d_1}}\|\widehat{V}^{(p)}-V\widehat{O}_V^{(p)}\|_{2,\max}\right)\mu\sqrt{\frac{r^4d_2\log d_1}{N_0p_0}} \\
        &\leq \frac{\sqrt{\frac{d_1}{d_2}}\|\widehat{U}^{(p)}-U\widehat{O}_U^{(p)}\|_{2,\max} + \|\widehat{V}^{(p)}-V\widehat{O}_V^{(p)}\|_{2,\max}}{40}.
    \end{align*}
    Then we have
    \begin{align*}
        &\eta\|\widehat{U}^{(p)}(\widehat{G}^{(p)}-\widehat{O}_U^{(p)\top}\Lambda\widehat{O}_V^{(p)})\widehat{R}_G^{(p)}(\widehat{\Lambda}^{(p)})^{-1} \|_{2,\max}\leq C_3\mu(\frac{r^{3/2}\sigma}{\lambda_{\min}}\frac{\sqrt{d_2}}{N_0p_0}\log d_1 + \frac{\sigma}{\lambda_{\min}}\sqrt{\frac{r^2d_2\log d_1}{N_0p_0d_1}}) \\
        &\quad + \frac{3\eta}{40}\left(\sqrt{\frac{d_1}{d_2}}\|\widehat{U}^{(p)}-U\widehat{O}_U^{(p)}\|_{2,\max} + \|\widehat{V}^{(p)}-V\widehat{O}_V^{(p)}\|_{2,\max}\right).
    \end{align*}
    Since $\|(\widehat{V}^{(p)}\widehat{O}_V^{(p)\top}-V)^{\top}\widehat{V}^{(p)}\|\leq \|\widehat{V}^{(p)}\widehat{O}_V^{(p)\top}-V\|^2$, we get
    \begin{align*}
        \|U\Lambda(\widehat{V}^{(p)}\widehat{O}_V^{(p)\top}-V)^{\top}\widehat{V}^{(p)}\widehat{R}_G^{(p)}(\widehat{\Lambda}^{(p)})^{-1}\|_{2,\max}&\leq 2\kappa\mu\sqrt{\frac{r}{d_1}}\|\widehat{V}^{(p)}\widehat{O}_V^{(p)\top}-V\|^2 \\ &\leq \frac{1}{40}\|\widehat{V}^{(p)}-V\widehat{O}_V^{(p)\top}\|_{2,\max}
    \end{align*}
    as long as $\|\widehat{V}^{(p)}\widehat{O}_V^{(p)\top}-V\|\lesssim 1/\kappa\mu\sqrt{r}\sqrt{d_2/d_1}$. Putting together the above bounds, we obtain
    \begin{align*}
        &\|\widehat{U}^{(p+0.5)}-U\widehat{O}_U^{(p)}\widehat{L}_G^{(p)}\|_{2,\max}\leq (1-\frac{9\eta}{10})\|\widehat{U}^{(p)}-U\widehat{O}_U^{(p)}\|_{2,\max} + C_3\mu(\frac{r^{3/2}\sigma}{\lambda_{\min}}\frac{\sqrt{d_2}}{N_0p_0}\log d_1 + \frac{\sigma}{\lambda_{\min}}\sqrt{\frac{r^2d_2\log d_1}{N_0p_0d_1}}) \\ &\quad + \frac{\eta}{10}\left(\sqrt{\frac{d_1}{d_2}}\|\widehat{U}^{(p)}-U\widehat{O}_U^{(p)}\|_{2,\max} + \|\widehat{V}^{(p)}-V\widehat{O}_V^{(p)}\|_{2,\max}\right) + \|\widehat{E}_V^{(p)}\|_{2,\max} + \|\widehat{E}_{\xi,V}^{(p)}\|_{2,\max}.
    \end{align*}
    Since $(\widehat{U}^{(p)}, \widehat{G}^{(p)}, \widehat{V}^{(p)})$ are independent with $\mathcal{D}_{2p+1}$ and $\widehat{U}^{(p)}, \widehat{V}^{(p)}$ are incoherent, for any $j\in [d_1]$, we have
    \begin{align*}
        \bigg\|\big\|e_j^{\top}\sum_{i=1}^{d_1}\sum_{q\in h_{it}} \xi_t^{iq} X_t^{iq}\widehat{V}^{(p)}\widehat{R}_G^{(p)}(\widehat{\Lambda}^{(p)})^{-1}\big\|\bigg\|_{\psi_2}\lesssim K\sigma\|\widehat{V}^{(p)}\|_{2,\max}\lambda_{\min}^{-1}\lesssim K\frac{\sigma}{\lambda_{\min}}\sqrt{\frac{r}{d_2}}
    \end{align*}
    and 
    \begin{align*}
        &\EE\left[(e_j^{\top}\sum_{i=1}^{d_1}\sum_{q\in h_{it}} \xi_t^{iq} X_t^{iq}\widehat{V}^{(p)}\widehat{R}_G^{(p)}(\widehat{\Lambda}^{(p)})^{-1})(e_j^{\top}\sum_{i=1}^{d_1}\sum_{q\in h_{it}}  \xi_t^{iq} X_t^{iq}\widehat{V}^{(p)}\widehat{R}_G^{(p)}(\widehat{\Lambda}^{(p)})^{-1})^{\top} \right] \\
        &\quad \leq \sigma^2Kp_0\|\widehat{V}^{(p)}\widehat{R}_G^{(p)}(\widehat{\Lambda}^{(p)})^{-1}\|_{2,\max}^2\lesssim \frac{\sigma^2Kp_0}{\lambda_{\min}^2}\frac{r}{d_2}.
    \end{align*}
    since $e_j^{\top}X_t^{iq}\neq 0$ only when $i=j$ and $|h_{it}|\sim B(K,p_0)$. Then by Bernstein inequality, as long as $T\gtrsim Kp_0^{-1}\log d_1$, with probability at least $1-d_1^{-10}$,
    \begin{align*}
        \|\widehat{E}_{\xi,V}^{(p)}\|_{2,\max}\lesssim \eta\frac{\sigma}{\lambda_{\min}}\sqrt{\frac{rd_2\log d_1}{N_0Kp}}.
    \end{align*}
    Notice that, $\alpha_d=d_2/d_1\geq K$, then the condition $T\gtrsim Kp_0^{-1}\log d_1$ can also be written as $T\gtrsim \alpha_dp_0^{-1}\log d_1$. Similarly, denote $\widehat{M}^{(p)}=\widehat{U}^{(p)}\widehat{G}^{(p)}\widehat{V}^{(p)\top}$, we have
    \begin{align*}
        \bigg\|\big\|e_j^{\top}\sum_{i=1}^{d_1}\sum_{q\in h_{it}} \inp{\widehat{M}^{(p)}-M}{X_t^{iq}} X_t^{iq}\widehat{V}^{(p)}\widehat{R}_G^{(p)}(\widehat{\Lambda}^{(p)})^{-1}\big\|\bigg\|_{\psi_2}\lesssim K\frac{\|\widehat{M}^{(p)}-M\|_{\max}}{\lambda_{\min}}\sqrt{\frac{r}{d_2}}
    \end{align*}
    and 
    \begin{align*}
        &\EE\left[(e_j^{\top}\sum_{i=1}^{d_1}\sum_{q\in h_{it}} \inp{\widehat{M}^{(p)}-M}{X_t^{iq}} X_t^{iq}\widehat{V}^{(p)}\widehat{R}_G^{(p)}(\widehat{\Lambda}^{(p)})^{-1})(e_j^{\top}\sum_{i=1}^{d_1}\sum_{q\in h_{it}} \inp{\widehat{M}^{(p)}-M}{X_t^{iq}}  X_t^{iq}\widehat{V}^{(p)}\widehat{R}_G^{(p)}(\widehat{\Lambda}^{(p)})^{-1})^{\top} \right] \\
        &\quad \leq \|\widehat{M}^{(p)}-M\|_{\max}^2(K^2p_0^2+Kp_0(1-p_0))\|\widehat{V}^{(p)}\widehat{R}_G^{(p)}(\widehat{\Lambda}^{(p)})^{-1}\|_{2,\max}^2\lesssim \frac{\sigma^2K^2p_0^2}{\lambda_{\min}^2}\frac{r}{d_2}.
    \end{align*}
    As long as $T\gtrsim p_0^{-1}\log d_1$, with probability at least $1-d_1^{-10}$,
    \begin{align*}
        \|\widehat{E}_V^{(p)}\|_{2,\max}\lesssim \eta\frac{\|\widehat{M}^{(p)}-M\|_{\max}}{\lambda_{\min}}\sqrt{\frac{rd_2\log d_1}{N_0p_0}}.
    \end{align*}
    Note that
    \begin{align*}
        &\|\widehat{M}^{(p)}-M\|_{\max}\leq \|(\widehat{U}^{(p)}-U\widehat{O}_U^{(p)})\widehat{G}^{(p)}\widehat{V}^{(p)\top}\|_{\max} + \|U(\widehat{O}_U^{(p)}\widehat{G}^{(p)}-\Lambda\widehat{O}_V^{(p)})\widehat{V}^{(p)\top}\|_{\max} \\ &\quad\quad  + \|U\Lambda(\widehat{V}^{(p)}\widehat{O}_V^{(p)}-V)^{\top}\|_{2,\max} \\
        &\quad \leq 2\mu\lambda_{\max}\left(\sqrt{\frac{r}{d_2}}\|\widehat{U}^{(p)}-U\widehat{O}_U^{(p)}\|_{2,\max} + \sqrt{\frac{r}{d_1}}\|\widehat{V}^{(p)}-V\widehat{O}_V^{(p)}\|_{2,\max}\right)  + \mu^2\sqrt{\frac{r^2}{d_1d_2}}\|\widehat{G}^{(p)}-\widehat{O}_U^{(p)}\Lambda\widehat{O}_V^{(p)}\|.
    \end{align*}
    Together with Lemma \ref{lemma:estimation2-1},
    \begin{align*}
        &\mu^2\sqrt{\frac{r^2}{d_1d_2}}\|\widehat{G}^{(p)}-\widehat{O}_U^{(p)}\Lambda\widehat{O}_V^{(p)}\|\leq C_3\mu^2(\frac{r^2\sigma}{N_0p_0}\log d_1 + \sqrt{\frac{\sigma^2 r^3\log d_1}{d_1N_0Kp_0}}) \\ &\quad + 2\lambda_{\max}\mu^2\sqrt{\frac{r^2}{d_1d_2}}(\|\widehat{U}^{(p)}-U\widehat{O}_U^{(p)}\|^2+\|\widehat{V}^{(p)}-V\widehat{O}_U^{(p)}\|^2) \\ &\quad + C_4\lambda_{\max}\left(\sqrt{\frac{1}{d_2}}\|\widehat{U}^{(p)}-U\widehat{O}_U^{(p)}\|_{2,\max} + \sqrt{\frac{1}{d_1}}\|\widehat{V}^{(p)}-V\widehat{O}_V^{(p)}\|_{2,\max}\right)\mu^3\sqrt{\frac{r^5\log d_1}{N_0p_0}}.
    \end{align*}
    As long as $T\gtrsim p_0^{-1}\mu^6r^4\log d_1$ and $\|\widehat{U}^{(p)}-U\widehat{O}_U^{(p)}\|+\|\widehat{V}^{(p)}-V\widehat{O}_U^{(p)}\|\lesssim \frac{1}{\mu^2r}$,
    \begin{align*}
        \|\widehat{M}^{(p)}-M\|_{\max}\leq \frac{\sigma}{\sqrt{Kp_0}}+2\mu\lambda_{\max}\left(\sqrt{\frac{r}{d_2}}\|\widehat{U}^{(p)}-U\widehat{O}_U^{(p)}\|_{2,\max} + \sqrt{\frac{r}{d_1}}\|\widehat{V}^{(p)}-V\widehat{O}_V^{(p)}\|_{2,\max}\right).
    \end{align*}
    Therefore, as long as $T\gtrsim p_0^{-1}\kappa^2\mu^6r^4\alpha_d\log d_1$, with probability at least $1-2d_1^{-10}$,
    \begin{align*}
        \|\widehat{E}_V^{(p)}\|_{2,\max} + \|\widehat{E}_{\xi,V}^{(p)}\|_{2,\max}\leq C_6\eta\frac{\sigma}{\lambda_{\min}}\sqrt{\frac{rd_2\log d_1}{N_0Kp_0}} + \frac{\eta}{40}\left(\sqrt{\frac{d_1}{d_2}}\|\widehat{U}^{(p)}-U\widehat{O}_U^{(p)}\|_{2,\max} + \|\widehat{V}^{(p)}-V\widehat{O}_V^{(p)}\|_{2,\max}\right).
    \end{align*}
    As a result, as long as $T\gtrsim p_0^{-1}\kappa^2\mu^6r^4\alpha_d\log d_1$, 
    \begin{align*}
        &\|\widehat{U}^{(p+0.5)}-U\widehat{O}_U^{(p)}\widehat{L}_G^{(p)}\|_{2,\max}\leq (1-\frac{9\eta}{10})\|\widehat{U}^{(p)}-U\widehat{O}_U^{(p)}\|_{2,\max} + C_6\eta\frac{\sigma}{\lambda_{\min}}\sqrt{\frac{rd_2\log d_1}{N_0Kp_0}}\\
        &\quad + \frac{\eta}{8}\left(\sqrt{\frac{d_1}{d_2}}\|\widehat{U}^{(p)}-U\widehat{O}_U^{(p)}\|_{2,\max} + \|\widehat{V}^{(p)}-V\widehat{O}_V^{(p)}\|_{2,\max}\right).
    \end{align*}
    Next, we investigate the singular values of $\widehat{U}^{(p+0.5)}-U\widehat{O}_U^{(p)}\widehat{L}_G^{(p)}$. Recall
    \begin{align*}
        \widehat{U}^{(p+0.5)}= \underbrace{\widehat{U}^{(p)}\widehat{L}_G^{(p)}-\eta(\widehat{U}^{(p)}\widehat{G}^{(p)}\widehat{V}^{(p)\top} - U\Lambda V^{\top})\widehat{V}^{(p)}\widehat{R}_G^{(p)}(\widehat{\Lambda}^{(p)})^{-1}}_{\mathcal{I}_1} + \underbrace{\widehat{E}_V^{(p)} + \widehat{E}_{\xi,V}^{(p)}}_{\mathcal{I}_2}. 
    \end{align*}
    By the independence between $(\widehat{U}^{(p)}, \widehat{G}^{(p)}, \widehat{V}^{(p)}, \widehat{L}_G^{(p)}, \widehat{R}_G^{(p)}, \widehat{\Lambda}^{(p)})$ and $\mathcal{D}_{2p+1}$, and notice that $\|\sum_{i=1}^{d_1} \sum_{q\in h_{it}} X_t^{iq}\| = \sqrt{\max_i |h_{it}|}\leq\sqrt{K}$, we have
    \begin{align*}
        \bigg\|\big\|\sum_{i=1}^{d_1}\sum_{q\in h_{it}} \xi_t^{iq} X_t^{iq}\widehat{V}^{(p)}\widehat{R}_G^{(p)}(\widehat{\Lambda}^{(p)})^{-1} \big\|\bigg\|_{\psi_2}\lesssim \frac{\sigma\sqrt{K}}{\lambda_{\min}}.
    \end{align*}
    Applying the Chernoff's bound for $|h_{it}|$ and taking a union bound over $d_1$ rows, we have with high probability, $\max_i |h_{it}|\lesssim Kp_0+\sqrt{Kp_0\log d_1} + \log d_1\lesssim Kp_0\log d_1$, then
    \begin{align*}
        &\bigg\|\EE\left[(\sum_{i=1}^{d_1}\sum_{q\in h_{it}} \xi_t^{iq} X_t^{iq}\widehat{V}^{(p)}\widehat{R}_G^{(p)}(\widehat{\Lambda}^{(p)})^{-1})(\sum_{i=1}^{d_1}\sum_{q\in h_{it}} \xi_t^{iq} X_t^{iq}\widehat{V}^{(p)}\widehat{R}_G^{(p)}(\widehat{\Lambda}^{(p)})^{-1})^{\top} \right]\bigg\| \\ 
        &\quad \leq \sigma^2\EE\left[\bigg\|\sum_{i=1}^{d_1}\sum_{q\in h_{it}} X_t^{iq}\widehat{V}^{(p)}\widehat{R}_G^{(p)}(\widehat{\Lambda}^{(p)})^{-1}\bigg\|^2\right]\lesssim \frac{\sigma^2Kp_0\log d_1}{\lambda_{\min}^2}.
    \end{align*}
    By matrix Bernstein inequality, with probability at least $1-d_1^{-10}$,
    \begin{align*}
        \|\widehat{E}_{\xi,V}^{(p)}\|\lesssim \eta\frac{\sigma}{\lambda_{\min}}\sqrt{\frac{d_2^2\log^2 d_1}{N_0Kp_0}}.
    \end{align*} 
    Similarly, with the same probability
    \begin{align*}
        \|\widehat{E}_V^{(p)}\|\lesssim \eta\frac{\|\widehat{M}^{(p)}-M\|_{\max}}{\lambda_{\min}}\sqrt{\frac{d_2^2\log^2 d_1}{N_0Kp_0}}.
    \end{align*}
    Combine with the previous result, with probability at least $1-2d_1^{-10}$,
    \begin{align*}
        \|\widehat{E}_V^{(p)} + \widehat{E}_{\xi,V}^{(p)}\|\lesssim \eta\frac{\sigma}{\lambda_{\min}}\sqrt{\frac{d_2^2\log^2 d_1}{N_0Kp_0}}  + \eta\sqrt{\frac{d_2^2\log^2 d_1}{N_0Kp_0}}\left(\sqrt{\frac{r}{d_2}}\|\widehat{U}^{(p)}-U\widehat{O}_U^{(p)}\|_{2,\max} + \sqrt{\frac{r}{d_1}}\|\widehat{V}^{(p)}-V\widehat{O}_V^{(p)}\|_{2,\max}\right).
    \end{align*}
    Note that the singular values of $\widehat{U}^{(p+0.5)}$ are the square root of eigenvalues of $\widehat{U}^{(p+0.5)\top}\widehat{U}^{(p+0.5)}$. We write
    \begin{align*}
        \widehat{U}^{(p+0.5)\top}\widehat{U}^{(p+0.5)}=\mathcal{I}_1^{\top}\mathcal{I}_1 + \mathcal{I}_2^{\top}\mathcal{I}_2 + \mathcal{I}_1^{\top}\mathcal{I}_2 + \mathcal{I}_2^{\top}\mathcal{I}_1.
    \end{align*}
    Since $\|\widehat{U}^{(p)\top}(\widehat{U}^{(p)}-U\widehat{O}_U^{(p)})\|\leq 2\|\widehat{U}^{(p)}-U\widehat{O}_U^{(p)}\|^2$ and $\|\widehat{V}^{(p)\top}(\widehat{V}^{(p)}-V\widehat{O}_V^{(p)})\|\leq 2\|\widehat{V}^{(p)}-V\widehat{O}_V^{(p)}\|^2$, by Lemma \ref{lemma:estimation2-1} and proof of Lemma \ref{lemma:estimation2}, we get
    \begin{align*}
        &\|\mathcal{I}_1^{\top}\mathcal{I}_1-I\|\leq 3(\kappa^2\eta^2+\kappa\eta)(\|\widehat{U}^{(p)}-U\widehat{O}_U^{(p)}\|^2 + \|\widehat{V}^{(p)}-V\widehat{O}_V^{(p)}\|^2) + 4\frac{\eta}{\lambda_{\min}}\|\widehat{O}_U^{(p)}\widehat{G}^{(p)} - \Lambda\widehat{O}_V^{(p)} \| \\
        &\quad\lesssim (\kappa^2\eta^2+\kappa\eta)(\|\widehat{U}^{(p)}-U\widehat{O}_U^{(p)}\|^2 + \|\widehat{V}^{(p)}-V\widehat{O}_V^{(p)}\|^2) + \eta\frac{\sigma}{\lambda_{\min}}\left( \frac{r}{N_0p_0}\sqrt{d_1d_2}\log d_1 + \sqrt{\frac{d_2r}{N_0Kp_0}\log d_1}\right) \\ &\quad\quad +  \eta\kappa\left(\sqrt{\frac{r}{d_2}}\|\widehat{U}^{(p)}-U\widehat{O}_U^{(p)}\|_{2,\max} + \sqrt{\frac{r}{d_1}}\|\widehat{V}^{(p)}-V\widehat{O}_V^{(p)}\|_{2,\max}\right)\sqrt{\frac{d_1d_2r^2\log d_1}{N_0p_0}}.
    \end{align*}
    Moreover, as long as $\|\widehat{U}^{(p)}-U\widehat{O}_U^{(p)}\| + \|\widehat{V}^{(p)}-V\widehat{O}_V^{(p)}\|\lesssim \frac{1}{\kappa}$, $\|\mathcal{I}_1\|\leq 2$, then $\|\mathcal{I}_1^{\top}\mathcal{I}_2 + \mathcal{I}_2^{\top}\mathcal{I}_1\|\leq 4\|\widehat{E}_V^{(p)} + \widehat{E}_{\xi,V}^{(p)}\|$. Finally, with probability at least $1-2d_1^{-10}$,
    \begin{align*}
        &\|\widehat{U}^{(p+0.5)\top}\widehat{U}^{(p+0.5)}-I\|\lesssim \eta\frac{\sigma}{\lambda_{\min}}\sqrt{\frac{d_2^2\log^2 d_1}{N_0Kp_0}} + (\kappa^2\eta^2+\kappa\eta)(\|\widehat{U}^{(p)}-U\widehat{O}_U^{(p)}\|^2 + \|\widehat{V}^{(p)}-V\widehat{O}_V^{(p)}\|^2) \\
        &\quad +  \eta\sqrt{\frac{d_2^2\log^2 d_1}{N_0Kp_0}}\left(\sqrt{\frac{r}{d_2}}\|\widehat{U}^{(p)}-U\widehat{O}_U^{(p)}\|_{2,\max} + \sqrt{\frac{r}{d_1}}\|\widehat{V}^{(p)}-V\widehat{O}_V^{(p)}\|_{2,\max}\right),
    \end{align*}
    implying that $\{|1-\lambda_{\min}(\widehat{U}^{(p+0.5)})|, |1-\lambda_{\max}(\widehat{U}^{(p+0.5)})|\}$ can be upper bounded by the above rate.
\end{proof}

\subsection{Proof of Lemma \ref{lemma:estimation3-1}}
\begin{proof}
    We first write
    \begin{align*}
        &\frac{1}{N_0\nu}\sum_{t\in \mathcal{D}_{2p}} \sum_{(i,j)\in g_{t}} \inp{\widehat{U}^{(p)}\widehat{G}^{(p)}\widehat{V}^{(p)\top} - U\Lambda V^{\top}}{X_t^{ij}}\widehat{U}^{(p)\top}X_t^{ij}\widehat{V}^{(p)} - \frac{1}{N_0\nu}\sum_{t\in \mathcal{D}_{2p}} \sum_{(i,j)\in h_{t}} \xi_t^{ij}\widehat{U}^{(p)\top}X_t^{ij}\widehat{V}^{(p)}=0
    \end{align*}
    where due to data splitting, $(\widehat{U}^{(p)},\widehat{V}^{(p)})$ are independent with $\mathcal{D}_{2p}$. Note that from proof of Lemma \ref{lemma:estimation1},
    \begin{align*}
        &\widehat{G}^{(p)}-\widehat{O}_U^{(p)\top}\Lambda\widehat{O}_V^{(p)}=(\widehat{G}^{(p)}-\widehat{O}_U^{(p)\top}\Lambda\widehat{O}_V^{(p)})  + \frac{1}{N_0\nu}\sum_{t\in \mathcal{D}_{2p}} \sum_{(i,j)\in g_{t}} \xi_t^{ij}\widehat{U}^{(p)\top}X_t^{ij}\widehat{V}^{(p)}\\ &\quad - \frac{1}{N_0\nu}\sum_{t\in \mathcal{D}_{2p}} \sum_{(i,j)\in g_{t}}  \inp{\widehat{G}^{(p)} - \widehat{O}_U^{(p)\top}\Lambda\widehat{O}_V^{(p)}}{\widehat{U}^{(p)\top}X_t^{ij}\widehat{V}^{(p)}}\widehat{U}^{(p)\top}X_t^{ij}\widehat{V}^{(p)} \\ &\quad - \frac{1}{N_0\nu}\sum_{t\in \mathcal{D}_{2p}} \sum_{(i,j)\in g_{t}}  \inp{\widehat{U}^{(p)}\widehat{O}_U^{(p)\top}\Lambda(V^{(p)}\widehat{O}_V^{(p)\top})^{\top}- U\Lambda V^{\top}}{X_t^{ij}}\widehat{U}^{(p)\top}X_t^{ij}\widehat{V}^{(p)}.
    \end{align*}
    Since $\|\widehat{U}^{(p)}\|\leq 2\mu\sqrt{r/d_1}$, $\|\widehat{V}^{(p)}\|\leq 2\mu\sqrt{r/d_2}$, then
    \begin{align*}
        \left\|\bigg\|\sum_{(i,j)\in g_{t}} \xi_t^{ig}\widehat{U}^{(p)\top}X_t^{ig}\widehat{V}^{(p)}\bigg\|\right\|_{\psi_2}\lesssim \sigma\frac{\mu^2r}{\sqrt{d_1d_2}}d_1=\sigma\mu^2r\sqrt{\frac{d_1}{d_2}}
    \end{align*}
    and 
    \begin{align*}
        &\left\|\EE\left[\bigg( \sum_{(i,j) \in g_{t}} \xi_t^{ij}\widehat{U}^{(p)\top}X_t^{ij}\widehat{V}^{(p)}\bigg)\bigg(\sum_{(i,j)\in g_{t}} \xi_t^{ij}\widehat{U}^{(p)\top}X_t^{ij}\widehat{V}^{(p)}\bigg)^{\top}\right]\right\| \\ &\quad = \left\|\EE\left[\sum_{(i,j)\in g_{t}}\xi_t^{ij2}\widehat{U}^{(p)\top}X_t^{ij}\widehat{V}^{(p)}\widehat{V}^{(p)\top}X_t^{ij\top}\widehat{U}^{(p)}\right]\right\|=\sigma^2r\nu
    \end{align*}
    by the independence of $\xi_t^{ij}$. By matrix Bernstein inequality (\cite{tropp2012user}), with probability at least $1-d_1^{-10}$,
    \begin{align*}
        \left\|\frac{1}{N_0\nu}\sum_{t\in \mathcal{D}_{2p}}\sum_{(i,j)\in g_{t}} \xi_t^{ij}\widehat{U}^{(p)\top}X_t^{ij}\widehat{V}^{(p)}\right\|\lesssim \frac{\sigma r}{N_0\nu}\sqrt{\frac{d_1}{d_2}}\log d_1 + \sqrt{\frac{\sigma^2r}{N_0\nu}\log d_1}. 
    \end{align*}
    Also from the proof of Lemma \ref{lemma:estimation1},
    \begin{align*}
        &\left\|(\widehat{G}^{(p)}-\widehat{O}_U^{(p)\top}\Lambda\widehat{O}_V^{(p)}) - \frac{1}{N_0\nu}\sum_{t\in \mathcal{D}_{2p}} \sum_{(i,j)\in g_{t}} \inp{\widehat{G}^{(p)} - \widehat{O}_U^{(p)\top}\Lambda\widehat{O}_V^{(p)}}{\widehat{U}^{(p)\top}X_t^{ij}\widehat{V}^{(p)}}\widehat{U}^{(p)\top}X_t^{ij}\widehat{V}^{(p)}\right\| \\
        &\quad \leq  \|\widehat{G}^{(p)}-\widehat{O}_U^{(p)\top}\Lambda\widehat{O}_V^{(p)}\|\max_{A\in \mathcal{N}_{1/3}(\mathcal{O}_r)} \left\|A - \frac{1}{N_0\nu}\sum_{t\in \mathcal{D}_{2p}} \sum_{(i,q)\in g_{t}} \inp{A}{\widehat{U}^{(p)\top}X_t^{ij}\widehat{V}^{(p)}}\widehat{U}^{(p)\top}X_t^{ij}\widehat{V}^{(p)}\right\|.
    \end{align*}
    For each $A\in \mathcal{N}_{1/3}(\mathcal{O}_r)$,
    \begin{align*}
        \|\sum_{(i,j)\in g_{t}} \inp{A}{\widehat{U}^{(p)\top}X_t^{ij}\widehat{V}^{(p)}}\widehat{U}^{(p)\top}X_t^i\widehat{V}^{(p)}\|&\leq \sum_{(i,j)\in g_{t}} \|\widehat{U}^{(p)\top}X_t^{ij}\widehat{V}^{(p)}\|_*\|\widehat{U}^{(p)\top}X_t^{ij}\widehat{V}^{(p)}\| \\ &\leq d_1 \|\widehat{U}^{(p)}\|_{2,\max}^2\|\widehat{V}^{(p)}\|_{2,\max}^2\leq \frac{\mu^2r^2}{d_2}
    \end{align*}
    where $\|\cdot\|_*$ denotes the matrix nuclear norm. Moreover, 
    \begin{align*}
        &\left\|\EE\left[\bigg(\sum_{(i,j)\in g_{t}} \inp{A}{\widehat{U}^{(p)\top}X_t^{ij}\widehat{V}^{(p)}}\widehat{U}^{(p)\top}X_t^{ij}\widehat{V}^{(p)}\bigg)\bigg(\sum_{(i,j)\in g_{t}} \inp{A}{\widehat{U}^{(p)\top}X_t^{ij}\widehat{V}^{(p)}}\widehat{U}^{(p)\top}X_t^{ij}\widehat{V}^{(p)}\bigg)^{\top}\right]\right\| \\
        &\quad \leq \max_{i,j}|\inp{A}{\widehat{U}^{(p)\top}e_ie_j^{\top}\widehat{V}^{(p)}}|^2d_1^2\max_{i,j,k,l}\|(\widehat{U}^{(p)\top}e_ie_j^{\top}\widehat{V}^{(p)})(\widehat{U}^{(p)\top}e_ke_l^{\top}\widehat{V}^{(p)})^{\top}\| \\
        &\quad \leq \fro{A}^2\max_{i,j}\fro{\widehat{U}^{(p)\top}e_ie_j^{\top}\widehat{V}^{(p)}}^2d_1^2\max_{i,j}\|\widehat{U}^{(p)\top}e_ie_j^{\top}\widehat{V}^{(p)}\|^2\leq \frac{\mu^4r^3}{d_2^2}.
    \end{align*}
    Then by matrix Bernstein inequality, with probability at least $1-d_1^{-10}$,
    \begin{align*}
        \left\|A - \frac{1}{N_0\nu}\sum_{t\in \mathcal{D}_{2p}} \sum_{(i,j)\in g_{t}} \inp{A}{\widehat{U}^{(p)\top}X_t^{ij}\widehat{V}^{(p)}}\widehat{U}^{(p)\top}X_t^{ij}\widehat{V}^{(p)}\right\|\lesssim \sqrt{\frac{r^3\log d_1}{N_0d_2^2\nu^2}} + \frac{r^2\log d_1}{N_0d_2\nu}.
    \end{align*}
   as long as $T\gtrsim r\log d_1$. This directly implies that
    \begin{align*}
        &\left\|(\widehat{G}^{(p)}-\widehat{O}_U^{(p)\top}\Lambda\widehat{O}_V^{(p)}) - \frac{1}{N_0\nu}\sum_{t\in \mathcal{D}_{2p}} \sum_{(i,j)\in g_{t}}  \inp{\widehat{G}^{(p)} - \widehat{O}_U^{(p)\top}\Lambda\widehat{O}_V^{(p)}}{\widehat{U}^{(p)\top}X_t^{ij}\widehat{V}^{(p)}}\widehat{U}^{(p)\top}X_t^{ij}\widehat{V}^{(p)}\right\| \\
        &\quad \lesssim  \|\widehat{G}^{(p)}-\widehat{O}_U^{(p)\top}\Lambda\widehat{O}_V^{(p)}\|\left(\sqrt{\frac{r^3\log d_1}{N_0d_2^2\nu^2}} + \frac{r^2\log d_1}{N_0d_2\nu}\right).
    \end{align*}
    Similarly, we first notice that 
    \begin{align*}
        &\left\|\EE\left[\sum_{(i,j)\in g_{t}}  \inp{\widehat{U}^{(p)}\widehat{O}_U^{(p)\top}\Lambda(V^{(p)}\widehat{O}_V^{(p)\top})^{\top}- U\Lambda V^{\top}}{X_t^{ij}}\widehat{U}^{(p)\top}X_t^{ij}\widehat{V}^{(p)}\right]\right\| \\ &\quad =\nu\left\|\widehat{U}^{(p)\top}(\widehat{U}^{(p)}\widehat{O}_U^{(p)\top}\Lambda(V^{(p)}\widehat{O}_V^{(p)\top})^{\top}- U\Lambda V^{\top})\widehat{V}^{(p)}\right\|.
    \end{align*}
    We also calculate the uniform bound
    \begin{align*}
        &\left\|\sum_{(i,j)\in g_{t}}  \inp{\widehat{U}^{(p)}\widehat{O}_U^{(p)\top}\Lambda(V^{(p)}\widehat{O}_V^{(p)\top})^{\top}- U\Lambda V^{\top}}{X_t^{ij}}\widehat{U}^{(p)\top}X_t^{ij}\widehat{V}^{(p)}\right\| \\ 
        &\quad \leq \|\widehat{U}^{(p)}\widehat{O}_U^{(p)\top}\Lambda(V^{(p)}\widehat{O}_V^{(p)\top})^{\top}- U\Lambda V^{\top}\|_{\max}^2d_1\max_{i,j}\|\widehat{U}^{(p)\top}e_ie_j^{\top}\widehat{V}^{(p)}\| \\ 
        &\quad \leq \|\widehat{U}^{(p)}\widehat{O}_U^{(p)\top}\Lambda(V^{(p)}\widehat{O}_V^{(p)\top})^{\top}- U\Lambda V^{\top}\|_{\max}\mu^2r\sqrt{\frac{d_1}{d_2}}
    \end{align*}
    and variance
    \begin{align*}
        &\left\|\EE\left[\bigg(\sum_{(i,j)\in g_{t}}  \inp{\widehat{U}^{(p)}\widehat{O}_U^{(p)\top}\Lambda(V^{(p)}\widehat{O}_V^{(p)\top})^{\top}- U\Lambda V^{\top}}{X_t^{ij}}\widehat{U}^{(p)\top}X_t^{ij}\widehat{V}^{(p)}\bigg) \right.\right.\\ &\left.\left.\quad \bigg(\sum_{(i,j)\in g_{t}}  \inp{\widehat{U}^{(p)}\widehat{O}_U^{(p)\top}\Lambda(V^{(p)}\widehat{O}_V^{(p)\top})^{\top}- U\Lambda V^{\top}}{X_t^{ij}}\widehat{U}^{(p)\top}X_t^{ij}\widehat{V}^{(p)}\bigg)^{\top}\right]\right\| \\
        &\quad\quad \leq \|\widehat{U}^{(p)}\widehat{O}_U^{(p)\top}\Lambda(V^{(p)}\widehat{O}_V^{(p)\top})^{\top}- U\Lambda V^{\top}\|_{\max}^2d_1^2\max_{i,j}\|\widehat{U}^{(p)\top}e_ie_j^{\top}\widehat{V}^{(p)}\|^2 \\
        &\quad\quad \leq \|\widehat{U}^{(p)}\widehat{O}_U^{(p)\top}\Lambda(V^{(p)}\widehat{O}_V^{(p)\top})^{\top}- U\Lambda V^{\top}\|_{\max}^2\frac{\mu^4r^2d_1}{d_2}
    \end{align*}
    By matrix Bernstein inequality, with probability at least $1-d_1^{-10}$, 
    \begin{align*}
        &\left\|\frac{1}{N_0\nu}\sum_{t\in \mathcal{D}_{2p}} \sum_{(i,j)\in g_{t}}  \inp{\widehat{U}^{(p)}\widehat{O}_U^{(p)\top}\Lambda(V^{(p)}\widehat{O}_V^{(p)\top})^{\top}- U\Lambda V^{\top}}{X_t^{ij}}\widehat{U}^{(p)\top}X_t^{iq}\widehat{V}^{(p)}\right\| \\
        &\quad \leq \left\|\widehat{U}^{(p)\top}(\widehat{U}^{(p)}\widehat{O}_U^{(p)\top}\Lambda(V^{(p)}\widehat{O}_V^{(p)\top})^{\top}- U\Lambda V^{\top})\widehat{V}^{(p)}\right\| \\ &\quad\quad + C_2\|\widehat{U}^{(p)}\widehat{O}_U^{(p)\top}\Lambda(V^{(p)}\widehat{O}_V^{(p)\top})^{\top}- U\Lambda V^{\top}\|_{\max}\sqrt{\frac{d_1r^2\log d_1}{N_0d_2\nu^2}}
    \end{align*}
    as long as $T\gtrsim r^3\log d_1$. Then following the last parts of the proof of Lemma \ref{lemma:estimation1}, as long as $T\gtrsim (d_2\nu)^{-1}r^3\log d_1$, with probability at least $1-3d_1^{-10}$,
    \begin{align*}
        &\|\widehat{G}^{(p)}-\widehat{O}_U^{(p)\top}\Lambda\widehat{O}_V^{(p)}\|\leq C_3\left( \frac{\sigma r}{N_0\nu}\sqrt{\frac{d_1}{d_2}}\log d_1 + \sqrt{\frac{\sigma^2r}{N_0\nu}\log d_1}\right) \\ &\quad + 2\lambda_{\max}(\|\widehat{U}^{(p)}-U\widehat{O}_U^{(p)}\|^2 + \|\widehat{V}^{(p)}-V\widehat{O}_V^{(p)}\|^2) \\ &\quad + C_4\lambda_{\max}\left(\sqrt{\frac{r}{d_2}}\|\widehat{U}^{(p)}-U\widehat{O}_U^{(p)}\|_{2,\max} + \sqrt{\frac{r}{d_1}}\|\widehat{V}^{(p)}-V\widehat{O}_V^{(p)}\|_{2,\max}\right)\sqrt{\frac{d_1r^2\log d_1}{N_0d_2\nu^2}}.
    \end{align*}
\end{proof}

\subsection{Proof of Lemma \ref{lemma:estimation3-2}}
\begin{proof}
    Recall that, 
    \begin{align*}
        &\widehat{U}^{(p+0.5)}-U\widehat{O}_U^{(p)}\widehat{L}_G^{(p)} = (\widehat{U}^{(p)}\widehat{L}_G^{(p)}-U\widehat{O}_U^{(p)}\widehat{L}_G^{(p)})(I-\eta\widehat{L}_G^{(p)\top}\widehat{O}_U^{(p)\top}\Lambda\widehat{O}_V^{(p)}\widehat{R}_G^{(p)}(\widehat{\Lambda}^{(p)})^{-1}) \\
        &\quad -\eta \widehat{U}^{(p)}(\widehat{G}^{(p)}-\widehat{O}_U^{(p)\top}\Lambda\widehat{O}_V^{(p)})\widehat{R}_G^{(p)}(\widehat{\Lambda}^{(p)})^{-1} - \eta U\Lambda(\widehat{V}^{(p)}\widehat{O}_V^{(p)\top}-V)^{\top}\widehat{V}^{(p)}\widehat{R}_G^{(p)}(\widehat{\Lambda}^{(p)})^{-1} + \widehat{E}_V^{(p)} + \widehat{E}_{\xi,V}^{(p)}.
    \end{align*}
    By Lemma \ref{lemma:estimation3-1}, 
    \begin{align*}
        &\|\widehat{\Lambda}^{(p)}-\widehat{L}_G^{(p)\top}\widehat{O}_U^{(p)\top}\Lambda\widehat{O}_V^{(p)}\widehat{R}_G^{(p)}\|\leq C_3(\frac{r\sigma}{N_0\nu}\sqrt{\frac{d_1}{d_2}}\log d_1 + \sqrt{\frac{\sigma^2 r\log d_1}{N_0\nu}}) \\ &\quad + 2\lambda_{\max}(\|\widehat{U}^{(p)}-U\widehat{O}_U^{(p)}\|^2+\|\widehat{V}^{(p)}-V\widehat{O}_U^{(p)}\|^2) \\ &\quad + C_4\lambda_{\max}\left(\sqrt{\frac{1}{d_2}}\|\widehat{U}^{(p)}-U\widehat{O}_U^{(p)}\|_{2,\max} + \sqrt{\frac{1}{d_1}}\|\widehat{V}^{(p)}-V\widehat{O}_V^{(p)}\|_{2,\max}\right)\mu\sqrt{\frac{r^3d_1\log d_1}{N_0d_2\nu^2}}
    \end{align*}
    which implies $\|\widehat{\Lambda}^{(p)}-\widehat{L}_G^{(p)\top}\widehat{O}_U^{(p)\top}\Lambda\widehat{O}_V^{(p)}\widehat{R}_G^{(p)}\|\leq \lambda_{\min}/20$ and $\lambda_{\min}(\widehat{\Lambda}^{(p)})\geq \lambda_{\min}/2$ as long as $\lambda_{\min}/\sigma \gtrsim \max\{\sqrt{d_1}/\sqrt{d_2}T\nu, \sqrt{1/T\nu}\}r\log d_1$, $T\gtrsim (d_2\nu)^{-2}r^3\log d_1$, $\|\widehat{U}^{(p)}\|_{2,\max}\leq 2\mu\sqrt{r/d_1}$, $\|\widehat{V}^{(p)}\|_{2,\max}\leq 2\mu\sqrt{r/d_2}$ and $\max\{\|\widehat{U}^{(p)}-U\widehat{O}_U^{(p)}\|, \|\widehat{V}^{(p)}-V\widehat{O}_V^{(p)}\|\}\lesssim 1/\sqrt{\kappa}$. 

    Since $\eta\leq 0.75$, by proof of Lemma \ref{lemma:estimation2}, we have
    \begin{align*}
        \|(\widehat{U}^{(p)}\widehat{L}_G^{(p)}-U\widehat{O}_U^{(p)}\widehat{L}_G^{(p)})(I-\eta\widehat{L}_G^{(p)\top}\widehat{O}_U^{(p)\top}\Lambda\widehat{O}_V^{(p)}\widehat{R}_G^{(p)}(\widehat{\Lambda}^{(p)})^{-1})\|_{2,\max}\leq (1-\frac{9\eta}{10})\|\widehat{U}^{(p)}-U\widehat{O}_U^{(p)}\|_{2,\max}.
    \end{align*}
    By $\|\widehat{U}^{(p)}\|_{2,\max}\leq 2\mu\sqrt{r/d_1}$, we have
    \begin{align*}
        &\eta\|\widehat{U}^{(p)}(\widehat{G}^{(p)}-\widehat{O}_U^{(p)\top}\Lambda\widehat{O}_V^{(p)})\widehat{R}_G^{(p)}(\widehat{\Lambda}^{(p)})^{-1} \|_{2,\max}\leq 2\eta\|\widehat{U}^{(p)}\|_{2,\max}\|\widehat{G}^{(p)}-\widehat{O}_U^{(p)\top}\Lambda\widehat{O}_V^{(p)}\|\lambda_{\min}^{-1} \\
        &\leq C_3\eta\mu(\frac{r^{3/2}\sigma}{\lambda_{\min}}\frac{1}{N_0\sqrt{d_2}\nu}\log d_1 + \frac{\sigma}{\lambda_{\min}}\sqrt{\frac{r^2\log d_1}{N_0d_1\nu}})  + C_4\eta\frac{\mu\kappa r^{1/2}}{\sqrt{d_1}}(\|\widehat{U}^{(p)}-U\widehat{O}_U^{(p)}\|^2+\|\widehat{V}^{(p)}-V\widehat{O}_U^{(p)}\|^2) \\ &\quad + C_5\eta\mu\kappa\left(\sqrt{\frac{1}{d_2}}\|\widehat{U}^{(p)}-U\widehat{O}_U^{(p)}\|_{2,\max} + \sqrt{\frac{1}{d_1}}\|\widehat{V}^{(p)}-V\widehat{O}_V^{(p)}\|_{2,\max}\right)\mu\sqrt{\frac{r^4\log d_1}{N_0d_2\nu^2}}.
    \end{align*}
    Again according to proof of Lemma \ref{lemma:estimation2}, if $T\gtrsim (d_1d_2)^{-1}\nu^{-2}\mu^4\kappa^2r^4\log d_1$, 
    \begin{align*}
        &C_5\mu\kappa\left(\sqrt{\frac{1}{d_2}}\|\widehat{U}^{(p)}-U\widehat{O}_U^{(p)}\|_{2,\max} + \sqrt{\frac{1}{d_1}}\|\widehat{V}^{(p)}-V\widehat{O}_V^{(p)}\|_{2,\max}\right)\mu\sqrt{\frac{r^4\log d_1}{N_0d_2\nu^2}} \\
        &\leq \frac{\sqrt{\frac{d_1}{d_2}}\|\widehat{U}^{(p)}-U\widehat{O}_U^{(p)}\|_{2,\max} + \|\widehat{V}^{(p)}-V\widehat{O}_V^{(p)}\|_{2,\max}}{40}.
    \end{align*}
    Then we have
    \begin{align*}
        &\eta\|\widehat{U}^{(p)}(\widehat{G}^{(p)}-\widehat{O}_U^{(p)\top}\Lambda\widehat{O}_V^{(p)})\widehat{R}_G^{(p)}(\widehat{\Lambda}^{(p)})^{-1} \|_{2,\max}\leq C_3\mu(\frac{r^{3/2}\sigma}{\lambda_{\min}}\frac{1}{N_0\sqrt{d_2}\nu}\log d_1 + \frac{\sigma}{\lambda_{\min}}\sqrt{\frac{r^2\log d_1}{N_0d_1\nu}}) \\
        &\quad + \frac{3\eta}{40}\left(\sqrt{\frac{d_1}{d_2}}\|\widehat{U}^{(p)}-U\widehat{O}_U^{(p)}\|_{2,\max} + \|\widehat{V}^{(p)}-V\widehat{O}_V^{(p)}\|_{2,\max}\right).
    \end{align*}
    Since $\|(\widehat{V}^{(p)}\widehat{O}_V^{(p)\top}-V)^{\top}\widehat{V}^{(p)}\|\leq \|\widehat{V}^{(p)}\widehat{O}_V^{(p)\top}-V\|^2$, we get
    \begin{align*}
        \|U\Lambda(\widehat{V}^{(p)}\widehat{O}_V^{(p)\top}-V)^{\top}\widehat{V}^{(p)}\widehat{R}_G^{(p)}(\widehat{\Lambda}^{(p)})^{-1}\|_{2,\max}&\leq 2\kappa\mu\sqrt{\frac{r}{d_1}}\|\widehat{V}^{(p)}\widehat{O}_V^{(p)\top}-V\|^2 \\ &\leq \frac{1}{40}\|\widehat{V}^{(p)}-V\widehat{O}_V^{(p)\top}\|_{2,\max}
    \end{align*}
    as long as $\|\widehat{V}^{(p)}\widehat{O}_V^{(p)\top}-V\|\lesssim 1/\kappa\mu\sqrt{r}\sqrt{d_2/d_1}$. Putting together the above bounds, we obtain
    \begin{align*}
        &\|\widehat{U}^{(p+0.5)}-U\widehat{O}_U^{(p)}\widehat{L}_G^{(p)}\|_{2,\max}\leq (1-\frac{9\eta}{10})\|\widehat{U}^{(p)}-U\widehat{O}_U^{(p)}\|_{2,\max} + C_3\mu(\frac{r^{3/2}\sigma}{\lambda_{\min}}\frac{1}{N_0\sqrt{d_2}\nu}\log d_1 + \frac{\sigma}{\lambda_{\min}}\sqrt{\frac{r^2\log d_1}{N_0d_1\nu}}) \\ &\quad + \frac{\eta}{10}\left(\sqrt{\frac{d_1}{d_2}}\|\widehat{U}^{(p)}-U\widehat{O}_U^{(p)}\|_{2,\max} + \|\widehat{V}^{(p)}-V\widehat{O}_V^{(p)}\|_{2,\max}\right) + \|\widehat{E}_V^{(p)}\|_{2,\max} + \|\widehat{E}_{\xi,V}^{(p)}\|_{2,\max}.
    \end{align*}
    Since $(\widehat{U}^{(p)}, \widehat{G}^{(p)}, \widehat{V}^{(p)})$ are independent with $\mathcal{D}_{2p+1}$ and $\widehat{U}^{(p)}, \widehat{V}^{(p)}$ are incoherent, for any $j\in [d_1]$, we have
    \begin{align*}
        \bigg\|\big\|e_j^{\top}\sum_{(i,q)\in h_{t}} \xi_t^{iq} X_t^{iq}\widehat{V}^{(p)}\widehat{R}_G^{(p)}(\widehat{\Lambda}^{(p)})^{-1}\big\|\bigg\|_{\psi_2}\lesssim \sigma\|\widehat{V}^{(p)}\|_{2,\max}\lambda_{\min}^{-1}\lesssim \frac{\sigma}{\lambda_{\min}}\sqrt{\frac{r}{d_2}}
    \end{align*}
    and 
    \begin{align*}
        &\EE\left[(e_j^{\top}\sum_{(i,j)\in g_{t}} \xi_t^{ij} X_t^{ij}\widehat{V}^{(p)}\widehat{R}_G^{(p)}(\widehat{\Lambda}^{(p)})^{-1})(e_j^{\top}\sum_{(i,j)\in g_{t}}  \xi_t^{ij} X_t^{ij}\widehat{V}^{(p)}\widehat{R}_G^{(p)}(\widehat{\Lambda}^{(p)})^{-1})^{\top} \right] \\
        &\quad = \sigma^2\fro{\widehat{V}^{(p)}\widehat{R}_G^{(p)}(\widehat{\Lambda}^{(p)})^{-1}}^2\leq \frac{\sigma^2r}{\lambda_{\min}^2}\nu.
    \end{align*}
    since $e_j^{\top}X_t^{iq}\neq 0$ only when $i=j$. Then by Bernstein inequality, as long as $T\gtrsim (d_2\nu)^{-1}\log d_1$, with probability at least $1-d_1^{-10}$,
    \begin{align*}
        \|\widehat{E}_{\xi,V}^{(p)}\|_{2,\max}\lesssim \eta\frac{\sigma}{\lambda_{\min}}\sqrt{\frac{r\log d_1}{N_0\nu}}.
    \end{align*}
    Similarly, with the same probability,
    \begin{align*}
        \|\widehat{E}_V^{(p)}\|_{2,\max}\lesssim \eta\frac{\|\widehat{M}^{(p)}-M\|_{\max}}{\lambda_{\min}}\sqrt{\frac{r\log d_1}{N_0\nu}}.
    \end{align*}
    Note that
    \begin{align*}
        &\|\widehat{M}^{(p)}-M\|_{\max}\leq \|(\widehat{U}^{(p)}-U\widehat{O}_U^{(p)})\widehat{G}^{(p)}\widehat{V}^{(p)\top}\|_{\max} + \|U(\widehat{O}_U^{(p)}\widehat{G}^{(p)}-\Lambda\widehat{O}_V^{(p)})\widehat{V}^{(p)\top}\|_{\max} \\ &\quad\quad  + \|U\Lambda(\widehat{V}^{(p)}\widehat{O}_V^{(p)}-V)^{\top}\|_{2,\max} \\
        &\quad \leq 2\mu\lambda_{\max}\left(\sqrt{\frac{r}{d_2}}\|\widehat{U}^{(p)}-U\widehat{O}_U^{(p)}\|_{2,\max} + \sqrt{\frac{r}{d_1}}\|\widehat{V}^{(p)}-V\widehat{O}_V^{(p)}\|_{2,\max}\right)  + \mu^2\sqrt{\frac{r^2}{d_1d_2}}\|\widehat{G}^{(p)}-\widehat{O}_U^{(p)}\Lambda\widehat{O}_V^{(p)}\|.
    \end{align*}
    Together with Lemma \ref{lemma:estimation3-1},
    \begin{align*}
        &\mu^2\sqrt{\frac{r^2}{d_1d_2}}\|\widehat{G}^{(p)}-\widehat{O}_U^{(p)}\Lambda\widehat{O}_V^{(p)}\|\leq C_3\mu^2(\frac{r^2\sigma}{N_0d_2\nu}\log d_1 + \sqrt{\frac{\sigma^2 r^3\log d_1}{N_0d_1d_2\nu}}) \\ &\quad + 2\lambda_{\max}\mu^2\sqrt{\frac{r^2}{d_1d_2}}(\|\widehat{U}^{(p)}-U\widehat{O}_U^{(p)}\|^2+\|\widehat{V}^{(p)}-V\widehat{O}_U^{(p)}\|^2) \\ &\quad + C_4\lambda_{\max}\left(\sqrt{\frac{1}{d_2}}\|\widehat{U}^{(p)}-U\widehat{O}_U^{(p)}\|_{2,\max} + \sqrt{\frac{1}{d_1}}\|\widehat{V}^{(p)}-V\widehat{O}_V^{(p)}\|_{2,\max}\right)\mu^3\sqrt{\frac{r^5\log d_1}{N_0d_2^2\nu^2}}.
    \end{align*}
    As long as $T\gtrsim (d_2\nu)^{-1}\mu^6r^4\log d_1$ and $\|\widehat{U}^{(p)}-U\widehat{O}_U^{(p)}\|+\|\widehat{V}^{(p)}-V\widehat{O}_U^{(p)}\|\lesssim \frac{1}{\mu^2r}$,
    \begin{align*}
        \|\widehat{M}^{(p)}-M\|_{\max}\leq \sigma +2\mu\lambda_{\max}\left(\sqrt{\frac{r}{d_2}}\|\widehat{U}^{(p)}-U\widehat{O}_U^{(p)}\|_{2,\max} + \sqrt{\frac{r}{d_1}}\|\widehat{V}^{(p)}-V\widehat{O}_V^{(p)}\|_{2,\max}\right).
    \end{align*}
    Therefore, as long as $T\gtrsim (d_1\nu)^{-1}\kappa^2\mu^6r^4\log d_1$, with probability at least $1-2d_1^{-10}$,
    \begin{align*}
        \|\widehat{E}_V^{(p)}\|_{2,\max} + \|\widehat{E}_{\xi,V}^{(p)}\|_{2,\max}\leq C_6\eta\frac{\sigma}{\lambda_{\min}}\sqrt{\frac{r\log d_1}{N_0\nu}} + \frac{\eta}{40}\left(\sqrt{\frac{d_1}{d_2}}\|\widehat{U}^{(p)}-U\widehat{O}_U^{(p)}\|_{2,\max} + \|\widehat{V}^{(p)}-V\widehat{O}_V^{(p)}\|_{2,\max}\right).
    \end{align*}
    As a result, 
    \begin{align*}
        &\|\widehat{U}^{(p+0.5)}-U\widehat{O}_U^{(p)}\widehat{L}_G^{(p)}\|_{2,\max}\leq (1-\frac{9\eta}{10})\|\widehat{U}^{(p)}-U\widehat{O}_U^{(p)}\|_{2,\max} + C_6\eta\frac{\sigma}{\lambda_{\min}}\sqrt{\frac{r\log d_1}{N_0\nu}}\\
        &\quad + \frac{\eta}{8}\left(\sqrt{\frac{d_1}{d_2}}\|\widehat{U}^{(p)}-U\widehat{O}_U^{(p)}\|_{2,\max} + \|\widehat{V}^{(p)}-V\widehat{O}_V^{(p)}\|_{2,\max}\right).
    \end{align*}
    Next, we investigate the singular values of $\widehat{U}^{(p+0.5)}-U\widehat{O}_U^{(p)}\widehat{L}_G^{(p)}$. Recall
    \begin{align*}
        \widehat{U}^{(p+0.5)}= \underbrace{\widehat{U}^{(p)}\widehat{L}_G^{(p)}-\eta(\widehat{U}^{(p)}\widehat{G}^{(p)}\widehat{V}^{(p)\top} - U\Lambda V^{\top})\widehat{V}^{(p)}\widehat{R}_G^{(p)}(\widehat{\Lambda}^{(p)})^{-1}}_{\mathcal{I}_1} + \underbrace{\widehat{E}_V^{(p)} + \widehat{E}_{\xi,V}^{(p)}}_{\mathcal{I}_2}. 
    \end{align*}
    By the independence between $(\widehat{U}^{(p)}, \widehat{G}^{(p)}, \widehat{V}^{(p)}, \widehat{L}_G^{(p)}, \widehat{R}_G^{(p)}, \widehat{\Lambda}^{(p)})$ and $\mathcal{D}_{2p+1}$, and notice that $\| \sum_{(i,q)\in h_{t}} X_t^{iq}\| =1$, then the same as proof of Lemma \ref{lemma:estimation2}, we have with probability at least $1-2d_1^{-10}$,
    \begin{align*}
        \|\widehat{E}_V^{(p)} + \widehat{E}_{\xi,V}^{(p)}\|\lesssim \eta\frac{\sigma}{\lambda_{\min}}\sqrt{\frac{d_2\log^2 d_1}{N_0\nu}}  + \eta\sqrt{\frac{d_2\log^2 d_1}{N_0\nu}}\left(\sqrt{\frac{r}{d_2}}\|\widehat{U}^{(p)}-U\widehat{O}_U^{(p)}\|_{2,\max} + \sqrt{\frac{r}{d_1}}\|\widehat{V}^{(p)}-V\widehat{O}_V^{(p)}\|_{2,\max}\right).
    \end{align*}
    Note that the singular values of $\widehat{U}^{(p+0.5)}$ are the square root of eigenvalues of $\widehat{U}^{(p+0.5)\top}\widehat{U}^{(p+0.5)}$. We write
    \begin{align*}
        \widehat{U}^{(p+0.5)\top}\widehat{U}^{(p+0.5)}=\mathcal{I}_1^{\top}\mathcal{I}_1 + \mathcal{I}_2^{\top}\mathcal{I}_2 + \mathcal{I}_1^{\top}\mathcal{I}_2 + \mathcal{I}_2^{\top}\mathcal{I}_1.
    \end{align*}
    Since $\|\widehat{U}^{(p)\top}(\widehat{U}^{(p)}-U\widehat{O}_U^{(p)})\|\leq 2\|\widehat{U}^{(p)}-U\widehat{O}_U^{(p)}\|^2$ and $\|\widehat{V}^{(p)\top}(\widehat{V}^{(p)}-V\widehat{O}_V^{(p)})\|\leq 2\|\widehat{V}^{(p)}-V\widehat{O}_V^{(p)}\|^2$, by Lemma \ref{lemma:estimation3-1} and proof of Lemma \ref{lemma:estimation2}, we get
    \begin{align*}
        &\|\mathcal{I}_1^{\top}\mathcal{I}_1-I\|\leq 3(\kappa^2\eta^2+\kappa\eta)(\|\widehat{U}^{(p)}-U\widehat{O}_U^{(p)}\|^2 + \|\widehat{V}^{(p)}-V\widehat{O}_V^{(p)}\|^2) + 4\frac{\eta}{\lambda_{\min}}\|\widehat{O}_U^{(p)}\widehat{G}^{(p)} - \Lambda\widehat{O}_V^{(p)} \| \\
        &\quad\lesssim (\kappa^2\eta^2+\kappa\eta)(\|\widehat{U}^{(p)}-U\widehat{O}_U^{(p)}\|^2 + \|\widehat{V}^{(p)}-V\widehat{O}_V^{(p)}\|^2) + \eta\frac{\sigma}{\lambda_{\min}}\left( \frac{r}{N_0\nu}\sqrt{\frac{d_1}{d_2}}\log d_1 + \sqrt{\frac{r}{N_0\nu}\log d_1}\right) \\ &\quad\quad +  \eta\kappa\left(\sqrt{\frac{r}{d_2}}\|\widehat{U}^{(p)}-U\widehat{O}_U^{(p)}\|_{2,\max} + \sqrt{\frac{r}{d_1}}\|\widehat{V}^{(p)}-V\widehat{O}_V^{(p)}\|_{2,\max}\right)\sqrt{\frac{d_1r^2\log d_1}{N_0d_2\nu^2}}.
    \end{align*}
    Moreover, as long as $\|\widehat{U}^{(p)}-U\widehat{O}_U^{(p)}\| + \|\widehat{V}^{(p)}-V\widehat{O}_V^{(p)}\|\lesssim \frac{1}{\kappa}$, $\|\mathcal{I}_1\|\leq 2$, then $\|\mathcal{I}_1^{\top}\mathcal{I}_2 + \mathcal{I}_2^{\top}\mathcal{I}_1\|\leq 4\|\widehat{E}_V^{(p)} + \widehat{E}_{\xi,V}^{(p)}\|$. Finally, with probability at least $1-2d_1^{-10}$,
    \begin{align*}
        &\|\widehat{U}^{(p+0.5)\top}\widehat{U}^{(p+0.5)}-I\|\lesssim \eta\frac{\sigma}{\lambda_{\min}}\sqrt{\frac{d_2\log^2 d_1}{N_0\nu}} + (\kappa^2\eta^2+\kappa\eta)(\|\widehat{U}^{(p)}-U\widehat{O}_U^{(p)}\|^2 + \|\widehat{V}^{(p)}-V\widehat{O}_V^{(p)}\|^2) \\
        &\quad +  \eta\sqrt{\frac{d_2\log^2 d_1}{N_0\nu}}\left(\sqrt{\frac{r}{d_2}}\|\widehat{U}^{(p)}-U\widehat{O}_U^{(p)}\|_{2,\max} + \sqrt{\frac{r}{d_1}}\|\widehat{V}^{(p)}-V\widehat{O}_V^{(p)}\|_{2,\max}\right),
    \end{align*}
    implying that $\{|1-\lambda_{\min}(\widehat{U}^{(p+0.5)})|, |1-\lambda_{\max}(\widehat{U}^{(p+0.5)})|\}$ can be upper bounded by the above rate.
\end{proof}

\subsection{Proof of Lemma \ref{lemma:estimation1-nonuniform}}
\begin{proof}
    Notice that by our previous notation, $X_t^i=e_ie_{j_t(i)}^{\top}$. By the definition of $\widehat{G}^{(p)}$ and $\mathscr{L}_{\mathcal{D}_{2p}}$, we first write
    \begin{align*}
        \frac{1}{N_0}\sum_{t\in \mathcal{D}_{2p}} \sum_{i=1}^{d_1} \frac{1}{p_{ij_t(i)}}\inp{\widehat{U}^{(p)}\widehat{G}^{(p)}\widehat{V}^{(p)\top} - U\Lambda V^{\top}}{X_t^i}\widehat{U}^{(p)\top}X_t^{i}\widehat{V}^{(p)} - \frac{1}{N_0}\sum_{t\in \mathcal{D}_{2p}} \sum_{i=1}^{d_1}\frac{1}{p_{ij_t(i)}}\xi_t^i\widehat{U}^{(p)\top}X_t^{i}\widehat{V}^{(p)}=0
    \end{align*}
    where due to data splitting, $(\widehat{U}^{(p)},\widehat{V}^{(p)})$ are independent with $\mathcal{D}_{2p}$. Note that
    \begin{align*}
        \widehat{U}^{(p)}\widehat{G}^{(p)}\widehat{V}^{(p)\top} - U\Lambda V^{\top}=\widehat{U}^{(p)}(\widehat{G}^{(p)} -\widehat{O}_U^{(p)\top}\Lambda \widehat{O}_V^{(p)})\widehat{V}^{(p)\top} + (\widehat{U}^{(p)}\widehat{O}_U^{(p)\top}\Lambda(V^{(p)}\widehat{O}_V^{(p)\top})^{\top}- U\Lambda V^{\top}).
    \end{align*}
    Then
    \begin{align*}
        &\widehat{G}^{(p)}-\widehat{O}_U^{(p)\top}\Lambda\widehat{O}_V^{(p)} =(\widehat{G}^{(p)}-\widehat{O}_U^{(p)\top}\Lambda\widehat{O}_V^{(p)}) \\ &\quad - \frac{1}{N_0}\sum_{t\in \mathcal{D}_{2p}} \sum_{i=1}^{d_1}\frac{1}{p_{ij_t(i)}} \inp{\widehat{G}^{(p)} - \widehat{O}_U^{(p)\top}\Lambda\widehat{O}_V^{(p)}}{\widehat{U}^{(p)\top}X_t^i\widehat{V}^{(p)}}\widehat{U}^{(p)\top}X_t^{i}\widehat{V}^{(p)} \\ &\quad - \frac{1}{N_0}\sum_{t\in \mathcal{D}_{2p}} \sum_{i=1}^{d_1} \frac{1}{p_{ij_t(i)}}\inp{\widehat{U}^{(p)}\widehat{O}_U^{(p)\top}\Lambda(V^{(p)}\widehat{O}_V^{(p)\top})^{\top}- U\Lambda V^{\top}}{X_t^i}\widehat{U}^{(p)\top}X_t^{i}\widehat{V}^{(p)}  \\ &\quad + \frac{1}{N_0}\sum_{t\in \mathcal{D}_{2p}} \sum_{i=1}^{d_1}\frac{1}{p_{ij_t(i)}}\xi_t^i\widehat{U}^{(p)\top}X_t^{i}\widehat{V}^{(p)}.
    \end{align*}
    Since $\|\widehat{U}^{(p)}\|\leq 2\mu\sqrt{r/d_1}$, $\|\widehat{V}^{(p)}\|\leq 2\mu\sqrt{r/d_2}$, 
    \begin{align*}
        \left\|\bigg\|\sum_{i=1}^{d_1}\frac{1}{p_{ij_t(i)}}\xi_t^i\widehat{U}^{(p)\top}X_t^{i}\widehat{V}^{(p)}\bigg\|\right\|_{\psi_2}\lesssim \sigma  \frac{\sqrt{d_1}\mu^2r}{p_{\min}\sqrt{d_2}}
    \end{align*}
    and 
    \begin{align*}
        &\left\|\EE\left[\bigg(\sum_{i=1}^{d_1}\frac{1}{p_{ij_t(i)}}\xi_t^i\widehat{U}^{(p)\top}X_t^{i}\widehat{V}^{(p)}\bigg)\bigg(\sum_{i=1}^{d_1}\frac{1}{p_{ij_t(i)}}\xi_t^i\widehat{U}^{(p)\top}X_t^{i}\widehat{V}^{(p)}\bigg)^{\top}\right]\right\| \\ &\quad = \left\|\EE\left[\sum_{i=1}^{d_1}\frac{1}{p_{ij_t(i)}^2}\xi_t^{i2}\widehat{U}^{(p)\top}X_t^{i}\widehat{V}^{(p)}\widehat{V}^{(p)\top}X_t^{i\top}\widehat{U}^{(p)}\right]\right\|\leq \frac{\sigma^2\mu^4r^2}{d_1d_2}\left(\sum_{i,j}\frac{1}{p_{ij}}\right) 
    \end{align*}
    by the independence of $\xi_t^i$. By matrix Bernstein inequality (\cite{tropp2012user}), with probability at least $1-d_1^{-10}$,
    \begin{align*}
        \left\|\frac{1}{N_0}\sum_{t\in \mathcal{D}_{2p}} \sum_{i=1}^{d_1}\frac{1}{p_{ij_t(i)}}\xi_t^i\widehat{U}^{(p)\top}X_t^{i}\widehat{V}^{(p)}\right\|\lesssim \frac{\sqrt{d_1}\sigma r}{N_0p_{\min}\sqrt{d_2}}\log d_1 + \sqrt{\frac{\sigma^2r^2}{d_1d_2N_0}\left(\sum_{i,j}\frac{1}{p_{ij}}\right)\log d_1}.
    \end{align*}
    Notice that
    \begin{align*}
        &\left\|(\widehat{G}^{(p)}-\widehat{O}_U^{(p)\top}\Lambda\widehat{O}_V^{(p)}) - \frac{1}{N_0}\sum_{t\in \mathcal{D}_{2p}} \sum_{i=1}^{d_1} \frac{1}{p_{ij_t(i)}}\inp{\widehat{G}^{(p)} - \widehat{O}_U^{(p)\top}\Lambda\widehat{O}_V^{(p)}}{\widehat{U}^{(p)\top}X_t^i\widehat{V}^{(p)}}\widehat{U}^{(p)\top}X_t^{i}\widehat{V}^{(p)}\right\| \\
        &\quad \leq \|\widehat{G}^{(p)}-\widehat{O}_U^{(p)\top}\Lambda\widehat{O}_V^{(p)}\|\sup_{A\in \RR^{r\times r}, \|A\|\leq 1} \left\|A - \frac{1}{N_0}\sum_{t\in \mathcal{D}_{2p}} \sum_{i=1}^{d_1} \frac{1}{p_{ij_t(i)}}\inp{A}{\widehat{U}^{(p)\top}X_t^i\widehat{V}^{(p)}}\widehat{U}^{(p)\top}X_t^i\widehat{V}^{(p)}\right\|.
    \end{align*}
    Denote $\mathcal{O}_r=\{A\in \RR^{r\times r},\|A\|\leq 1\}$ and $\mathcal{N}_{1/3}(\mathcal{O}_r)$ the $1/3$-net of $\mathcal{O}_r$, i.e., for any $A\in \mathcal{O}_r$, these exists $A_0\in \mathcal{N}_{1/3}(\mathcal{O}_r)$ so that $\|A-A_0\|\leq 1/3$. It is well-known by (\cite{pajor1998metric, koltchinskii2015optimal}) that $\text{Card}(\mathcal{N}_{1/3}(\mathcal{O}_r))\leq 3^{C_2r^2}$ for some constant $C_2>0$. By definition of $\mathcal{N}_{1/3}(\mathcal{O}_r)$,
    \begin{align*}
        &\left\|(\widehat{G}^{(p)}-\widehat{O}_U^{(p)\top}\Lambda\widehat{O}_V^{(p)}) - \frac{1}{N_0}\sum_{t\in \mathcal{D}_{2p}} \sum_{i=1}^{d_1} \frac{1}{p_{ij_t(i)}}\inp{\widehat{G}^{(p)} - \widehat{O}_U^{(p)\top}\Lambda\widehat{O}_V^{(p)}}{\widehat{U}^{(p)\top}X_t^i\widehat{V}^{(p)}}\widehat{U}^{(p)\top}X_t^{i}\widehat{V}^{(p)}\right\| \\
        &\quad \leq  \|\widehat{G}^{(p)}-\widehat{O}_U^{(p)\top}\Lambda\widehat{O}_V^{(p)}\|\max_{A\in \mathcal{N}_{1/3}(\mathcal{O}_r)} \left\|A - \frac{1}{N_0}\sum_{t\in \mathcal{D}_{2p}} \sum_{i=1}^{d_1} \frac{1}{p_{ij_t(i)}}\inp{A}{\widehat{U}^{(p)\top}X_t^i\widehat{V}^{(p)}}\widehat{U}^{(p)\top}X_t^i\widehat{V}^{(p)}\right\|.
    \end{align*}
    For each $A\in \mathcal{N}_{1/3}(\mathcal{O}_r)$,
    \begin{align*}
        \|\sum_{i=1}^{d_1}\frac{1}{p_{ij_t(i)}}\inp{A}{\widehat{U}^{(p)\top}X_t^i\widehat{V}^{(p)}}\widehat{U}^{(p)\top}X_t^i\widehat{V}^{(p)}\|&\leq \sum_{i=1}^{d_1}\frac{1}{p_{ij_t(i)}}\|\widehat{U}^{(p)\top}X_t^i\widehat{V}^{(p)}\|_*\|\widehat{U}^{(p)\top}X_t^i\widehat{V}^{(p)}\| \\ &\leq \frac{d_1}{p_{\min}} \|\widehat{U}^{(p)}\|_{2,\max}^2\|\widehat{V}^{(p)}\|_{2,\max}^2\leq \frac{\mu^2r^2}{p_{\min}d_2}
    \end{align*}
    where $\|\cdot\|_*$ denotes the matrix nuclear norm. Moreover,
    \begin{align*}
        &\left\|\EE\left[\bigg(\sum_{i=1}^{d_1} \frac{1}{p_{ij_t(i)}}\inp{A}{\widehat{U}^{(p)\top}X_t^i\widehat{V}^{(p)}}\widehat{U}^{(p)\top}X_t^i\widehat{V}^{(p)}\bigg)\bigg(\sum_{i=1}^{d_1}\frac{1}{p_{ij_t(i)}} \inp{A}{\widehat{U}^{(p)\top}X_t^i\widehat{V}^{(p)}}\widehat{U}^{(p)\top}X_t^i\widehat{V}^{(p)}\bigg)^{\top}\right]\right\| \\
        &\quad \leq \EE\left[\left\| \sum_{i=1}^{d_1} \frac{1}{p_{ij_t(i)}}\inp{A}{\widehat{U}^{(p)\top}X_t^i\widehat{V}^{(p)}}\widehat{U}^{(p)\top}X_t^i\widehat{V}^{(p)} \right\|^2\right] \\
        &\quad \leq d_1\max_{i,j}|\inp{A}{\widehat{U}^{(p)\top}e_ie_j^{\top}\widehat{V}^{(p)}}|^2 \sum_{i=1}^{d_1}\sum_{j=1}^{d_2}\frac{1}{p_{ij}}\|(\widehat{U}^{(p)\top}e_ie_j^{\top}\widehat{V}^{(p)})(\widehat{U}^{(p)\top}e_ie_j^{\top}\widehat{V}^{(p)})^{\top}\| \\
        &\quad \leq d_1\fro{A}^2\max_{i,j}\fro{\widehat{U}^{(p)\top}e_ie_j^{\top}\widehat{V}^{(p)}}^2\sum_{i=1}^{d_1}\sum_{j=1}^{d_2}\frac{1}{p_{ij}}\|\widehat{U}^{(p)}\|_{2,\max}^2\|\widehat{V}^{(p)}\|_{2,\max}^2   \leq \frac{\mu^8r^4}{d_1d_2^2}\left(\sum_{i=1}^{d_1}\sum_{j=1}^{d_2}\frac{1}{p_{ij}}\right),
    \end{align*}
    where the last inequality comes from the incoherence property. Then by matrix Bernstein inequality, with probability at least $1-d_1^{-10}$,
    \begin{align*}
        \left\|A - \frac{d_2}{N_0}\sum_{t\in \mathcal{D}_{2p}} \sum_{i=1}^{d_1} \inp{A}{\widehat{U}^{(p)\top}X_t^i\widehat{V}^{(p)}}\widehat{U}^{(p)\top}X_t^i\widehat{V}^{(p)}\right\|\lesssim \sqrt{\frac{r^4\log d_1}{N_0d_1d_2^2}\left(\sum_{i=1}^{d_1}\sum_{j=1}^{d_2}\frac{1}{p_{ij}}\right)}
    \end{align*}
    as long as $p_{\min}\lesssim \sqrt{N_0d_1^{-1}\sum_{i,j}p_{ij}^{-1}\log d_1}$. This directly implies that
    \begin{align*}
        &\left\|(\widehat{G}^{(p)}-\widehat{O}_U^{(p)\top}\Lambda\widehat{O}_V^{(p)}) - \frac{1}{N_0}\sum_{t\in \mathcal{D}_{2p}} \sum_{i=1}^{d_1}  \frac{1}{p_{ij_t(i)}}\inp{\widehat{G}^{(p)} - \widehat{O}_U^{(p)\top}\Lambda\widehat{O}_V^{(p)}}{\widehat{U}^{(p)\top}X_t^i\widehat{V}^{(p)}}\widehat{U}^{(p)\top}X_t^{i}\widehat{V}^{(p)}\right\| \\
        &\quad \lesssim  \|\widehat{G}^{(p)}-\widehat{O}_U^{(p)\top}\Lambda\widehat{O}_V^{(p)}\|\sqrt{\frac{r^4\log d_1}{N_0d_1d_2^2}\left(\sum_{i=1}^{d_1}\sum_{j=1}^{d_2}\frac{1}{p_{ij}}\right)}.
    \end{align*}
    Similarly, we first notice that 
    \begin{align*}
        &\left\|\EE\left[\sum_{i=1}^{d_1}  \frac{1}{p_{ij_t(i)}}\inp{\widehat{U}^{(p)}\widehat{O}_U^{(p)\top}\Lambda(V^{(p)}\widehat{O}_V^{(p)\top})^{\top}- U\Lambda V^{\top}}{X_t^i}\widehat{U}^{(p)\top}X_t^{i}\widehat{V}^{(p)}\right]\right\| \\ &\quad =\left\|\widehat{U}^{(p)\top}(\widehat{U}^{(p)}\widehat{O}_U^{(p)\top}\Lambda(V^{(p)}\widehat{O}_V^{(p)\top})^{\top}- U\Lambda V^{\top})\widehat{V}^{(p)}\right\|.
    \end{align*}
    We also calculate the uniform bound
    \begin{align*}
        &\left\|\sum_{i=1}^{d_1} \frac{1}{p_{ij_t(i)}}\inp{\widehat{U}^{(p)}\widehat{O}_U^{(p)\top}\Lambda(V^{(p)}\widehat{O}_V^{(p)\top})^{\top}- U\Lambda V^{\top}}{X_t^i}\widehat{U}^{(p)\top}X_t^{i}\widehat{V}^{(p)}\right\| \\ 
        &\quad \leq \frac{d_1}{p_{\min}}\|\widehat{U}^{(p)}\widehat{O}_U^{(p)\top}\Lambda(V^{(p)}\widehat{O}_V^{(p)\top})^{\top}- U\Lambda V^{\top}\|_{\max}^2 \max_{k,l}\|\widehat{U}^{(p)\top}e_ke_l^{\top}\widehat{V}^{(p)}\| \\ 
        &\quad \leq \|\widehat{U}^{(p)}\widehat{O}_U^{(p)\top}\Lambda(V^{(p)}\widehat{O}_V^{(p)\top})^{\top}- U\Lambda V^{\top}\|_{\max}\mu^2r\sqrt{\frac{d_1}{p_{\min}^2d_2}}
    \end{align*}
    and variance
    \begin{align*}
        &\left\|\EE\left[\bigg(\sum_{i=1}^{d_1}\frac{1}{p_{ij_t(i)}} \inp{\widehat{U}^{(p)}\widehat{O}_U^{(p)\top}\Lambda(V^{(p)}\widehat{O}_V^{(p)\top})^{\top}- U\Lambda V^{\top}}{X_t^i}\widehat{U}^{(p)\top}X_t^{i}\widehat{V}^{(p)}\bigg) \right.\right.\\ &\left.\left.\quad \bigg(\sum_{i=1}^{d_1} \frac{1}{p_{ij_t(i)}}\inp{\widehat{U}^{(p)}\widehat{O}_U^{(p)\top}\Lambda(V^{(p)}\widehat{O}_V^{(p)\top})^{\top}- U\Lambda V^{\top}}{X_t^i}\widehat{U}^{(p)\top}X_t^{i}\widehat{V}^{(p)}\bigg)^{\top}\right]\right\| \\
        &\quad\quad \leq \EE\left[\left\| \sum_{i=1}^{d_1}\frac{1}{p_{ij_t(i)}} \inp{\widehat{U}^{(p)}\widehat{O}_U^{(p)\top}\Lambda(V^{(p)}\widehat{O}_V^{(p)\top})^{\top}- U\Lambda V^{\top}}{X_t^i}\widehat{U}^{(p)\top}X_t^{i}\widehat{V}^{(p)} \right\|^2\right]\\
        &\quad\quad \leq d_1\|\widehat{U}^{(p)}\widehat{O}_U^{(p)\top}\Lambda(V^{(p)}\widehat{O}_V^{(p)\top})^{\top}- U\Lambda V^{\top}\|_{\max}^2\sum_{i=1}^{d_1}\sum_{j=1}^{d_2}\frac{1}{p_{ij}}\|\widehat{U}^{(p)\top}e_ie_j^{\top}\widehat{V}^{(p)}\|^2 \\
        &\quad\quad \leq \|\widehat{U}^{(p)}\widehat{O}_U^{(p)\top}\Lambda(V^{(p)}\widehat{O}_V^{(p)\top})^{\top}- U\Lambda V^{\top}\|_{\max}^2\frac{\mu^4r^2}{d_2}\left(\sum_{i=1}^{d_1}\sum_{j=1}^{d_2}\frac{1}{p_{ij}}\right).
    \end{align*}
    By matrix Bernstein inequality, as long as $p_{\min}^{-1}\lesssim \sqrt{N_0d_1^{-1}\sum_{i,j}p_{ij}^{-1}/\log d_1}$, with probability at least $1-d_1^{-10}$, 
    \begin{align*}
        &\left\|\frac{1}{N_0}\sum_{t\in \mathcal{D}_{2p}} \sum_{i=1}^{d_1}\frac{1}{p_{ij_t(i)}} \inp{\widehat{U}^{(p)}\widehat{O}_U^{(p)\top}\Lambda(V^{(p)}\widehat{O}_V^{(p)\top})^{\top}- U\Lambda V^{\top}}{X_t^i}\widehat{U}^{(p)\top}X_t^{i}\widehat{V}^{(p)}\right\| \\
        &\quad \leq \left\|\widehat{U}^{(p)\top}(\widehat{U}^{(p)}\widehat{O}_U^{(p)\top}\Lambda(V^{(p)}\widehat{O}_V^{(p)\top})^{\top}- U\Lambda V^{\top})\widehat{V}^{(p)}\right\| \\ &\quad\quad + C_2\|\widehat{U}^{(p)}\widehat{O}_U^{(p)\top}\Lambda(V^{(p)}\widehat{O}_V^{(p)\top})^{\top}- U\Lambda V^{\top}\|_{\max}\sqrt{\frac{r^2\log d_1}{d_2N_0}\left(\sum_{i=1}^{d_1}\sum_{j=1}^{d_2}\frac{1}{p_{ij}}\right)}.
    \end{align*}
    Note that 
    \begin{align*}
        &\|\widehat{U}^{(p)}\widehat{O}_U^{(p)\top}\Lambda(V^{(p)}\widehat{O}_V^{(p)\top})^{\top}- U\Lambda V^{\top}\|_{\max} \\&\quad \leq 3\lambda_{\max}\mu\left(\sqrt{\frac{r}{d_2}}\|\widehat{U}^{(p)}-U\widehat{O}_U^{(p)}\|_{2,\max} + \sqrt{\frac{r}{d_1}}\|\widehat{V}^{(p)}-V\widehat{O}_V^{(p)}\|_{2,\max}\right).
    \end{align*}
    By the differential property of Grassmannians (\cite{edelman1998geometry,keshavan2010matrix,xia2019polynomial}),
    \begin{align*}
        &\left\|\widehat{U}^{(p)\top}(\widehat{U}^{(p)}\widehat{O}_U^{(p)\top}\Lambda(V^{(p)}\widehat{O}_V^{(p)\top})^{\top}- U\Lambda V^{\top})\widehat{V}^{(p)}\right\| \\
        &\quad \leq \lambda_{\max}\|\widehat{U}^{(p)\top}(\widehat{U}^{(p)}\widehat{O}_U^{(p)\top}-U)\| + \lambda_{\max}\|\widehat{V}^{(p)\top}(\widehat{V}^{(p)}\widehat{O}_V^{(p)\top}-V)\| \\
        &\quad \leq 2\lambda_{\max}(\|\widehat{U}^{(p)}-U\widehat{O}_U^{(p)}\|^2 + \|\widehat{V}^{(p)}-V\widehat{O}_V^{(p)}\|^2).
    \end{align*}
    As a result, with probability at least $1-3d_1^{-10}$,
    \begin{align*}
        &\|\widehat{G}^{(p)}-\widehat{O}_U^{(p)\top}\Lambda\widehat{O}_V^{(p)}\|\leq C_3\left(\frac{\sqrt{d_1}\sigma r}{N_0p_{\min}\sqrt{d_2}}\log d_1 + \sqrt{\frac{\sigma^2r^2}{d_1d_2N_0}\left(\sum_{i=1}^{d_1}\sum_{j=1}^{d_2} \frac{1}{p_{ij}}\right)\log d_1}\right) \\ &\quad + 2\lambda_{\max}(\|\widehat{U}^{(p)}-U\widehat{O}_U^{(p)}\|^2 + \|\widehat{V}^{(p)}-V\widehat{O}_V^{(p)}\|^2) \\ &\quad + C_4\lambda_{\max}\left(\sqrt{\frac{r}{d_2}}\|\widehat{U}^{(p)}-U\widehat{O}_U^{(p)}\|_{2,\max} + \sqrt{\frac{r}{d_1}}\|\widehat{V}^{(p)}-V\widehat{O}_V^{(p)}\|_{2,\max}\right)\sqrt{\frac{r^2\log d_1}{d_2N_0}\left(\sum_{i=1}^{d_1}\sum_{j=1}^{d_2}\frac{1}{p_{ij}}\right)}.
    \end{align*}
\end{proof}

\subsection{Proof of Lemma \ref{lemma:estimation2-nonuniform}}
\begin{proof}
    Recall that
    \begin{align*}
        &\widehat{U}^{(p+0.5)}-U\widehat{O}_U^{(p)}\widehat{L}_G^{(p)} = (\widehat{U}^{(p)}\widehat{L}_G^{(p)}-U\widehat{O}_U^{(p)}\widehat{L}_G^{(p)})(I-\eta\widehat{L}_G^{(p)\top}\widehat{O}_U^{(p)\top}\Lambda\widehat{O}_V^{(p)}\widehat{R}_G^{(p)}(\widehat{\Lambda}^{(p)})^{-1}) \\
        &\quad -\eta \widehat{U}^{(p)}(\widehat{G}^{(p)}-\widehat{O}_U^{(p)\top}\Lambda\widehat{O}_V^{(p)})\widehat{R}_G^{(p)}(\widehat{\Lambda}^{(p)})^{-1} - \eta U\Lambda(\widehat{V}^{(p)}\widehat{O}_V^{(p)\top}-V)^{\top}\widehat{V}^{(p)}\widehat{R}_G^{(p)}(\widehat{\Lambda}^{(p)})^{-1} + \widehat{E}_V^{(p)} + \widehat{E}_{\xi,V}^{(p)}.
    \end{align*}
    By Lemma \ref{lemma:estimation1-nonuniform}, 
    \begin{align*}
         &\|\widehat{\Lambda}^{(p)}-\widehat{L}_G^{(p)\top}\widehat{O}_U^{(p)\top}\Lambda\widehat{O}_V^{(p)}\widehat{R}_G^{(p)}\|\leq C_3\left(\frac{\sqrt{d_1}\sigma r}{N_0p_{\min}\sqrt{d_2}}\log d_1 + \sqrt{\frac{\sigma^2r^2}{d_1d_2N_0}\left(\sum_{i=1}^{d_1}\sum_{j=1}^{d_2}\frac{1}{p_{ij}}\right)\log d_1}\right) \\ &\quad + 2\lambda_{\max}(\|\widehat{U}^{(p)}-U\widehat{O}_U^{(p)}\|^2 + \|\widehat{V}^{(p)}-V\widehat{O}_V^{(p)}\|^2) \\ &\quad + C_4\lambda_{\max}\left(\sqrt{\frac{r}{d_2}}\|\widehat{U}^{(p)}-U\widehat{O}_U^{(p)}\|_{2,\max} + \sqrt{\frac{r}{d_1}}\|\widehat{V}^{(p)}-V\widehat{O}_V^{(p)}\|_{2,\max}\right)\sqrt{\frac{r^2\log d_1}{d_2N_0}\left(\sum_{i=1}^{d_1}\sum_{j=1}^{d_2}\frac{1}{p_{ij}}\right)}.
    \end{align*}
    which implies $\|\widehat{\Lambda}^{(p)}-\widehat{L}_G^{(p)\top}\widehat{O}_U^{(p)\top}\Lambda\widehat{O}_V^{(p)}\widehat{R}_G^{(p)}\|\leq \lambda_{\min}/20$ and $\lambda_{\min}(\widehat{\Lambda}^{(p)})\geq \lambda_{\min}/2$ as long as $\lambda_{\min}/\sigma \gtrsim \max\{\sqrt{d_1}/(p_{\min}\sqrt{d_2}N_0), \sqrt{(\sum_{i,j}p_{ij}^{-1})/d_1d_2N_0}\}r\log d_1$, $N_0\gtrsim r^3(\sum_{i,j}p_{ij}^{-1})\log d_1/(d_1d_2^2)$, $\|\widehat{U}^{(p)}\|_{2,\max}\leq 2\mu\sqrt{r/d_1}$, $\|\widehat{V}^{(p)}\|_{2,\max}\leq 2\mu\sqrt{r/d_2}$ and $\max\{\|\widehat{U}^{(p)}-U\widehat{O}_U^{(p)}\|, \|\widehat{V}^{(p)}-V\widehat{O}_V^{(p)}\|\}\lesssim 1/\sqrt{\kappa}$. 

    Since $\eta= 0.75$, we have
    \begin{align*}
        &\|(\widehat{U}^{(p)}\widehat{L}_G^{(p)}-U\widehat{O}_U^{(p)}\widehat{L}_G^{(p)})(I-\eta\widehat{L}_G^{(p)\top}\widehat{O}_U^{(p)\top}\Lambda\widehat{O}_V^{(p)}\widehat{R}_G^{(p)}(\widehat{\Lambda}^{(p)})^{-1})\|_{2,\max} \\
        &\quad \leq \|\widehat{U}^{(p)}\widehat{L}_G^{(p)}-U\widehat{O}_U^{(p)}\widehat{L}_G^{(p)}\|_{2,\max} \|I-\eta\widehat{L}_G^{(p)\top}\widehat{O}_U^{(p)\top}\Lambda\widehat{O}_V^{(p)}\widehat{R}_G^{(p)}(\widehat{\Lambda}^{(p)})^{-1}\| \\
        &\quad \leq \|\widehat{U}^{(p)}-U\widehat{O}_U^{(p)}\|_{2,\max}(1-\eta) + \|\widehat{U}^{(p)}-U\widehat{O}_U^{(p)}\|_{2,\max}\eta\|\widehat{\Lambda}^{(p)}-\widehat{L}_G^{(p)\top}\widehat{O}_U^{(p)\top}\Lambda\widehat{O}_V^{(p)}\widehat{R}_G^{(p)}\|\|(\widehat{\Lambda}^{(p)})^{-1}\| \\
        &\quad \leq (1-\eta)\|\widehat{U}^{(p)}-U\widehat{O}_U^{(p)}\|_{2,\max} + 2\lambda_{\min}^{-1}\|\widehat{U}^{(p)}-U\widehat{O}_U^{(p)}\|_{2,\max}\eta\|\widehat{\Lambda}^{(p)}-\widehat{L}_G^{(p)\top}\widehat{O}_U^{(p)\top}\Lambda\widehat{O}_V^{(p)}\widehat{R}_G^{(p)}\|.
    \end{align*}
    Then we can obtain
    \begin{align*}
        \|(\widehat{U}^{(p)}\widehat{L}_G^{(p)}-U\widehat{O}_U^{(p)}\widehat{L}_G^{(p)})(I-\eta\widehat{L}_G^{(p)\top}\widehat{O}_U^{(p)\top}\Lambda\widehat{O}_V^{(p)}\widehat{R}_G^{(p)}(\widehat{\Lambda}^{(p)})^{-1})\|_{2,\max}\leq (1-\frac{9\eta}{10})\|\widehat{U}^{(p)}-U\widehat{O}_U^{(p)}\|_{2,\max}.
    \end{align*}
    By $\|\widehat{U}^{(p)}\|_{2,\max}\leq 2\mu\sqrt{r/d_1}$, we have
    \begin{align*}
        &\eta\|\widehat{U}^{(p)}(\widehat{G}^{(p)}-\widehat{O}_U^{(p)\top}\Lambda\widehat{O}_V^{(p)})\widehat{R}_G^{(p)}(\widehat{\Lambda}^{(p)})^{-1} \|_{2,\max}\leq 2\eta\|\widehat{U}^{(p)}\|_{2,\max}\|\widehat{G}^{(p)}-\widehat{O}_U^{(p)\top}\Lambda\widehat{O}_V^{(p)}\|\lambda_{\min}^{-1} \\
        &\leq C_3\eta\mu\left(\frac{r^{3/2}\sigma}{\lambda_{\min}}\frac{1}{\sqrt{d_2}p_{\min}N_0}\log d_1 + \frac{\sigma}{\lambda_{\min}}\sqrt{\frac{r^2\log d_1}{N_0d_1^2d_2}\left(\sum_{i=1}^{d_1}\sum_{j=1}^{d_2}\frac{1}{p_{ij}}\right)}\right)  \\ &\quad + C_4\eta\frac{\mu\kappa r^{1/2}}{\sqrt{d_1}}(\|\widehat{U}^{(p)}-U\widehat{O}_U^{(p)}\|^2+\|\widehat{V}^{(p)}-V\widehat{O}_U^{(p)}\|^2) \\ &\quad + C_5\eta\mu\kappa\left(\sqrt{\frac{1}{d_2}}\|\widehat{U}^{(p)}-U\widehat{O}_U^{(p)}\|_{2,\max} + \sqrt{\frac{1}{d_1}}\|\widehat{V}^{(p)}-V\widehat{O}_V^{(p)}\|_{2,\max}\right)\mu\sqrt{\frac{r^4\log d_1}{d_1d_2N_0}\left(\sum_{i=1}^{d_1}\sum_{j=1}^{d_2}\frac{1}{p_{ij}}\right)}.
    \end{align*}
    Observe that 
    \begin{align*}
        &C_4\frac{\mu\kappa r^{1/2}}{\sqrt{d_1}}\|\widehat{U}^{(p)}-U\widehat{O}_U^{(p)}\|^2= C_4\mu\kappa r^{1/2} \|\widehat{U}^{(p)}-U\widehat{O}_U^{(p)}\|\frac{\|\widehat{U}^{(p)}-U\widehat{O}_U^{(p)}\|}{\sqrt{d_1}} \\
        &\quad \leq C_4\mu\kappa r^{1/2} \|\widehat{U}^{(p)}-U\widehat{O}_U^{(p)}\|\frac{\fro{\widehat{U}^{(p)}-U\widehat{O}_U^{(p)}}}{\sqrt{d_1}} \leq \frac{\|\widehat{U}^{(p)}-U\widehat{O}_U^{(p)}\|_{2,\max}}{20\sqrt{d_2/d_1}}
    \end{align*}
    if $\|\widehat{U}^{(p)}-U\widehat{O}_U^{(p)}\|\lesssim 1/\kappa\mu\sqrt{rd_2/d_1}$. Similarly, we have
    \begin{align*}
        C_4\frac{\mu\kappa r^{1/2}}{\sqrt{d_1}}\|\widehat{V}^{(p)}-V\widehat{O}_V^{(p)}\|^2\leq \frac{\|\widehat{V}^{(p)}-V\widehat{O}_V^{(p)}\|_{2,\max}}{20}.
    \end{align*}
    Therefore, 
    \begin{align*}
        C_4\frac{\mu\kappa r^{1/2}}{\sqrt{d_1}}(\|\widehat{U}^{(p)}-U\widehat{O}_U^{(p)}\|^2+\|\widehat{V}^{(p)}-V\widehat{O}_U^{(p)}\|^2)\leq \frac{\sqrt{\frac{d_1}{d_2}}\|\widehat{U}^{(p)}-U\widehat{O}_U^{(p)}\|_{2,\max} + \|\widehat{V}^{(p)}-V\widehat{O}_V^{(p)}\|_{2,\max} }{20}.
    \end{align*}
    Moreover, if $N_0\gtrsim \mu^4\kappa^2r^4(d_1^2d_2)^{-1}(\sum_{i,j} p_{ij}^{-1})\log d_1$,
    \begin{align*}
        &C_5\mu\kappa\left(\sqrt{\frac{1}{d_2}}\|\widehat{U}^{(p)}-U\widehat{O}_U^{(p)}\|_{2,\max} + \sqrt{\frac{1}{d_1}}\|\widehat{V}^{(p)}-V\widehat{O}_V^{(p)}\|_{2,\max}\right)\mu\sqrt{\frac{r^4d_2\log d_1}{N_0}} \\
        &\leq \frac{\sqrt{\frac{d_1}{d_2}}\|\widehat{U}^{(p)}-U\widehat{O}_U^{(p)}\|_{2,\max} + \|\widehat{V}^{(p)}-V\widehat{O}_V^{(p)}\|_{2,\max}}{40}.
    \end{align*}
    Then we have
    \begin{align*}
        &\eta\|\widehat{U}^{(p)}(\widehat{G}^{(p)}-\widehat{O}_U^{(p)\top}\Lambda\widehat{O}_V^{(p)})\widehat{R}_G^{(p)}(\widehat{\Lambda}^{(p)})^{-1} \|_{2,\max} \\ &\quad\leq C_3\eta\mu\left(\frac{r^{3/2}\sigma}{\lambda_{\min}}\frac{1}{\sqrt{d_2}p_{\min}N_0}\log d_1 + \frac{\sigma}{\lambda_{\min}}\sqrt{\frac{r^2\log d_1}{N_0d_1^2d_2}\left(\sum_{i=1}^{d_1}\sum_{j=1}^{d_2}\frac{1}{p_{ij}}\right)}\right)\\
        &\quad\quad + \frac{3\eta}{40}\left(\sqrt{\frac{d_1}{d_2}}\|\widehat{U}^{(p)}-U\widehat{O}_U^{(p)}\|_{2,\max} + \|\widehat{V}^{(p)}-V\widehat{O}_V^{(p)}\|_{2,\max}\right).
    \end{align*}
    Since $\|(\widehat{V}^{(p)}\widehat{O}_V^{(p)\top}-V)^{\top}\widehat{V}^{(p)}\|\leq \|\widehat{V}^{(p)}\widehat{O}_V^{(p)\top}-V\|^2$, we get
    \begin{align*}
        \|U\Lambda(\widehat{V}^{(p)}\widehat{O}_V^{(p)\top}-V)^{\top}\widehat{V}^{(p)}\widehat{R}_G^{(p)}(\widehat{\Lambda}^{(p)})^{-1}\|_{2,\max}&\leq 2\kappa\mu\sqrt{\frac{r}{d_1}}\|\widehat{V}^{(p)}\widehat{O}_V^{(p)\top}-V\|^2 \\ &\leq \frac{1}{40}\|\widehat{V}^{(p)}-V\widehat{O}_V^{(p)\top}\|_{2,\max}
    \end{align*}
    as long as $\|\widehat{V}^{(p)}\widehat{O}_V^{(p)\top}-V\|\lesssim 1/\kappa\mu\sqrt{r}\sqrt{d_2/d_1}$. Putting together the above bounds, we obtain
    \begin{align*}
        &\|\widehat{U}^{(p+0.5)}-U\widehat{O}_U^{(p)}\widehat{L}_G^{(p)}\|_{2,\max}\leq (1-\frac{9\eta}{10})\|\widehat{U}^{(p)}-U\widehat{O}_U^{(p)}\|_{2,\max} \\ &\quad + C_3\eta\mu\left(\frac{r^{3/2}\sigma}{\lambda_{\min}}\frac{1}{\sqrt{d_2}p_{\min}N_0}\log d_1 + \frac{\sigma}{\lambda_{\min}}\sqrt{\frac{r^2\log d_1}{N_0d_1^2d_2}\left(\sum_{i=1}^{d_1}\sum_{j=1}^{d_2}\frac{1}{p_{ij}}\right)}\right) \\ &\quad + \frac{\eta}{10}\left(\sqrt{\frac{d_1}{d_2}}\|\widehat{U}^{(p)}-U\widehat{O}_U^{(p)}\|_{2,\max} + \|\widehat{V}^{(p)}-V\widehat{O}_V^{(p)}\|_{2,\max}\right) + \|\widehat{E}_V^{(p)}\|_{2,\max} + \|\widehat{E}_{\xi,V}^{(p)}\|_{2,\max}.
    \end{align*}
    Since $(\widehat{U}^{(p)}, \widehat{G}^{(p)}, \widehat{V}^{(p)})$ are independent with $\mathcal{D}_{2p+1}$ and $\widehat{U}^{(p)}, \widehat{V}^{(p)}$ are incoherent, and $\|\sum_{i=1}^{d_1} X_t^i\|=1$, we have
    \begin{align*}
        \bigg\|\big\|e_j^{\top}\sum_{i=1}^{d_1}\frac{1}{p_{ij_t(i)}} \xi_t^i X_t^{i}\widehat{V}^{(p)}\widehat{R}_G^{(p)}(\widehat{\Lambda}^{(p)})^{-1}\big\|\bigg\|_{\psi_2}\lesssim p_{\min}^{-1}\sigma\|\widehat{V}^{(p)}\|_{2,\max}\lambda_{\min}^{-1}\lesssim \frac{\sigma}{p_{\min}\lambda_{\min}}\sqrt{\frac{r}{d_2}}
    \end{align*}
    and 
    \begin{align*}
        &\EE\left[(e_j^{\top}\sum_{i=1}^{d_1}\frac{1}{p_{ij_t(i)}} \xi_t^i X_t^{i}\widehat{V}^{(p)}\widehat{R}_G^{(p)}(\widehat{\Lambda}^{(p)})^{-1})(e_j^{\top}\sum_{i=1}^{d_1}\frac{1}{p_{ij_t(i)}} \xi_t^i X_t^{i}\widehat{V}^{(p)}\widehat{R}_G^{(p)}(\widehat{\Lambda}^{(p)})^{-1})^{\top} \right] \\
        &\quad \leq \sigma^2\EE\left[\|(p_{jj_t(j)})^{-1}e_j^{\top} X_t^{j}\widehat{V}^{(p)}\widehat{R}_G^{(p)}(\widehat{\Lambda}^{(p)})^{-1}\|^2\right]=\sigma^2 \sum_{k=1}^{d_2} \frac{1}{p_{jk}}\|e_k^{\top}\widehat{V}^{(p)}\widehat{R}_G^{(p)}(\widehat{\Lambda}^{(p)})^{-1}\|^2\\
        &\lesssim \frac{\sigma^2}{\lambda_{\min}^2}\frac{r}{p_{\min}},
    \end{align*}
    since $e_j^{\top}X_t^i\neq 0$ only when $i=j$. As long as $(p_{\min})^{-1}\log d_1\lesssim N_0d_2$, by Bernstein inequality, with probability at least $1-d_1^{-10}$,
    \begin{align*}
        \|\widehat{E}_{\xi,V}^{(p)}\|_{2,\max}\lesssim \eta\frac{\sigma}{\lambda_{\min}}\sqrt{\frac{r\log d_1}{p_{\min}N_0}}.
    \end{align*}
    Similarly, with the same probability
    \begin{align*}
        \|\widehat{E}_V^{(p)}\|_{2,\max}\lesssim \eta\frac{\|\widehat{M}^{(p)}-M\|_{\max}}{\lambda_{\min}}\sqrt{\frac{r\log d_1}{p_{\min}N_0}}
    \end{align*}
    where $\widehat{M}^{(p)}=\widehat{U}^{(p)}\widehat{G}^{(p)}\widehat{V}^{(p)\top}$. Note that
    \begin{align*}
        &\|\widehat{M}^{(p)}-M\|_{\max}\leq \|(\widehat{U}^{(p)}-U\widehat{O}_U^{(p)})\widehat{G}^{(p)}\widehat{V}^{(p)\top}\|_{\max} + \|U(\widehat{O}_U^{(p)}\widehat{G}^{(p)}-\Lambda\widehat{O}_V^{(p)})\widehat{V}^{(p)\top}\|_{\max} \\ &\quad\quad  + \|U\Lambda(\widehat{V}^{(p)}\widehat{O}_V^{(p)}-V)^{\top}\|_{2,\max} \\
        &\quad \leq 2\mu\lambda_{\max}\left(\sqrt{\frac{r}{d_2}}\|\widehat{U}^{(p)}-U\widehat{O}_U^{(p)}\|_{2,\max} + \sqrt{\frac{r}{d_1}}\|\widehat{V}^{(p)}-V\widehat{O}_V^{(p)}\|_{2,\max}\right)  + \mu^2\sqrt{\frac{r^2}{d_1d_2}}\|\widehat{G}^{(p)}-\widehat{O}_U^{(p)}\Lambda\widehat{O}_V^{(p)}\|.
    \end{align*}
    Together with Lemma \ref{lemma:estimation1-nonuniform},
    \begin{align*}
        &\mu^2\sqrt{\frac{r^2}{d_1d_2}}\|\widehat{G}^{(p)}-\widehat{O}_U^{(p)}\Lambda\widehat{O}_V^{(p)}\|\leq C_3\mu^2\left({r^{3/2}\sigma}\frac{1}{d_2p_{\min}N_0}\log d_1 + \sigma\sqrt{\frac{r^2\log d_1}{N_0d_1^2d_2^2}\left(\sum_{i=1}^{d_1}\sum_{j=1}^{d_2}\frac{1}{p_{ij}}\right)}\right) \\ &\quad + 2\frac{\mu^2 r\lambda_{\max}}{\sqrt{d_1d_2}}\left(\|\widehat{U}^{(p)}-U\widehat{O}_U^{(p)}\|^2 + \|\widehat{V}^{(p)}-V\widehat{O}_V^{(p)}\|^2\right) \\ &\quad + C_4\lambda_{\max}\left(\sqrt{\frac{1}{d_2}}\|\widehat{U}^{(p)}-U\widehat{O}_U^{(p)}\|_{2,\max} + \sqrt{\frac{1}{d_1}}\|\widehat{V}^{(p)}-V\widehat{O}_V^{(p)}\|_{2,\max}\right)\mu^3\sqrt{\frac{r^2\log d_1}{d_1d_2^2N_0}\left(\sum_{i=1}^{d_1}\sum_{j=1}^{d_2}\frac{1}{p_{ij}}\right)}.
    \end{align*}
    As long as $N_0\gtrsim (\sum_{i,j}p_{ij}^{-1})(d_1d_2^2)^{-1}\mu^6r^4\log d_1$ and $N_0\gtrsim (p_{\min}d_2)^{-1}\log d_1$,
    \begin{align*}
        \|\widehat{M}^{(p)}-M\|_{\max}\leq \sigma+2\mu\lambda_{\max}\left(\sqrt{\frac{r}{d_2}}\|\widehat{U}^{(p)}-U\widehat{O}_U^{(p)}\|_{2,\max} + \sqrt{\frac{r}{d_1}}\|\widehat{V}^{(p)}-V\widehat{O}_V^{(p)}\|_{2,\max}\right).
    \end{align*}
    Therefore, with probability at least $1-2d_1^{-10}$,
    \begin{align*}
        \|\widehat{E}_V^{(p)}\|_{2,\max} + \|\widehat{E}_{\xi,V}^{(p)}\|_{2,\max}\leq C_6\eta\frac{\sigma}{\lambda_{\min}}\sqrt{\frac{r\log d_1}{p_{\min}N_0}} + \frac{\eta}{40}\left(\sqrt{\frac{d_1}{d_2}}\|\widehat{U}^{(p)}-U\widehat{O}_U^{(p)}\|_{2,\max} + \|\widehat{V}^{(p)}-V\widehat{O}_V^{(p)}\|_{2,\max}\right).
    \end{align*}
    As a result,
    \begin{align*}
        &\|\widehat{U}^{(p+0.5)}-U\widehat{O}_U^{(p)}\widehat{L}_G^{(p)}\|_{2,\max}\leq (1-\frac{9\eta}{10})\|\widehat{U}^{(p)}-U\widehat{O}_U^{(p)}\|_{2,\max} + C_6\eta\frac{\sigma}{\lambda_{\min}}\sqrt{\frac{r\log d_1}{p_{\min}N_0}}\\
        &\quad + \frac{\eta}{8}\left(\sqrt{\frac{d_1}{d_2}}\|\widehat{U}^{(p)}-U\widehat{O}_U^{(p)}\|_{2,\max} + \|\widehat{V}^{(p)}-V\widehat{O}_V^{(p)}\|_{2,\max}\right).
    \end{align*}
    Next, we investigate the singular values of $\widehat{U}^{(p+0.5)}-U\widehat{O}_U^{(p)}\widehat{L}_G^{(p)}$. Recall
    \begin{align*}
        \widehat{U}^{(p+0.5)}= \underbrace{\widehat{U}^{(p)}\widehat{L}_G^{(p)}-\eta(\widehat{U}^{(p)}\widehat{G}^{(p)}\widehat{V}^{(p)\top} - U\Lambda V^{\top})\widehat{V}^{(p)}\widehat{R}_G^{(p)}(\widehat{\Lambda}^{(p)})^{-1}}_{\mathcal{I}_1} + \underbrace{\widehat{E}_V^{(p)} + \widehat{E}_{\xi,V}^{(p)}}_{\mathcal{I}_2}. 
    \end{align*}
    By the independence between $(\widehat{U}^{(p)}, \widehat{G}^{(p)}, \widehat{V}^{(p)}, \widehat{L}_G^{(p)}, \widehat{R}_G^{(p)}, \widehat{\Lambda}^{(p)})$ and $\mathcal{D}_{2p+1}$, and notice that $\|X_t\|=1$, we have
    \begin{align*}
        \bigg\|\big\|\sum_{i=1}^{d_1}\frac{1}{p_{ij_t(i)}}\xi_t^i X_t^{i}\widehat{V}^{(p)}\widehat{R}_G^{(p)}(\widehat{\Lambda}^{(p)})^{-1} \big\|\bigg\|_{\psi_2}\lesssim p_{\min}^{-1}\frac{\sigma}{\lambda_{\min}}
    \end{align*}
    and
    \begin{align*}
        &\bigg\|\EE\left[(\sum_{i=1}^{d_1}\frac{1}{p_{ij_t(i)}}\xi_t^i X_t^{i}\widehat{V}^{(p)}\widehat{R}_G^{(p)}(\widehat{\Lambda}^{(p)})^{-1})(\sum_{i=1}^{d_1}\frac{1}{p_{ij_t(i)}}\xi_t^i X_t^{i}\widehat{V}^{(p)}\widehat{R}_G^{(p)}(\widehat{\Lambda}^{(p)})^{-1})^{\top} \right]\bigg\| \\ 
        &\quad\leq \EE\left[\fro{\sum_{i=1}^{d_1}\frac{1}{p_{ij_t(i)}}\xi_t^i X_t^{i}\widehat{V}^{(p)}\widehat{R}_G^{(p)}(\widehat{\Lambda}^{(p)})^{-1}}^2  \right]\\
        &\quad \leq \sigma^2  \sum_{i=1}^{d_1}\EE[\|\frac{1}{p_{ij_t(i)}}e_{j_t(i)}\widehat{V}^{(p)}\widehat{R}_G^{(p)}(\widehat{\Lambda}^{(p)})^{-1}\|^2]\\
        &\quad \leq \sigma^2 \sum_{i=1}^{d_1}\sum_{j=1}^{d_2} \frac{1}{p_{ij}}\|e_j^{\top}\widehat{V}^{(p)}\widehat{R}_G^{(p)}(\widehat{\Lambda}^{(p)})^{-1}\|^2\lesssim \frac{\sigma^2}{d_2\lambda_{\min}^2}\left(\sum_{i=1}^{d_1}\sum_{j=1}^{d_2}\frac{1}{p_{ij}}\right).
    \end{align*}
    By matrix Bernstein inequality, as long as $p_{\min}^{-1}\log d_1\lesssim \sqrt{d_2^{-1}N_0(\sum_{i=1}^{d_1}\sum_{j=1}^{d_2}(p_{ij}^{-1}))\log d_1}$, with probability at least $1-d_1^{-10}$,
    \begin{align*}
        \|\widehat{E}_{\xi,V}^{(p)}\|\lesssim \eta\frac{\sigma}{\lambda_{\min}}\sqrt{\frac{\log d_1}{d_2N_0}\left(\sum_{i=1}^{d_1}\sum_{j=1}^{d_2}\frac{1}{p_{ij}}\right)}.
    \end{align*} 
    Similarly, with the same probability
    \begin{align*}
        \|\widehat{E}_V^{(p)}\|\lesssim \eta\frac{\|\widehat{M}^{(p)}-M\|_{\max}}{\lambda_{\min}}\sqrt{\frac{\log d_1}{d_2N_0}\left(\sum_{i=1}^{d_1}\sum_{j=1}^{d_2}\frac{1}{p_{ij}}\right)}.
    \end{align*}
    Combine with the previous result, with probability at least $1-2d_1^{-10}$,
    \begin{align*}
        &\|\widehat{E}_V^{(p)} + \widehat{E}_{\xi,V}^{(p)}\|\lesssim \eta\frac{\sigma}{\lambda_{\min}}\sqrt{\frac{\log d_1}{d_2N_0}\left(\sum_{i=1}^{d_1}\sum_{j=1}^{d_2}\frac{1}{p_{ij}}\right)} \\ &\quad + \eta\sqrt{\frac{\log d_1}{d_2N_0}\left(\sum_{i=1}^{d_1}\sum_{j=1}^{d_2}\frac{1}{p_{ij}}\right)}\left(\sqrt{\frac{r}{d_2}}\|\widehat{U}^{(p)}-U\widehat{O}_U^{(p)}\|_{2,\max} + \sqrt{\frac{r}{d_1}}\|\widehat{V}^{(p)}-V\widehat{O}_V^{(p)}\|_{2,\max}\right).
    \end{align*}
    Note that the singular values of $\widehat{U}^{(p+0.5)}$ are the square root of eigenvalues of $\widehat{U}^{(p+0.5)\top}\widehat{U}^{(p+0.5)}$. We write
    \begin{align*}
        \widehat{U}^{(p+0.5)\top}\widehat{U}^{(p+0.5)}=\mathcal{I}_1^{\top}\mathcal{I}_1 + \mathcal{I}_2^{\top}\mathcal{I}_2 + \mathcal{I}_1^{\top}\mathcal{I}_2 + \mathcal{I}_2^{\top}\mathcal{I}_1.
    \end{align*}
    Since $\|\widehat{U}^{(p)\top}(\widehat{U}^{(p)}-U\widehat{O}_U^{(p)})\|\leq 2\|\widehat{U}^{(p)}-U\widehat{O}_U^{(p)}\|^2$ and $\|\widehat{V}^{(p)\top}(\widehat{V}^{(p)}-V\widehat{O}_V^{(p)})\|\leq 2\|\widehat{V}^{(p)}-V\widehat{O}_V^{(p)}\|^2$, by Lemma \ref{lemma:estimation1-nonuniform}, we get
    \begin{align*}
        &\|\mathcal{I}_1^{\top}\mathcal{I}_1-I\|\leq 3\kappa\eta(\|\widehat{U}^{(p)}-U\widehat{O}_U^{(p)}\|^2 + \|\widehat{V}^{(p)}-V\widehat{O}_V^{(p)}\|^2) + 2\frac{\eta}{\lambda_{\min}}\|\widehat{O}_U^{(p)}\widehat{G}^{(p)} - \Lambda\widehat{O}_V^{(p)} \| \\ &\quad\quad + 2\eta^2(\kappa^2\|\widehat{U}^{(p)}-U\widehat{O}_U^{(p)}\|^2 + \kappa^2\|\widehat{V}^{(p)}-V\widehat{O}_V^{(p)}\|^2 + \lambda_{\min}^{-2}\|\widehat{O}_U^{(p)}\widehat{G}^{(p)} - \Lambda\widehat{O}_V^{(p)} \|^2) \\
        &\quad\leq 3(\kappa^2\eta^2+\kappa\eta)(\|\widehat{U}^{(p)}-U\widehat{O}_U^{(p)}\|^2 + \|\widehat{V}^{(p)}-V\widehat{O}_V^{(p)}\|^2) + 4\frac{\eta}{\lambda_{\min}}\|\widehat{O}_U^{(p)}\widehat{G}^{(p)} - \Lambda\widehat{O}_V^{(p)} \| \\
        &\quad\lesssim (\kappa^2\eta^2+\kappa\eta)(\|\widehat{U}^{(p)}-U\widehat{O}_U^{(p)}\|^2 + \|\widehat{V}^{(p)}-V\widehat{O}_V^{(p)}\|^2) \\ &\quad\quad + \eta\frac{\sigma}{\lambda_{\min}}\left(\frac{\sqrt{d_1} r}{N_0p_{\min}\sqrt{d_2}}\log d_1 + \sqrt{\frac{r^2}{d_1d_2N_0}\left(\sum_{i=1}^{d_1}\sum_{j=1}^{d_2} \frac{1}{p_{ij}}\right)\log d_1}\right) \\ &\quad\quad +  \eta\kappa\left(\sqrt{\frac{r}{d_2}}\|\widehat{U}^{(p)}-U\widehat{O}_U^{(p)}\|_{2,\max} + \sqrt{\frac{r}{d_1}}\|\widehat{V}^{(p)}-V\widehat{O}_V^{(p)}\|_{2,\max}\right)\sqrt{\frac{r^2\log d_1}{d_2N_0}\left(\sum_{i=1}^{d_1}\sum_{j=1}^{d_2}\frac{1}{p_{ij}}\right)}.
    \end{align*}
    Moreover, as long as $\|\widehat{U}^{(p)}-U\widehat{O}_U^{(p)}\| + \|\widehat{V}^{(p)}-V\widehat{O}_V^{(p)}\|\lesssim \frac{1}{\kappa}$, $\|\mathcal{I}_1\|\leq 2$, then $\|\mathcal{I}_1^{\top}\mathcal{I}_2 + \mathcal{I}_2^{\top}\mathcal{I}_1\|\leq 4\|\widehat{E}_V^{(p)} + \widehat{E}_{\xi,V}^{(p)}\|$. Finally, with probability at least $1-2d_1^{-10}$,
    \begin{align*}
        &\|\widehat{U}^{(p+0.5)\top}\widehat{U}^{(p+0.5)}-I\|\lesssim \eta\frac{\sigma}{\lambda_{\min}}\sqrt{\frac{\log d_1}{d_2N_0}\left(\sum_{i=1}^{d_1}\sum_{j=1}^{d_2}\frac{1}{p_{ij}}\right)} + (\kappa^2\eta^2+\kappa\eta)(\|\widehat{U}^{(p)}-U\widehat{O}_U^{(p)}\|^2 + \|\widehat{V}^{(p)}-V\widehat{O}_V^{(p)}\|^2) \\
        &\quad +  \eta\sqrt{\frac{\log d_1}{d_2N_0}\left(\sum_{i=1}^{d_1}\sum_{j=1}^{d_2}\frac{1}{p_{ij}}\right)}\left(\sqrt{\frac{r}{d_2}}\|\widehat{U}^{(p)}-U\widehat{O}_U^{(p)}\|_{2,\max} + \sqrt{\frac{r}{d_1}}\|\widehat{V}^{(p)}-V\widehat{O}_V^{(p)}\|_{2,\max}\right),
    \end{align*}
    implying that $\{|1-\lambda_{\min}(\widehat{U}^{(p+0.5)})|, |1-\lambda_{\max}(\widehat{U}^{(p+0.5)})|\}$ can be upper bounded by the above rate.
\end{proof}

\end{sloppypar}

\end{document}